\documentclass[printer]{aa}
\UseRawInputEncoding 
\usepackage[utf8]{inputenc}
\usepackage{pdflscape}
\usepackage{xcolor}
\usepackage{mathtools}
\usepackage{lscape}
 \usepackage[hidelinks]{hyperref}
   \hypersetup{
      colorlinks = true,
      citecolor ={cyan}
   }
\usepackage{graphicx}
\newcommand{\ergs}{${\rm erg}\,{\rm s}^{-1}$}

\newcommand{\kms}{km\,${\rm s}^{-1}$}

\newcommand{\HeI}[1]{He\,{\sc i}\,$\lambda {#1}$}
\newcommand{\HeII}[1]{He\,{\sc ii}\,$\lambda {#1}$}

\usepackage{txfonts}

\begin{document}

   \title{The Tarantula Massive Binary Monitoring VI: Characterisation of hidden companions in 51 single-lined O-type binaries}
   \subtitle{A flat mass-ratio distribution and black-hole binary candidates} 
   \author{
  T.\ Shenar\inst{1,2}
    \and H.\ Sana\inst{2}    
    \and L.\ Mahy\inst{3, 2}  
    \and J.\ Ma\'{\i}z Apell\'aniz\inst{9}       
    \and Paul A.\ Crowther$^{4}$   
    \and M.\ Gromadzki$^{12}$
    \and A.\ Herrero$^{5, 6}$    
    \and N.\ Langer\inst{7, 8}        
    \and P.\ Marchant\inst{2}    
    \and Fabian R.\ N.\ Schneider$^{10, 11}$    
    \and K.\ Sen$^{7, 8}$    
    \and I.\ Soszy{\'n}ski$^{12}$
    \and S.\ Toonen$^{1}$    
          }
          
   \institute{
   \inst{1}{Anton Pannekoek Institute for Astronomy, Science Park 904, 1098 XH, Amsterdam, The Netherlands}; \\\email{tomer.shenar@uva.be} \\
  \inst{2}{Institute of Astronomy, KU Leuven, Celestijnenlaan 200D, 3001 Leuven, Belgium} \\
  \inst{3}{Royal Observatory of Belgium, Avenue circulaire/Ringlaan 3, B-1180 Brussels, Belgium} \\
  \inst{4}{Department of Physics \& Astronomy, Hounsfield Road, University of Sheffield, Sheffield, S3 7RH United Kingdom} \\  
  \inst{5}{Instituto de Astrof\'isica de Canarias,E-38 200 La Laguna, Tenerife, Spain} \\  
  \inst{6}{Dpto.\ Astrof\'isica, Universidad de La Laguna, E-38\,205 La Laguna, Tenerife, Spain} \\  
  \inst{7}{Argelander-Institut f\"ur Astronomie, Universit\"at Bonn, Auf dem H\"ugel 71, 53121 Bonn, Germany} \\  
  \inst{8}{Max-Planck-Institut f\"ur Radioastronomie, Auf dem Hügel 69, 53121 Bonn, Germany} \\   
  \inst{9}{Centro de Astrobiolog\'{\i}a (CAB), CSIC-INTA, Campus ESAC, Camino bajo del castillo, E-28\,692 Villanueva de la Ca\~nada, Madrid, Spain} \\
  \inst{10}{Heidelberger Institut f\"ur Theoretische Studien, Schloss-Wolfsbrunnenweg 35, 69118 Heidelberg, Germany} \\
  \inst{11}{Astronomisches Rechen-Institut, Zentrum f\"ur Astronomie der Universit\"at Heidelberg, M\"onchhofstr. 12-14, 69120 Heidelberg, Germany}   \\
  \inst{12}{Astronomical Observatory, University of Warsaw, Al. Ujazdowskie 4, 00-478 Warszawa, Poland} 
              }

  \date{Received ?; accepted ?}


  \abstract
   {Massive binaries hosting a black hole (OB+BH) represent a critical phase in the production of BH mergers in the context of binary evolution. In spite of this, such systems have so far largely avoided detection. Single-lined spectroscopic (SB1) O-type binaries are ideal objects to search  for elusive BH companions.  Moreover, SB1 binaries hosting two main sequence stars probe a regime of more extreme mass ratios and longer periods compared to double-lined binaries (SB2), and they are thus valuable for establishing  the natal mass ratio distribution of massive stars.}
   {We characterise the hidden companions in 51 SB1 O-type and evolved B-type binaries identified in the Large Magellanic Cloud (LMC) in the framework of the VLT-FLAMES Tarantula Survey (VFTS) and its follow-up, the Tarantula Massive Binary Monitoring (TMBM). The binaries cover periods between a few days to years ($0<\log P<3\,$[d]). Our goals are to hunt for BHs and  sample the low-mass end of the mass-ratio distribution.}
   {To uncover the hidden companions, we implemented the shift-and-add grid disentangling algorithm using 32 epochs of spectroscopy acquired in the framework of TMBM with the FLAMES spectrograph, allowing us to detect companions contributing as little as $\approx 1-2\%$ to the visual flux. We further analysed OGLE photometric data for the presence of eclipses or ellipsoidal variations. }
   {Out of the 51 SB1 systems, 43 (84\%) are found to have non-degenerate stellar companions, of which 28 are confident detections and 15 are less certain (SB1: or SB2:). Of these 43 targets, one is found to be a triple (VFTS~64),  and two are found to be quadruples (VFTS~120, 702). Our sample includes a total of eight eclipsing binaries.   The remaining eight targets (16\%) retain an SB1 classification. We modelled the mass-ratio distribution as  $f(q)\propto q^\kappa$, and 
   derived  $\kappa$ through a Bayesian approach. We used  mass-ratio constraints from previously known SB2 binaries, newly uncovered SB2 binaries,  and SB1 binaries, while accounting for binary detection bias. We found $\kappa = 0.2\pm 0.2$ for the entire sample and   $\kappa = -0.2 \pm 0.2$ when excluding binaries with periods shorter than 10\,d. In contrast, $\kappa = 1.2\pm 0.5$ was retrieved for tight binaries ($P<10\,$d), and it is proposed here to be a consequence of binary interactions. Aside from the unambiguous O+BH binary VFTS~243, which was analysed in detail in a separate paper, we identified two additional  OB+BH candidates: VFTS~514 and 779.
   }
   {Our study firmly establishes a virtually flat natal mass-ratio distribution ($\kappa = 0$) for O-type stars at LMC metallicity, covering the entire mass-ratio range ($0.05 < q < 1$) and periods in the range $0 < \log P < 3\,$[d]. The nature  of the  OB+BH candidates should be verified through future monitoring, but the frequency of OB+BH candidates is generally in line with recent predictions at LMC metallicity.   }

   \titlerunning{Characterisation of hidden companions in SB1 binaries in the Tarantula}
   \authorrunning{T. Shenar et al.}

   \maketitle
%
%

\section{Introduction}
\label{sec:intro}

The number of  LIGO/VIRGO gravitational-wave (GW) detections originating in coalescing neutron stars (NSs) and black holes (BHs) is rapidly growing. The number of merger events is approaching the 100 mark \citep{Abbott2019, Abbott2021_third}, and it is expected to grow by orders of magnitude in the coming years. There is an overwhelming international effort aimed at constraining the pathways and rates of these merger events \citep[see][for a recent review]{Mandel2022}. This endeavour is intricately tied with our understanding of the progenitors of BHs and NSs --  massive stars -- which are born with masses higher than $8\,M_\odot$. 


A central problem in the field of massive stars concerns the impact and incidence of stellar multiplicity \citep[e.g.][]{Paczynski1967, Vanbeveren1998, Langer2012, Sana2012}. 
A fundamental input for binary population syntheses in this context is the initial mass ratio distribution of the binary components in massive binaries.
In principle, direct measurements of the mass ratios of spectroscopic binaries can only be obtained from double-lined spectroscopic binaries (SB2).
However, SB2 binaries are, by definition, systems in which both stars can be observed, and they are hence biased towards  mass ratios close to unity. Moreover, SB2 binaries are most easily spotted when the spectral lines are not blended. For this to happen,  the binary components need to  exhibit appreciable radial-velocity (RV) amplitudes, which again introduces a bias towards short periods and high-mass companions. Short-period ($P\lesssim 10\,$d) binaries specifically entail the danger of having undergone an interaction in the past, as testified by short-period Algol-like systems \citep[e.g.][]{Mahy2020a, Mahy2020b, Janssens2021, Sen2022}. In contrast, single-lined spectroscopic binaries (SB1) are dominated by binaries with more extreme mass ratios and longer periods, where a past binary interaction can be excluded in the case of two non-degenerate components. They, therefore, provide a vital extension to SB2 binaries for a derivation of the natal mass ratio distributions. 

Furthermore, SB1 binaries  contain an additional population: that of post-interaction binaries \citep{deMink2014}. Of special interest are OB+BH binaries. Such binaries represent an intermediate evolutionary phase that appears in the majority of  evolutionary sequences leading to the formation of BH+BH mergers \citep{Belczynski2002, Giacobbo2018, Kruckow2018,  Marchant2019}.   As such, OB+BH binaries provide crucial constraints on binary evolution, supernova physics, and the presence of BH kicks  \citep[e.g.][]{Fryer2012, Vanbeveren2020, Sen2021}. \citet{Langer2020} predicted that about 2 to 3\% of massive stars should host a BH companion, implying that hundreds of such binaries are lurking in Galactic and extragalactic OB-star catalogues. And yet, such binaries are sparsely known, and the ones observed are primarily  high-mass X-ray binaries \citep[HMXBs,][]{ Corral-Santana2016}. However, the X-ray irradiation, mass transfer, and tidal interaction in X-ray binaries makes their  analysis and interpretation complex (e.g. \citealt{GimenezGarcia2016}, \citealt{Hainich2020}, \citealt{Ramachandran2022}). The strong tidal circularisation in these binaries  also blurs information stored in their orbit regarding the ejecta mass and kick of the BH progenitor. Identifying X-ray quiescent O+BH binaries is thus highly desirable. Indeed, the HMXB population is expected to represent just a small fraction of  a much larger population of X-ray quiescent binaries hosting a BH \citep{Langer2020, Sen2021}.

A few reports of quiescent OB+BH systems have recently emerged (e.g. \object{MWC~656}: \citealt{Casares2014}; \object{LB-1}: \citealt{Liu2019}; \object{HR~6819}: \citealt{Rivinius2020}; \object{NGC 1850 BH1}: \citealt{Saracino2022}; \object{NGC2004\#115}: \citealt{Lennon2021}; \object{HD~96670}: \citealt{Gomez2021}). However, most of these discoveries were soon thereafter challenged \citep[e.g.][]{Abdul-Masih2020,  Shenar2020LB1, Bodensteiner2020HR,  El-Badry2021Lennon, El-Badry2021NGC1850, Frost2022}, and the presence of a BH in MWC~656 and HD~96670 still requires  validation.
In the upcoming year, \textit{Gaia} is expected to uncover hundreds of OB+BH binaries in our Galaxy relying on accurate astrometry \citep{Breivik2017, Mashian2017, Janssens2022}. However, to uncover these populations at subsolar metallicities in neighbouring galaxies such as the Small and Large Magellanic Clouds (SMC, LMC), spectroscopy of SB1 binaries is the method of choice. \citet{Shenar2022VFTS243} identified VFTS~243 in the LMC as an O+BH binary as part of the study put forth here. Recently, a  Galactic O+BH system was also reported  (\object{HD~130298}, Mahy et al.\ 2022, accepted to A\&A). 

Owing to its well constrained distance ($d = 49.97\,$\,kpc, \citealt{Pietrzynski2013}), modest reddening \citep{Maiz2014}, ongoing star formation, and sub-solar metallicity content ($Z\approx 0.5\,Z_\odot$), the \object{Tarantula nebula} (\object{30 Doradus}) in the LMC provides an ideal laboratory for investigating massive binaries. It hosts a rich population of massive OB-type stars \citep{Walborn1997, Walborn2014}, including the most massive stars and binaries reported to date \citep{deKoter1997, Crowther2010, Hainich2014, Shenar2017, Shenar2021, Bestenlehner2022, Brands2022}. 
The VLT-FLAMES Tarantula Survey \citep[VFTS,][]{Evans2011} provided a homogeneous, multi-epoch spectroscopic dataset aimed to establish the stellar and binary properties of the massive-star content in this region. Among the $\approx 800$ targets included in the VFTS, 360 were classified as O-type \citep{Evans2011, Walborn2014}. Of those, 116  (32\%)  were  found to exhibit large radial velocity (RV) variations ($\Delta {\rm RV} > 20\,$\kms), making them excellent  binary candidates \citep{Sana2013}. The Tarantula Massive Binary Monitoring (TMBM, PI: Sana, programme IDs: 090.D-0323, 092.D-0136) campaign acquired 32 additional epochs for the majority of the O-type binary candidates, as well as a handful of evolved B-type stars allocated to spare fibres. In total, 93 binaries were studied by \citet{Almeida2017}, resulting in orbital solutions for 31 SB2 and 51 SB1 systems. The SB2s were recently analysed by \citet{Mahy2020b, Mahy2020a}.

This paper aims to  characterise the hidden companions in  a sample of O-type and evolved B-type SB1 binaries in the TMBM sample utilising  multi-epoch spectroscopy (Sect.\,\ref{sec:data}). Our goal is twofold: to probe the lower-end of the mass ratio distribution, and to hunt for OB+BH candidates.  This is achieved through the implementation of spectral disentangling, described in Sect.\,\ref{sec:analysis} and Appendix\,\ref{sec:appregg}. The results are presented in Sect.\,\ref{sec:results} and discussed in Sect.\,\ref{sec:disc}. We provide our final conclusions regarding the mass-ratio distribution and OB+BH population in this sample in Sect.\,\ref{sec:conclusions}.

\renewcommand{\arraystretch}{1.3}

\begin{table*}[!h]
\centering
\tiny
\caption{Main properties of the sample of 51 O-type SB1 binaries analysed in this work. }
\resizebox{\textwidth}{!}{\begin{tabular}{lllccccccccccc}\hline \hline
VFTS  & SpT (old)\,\tablefootmark{a, b} & SpT (new)\,\tablefootmark{c}  & $P\,$\tablefootmark{a}     & epochs & mean S/N & $f$\,\tablefootmark{a}  & $f_1/f_{\rm tot}(V)$\,\tablefootmark{d} & $T_{\rm eff, 1}\,$\tablefootmark{e} & $\log L_1\,$\tablefootmark{e} & $M_{1, {\rm SpT}}\,\tablefootmark{f}$ & $M_{1, {\rm ev}}\,\tablefootmark{g}$ & $M_{\rm min, 2}\,\tablefootmark{h}$ & $\log L_X\,\tablefootmark{i}$\\  
     &    &     &  d       &  &  & $M_\odot$  & &  K & $[{\rm erg}\,{\rm s}^{-1}]$ &  $M_\odot$ & $M_\odot$ & $M_\odot$ & $[{\rm erg}\,{\rm s}^{-1}]$\\ 
 \hline   
64 & O7.5 II(f) & O8 II:(f) + (O7~V+O7~V) & 903$\pm$4&37 & 71 & 10.7$\pm$0.4 &0.55 & 33300 & 5.07 & $ 28 \pm 5$  & $20^{+4}_{-3}$ & $34.8_{-3.4}^{+3.3} $ & <32.51\\ 
73 & O9.5 III & O9.5 IV + B: & 150.60$\pm$0.13&32 & 27 & 0.288$\pm$0.024 &$\approx 1$ &31850 & 5.00 & $ 19 \pm 3$  & $17^{+3}_{-2}$ & $5.6_{-0.8}^{+0.8} $ & -\\ 
86 & O9 III((n)) & O9.5 III + O8~IV:n & 182.95$\pm$0.14&32 & 84 & 1.0$\pm$0.5\,\tablefootmark{l} &0.56 & 30800 & 5.01 & $ 20 \pm 5$  & $17^{+3}_{-3}$ & $9.2_{-3.7}^{+3.5} $ & -\\ 
93 & O9.2 III-IV & O8.5 V + B0.2:~V & 250.13$\pm$0.33&32 & 67 & 0.051$\pm$0.013\,\tablefootmark{l} &0.79 & 34500\,\tablefootmark{j} & 5.02\,\tablefootmark{j} & $ 21 \pm 3$  & $21^{+4}_{-3}$ & $3.1_{-0.5}^{+0.5} $ & <32.22\\ 
120 & O9.5 IV: & (O9.2 IV + B) + (O9.5~V + O9.5~V) & 15.6546$\pm$0.0011&32 & 75 & 0.0217$\pm$0.0012 &0.35 & 32350 & 4.69 & $ 19 \pm 3$  & $16^{+2}_{-2}$ & $2.1_{-0.3}^{+0.3} $ & <32.28\\ \vspace{0.3cm} 
171 & O8 II-III(f) & O8.5 III:(f) + B1.5:~V & 677.0$\pm$0.8&32 & 116 & 0.0659$\pm$0.0029 &0.92 & 34250\,\tablefootmark{j} & 5.43\,\tablefootmark{j} & $ 21 \pm 4$  & $23^{+7}_{-4}$ & $3.4_{-0.5}^{+0.5} $ & -\\ 
184 & O6.5 Vnz & O6.5 Vn + OB: & 32.128$\pm$0.022&32 & 65 & 0.0055$\pm$0.0015 &$\approx 1$ &38900 & 4.91 & $ 30 \pm 6$  & $26^{+5}_{-3}$ & $1.7_{-0.4}^{+0.4} $ & -\\ 
191 & O9.5 V & O9.2 V + O9.7:~V: & 358.9$\pm$0.8&32 & 50 & 0.42$\pm$0.13 &0.87 & 33400 & 4.81 & $ 20 \pm 3$  & $18^{+3}_{-2}$ & $6.6_{-1.5}^{+1.5} $ & -\\ 
201 & O9.7 V + sec & O9.7 V  + B1.5:~V & 15.3270$\pm$0.0020&32 & 29 & 0.55$\pm$0.19\,\tablefootmark{l} &0.70 & 32500 & 4.46 & $ 19 \pm 3$  & $15^{+2}_{-2}$ & $7.1_{-1.7}^{+1.6} $ & -\\ 
225 & B0.7-1III-II & B0.7 III & 8.2337$\pm$0.0004&32 & 62 & 0.0211$\pm$0.0005 &$\approx 1$ &24500 & 4.53 & $ 15 \pm 3$  & $12^{+2}_{-2}$ & $1.8_{-0.3}^{+0.3} $ & -\\ \vspace{0.3cm} 
231 & O9.7 IV:(n) + sec & O9.7 V  + B1.5:~V & 7.92911$\pm$0.00022&32 & 39 & 0.61$\pm$0.21\,\tablefootmark{l} &0.67 & 31450 & 4.42 & $ 19 \pm 3$  & $14^{+2}_{-2}$ & $7.4_{-1.7}^{+1.6} $ & -\\ 
243 & O7 V(n)((f)) & O7 V:(n)((f)) & 10.4031$\pm$0.0004&37 & 61 & 0.581$\pm$0.028 &$\approx 1$ &36000\,\tablefootmark{k} & 5.20\,\tablefootmark{k} & $ 27 \pm 4$  & $26^{+2}_{-2}$ & $9.1_{-1.0}^{+1.0} $ & <32.15\\ 
256 & O7.5-8 V((n))z & O7.5 V: + OB: & 246.0$\pm$0.5&32 & 83 & 0.085$\pm$0.010 &$\approx 1$ &36900 & 4.98 & $ 25 \pm 3$  & $22^{+4}_{-3}$ & $4.1_{-0.5}^{+0.5} $ & -\\ 
277 & O9 V & O9 V + B1.5:~V & 240.42$\pm$0.13&32 & 77 & 0.33$\pm$0.15 &0.96 & 33900 & 5.00 & $ 20 \pm 3$  & $19^{+4}_{-2}$ & $6.1_{-1.8}^{+1.7} $ & <32.03\\ 
314 & O9.7 IV:(n) + sec & O9.7 V(n)  + B & 2.550786$\pm$0.000005\,\tablefootmark{l} &32 & 44 & 0.345$\pm$0.012\,\tablefootmark{l} &$\approx 1$ &32500 & 4.58 & $ 19 \pm 3$  & $17^{+3}_{-2}$ & $5.9_{-0.6}^{+0.6} $ & -\\ \vspace{0.3cm} 
318 & O((n))p & O9.5  V + O9.2~V & 14.0043$\pm$0.0029&23 & 34 & 0.0181$\pm$0.0026 &0.67 & 32900 & 4.07 & $ 20 \pm 3$  & $15^{+2}_{-2}$ & $2.0_{-0.3}^{+0.3} $ & -\\ 
329 & O9.7 II-III(n) & O9.5 V(n) + B1:~V: & 7.0491$\pm$0.0004&32 & 64 & 0.43$\pm$0.11\,\tablefootmark{l} &0.78 & 32900 & 4.64 & $ 20 \pm 3$  & $18^{+3}_{-2}$ & $6.6_{-1.3}^{+1.3} $ & -\\ 
332 & O9.2 II-III & O9 III + O9.2~V & 1025$\pm$9&37 & 106 & 10$\pm$6\,\tablefootmark{l} &0.58 & 31800 & 5.19 & $ 20 \pm 5$  & $18^{+5}_{-3}$ & $29.0_{-12.6}^{+12.0} $ & <31.99\\ 
333 & O8 II-III((f)) & O9 II((f)) + O6.5~V: & 980.1$\pm$1.5&37 & 272 & 2.4$\pm$0.4\,\tablefootmark{l} &0.76 & 33800 & 5.88 & $ 20 \pm 6$  & $32^{+15}_{-8}$ & $14.2_{-3.4}^{+3.4} $ & 32.20\\ 
350 & O8 V & O8.5 V + O9.5~V & 69.570$\pm$0.005&32 & 82 & 2.06$\pm$0.33\,\tablefootmark{l} &0.75 & 34900 & 5.12 & $ 21 \pm 3$  & $21^{+4}_{-3}$ & $13.5_{-1.9}^{+1.9} $ & <31.86\\ \vspace{0.3cm} 
386 & O9 IV(n) & O9 V(n) + B1~V: & 20.451$\pm$0.020\,\tablefootmark{l} &32 & 114 & 0.060$\pm$0.012\,\tablefootmark{l} &0.84 & 32900 & 5.32 & $ 20 \pm 3$  & $22^{+5}_{-4}$ & $3.2_{-0.6}^{+0.6} $ & <32.02\\ 
390 & O5-6 V(n)((fc))z & O5.5 V:((fc)) + O9.7: V: & 21.9059$\pm$0.0002\,\tablefootmark{l} &32 & 54 & 0.505$\pm$0.033\,\tablefootmark{l} &0.85 & 40900 & 5.18 & $ 33 \pm 5$  & $31^{+6}_{-5}$ & $9.7_{-1.2}^{+1.2} $ &  <31.95\\ 
404 & O3.5 V(n)((fc)) & O3.5:   V:((fc)) + O5~V: & 145.76$\pm$0.08&32 & 110 & 4.8$\pm$1.7\,\tablefootmark{l} &0.58 & 47000 & 5.55 & $ 45 \pm 9$  & $45^{+11}_{-8}$ & $29.9_{-8.2}^{+8.0} $ & 32.84\\ 
409 & O4 V((f))z & O3.5: V:((f)) + B: & 22.1909$\pm$0.0012&32 & 48 & 0.162$\pm$0.008 &$\approx 1$ &47000 & 5.94 & $ 45 \pm 9$  & $53^{+15}_{-12}$ & $7.6_{-1.2}^{+1.1} $ & <31.88\\ 
429 & O7.5-8 V & O7 V: + B1:~V: & 30.0450$\pm$0.0003\,\tablefootmark{l} &32 & 78 & 1.40$\pm$0.04\,\tablefootmark{l} &0.90 & 37900 & 5.09 & $ 27 \pm 4$  & $24^{+4}_{-3}$ & $13.1_{-1.3}^{+1.3} $ & <32.07\\ \vspace{0.3cm} 
440 & O6-6.5 II(f) & O6: V:(f) + O8~V & 1019$\pm$9&32 & 144 & 0.148$\pm$0.025 &0.84 & 33800\,\tablefootmark{j} & 5.63\,\tablefootmark{j} & $ 31 \pm 5$  & $25^{+10}_{-5}$ & $5.8_{-1.0}^{+1.0} $ & <32.01\\ 
441 & O9.5 V & O9.2 V + B0.5~V & 6.86858$\pm$0.00022&32 & 73 & 0.26$\pm$0.05\,\tablefootmark{l} &0.79 & 33400 & 4.77 & $ 20 \pm 3$  & $18^{+3}_{-2}$ & $5.4_{-1.0}^{+1.0} $ & -\\ 
475 & O9.7 III & O9.7 V + B0~V & 4.05424$\pm$0.00012&32 & 28 & 0.6$\pm$0.4\,\tablefootmark{l} &0.65 & 32500 & 4.72 & $ 19 \pm 3$  & $17^{+3}_{-2}$ & $7.2_{-3.4}^{+2.6} $ & -\\ 
479 & O4-5 V((fc))z & O4.5 V((fc))z + B: & 14.7254$\pm$0.0009&32 & 47 & 0.509$\pm$0.024 &$\approx 1$ &42900 & 5.14 & $ 39 \pm 6$  & $32^{+6}_{-5}$ & $10.8_{-1.3}^{+1.3} $ & <31.91\\ 
481 & O8.5 III & O8.5 V + O9.7:~V: & 141.823$\pm$0.009&32 & 128 & 1.57$\pm$0.24 &0.90 & 34900 & 5.29 & $ 21 \pm 3$  & $22^{+5}_{-4}$ & $12.0_{-1.6}^{+1.6} $ & <32.18\\ \vspace{0.3cm} 
514 & O9.7 III & O9.7 V & 184.92$\pm$0.11&32 & 55 & 0.175$\pm$0.010 &$\approx 1$ &32500 & 4.44 & $ 19 \pm 3$  & $15^{+2}_{-2}$ & $4.5_{-0.5}^{+0.5} $ & -\\ 
532 & O3 V(n)((f*))z+OB & O3.5: V:((f*)) + B~III& 5.796223$\pm$0.000002\,\tablefootmark{l} &32 & 79 & 0.022$\pm$0.004\,\tablefootmark{l} &$\approx 1$ &44700\,\tablefootmark{m} & 5.74\,\tablefootmark{m} & $ 45 \pm 9$  & $42^{+11}_{-8}$ & $3.7_{-0.7}^{+0.7} $ & 31.92\\ 
603 & O4 III(fc) & O4 III:(fc) + OB: & 1.756777$\pm$0.000024&32 & 107 & 0.000265$\pm$0.000020 &$\approx 1$ &42200\,\tablefootmark{m} & 5.98\,\tablefootmark{m} & $ 61 \pm 10$  & $41^{+16}_{-8}$ & $1.0_{-0.1}^{+0.1} $ & 31.95\\ 
613 & O8.5 Vz & O9  V + O7.5~V & 69.16$\pm$0.04&32 & 40 & 1.7$\pm$1.2\,\tablefootmark{l} &0.51 & 33900 & 4.64 & $ 20 \pm 3$  & $17^{+3}_{-2}$ & $12.0_{-5.4}^{+4.5} $ & -\\ 
619 & O7-8 V(n) & O8: V & 14.5043$\pm$0.0026&32 & 42 & 0.074$\pm$0.008 &$\approx 1$ &35900 & 4.86 & $ 23 \pm 3$  & $20^{+3}_{-3}$ & $3.8_{-0.4}^{+0.4} $ & -\\ \vspace{0.3cm} 
631 & O9.7 III(n) & O9.7 V & 5.37487$\pm$0.00018&32 & 44 & 0.064$\pm$0.004 &$\approx 1$ &32500 & 4.75 & $ 19 \pm 3$  & $17^{+3}_{-2}$ & $3.1_{-0.4}^{+0.4} $ & -\\ 
645 & O9.5 V((n)) & O9.5 V & 12.5458$\pm$0.0016&37 & 35 & 0.038$\pm$0.009 &$\approx 1$ &32900 & 4.68 & $ 20 \pm 3$  & $17^{+2}_{-2}$ & $2.6_{-0.5}^{+0.5} $ & -\\ 
657 & O7-8 II(f) & O7 II:(f) + OB: & 63.466$\pm$0.008&32 & 47 & 0.39$\pm$0.04 &$\approx 1$ &35300 & 5.52 & $ 35 \pm 15$  & $25^{+7}_{-5}$ & $9.2_{-2.8}^{+2.6} $ & <31.83\\ 
702 & O8 V(n) & (O8 V(n) + OB) + (OB+OB) & 1.981440$\pm$0.000020\,\tablefootmark{l} &32 & 33 & 0.244$\pm$0.021\,\tablefootmark{l} &$\approx 1$ &35900 & 5.02 & $ 23 \pm 3$  & $21^{+4}_{-3}$ & $5.9_{-0.6}^{+0.6} $ & <31.88\\ 
733 & O9.7p & O7.5 V + B1~II & 5.922078$\pm$0.000005&32 & 96 & 0.038$\pm$0.005\,\tablefootmark{l} &0.44 & 36900 & 5.23 & $ 25 \pm 3$  & $24^{+5}_{-3}$ & $3.1_{-0.4}^{+0.4} $ & <31.95\\ \vspace{0.3cm} 
736 & O9.5 V & O9.5 V + B: & 68.800$\pm$0.021&32 & 48 & 0.105$\pm$0.006 &$\approx 1$ &32900 & 4.70 & $ 20 \pm 3$  & $17^{+2}_{-2}$ & $3.9_{-0.5}^{+0.5} $ & -\\ 
743 & O9.5 V((n)) & O9.5 V((n)) & 14.9473$\pm$0.0009&32 & 84 & 0.0199$\pm$0.0014 &$\approx 1$ &32900 & 4.72 & $ 20 \pm 3$  & $17^{+3}_{-2}$ & $2.1_{-0.3}^{+0.3} $ & -\\ 
750 & O9.5 IV & O9.5 V + B: & 417$\pm$8&37 & 59 & 0.27$\pm$0.08 &$\approx 1$ &32900 & 4.64 & $ 20 \pm 3$  & $17^{+2}_{-2}$ & $5.6_{-1.2}^{+1.2} $ & -\\ 
769 & O9.7 II-III & O9.7 V & 2.365648$\pm$0.000016&32 & 48 & 0.0166$\pm$0.0010 &$\approx 1$ &32500 & 4.74 & $ 19 \pm 3$  & $17^{+2}_{-2}$ & $1.9_{-0.2}^{+0.2} $ & -\\ 
779 & B1 II-Ib & B1 II-III & 59.945$\pm$0.025&32 & 54 & 0.181$\pm$0.004 &$\approx 1$ &23500\,\tablefootmark{n} & 4.73\,\tablefootmark{n} & $ 14 \pm 4$  & $12^{+3}_{-2}$ & $3.9_{-0.7}^{+0.6} $ & -\\ \vspace{0.3cm} 
802 & O7.5 Vz & O7 V: + O8~Vn & 181.88$\pm$0.04&23 & 129 & 3.35$\pm$0.28\,\tablefootmark{l} &0.64 & 37900 & 5.05 & $ 27 \pm 4$  & $24^{+4}_{-3}$ & $19.3_{-2.3}^{+2.3} $ & <32.34\\ 
810 & O9.7 V + B1: V: & O9.7 V  + B1~V & 15.6886$\pm$0.0006&32 & 34 & 0.81$\pm$0.20\,\tablefootmark{l} &0.71 & 32500 & 4.36 & $ 19 \pm 3$  & $15^{+2}_{-2}$ & $8.4_{-1.6}^{+1.6} $ & -\\ 
812 & O4-5 V((fc)) & O4 V((fc)) & 17.28443$\pm$0.00035&32 & 88 & 0.0664$\pm$0.0032 &$\approx 1$ &43900 & 5.48 & $ 40 \pm 7$  & $37^{+8}_{-6}$ & $5.1_{-0.7}^{+0.7} $ & <32.14\\ 
827 & B1.5 Ib & B1.5 III & 43.221$\pm$0.017&32 & 58 & 0.0662$\pm$0.0027 &$\approx 1$ &21000\,\tablefootmark{n} & 5.03\,\tablefootmark{n} & $ 13 \pm 4$  & $13^{+3}_{-3}$ & $2.5_{-0.5}^{+0.5} $ & <32.48\\ 
829 & B1.5-2 II & B1.5 III & 202.9$\pm$0.9&32 & 56 & 0.037$\pm$0.006 &$\approx 1$ &20500\,\tablefootmark{n} & 4.78\,\tablefootmark{n} & $ 13 \pm 4$  & $11^{+3}_{-2}$ & $2.0_{-0.5}^{+0.5} $ & -\\ \vspace{0.3cm} 
887 & O9.5 II-IIIn & O9.7: V: + O9.5:~V & 2.672807$\pm$0.000035&32 & 65 & 0.32$\pm$0.06\,\tablefootmark{l} &0.55 & 32500 & 4.63 & $ 19 \pm 3$  & $16^{+2}_{-2}$ & $5.7_{-0.9}^{+0.9} $ & -\\ 
\hline
\end{tabular}}
\tablefoot{
Provided are VFTS identifiers, previous spectral types (adopted from \citealt{Almeida2017}), new spectral types from our study, and orbital periods (from \citealt{Almeida2017}, possibly updated in our study). We also provide the number of epochs used, mean S/N of spectra, mass function, estimates for the mass of the primary from spectral-type calibration and evolution tracks, and the mass range of the companion $M_2$.
\tablefoottext{a}{\citet{Almeida2017}, unless revised here}, 
\tablefoottext{b}{\citet{Walborn2014}},
\tablefoottext{c}{Derived using disentangled spectra. colon (:) signifies uncertain classification, see \cite{Sota2011} and \cite{Walborn2014} for meaning of remaining qualifiers},
\tablefoottext{d}{Flux ratio of the primary in the visual (see Sect.\,\ref{subsec:lcs}). Typical errors are 10\%. }, 
\tablefoottext{e}{Unless stated otherwise, using spectral-type calibrations from \citet{Doran2013} and available photometry (see text for details). Representative errors are $2000\,K$ (1 spectral type ordinate) and $0.3\,$dex (including uncertainties on SpT, luminosity class, light ratios, and reddening)  on $T_{\rm eff}$ and $\log L$, respectively.  }, 
\tablefoottext{f}{Mean evolutionary mass of all apparently-single stars with similar  spectral types in \cite{Schneider2018}.}, 
\tablefoottext{g}{Evolutionary mass obtained with the BONNSAI tool \citep{Brott2011, Koehler2015, Schneider2014} using $T_{\rm eff}$ and $\log L$ as input.}, 
\tablefoottext{h}{$M_{\rm min, 2}$ is evaluated from mass function using $M_{\rm 1, SpT}$. Errors represent 1$\sigma$ (68\% confidence interval). }, 
\tablefoottext{i}{X-ray detection with the Chandra Visionary programme T-ReX \citep{Crowther2022TREX}. Targets with no entry were excluded from the X-ray analysis, either because they have $\log L < 5.0$\,[\ergs], or due to complex diffuse background contamination.},
\tablefoottext{j}{\citet{Ramirez-Agudelo2017} },
\tablefoottext{k}{\cite{Shenar2022VFTS243}},
\tablefoottext{l}{Revised here},
\tablefoottext{m}{\citet{Bestenlehner2014} },
\tablefoottext{n}{\citet{McEvoy2015} }
}
\label{tab:Sample}\end{table*}

\section{Sample and observations}
\label{sec:data}


Our sample comprises the 51 SB1 binaries identified by \citet{Almeida2017} in the TMBM sample. In Table\,\ref{tab:Sample}, we provide the list of the 51 targets, along with their previous spectral types (SpT), new SpTs derived from the disentangled spectra (Sect.\,\ref{sec:analysis}), and the orbital periods. We note that \citet{Almeida2017} defined SB1 binaries as systems for which RVs could not be retrieved for a secondary source, which could either mean that the secondary is too faint to be detected, or that severe line blending prevents a trustworthy measurement of the RVs. Hence, for a few targets in the sample, the original SpTs by \citet{Almeida2017} already suggest the presence of a non-degenerate secondary.

Our analysis relies on 32 epochs of optical spectra acquired with the Fibre Large Array Multi Element Spectrograph (FLAMES/GIRAFFE)  mounted on UT2 of the Very Large Telescope (VLT). The data and their reduction are described in detail in \citet{Almeida2017}; here, we only repeat the essentials.  Data were acquired between Oct 2012 and Mar 2014 with the L427.2 (LR02) grating, which covers the wavelength range 3964--4567\,\AA~at a resolving power of $R = 6400$ and rebinned sampling of 0.2\,\AA. For each target, 32 epochs were secured, with the exception of VFTS~318 and 802, for which only 18 epochs are available. For highly eccentric or long-period ($P \gtrsim$1.5\,yr) binaries, we also use five additional epochs from the original VFTS campaign, described in \citet{Evans2011}. Table\,\ref{tab:Sample} provides the total number of epochs and average signal-to-noise ratios (S/N), which are typically of the order of 50-100 per wavelength element. 

In Table\,\ref{tab:Sample}, we also provide the mass functions $f$ for each systems. Their calculation is primarily based on the results obtained by \citet{Almeida2017}, though we use updated orbital parameters obtained in our study when applicable.  The estimated flux ratios of the primaries in the visual $f_1 / f_{\rm tot} (V)$ are also given (see Sect.\,\ref{subsec:lcs}). 
Table\,\ref{tab:Sample} also lists the effective temperatures $T_{\rm eff, 1}$ and bolometric luminosities $\log L_1$ of the primaries. Generally, the effective temperature is obtained using SpT calibrations by \citet{Doran2013}. For $\log L_1$,  we use bolometric corrections from \citet{Doran2013}, $V$-band magnitudes and $B-V$ colours from \citet{Evans2011}, and reddening obtained from the difference between the observed colours and SpT-dependent colour (typically $B-V_0 \approx -0.31\pm 0.01$, \citealt{Conti2008}), assuming a relative extinction of $R_V = 3.5\,$. We also account for the flux ratio of the primary. We implicitly assume here that the observed colours are not affected by a potential secondary source, which is valid given the early spectral types of the primaries and their companions.  For the few targets that were analysed as single stars in previous studies \citep{Bestenlehner2014, McEvoy2015, Ramirez-Agudelo2017}, we adopted previously published $T_{\rm eff}$ and $\log L$ unless the secondary is found to contribute more than 20\% to the visual flux.  

We compute two independent estimates for the mass of the primaries: once based on SpT calibrations ($M_{\rm 1, SpT}$) and once based on evolution models $(M_{\rm 1, ev})$. To estimate  $M_{\rm 1, SpT}$,  we calculate the mean of the evolutionary masses  of all apparently-single stars in the VFTS  sample \citep{Schneider2018} sharing a similar  spectral types (i.e. differing by not more than one subtype and luminosity class ordinate). For the evolutionary masses, we use  $T_{\rm eff}$ and $\log L$ estimates as input in the BONNSAI tool\footnote{https://www.astro.uni-bonn.de/stars/bonnsai/} \citep{Schneider2014} using tracks from \citet{Brott2011} and \citet{Koehler2015}. As priors, we used the Salpter initial mass function and the initial rotational velocity distribution from \citet{Ramirez-Agudelo2013}. While some discrepancies between $M_{\rm 1, SpT}$ and $(M_{\rm 1, ev})$ are observed,  they all agree within $1\sigma$, and no obvious trends are present. This is not entirely surprising, since we are comparing evolutionary masses computed by two different means, which suffer similar systematics (unlike comparing spectroscopic or dynamical masses,  e.g. \citealt{Herrero1992}) . As $M_{\rm 1, ev}$ suffers from more assumptions than $M_{\rm 1, ev}$, we adopt $M_1 =M_{\rm 1, SpT}$ for the analysis, but using $M_{\rm 1, ev}$ would not impact our conclusions. From the binary mass function and the estimates on the masses of the primaries, constraints on the mass of the secondary $M_2$ can be derived, which are also provided in Table\,\ref{tab:Sample}. The derivation of these constraints is described in Sect.\,\ref{sec:results}.  We note that low-mass bloated stripped stars could be contaminating the sample \citep{Irrgang2020, Shenar2020LB1}, and to rule this out, spectroscopic analyses will be needed. We discuss this in Sect.\,\ref{subsec:BHcandidatesDisc}.  

Finally, Table\,\ref{tab:Sample} provides constraints on intrinsic X-ray luminosities obtained with the Chandra
Visionary programme T-ReX \citet{Crowther2022TREX}. Evidently, only four targets (VFTS~333, 404, 532, 603) have finite detections, all of which smaller than $10^{33}\,$\ergs and lying below or close to  the canonical value $L_x / L \approx -7 $ \citep{Pallavicini1981, Sana2006}. This stands in contrast to the only HMXB candidate in the 30 Dor region,  \object{VFTS~399} ($\log L_X = 34.7\,$[\ergs], \citealt{Clark2015}). The few X-ray detections are discussed in Appendix\,\ref{sec:indiv}.

In addition to the spectroscopy, we also make use of $I$-band photometry obtained between Oct 2001 and Mar 2020 with the Optical Gravitational Lensing Experiment data reductions III and IV \citep[OGLE,][]{Udalski2003, Udalski2008, Udalski2015}.

\section{Analysis}
\label{sec:analysis}

\subsection{Spectral disentangling}
\label{subsec:specdis}

To investigate whether non-degenerate secondaries are present in our sample, we utilise the method of spectral disentangling \citep{Bagnuolo1991, Simon1994, Hadrava1995, Ilijic2004, Sablowski2019}. By exploiting the information stored in the spectral-line variability of a spectral time series of a binary, disentangling allows one to separate the component spectra in a binary and derive the orbital parameters. There are different algorithms of spectral disentangling, operating in either wavelength space or Fourier space. Here, we utilise the iterative shift-and-add scheme \citep{Marchenko1998, Gonzalez2006, Mahy2010, Shenar2017}. This technique generally yields comparable results to  Fourier disentangling \citep[e.g.][]{Shenar2020LB1}, and, in some cases,  appears to be more robust \citep[e.g.][]{Bodensteiner2020HR}. The method of spectral disentangling is not limited to a specific types of stars: it has been implemented on O-type stars \citep[e.g.][]{Martins2012, Mahy2020a}, B and Be stars \citep{Hensberge2000, Bisikalo2000, Saad2005}, lower-mass stars \citep[e.g.][]{Maceroni2014}, and even Wolf-Rayet stars \citep[e.g.][]{Shenar2017, Shenar2018, Shenar2019, Shenar2021}. The underlying assumption is that the variability of the spectral lines is dominated by Doppler shifts of one or more sources (see also Appendix\,\ref{sec:appregg}.)

The  spectral separation procedure assumes knowledge of the RVs of both components, $RV_{i}$ and $RV_{2, i}$, where $i \in {1, ..., N}$ runs over all available epochs of observation. Equivalently, the orbital parameters can be assumed, from which the RVs may be computed.   The procedure is iterative. Starting with an approximation for the secondary's spectrum at the $j^{\rm th}$ iteration, $B_{j}$, the approximation for the primary's spectrum at the $j^{\rm th}$ iteration, $A_{j}$ is computed by shifting-and-subtracting $B_{j}$ from all observations according to the available RVs, and co-adding the residual spectra in the frame of reference of the primary. The same procedure is then used to compute $B_{j+1}$ from $A_{\rm j}$. The first iteration assumes a flat spectrum for the (fainter) secondary, unless otherwise stated, such that the first approximation for the primary is simply the co-added spectrum. 
The convergence rate of the separation algorithm and hence the number of iterations depends on the RV amplitudes and line profiles, and may range between a few to hundreds, depending on the line profiles, RV amplitudes, and desired level of accuracy. 

In many applications, however, the RVs of one or both components are not constrained. In this case, one can perform the disentangling across the $K_1, K_2$ axes  (grid disentangling), where $K_1, K_2$ are the RV semi-amplitudes of both components. In this context, specific spectral lines (or sets of lines) are disentangled. By computing $\chi^2(K_1, K_2)$ from the residual spectra:

\begin{equation}
   \chi^2(K_1, K_2) = \frac{1}{N_{\lambda} (N_{\rm epochs} - 2)}\,\sum_{i=1}^{N_{\rm epochs}}\,\sum_{k=1}^{N_{\lambda}} \frac{ \left(A_{i,k} + B_{i,k} - O_{i,k}\right)^2}{\sigma_i^2},
\label{eq:chi2}
\end{equation}
one can minimise $\chi^2$ to retrieve the $K_1, K_2$ values that best reproduce the data, and separate the composite spectra with these values. In Eq.\,(\ref{eq:chi2}), $N_{\lambda}$ is the number of wavelength bins in the selected range, $N_{\rm epochs}$ is the number of epochs (cf.\ Table\,\ref{tab:Sample}). $A_{i,k}, B_{i,k}$ are the disentangled spectra obtained for $K_1, K_2$, evaluated at $\lambda_k$, and  shifted to the appropriate RVs of the $i^{\rm th}$ epoch. $O_i$ is the observed $i^{\rm th}$ spectrum,  $\sigma_i$ is the S/N in the continuum region in the vicinity of the spectral line, and $\nu = N_{\lambda} (N_{\rm epochs} - 2)$ is the number of degrees of freedom, since each pixel in each of the two disentangled spectra is considered a free variable.  Naturally, one may extend this to other orbital parameters, such as the eccentricity ($e$) or the orbital period ($P$), albeit this becomes computationally expensive for the shift-and-add algorithm. However, since these parameters are typically well constrained in SB1 binaries, they can be kept fixed in our study, unless clear deviations are seen that require a revision of the parameters.

It is well known that the shift-and-add technique often results in apparent "emission wings" in the disentangled spectra, which are not of astrophysical origin, but are  intrinsic to the method \citep{Quintero2020}. Removing these features typically requires very large number of iterations, making the disentangling procedure inefficient. A workaround is to  enforce the disentangled spectra to lie below the continuum for absorption-line spectra, unless indications exist that the spectrum contains emission \citep{Shenar2019, Quintero2020}. This assumption is adopted throughout the paper, unless otherwise stated.

The method is thoroughly tested by creating mock data of different synthetic binaries, discussed in detail in Appendix\,\ref{sec:appregg}. Generally, the method works well in terms of reproducing the input spectra and RV semi-amplitudes whenever the semi-amplitudes are larger than half the resolution element ($K_1, K_2 \gtrsim 20\,$\kms). Fortunately, this holds for the majority of targets. Depending on the S/N, we can retrieve companion spectra down to flux contributions of \mbox{$f_2 / f_{\rm tot} \approx 1-3\%$}, noting that the limit strongly depends on the data quality and the rotational velocity of the companion. In principle, pushing to fainter companions could have been possible using the strong Balmer lines. However, the strong nebular contamination in the TMBM data often make Balmer lines unusable for this purpose (see Sect.\,\ref{subsec:neblines}). 

Finally, we highlight a few facts that appear to hold generally: (i) it is  easier to detect the presence of a companion than to determine $K_2$. Therefore, in some cases, we can confidently conclude the presence of a non-degenerate companion even if we cannot determine  $K_2$. (ii) Due to the large number of degrees of freedom, errors on $K_1, K_2$ from disentangling are  larger than those obtained from orbital fitting.  (iii) The fainter the secondary,  the less of an impact $K_2$ has on the disentangled spectra. In contrast, even small deviations in $K_1$ can result in spurious features in the disentangled spectrum of the secondary. This typically  results in a secondary spectrum that mimics that  of the primary, which can lead to an erroneous detection of a non-degenerate companion. We critically assess these facts throughout our analysis, and discuss this for the individual targets in Appendix\,\ref{sec:indiv}.

\subsection{Identifying non-degenerate companions}
\label{subsec:NonDegComp}

In the absence of eclipses,  the disentangled spectra can only be retrieved up to a scaling factor that depends on the light contribution of the two components to the total flux, $l_1 = f_1 / (f_1 + f_2)$ and $l_2 = 1-l_1 = f_2 / (f_1 + f_2)$. The fainter the companion is, the larger the amplification factor $1/l_2$ is.
Fortunately, the spectral typing does not depend on these scaling factors. 

It is important to realise that the method of disentangling, by nature, always returns a disentangled spectrum for a secondary star. It is then up to the user to decide whether the spectrum is "flat", or whether it is of stellar origin. 
Whether or not the companion is classified as  "non-degenerate" is left as a subjective decision, based on a visual inspection of its spectrum after applying the scaling, as well as comparison to spectral models. We generally avoid relying on Balmer lines due to the strong nebular contamination (Sect.\,\ref{subsec:neblines}), unless stated otherwise in Appendix\,\ref{sec:indiv}. We give much weight to the presence of He\,{\sc i} and He\,{\sc ii} lines (depending on the level of nebular contamination). In most cases, when a non-degenerate companion is reported, the stellar spectrum of the secondary is evident, especially upon magnification through re-scaling.

In quite a few cases, it is not clear whether the observed features are of stellar origin, or whether they are caused by nebular contamination (Sect.\,\ref{subsec:neblines})  or sources of intrinsic variability (Appendix\,\ref{sec:appregg}). In these cases, we mark the systems as "uncertain SB2" or "uncertain SB1" (SB2:, SB1:), depending on the "degree of belief" we attribute to the spectral features of the secondary, such as their line profiles and widths, their appearance with respect to those of the primary, and the overall level of nebular contamination. Our reasoning for each target is individually discussed in Appendix\,\ref{sec:indiv}.
While automated criteria could be developed to identify and classify the binaries, the presence of nebular lines makes it difficult to establish suitable objective criteria. Given the limited sample size, we prefer to keep the classification subjective, since developing an automated algorithm in this context would anyhow boil down to a subjective choice of  criteria. The disentangled spectra are presented in Appendix\,\ref{sec:DisSpec} and are available online\footnote{Disentangled spectra are available at the Centre de Donn\'ees astronomiques (CDS)  via anonymous ftp to cdsarc.u-strasbg.fr (130.79.128.5)} for the independent inspection of the reader.

\subsection{Spectral classification and light ratios}
\label{subsec:lcs}

For the spectral classification, we use the Marxist Ghost Buster (MGB) code \citep{Maizetal12,Maizetal15b} on the disentangled spectra, which allows the user to interactively compare the observed spectra with the spectral type-luminosity class 2-D grid of spectral standards. 
We use the latest spectral grid derived from the Galactic O-Star Spectroscopic Survey \citep[GOSSS][]{Maizetal11} that includes stars as late as A0 (Ma\'{\i}z Apell\'aniz et al. in prep.). As the GOSSS grid has a spectral resolution 2500, we degrade our spectra to that value.


Spectral classification of hot stars in the blue-violet is usually done in the 3900-5100~\AA\ range. However, here we rely primarily on the disentangled spectra, which only extends down to 4560~\AA.  The primary luminosity classification criteria and various qualifiers  for most O stars involve \HeII{4686} and neighbouring lines, which are only present in the non-disentangled spectra.  The 'z' qualifier (\HeII{4686} in absorption and stronger than \HeI{4471} and \HeII{4542}) cannot be established as that depends sensitively on the ratio of these lines, and we therefore omit it from the classification scheme  \citep{Ariaetal16,Maizetal16}. As we are also missing the 4630-4660~\AA\ region in the disentangled spectra, the f and c qualifiers (involving the presence of N and C emission lines in this region) cannot be determined \citep{Walbetal10a,Sota2011}. In this case, we adopted these qualifiers  from the original classifications of \citet{Walborn2014}, implicitly assuming that the features seen in the non-disentangled spectra belong to the primary. We therefore note that the spectral types provided here reflect those that best match our data (disentangled and non-disentangled combined), and may be revised in the future, when disentangling of the full 3900-5100~\AA\ range might become available.

To obtain estimates for the light ratios and the correct scaling of the disentangled spectra, we  rely on calibrations between spectral types and equivalent widths of \HeII4200, 4542, and \HeI4026, 4144, 4388, and 4471, as long as the equivalent widths are significant and not strongly contaminated by nebular lines. We use equivalent-width formulae provided by Sana et al.\ (in prep.), which are reminiscent of existing quantitative classification schemes by \citet{Conti1971} and \citet{Conti1974}, but using modern data. If two non-degenerate stars are retrieved from disentangling, we compute the light ratio as a weighted mean between the light ratios obtained from the primary's scaling and secondary's scaling. If the spectral subtype of the secondary cannot be determined, we adopt a light ratio of 5\% for putative non-degenerate companions and 3\% otherwise.

\subsection{Impact of nebular lines}
\label{subsec:neblines}

Only a limited number of spectral lines are available in the data with sufficient S/N to be  useful for the disentangling: the Balmer lines H$\delta$ and H$\gamma$, the \HeI4026, 4144, 4388, and 4471 lines,  the \HeII4200, 4542 lines, and rarely metal lines such as Mg\,{\sc ii}\,$\lambda 4481$ and Si\,{\sc iii}\,$\lambda 4553$,
due to the limited S/N. Of those, only the He\,{\sc ii} lines are free of nebular contamination, but  as we are mostly hunting for faint B-type companions, He\,{\sc ii} lines are of limited use. While sky spectra were subtracted to reduce the nebular contamination, strong residuals remain, likely due to the variability of the emission across the Tarantula. This problem worsens by the fact that the nebular residuals are not constant, but vary from epoch to epoch in strength and shape in a non-trivial manner,  presumably due to the varying weather and seeing conditions between the different epochs.
At the moderate resolution of our data ($\Delta \lambda \approx 0.7$\,\AA\ ), the nebular lines cover a significant part of the stellar profiles, and cannot be removed by simple interpolation without significantly biasing the data. To partly account for the nebular contamination, we extended the shift-and-add algorithm to three components, where component C represents the nebular lines \citep[see also][]{Abdul-Masih2019}. Component C is enforced to be static, lie above the continuum, and achieve non-vanishing values only in regions overlapping with the nebular emission. To account for the varying strengths of the nebular lines, we scale the component C with the equivalent widths of the emission cores of H$\gamma$.

\begin{figure}
\centering
\includegraphics[width=.5\textwidth]{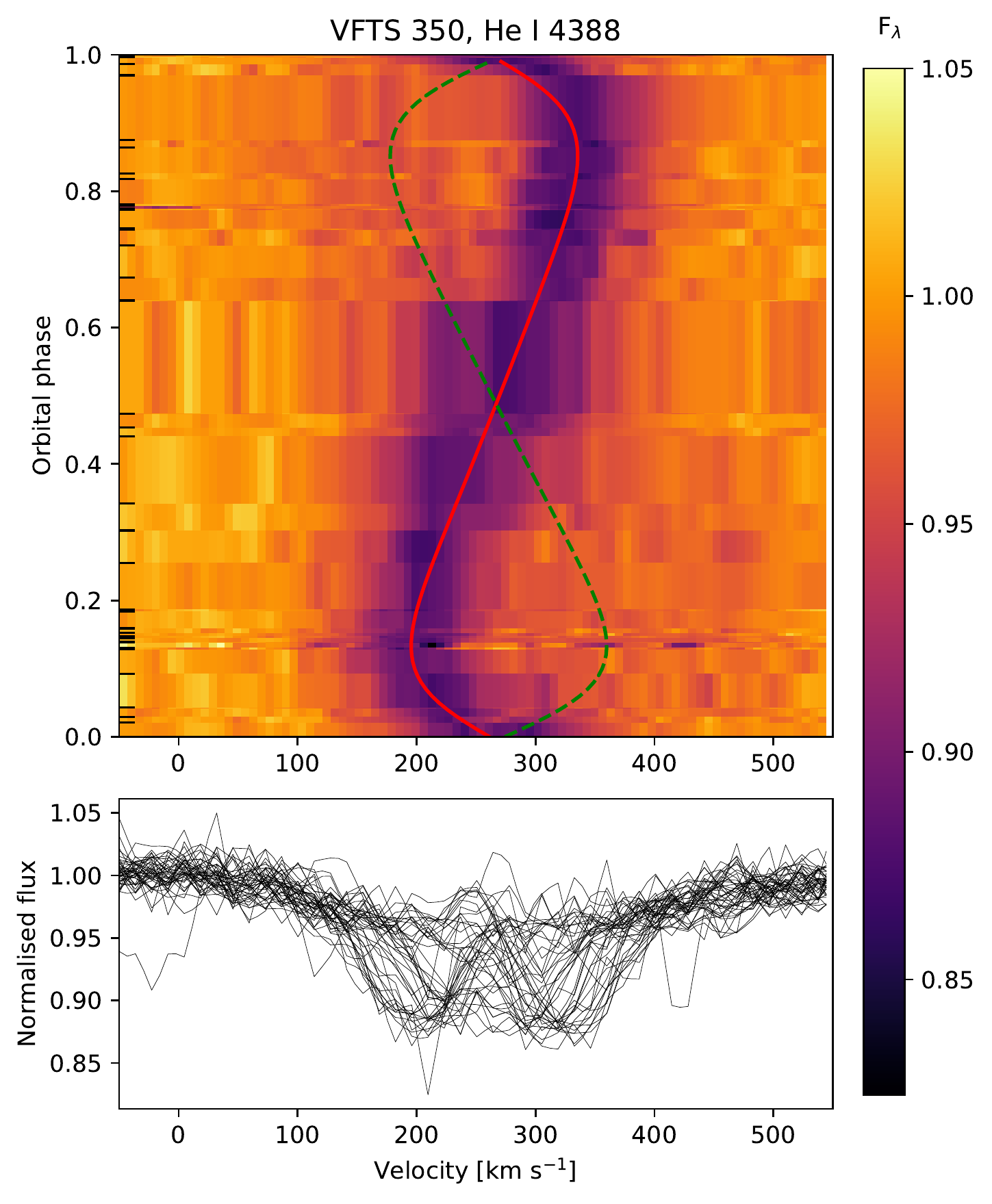}
\caption{Dynamical spectrum of VFTS~350, \HeI4388 line, phased with the orbital period $P=69.57\,$d (upper panel), and the individual spectra (lower panel). Also plotted are the RV curves of the O9.5~V primary (red solid line) and the hidden B0~V secondary (dashed green line).} \label{fig:VFTS350dyn}
\end{figure}

\begin{figure}
\centering
\includegraphics[width=.5\textwidth]{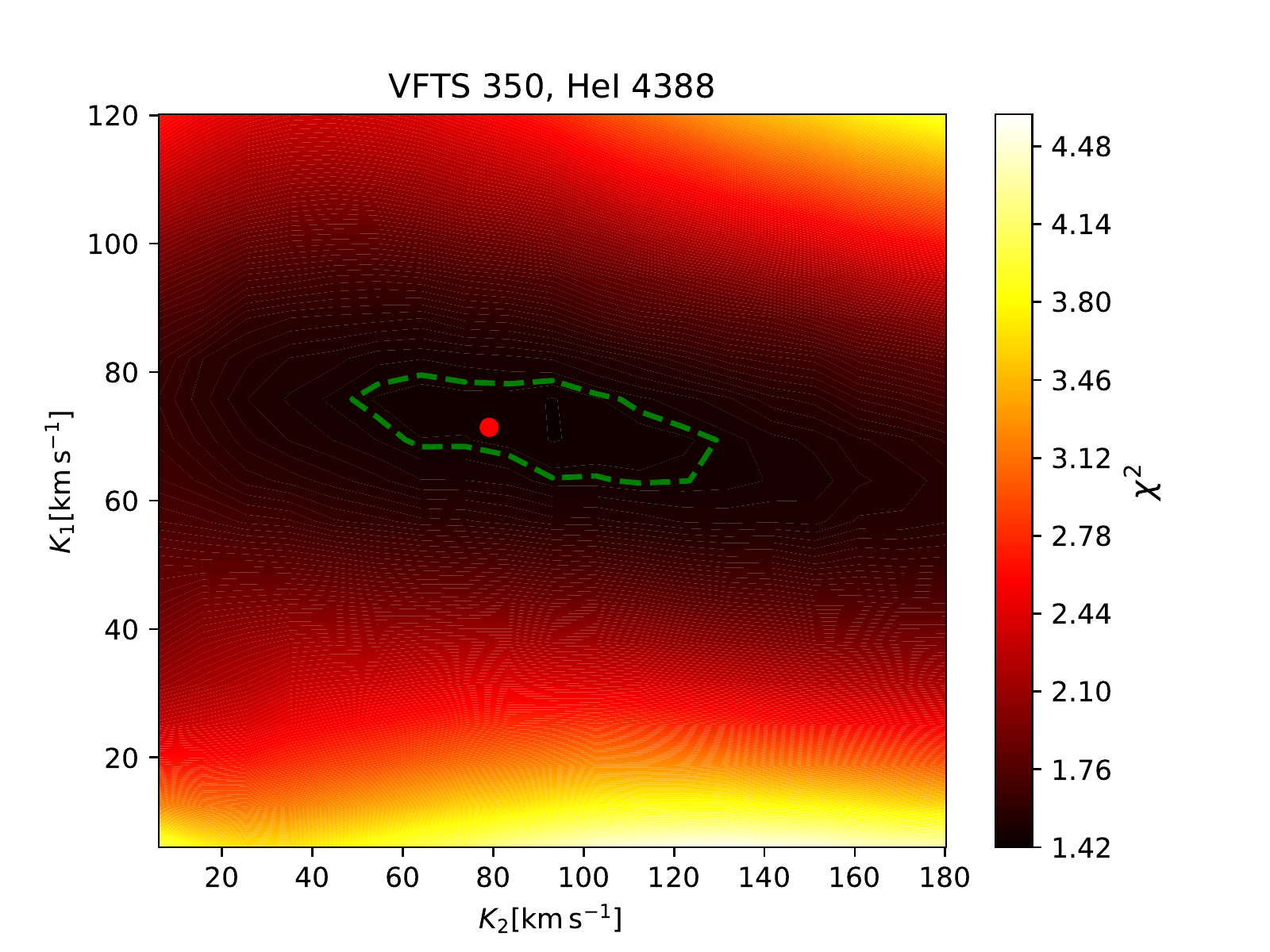}
\caption{$\chi^2$ map of the residuals of VFTS~350 (Eq.\,\ref{eq:chi2}), showing the 1$\sigma$ contour as well (green dashed lined). The minimum is at $K_1 = 72 \pm 8$ and $K_2 = 79 \pm 28$\,\kms, while the measurement considering multiple lines is $K_1= 70.3\pm3.7$\,\kms~and $K_2 = 91\pm20\,$\kms.    } \label{fig:chi2350}
\end{figure}

While our method for removing the residual nebular-line contamination works reasonably well (see examples in Sect.\,\ref{subsec:examples} and Appendix\,\ref{sec:indiv}),  it does not fully remove the impact of nebular lines. This is because the nebular lines vary in shape and strength, and are at times saturated or over-subtracted. This is especially important when the nebular contamination is very strong (typically in Balmer lines and often in the \HeI4471 line), or when the companion is faint, such that the nebular contamination dominates over the secondary's features. 
Generally, not accounting for nebular contamination leads to spurious features that are especially apparent in the amplified spectra of the faint secondaries. These features can appear confusingly similar to actual stellar features, though they often exhibit strong asymmetries. As we are not able to fully remove the impact of nebular lines, for some systems, our results are ambiguous in terms of the presence of a non-degenerate companion (see Sect.\,\ref{subsec:NonDegComp}).

\subsection{Examples: VFTS 350 (O+O) and VFTS~779 (B+BH?)}
\label{subsec:examples}

We illustrate the method by using VFTS~350 and 779 as examples. VFTS~350 was classified O8~V by \citet{Walborn2014}, revised to O8.5~V+O9.5~V here. \citet{Almeida2017} derived period of $P=69.57\,$d and an eccentricity of $e = 0.351$ for this system, which they classified SB1. Indeed, inspection of the dynamical spectrum of VFTS~350 (Fig.\,\ref{fig:VFTS350dyn}) does not readily reveal indications for a non-degenerate secondary. However, an implementation of grid disentangling of lines such as \HeI4388 (Fig.\,\ref{fig:chi2350}) reveals a well constrained minimum of $\chi^2(K_1, K_2)$. We perform a similar analysis for all strong helium lines, avoiding \HeI4471 and the Balmer lines due to strong nebular contamination. We obtain the final $K_1, K_2$ values by computing the weighted mean of all measurements, amounting to $K_1 = 70.3 \pm 3.7\,$\kms, and $K_2 = 91 \pm 20\,$\kms.  We note that the $K_1$ value derived here exceeds the value of $60.2\,$\kms~derived by \citet{Almeida2017}. This is a typical bias seen when spectral lines are blended by two components \citep{Bodensteiner2021, Banyard2022}.

\begin{figure}
\centering
\includegraphics[width=.5\textwidth]{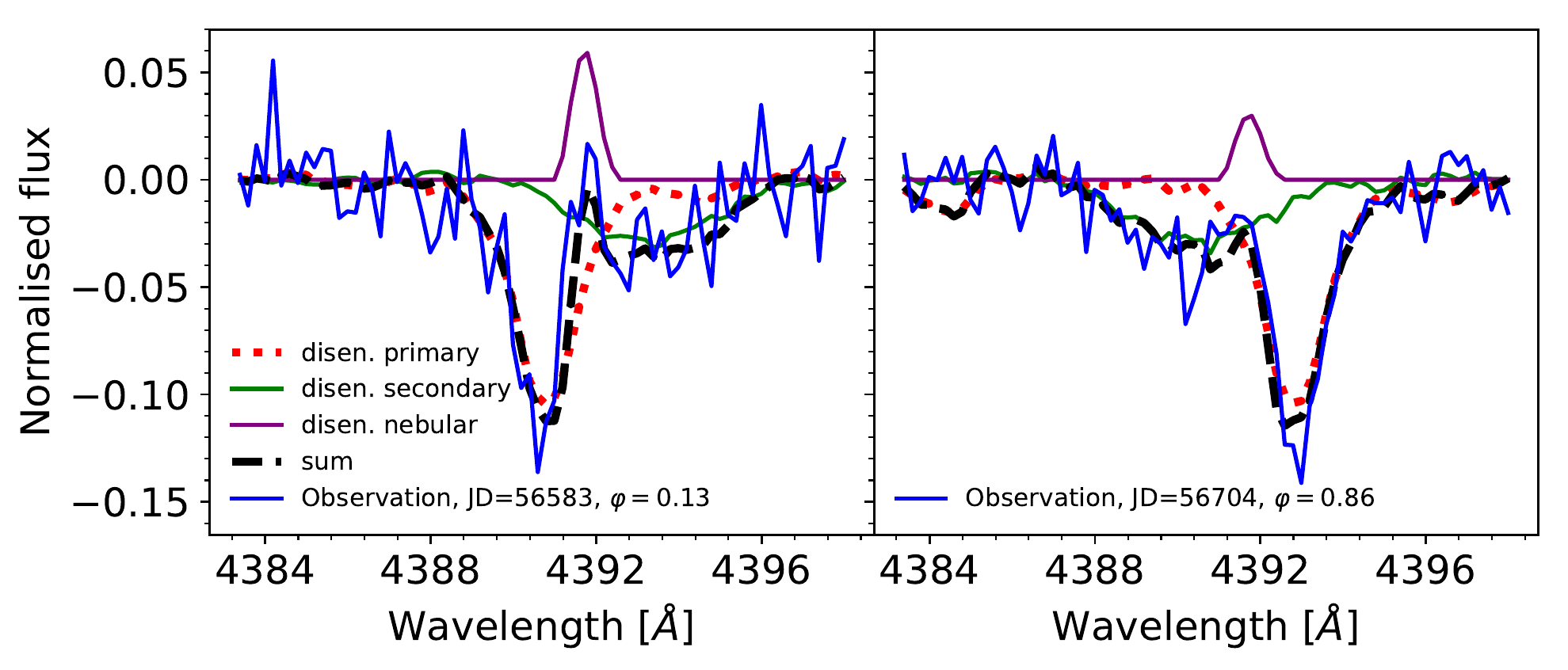}
\caption{Disentangling of the \HeI4388 line of \object{VFTS~350}. Shown are the normalised spectra of \object{VFTS~350}, vertically shifted by -1, at RV extremes  (phases $\varphi = 0.13$ and 0.86, left and right panels, noisy blue line), compared to the disentangled spectrum of the primary (dotted red line), secondary (solid green line), nebula (solid purple line), and their sum (dashed black line). Dates in the legend are given as JD - 2400000. The disentangling is performed for the derived  $K_1 =70\,$\kms~and $K_2 = 91\,$\kms~(see text). 
The disentangled spectra are not scaled by the light ratio in this plot. } \label{fig:VFTS350EXTexmp}
\end{figure}

\begin{figure}
\centering
\includegraphics[width=.5\textwidth]{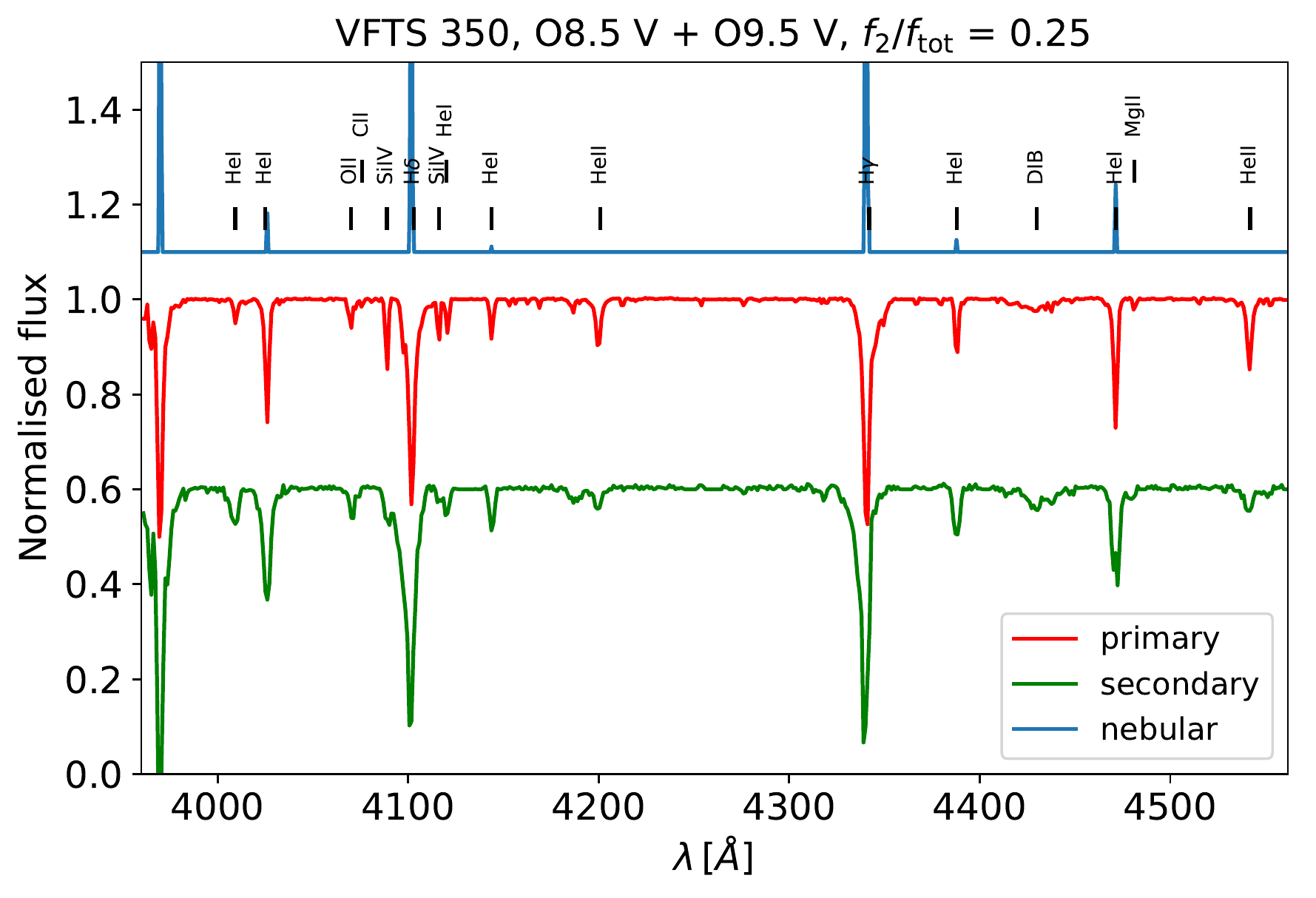}
\caption{Disentangled primary (red line, middle), secondary (green line, bottom), and nebular (blue line, top) spectra of VFTS~350. The spectra were binned at $\Delta \lambda = 1\,\AA$, or $R \approx 4000$ and shifted vertically for clarity. The component spectra are corrected for line dilution. } \label{fig:DISSPEC350}
\end{figure}

Figure\,\ref{fig:VFTS350EXTexmp} shows a comparison between the observations and the disentangled spectra for the derived $K_1, K_2$ values at two epochs corresponding to the RV extremes. Upon inspection of the line variability in these two epochs, the presence of a non-degenerate companion becomes readily apparent.  Evidently, the secondary exhibits broader lines, making it more  difficult to spot. 
The full disentangled spectra, corrected for line dilution, are shown in Fig.\,\ref{fig:DISSPEC350}. Relying on \citet{Sota2011}, we classify the secondary as O9.5~V, and estimated its flux contribution at 25\%.

\begin{figure}
\centering
\includegraphics[width=.5\textwidth]{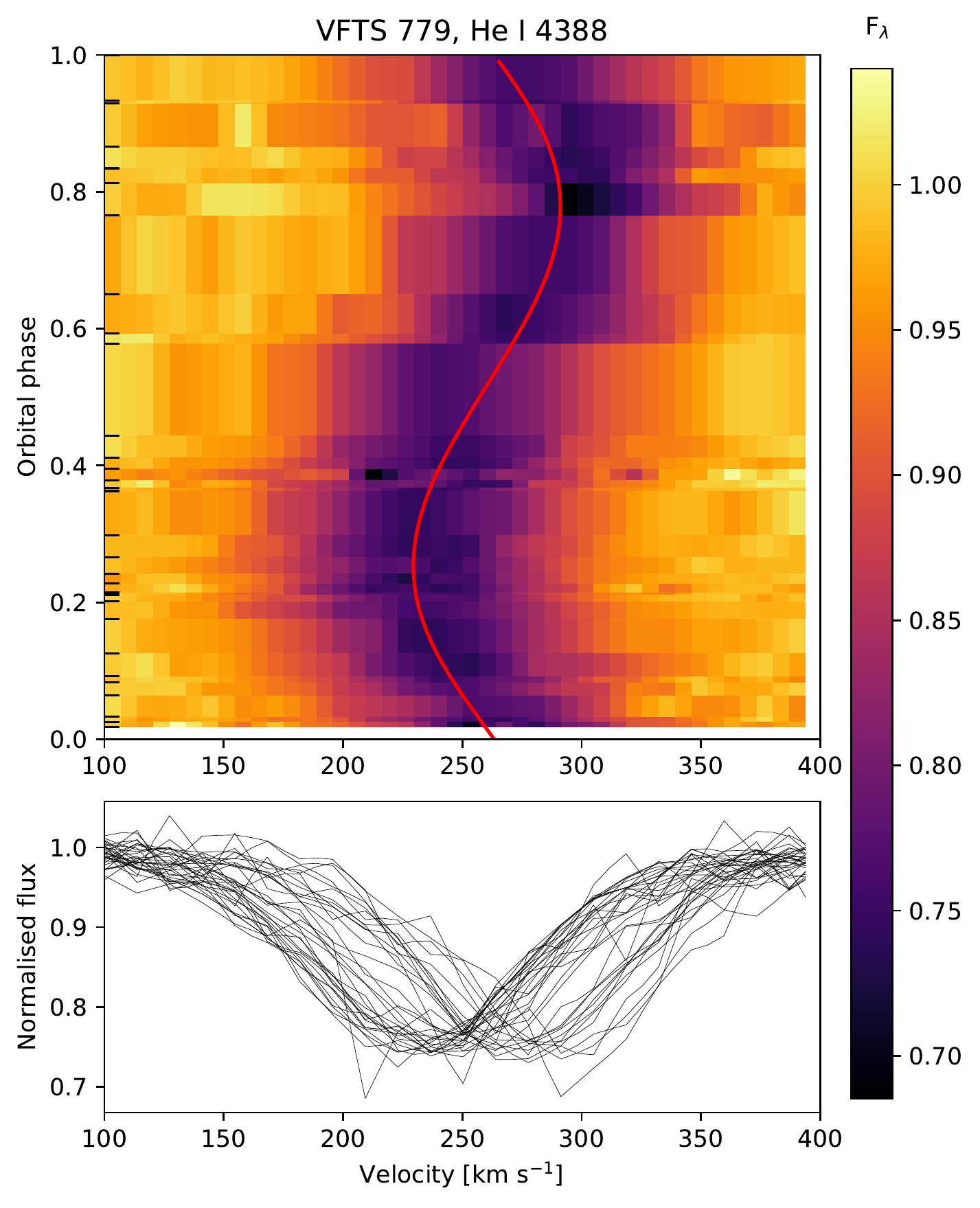}
\caption{As Fig.\,\ref{fig:VFTS350dyn}, but for VFTS~779, a B1~II-III SB1 binary in which no luminous companion is identified.  } \label{fig:VFTS779dyn}
\end{figure}

\begin{figure}
\centering
\includegraphics[width=.5\textwidth]{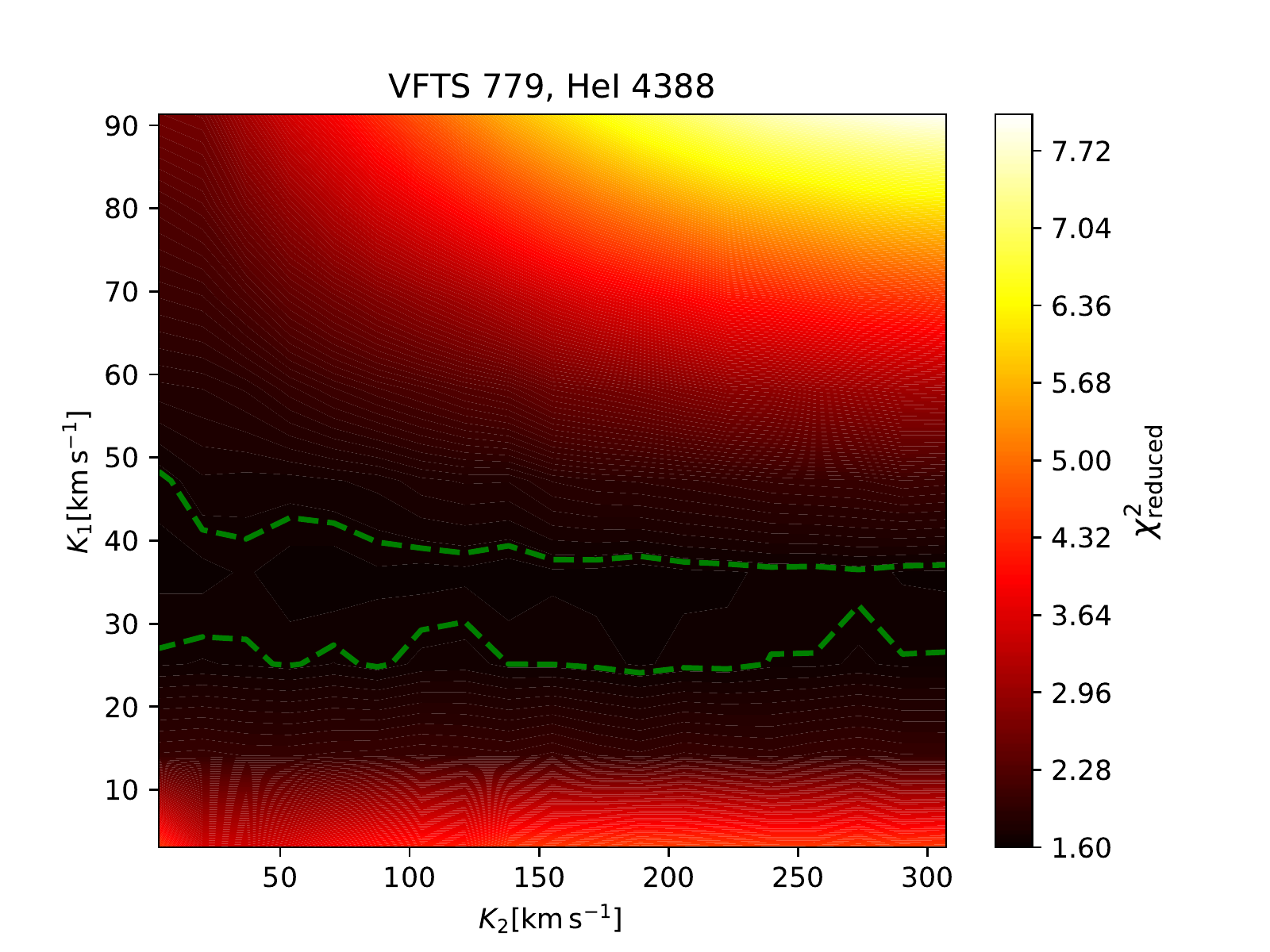}
\caption{As Fig.\,\ref{fig:chi2350}, but for  VFTS~779. $K_1$ is measured at $K_1 = 35.5 \pm 4.0\,$\kms~(consistent with the final measurement of $K_1=30.8\,$\kms), but $K_2$ is not constrained.} \label{fig:chi2779}
\end{figure}

\begin{figure}[!h]
\centering
\includegraphics[width=.5\textwidth]{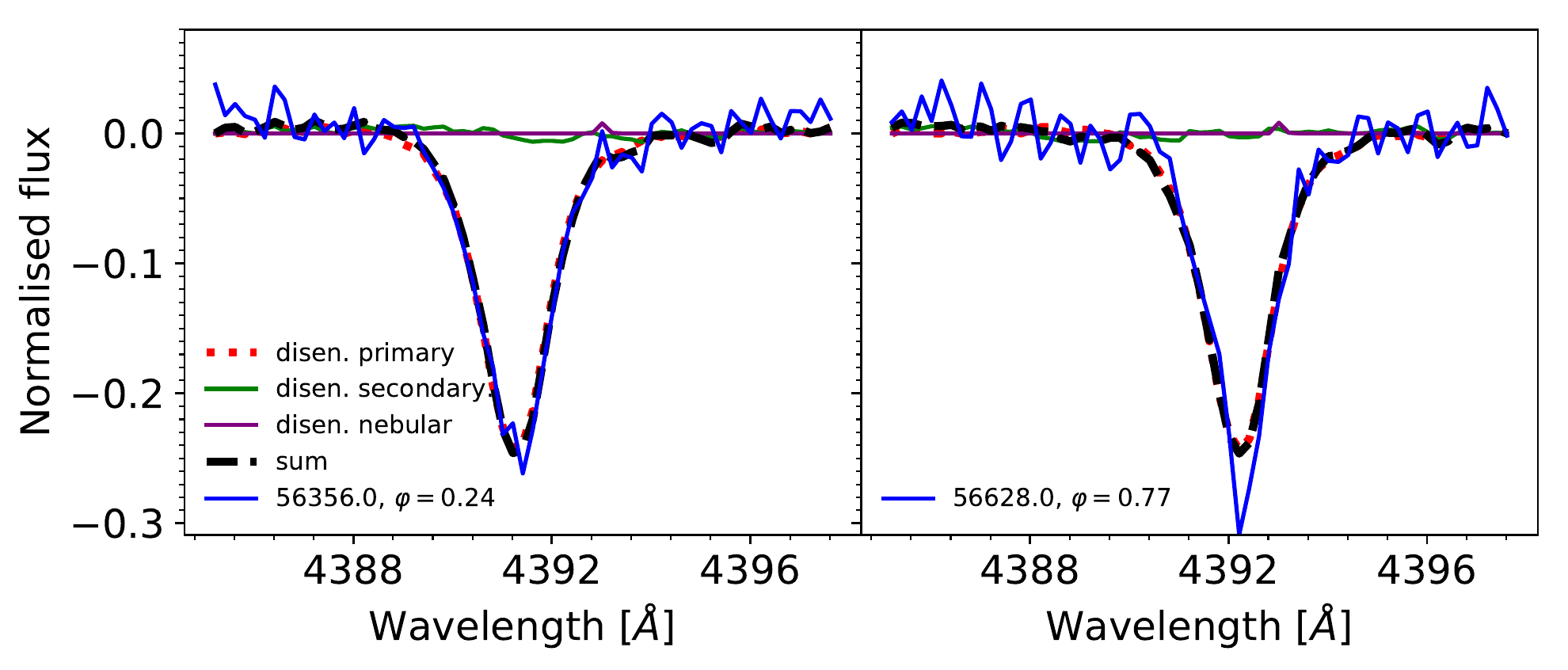}
\caption{As Fig.\,\ref{fig:VFTS350EXTexmp}, but for  \object{VFTS~779},  as derived for  $K_1 =31\,$\kms and $K_2 = 3 \times K_1 = 93\,$\kms. } \label{fig:VFTS779EXTexmp}
\end{figure}

\begin{figure}
\centering
\includegraphics[width=.5\textwidth]{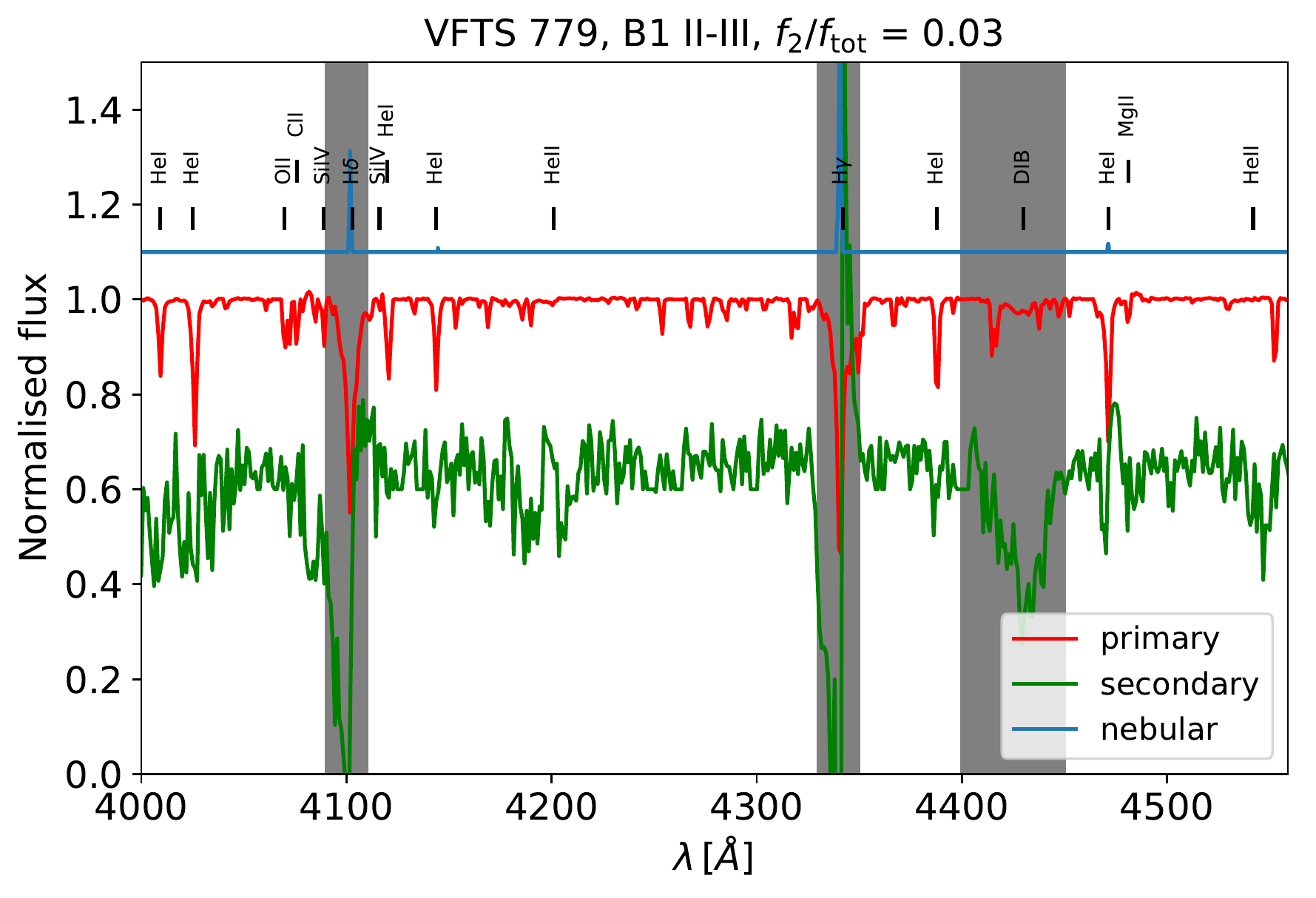}
\caption{As Fig.\,\ref{fig:DISSPEC350}, but for VFTS~779. The diffuse interstellar band (DIB) at $\approx 4410\,\AA$ and the Balmer lines, which suffer strong nebular contamination, are greyed out. No other significant stellar features within the S/N are observed.  } \label{fig:DISSPEC779}
\end{figure}

We now repeat this exercise for VFTS~779, which  was classified  B1~II-Ib by \citet{Walborn2014} and revised to B1~II-III here. \citet{Almeida2017} determined it to be an SB1 binary with $P=59.9\,$d and a near-circular orbit. A dynamical spectrum is shown in Fig.\,\ref{fig:VFTS779dyn}. Like for VFTS~350, a companion is not readily seen in the dynamical spectrum. In Fig.\,\ref{fig:chi2779}, we show the $\chi^2(K_1, K_2)$ map obtained for the \HeI4338 line. While $K_1$ can be well constrained, $K_2$ is fully unconstrained.  Indeed, when inspecting the spectral variability of this line at RV extremes (Fig.\,\ref{fig:VFTS779EXTexmp}), a companion cannot be identified. The spectral appearance of the disentangled spectra is virtually independent of the value of $K_2$. As the spectral separation still requires a $K_2$ value,  we  adopt the arbitrary value of $K_2 = 3\times K_1 = 93\,$\kms~(i.e. assuming the companion is three times less massive).  In Fig.\,\ref{fig:DISSPEC779}, the full disentangled spectra are shown. To enhance the spectral features of a putative secondary, we adopt an extreme light ratio of $l_2 = f_2 / f_{\rm tot} = 0.03$ in the visual. Evidently, the disentangled spectrum is virtually flat, with the exception of features originating in diffuse interstellar bands (DIBs) and nebular contamination.

VFTS~779 therefore does not have an identified companion. With a minimum secondary mass of $M_2 > 3.9\pm0.6\,M_\odot$ (Table\,\ref{tab:Sample}), this makes VFTS~779 a prime OB+BH candidate. However, given the that the system can accommodate a secondary as light as $\approx 3\,M_\odot$, we cannot rule out that it is a faint non-degenerate companion (main sequence star or helium star), which would contribute less than $1\%$ to the flux. Indeed, VFTS~243 is the only candidate for which non-degenerate companions can be ruled out \citep{Shenar2022VFTS243}. 

\subsection{Light curve analysis}
\label{subsec:lcanalysis}

The OGLE light curves are used to search for photometric variability in the  systems, with focus on eclipses or ellipsoidal variations. 
We analysed available light curves using the Period04 \citep{Lenz2005} tool, which performs a Lomb-Scargle analysis on the data. Peak frequencies are refined using through phase dispersion minimisation \citep[pwkit pdm][]{Stellingwerf1978, Schwarzenberg-Czerny1997}, and the light curves are phased using these periods and inspected. If a notable signal is observed, we present the light curve and discuss it in  Appendix\,\ref{sec:indiv}. While not fundamental for our study, the light curves help uncover several eclipsing binaries and high-order multiples. An in-depth analysis of the light curves is beyond the scope of our study.

\section{Results}
\label{sec:results}

We perform grid disentangling (1D or 2D) for all targets in the sample.  A detailed account of our analysis for each system is given in Appendix\,\ref{sec:indiv}, and the scaled disentangled spectra are presented in Appendix\,\ref{sec:DisSpec}. Table\,\ref{tab:SampleFin} provides the final orbital parameters, mass ratios $q$, estimated secondary masses $M_2$, and classifications. 
Figure\,\ref{fig:samplesolved} summarises our findings visually by showing, for each system, the 68\% confidence interval on the mass of the secondary, $M_2$. The derivation of these mass ranges is described in Sect.\,\ref{subsec:MonteCarlo}.

Using spectral disentangling, we can recover non-degenerate companions for the majority of our sample. In total, among the original 51 SB1 binaries, non-degenerate companions were found in 43 systems (84\%), of which 28  are considered certain and 15 less certain. For the majority of these new SB2 systems, $K_2$ RV semi-amplitudes could be derived, albeit often with substantial associated uncertainties. The hidden companions are typically early B-type stars, with some exceptions.  
As Fig.\,\ref{fig:samplesolved} illustrates, almost all companions with minimum masses above $\approx 5\,M_\odot$ could be recovered through disentangling.  In the remaining eight SB1 binaries (16\%), non-degenerate companions could not be uncovered, including the confirmed O+BH binary VFTS~243 \citep{Shenar2022VFTS243} and two additional OB+BH candidates, VFTS~514 and VFTS~779. We discuss this sample in Sect.\,\ref{subsec:BHcandidatesDisc}.

\subsection{Derivation of mass ratios and mass ranges}
\label{subsec:MonteCarlo}

\begin{figure}[!h]
\begin{tabular}{l}
\includegraphics[width=.485\textwidth]{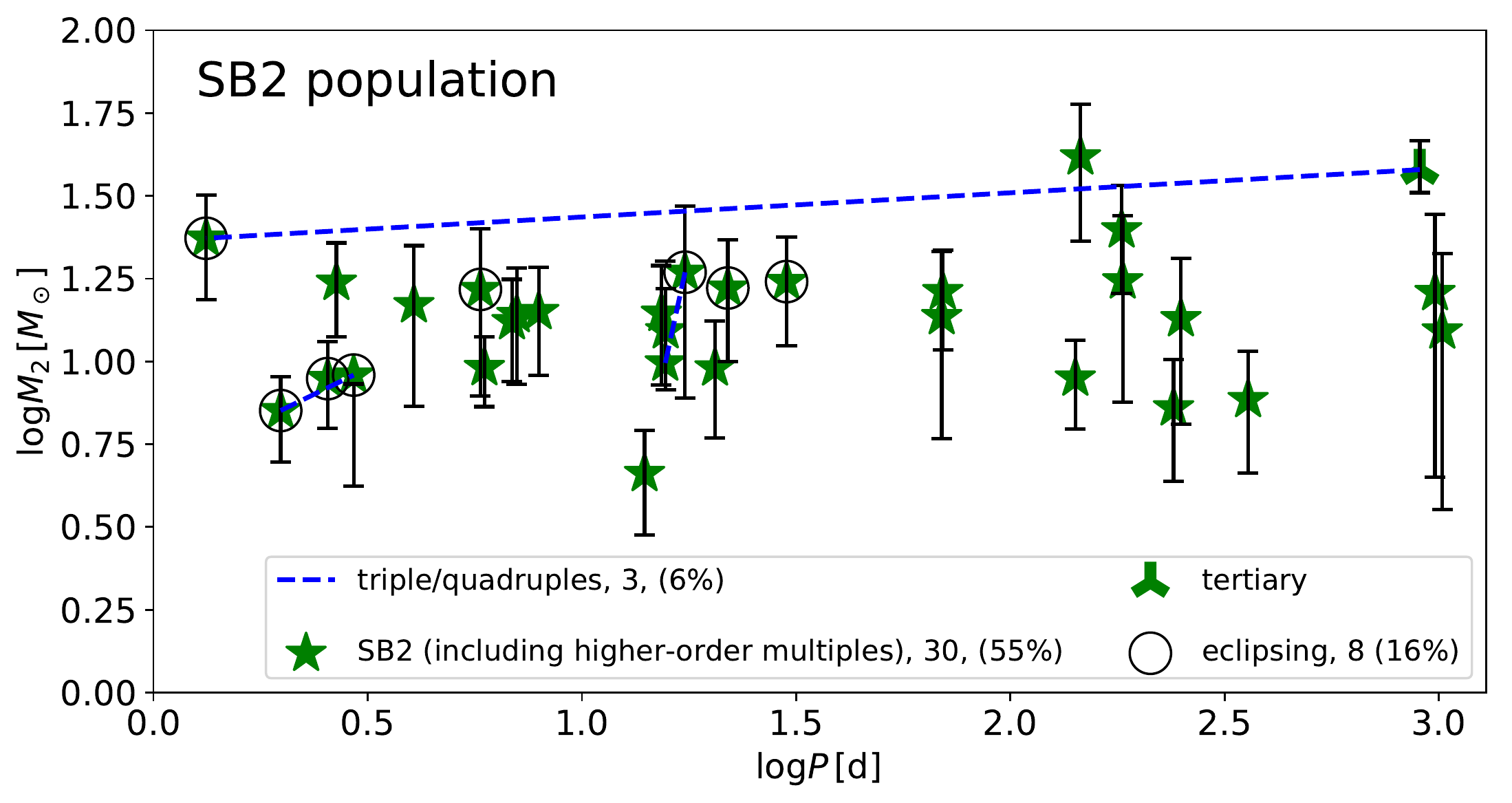}\\
\includegraphics[width=.485\textwidth]{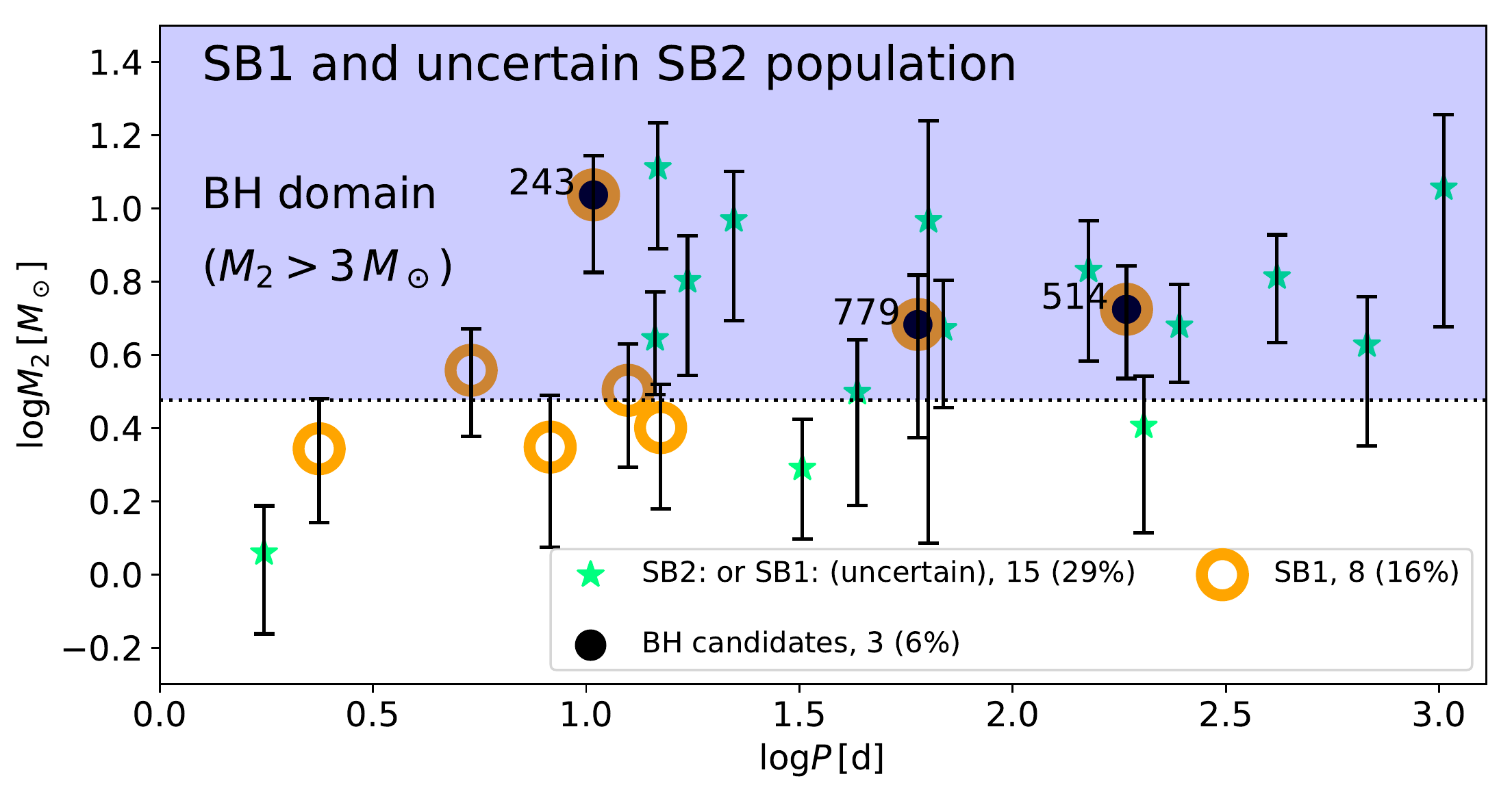}
\end{tabular}
    \caption{Nature of the hidden companions in the 51 SB1 binaries analysed here. Upper panel: Masses of the previously "hidden" companions in the newly uncovered SB2 systems  as a function of the orbital period. In total, 28 out of the  51 SB1 systems in our sample are  unambiguously found to host non-degenerate companions, though 31 symbols are plotted due to the presence of higher-order multiples. For the tertiary of VFTS~64, the outer period and inner binary mass are denoted. For the quadruples (VFTS~120 and 702), only the binary pairs are considered, since the outer period is not known, and a primary mass of $M_1=20\pm10\,M_\odot$ is assumed for VFTS~702B for plotting purposes.   The symbols mark the medians of the distributions, and the error bars denote the 68\% confidence intervals (see Sect.\,\ref{subsec:MonteCarlo}).  Lower Panel: as upper panel, but for the SB1 and uncertain SB1 and SB2 systems (SB1:, SB2:), which could in principle host a BH.  Shaded region depicts the mass regime $M_2 > 3\,M_\odot$ in which BHs are expected to reside. Along with the confirmed O+BH binary VFTS~243, two other OB+BH candidates are highlighted (VFTS~514 and 779). }
    \label{fig:samplesolved} 
\end{figure}

For systems in which $K_2$ could be derived, the computation of $q$ and $M_2$ along with their 68\% confidence intervals  follows directly from the inferred mass ratio ($K_1/K_2$) and the primary mass, for which we adopt $M_1 = M_{\rm 1, SpT}$ with the associated errors.
For systems for which $K_2$ could not be constrained (including all SB1 systems), the computation is more involved. First, we  asses the possible range of orbital inclinations for these systems. We assume a maximum inclination of 90$^\circ$ for all targets. The minimum inclination $i_{\rm crit}$ is estimated by equating the measured equatorial rotation velocity $\varv \sin i / \sin i_{\rm crit}$ with the critical rotation velocity $\varv_{\rm eq, crit}$ of the luminous components \citep{Mahy2022BH}. For simplicity, we adopt a typical value of the critical equatorial velocity as 500\,\kms~\citep[e.g.][]{Townsend2004}, such that:

\begin{equation}
    \sin i_{\rm crit} = \frac{\max(\varv \sin i_1, \varv \sin i_2)}{\varv_{\rm eq, crit}} \approx \frac{1}{500}\max(\varv \sin i_1, \varv \sin i_2).
\label{eq:icrit2}
\end{equation}

While we could estimate the critical rotation individually, the errors in spectral-type calibration and disentangling procedure easily dominate over the  variance of $\varv_{\rm crit}$ over the relevant spectral range. To estimate $\varv \sin i$, we measure the full-widths half-maxima (FWHM) of the \HeI4388~and \HeII4542 lines of all spectra through Gaussian fitting, and use $\varv \sin i$-FWHM calibrations provided by \citet{Ramirez-Agudelo2015} to obtain $\varv \sin i$. We only accept measurements for which the peak of the Gaussian exceeds the S/N by $6\sigma$ (i.e. the half-maximum is at least 3$\sigma$ below the continuum). We provide the computed $\varv \sin i$ and $i_{\rm crit}$ values in Appendix\,\ref{sec:vsini}. We note that the method neglects other sources of star-dependent broadening mechanisms (pressure broadening, micro- and macroturbulence), such that the $\varv \sin i$ values should only be considered approximate. The method generally results in poor lower bounds on the inclination of the order of $10^\circ$, and is therefore hardly constraining. We note that this method may not be valid for binaries containing degenerate secondaries due to kicks, which could cause the stellar spin to mis-align with the orbital spin. However, since the method only impacts the upper bound on the mass, it does not have important consequences in the context of identifying OB+BH candidates. 

With the mass functions, estimated primary masses ($M_1 = M_{\rm 1, SpT}$), and inclination ranges,  a probability density for $q$ can be computed for systems without a $K_2$ measurement.  To do so, we  perform a Monte Carlo simulation.
We ignore errors on the mass function, which are negligible in this context. We sample a Gaussian distribution for the mass of the primary using the values in Table\,\ref{tab:Sample}. We sample the inclination in the range $i \in [i_{\rm crit}, 90^\circ]$ (see Eq.\,\ref{eq:icrit2}), following the distribution $P(i) \propto \sin i$, which results from the projection of randomly oriented orbital orientations. For given $i$ and $M_1$, the mass of the secondary $M_2$ and the mass ratio $q$ can then be computed. For each SB1 system (or SB2 for which $K_2$ is not known), we thus obtain probability densities $P(M_2)$ and $P(q)$. Using these probability densities, we calculate the medians $\tilde{M_2}$ and $\tilde{q}$ and the 68\% confidence intervals. These values are given in Table\,\ref{tab:SampleFin}, and shown in Fig.\,\ref{fig:samplesolved}.

\subsection{High-order multiples}

By combining spectroscopic and photometric information, three systems are  found to be higher-order multiples.  Hence, Table\,\ref{tab:SampleFin} provides 56 entries in total instead of 51. VFTS~64 is a triple system consisting of an outer O7.5~II tertiary orbiting an eclipsing O6+O6 contact binary ($P_{\rm out} = 903\,$d, $P_{\rm in} = 1.3\,$d). VFTS~120 is found to be a quadruple system comprising an O9.5~IV + B0~V binary with a spectroscopic period of 15.65\,d and an eclipsing B+B binary with an photometric period of  $P=17.41\,$d. Finally, VFTS~702 is found to be a quadruple system owing to the presence of two distinct eclipse periods (1.98\,d and 2.93\,d), but the quality of the spectra does not enable us to determine the spectroscopic properties of the system unambiguously. Since the outer periods of both quadruples could not be determined, Fig.\,\ref{fig:samplesolved} includes 54 entries in total. 

Given that high-order multiples were only found owing to eclipses, it is likely that additional ones are present in the sample.  The FLAMES and OGLE angular resolutions (0.6'' and 0.26'', respectively) at LMC distance amount to separations of the order of tens of thousands of au, and orbital periods extending up to $\log P \lesssim 8\,$[d]. In this range, the high-order multiplicity fraction is expected to be larger than 50\% \citep{Moe2017}.
However, spectroscopy and photometry are not efficient techniques for establishing the multiplicity of stars in the range $\log P \gtrsim 3-4\,$[d], where instead interferometry or direct imaging are needed, both of which are challenging in the LMC. We therefore cannot rule out that higher-order multiples contaminate our results. However, bright companions that do not trace the orbital period tend to result in $K_2$ amplitudes of the order of 0\,\kms~(see, e.g. VFTS~64, Appendix\,\ref{sec:indiv}), and such rare cases are discussed in Appendix\,\ref{sec:indiv}. Relatively faint companions would be interesting to find, but have no implication on our analysis. Hence, it is unlikely that higher-order multiplicity impacts our results.

\begin{table*}[!h]
\centering
\tiny
\caption{Orbital parameters and companion mass ranges for the 51 O-type SB1 binaries considered in this work.}
\resizebox{\textwidth}{!}{\begin{tabular}{llcccccccccl}\hline \hline
VFTS & SpT\,\tablefootmark{a} &  $P_{\rm orb}$\,\tablefootmark{b} & $T_0$\,\tablefootmark{b}  & $e$\,\tablefootmark{b} & $\omega$\,\tablefootmark{b} & $K_1$\,\tablefootmark{b} &  $K_2$\,\tablefootmark{a} & $q$\,\tablefootmark{c} & $P$(BH|co)\tablefootmark{d} & $M_2$\,\tablefootmark{c} & comment \\ 
  
 &  &  d & HJD-2400000 &  & &  \kms &  \kms &   &   & $M_\odot$ &  \\ 
\hline  
64 & O8 II:(f) + (O6~V+O6~V) & 903$\pm$4 & 56349.5$\pm$2.3 & 0.528 $\pm$ 0.010 & 351.3 $\pm$  1.3 & 57.2 $\pm$ 0.6 & - & $0.57^{+0.37}_{-0.35}$ &  - &$38.9^{+8.4}_{-9.4}$ & triple\\ 
64A & O7 Vn: + O7~Vn & 1.327181$\pm$0.000003\,\tablefootmark{a} &56350.633$\pm$0.010\,\tablefootmark{a} &0\,\tablefootmark{e} & 90\,\tablefootmark{f} & 282.0 $\pm$ 36.0\,\tablefootmark{a} & 323 $\pm$ 33 & 0.87$\pm$0.14 &  - &$23.5\pm8.2$ & SB2, contact, ecl.\\ 
73 & O9.5 IV + B: & 150.60$\pm$0.13 & 54940$\pm$4 & 0.203 $\pm$ 0.031 & -12.5 $\pm$  9.0 & 27.0 $\pm$ 0.7 & - & $0.35^{+0.11}_{-0.1}$ &  - &$6.8^{+3.0}_{-2.5}$ & SB2:\\ 
86 & O9.5 III + O8~IV:n & 182.95$\pm$0.14 & 55026.2$\pm$1.3 & 0.514 $\pm$ 0.030 & 344.9 $\pm$  4.2 & 43.0 $\pm$ 8.0\,\tablefootmark{a} & 48 $\pm$ 10 & 0.90$\pm$0.25 &  - &$17.5\pm10.0$ & SB2\\ 
93 & O8.5 V + B0.2:~V & 250.13$\pm$0.33 & 54881$\pm$4 & 0.203 $\pm$ 0.027 & 224.8 $\pm$  8.5 & 12.8 $\pm$ 1.1\,\tablefootmark{a} & 20 $\pm$ 9 & 0.64$\pm$0.29 &  - &$13.5\pm7.0$ & SB2\\ \vspace{0.3cm} 
120 & (O9.2 IV + B) + (O9.5~V + O9.5~V) & - & - & - & - & - & - & - & - & quad.\\ 
120A & B + B0~V & 15.6546$\pm$0.0011\,\tablefootmark{a} &54802.88$\pm$0.18\,\tablefootmark{a} &0.280 $\pm$ 0.015 & 47.8 $\pm$  4.0 & 93.0 $\pm$ 17.0\,\tablefootmark{a} & 92 $\pm$ 13 & 0.99$\pm$0.23 &  - & - & SB2\\ 
120B & O9.5 V + B & 17.4098$\pm$0.0005\,\tablefootmark{a} &54813.02$\pm$0.05\,\tablefootmark{a} &0\,\tablefootmark{e} & 90\,\tablefootmark{f} & 20.0 $\pm$ 7.0\,\tablefootmark{a} & 21 $\pm$ 7 & 0.95$\pm$0.46 &  - &$18.6\pm10.8$ & SB2, ecl.\\ 
171 & O8.5 III:(f) + B1.5:~V & 677.0$\pm$0.8 & 55535.2$\pm$1.8 & 0.555 $\pm$ 0.011 & 77.4 $\pm$  2.1 & 11.8 $\pm$ 0.1 & - & $0.21^{+0.07}_{-0.07}$ &  - &$4.1^{+1.8}_{-1.7}$ & SB2:\\ 
184 & O6.5 Vn + OB: & 32.128$\pm$0.022 & 54873.4$\pm$2.9 & 0.200 $\pm$ 0.075 & 53.1 $\pm$  40.0 & 12.1 $\pm$ 1.1 & - & $0.07^{+0.02}_{-0.03}$ &  - &$2.0^{+0.7}_{-0.7}$ & SB2:\\ \vspace{0.3cm} 
191 & O9.2 V + O9.7:~V: & 358.9$\pm$0.8 & 54972$\pm$19 & 0.220 $\pm$ 0.070 & 236.6 $\pm$  18.9 & 23.0 $\pm$ 2.4 & - & $0.41^{+0.12}_{-0.09}$ &  - &$7.7^{+2.5}_{-2.2}$ & SB2\\ 
201 & O9.7 V  + B1.5:~V & 15.3270$\pm$0.0020 & 54872.51$\pm$0.28 & 0.463 $\pm$ 0.041 & 5.5 $\pm$  3.4 & 79.0 $\pm$ 9.0\,\tablefootmark{a} & 105 $\pm$ 26 & 0.75$\pm$0.21 &  - &$14.0\pm5.5$ & SB2\\ 
225 & B0.7 III & 8.2337$\pm$0.0004 & 54850.25$\pm$0.34 & 0.021 $\pm$ 0.008 & 319.7 $\pm$  13.4 & 29.2 $\pm$ 0.2 & - & $0.15^{+0.05}_{-0.06}$ & 23\% & $2.2^{+1.0}_{-0.8}$ & SB1\\ 
231 & O9.7 V  + B1.5:~V & 7.92911$\pm$0.00022 & 54837.36$\pm$0.10 & 0.406 $\pm$ 0.037 & 200.5 $\pm$  5.4 & 99.0 $\pm$ 11.0\,\tablefootmark{a} & 130 $\pm$ 25 & 0.76$\pm$0.17 &  - &$14.1\pm5.1$ & SB2\\ 
243 & O7 V:(n)((f)) & 10.4031$\pm$0.0004 & 54870.7$\pm$1.5 & 0.017 $\pm$ 0.012 & 66.0 $\pm$  53 & 81.4 $\pm$ 1.3 & - & $0.41^{+0.11}_{-0.1}$ & 100\% & $10.9^{+3.8}_{-3.1}$ & SB1, BH\\ \vspace{0.3cm} 
256 & O7.5 V: + OB: & 246.0$\pm$0.5 & 55065.6$\pm$2.2 & 0.629 $\pm$ 0.024 & 75.1 $\pm$  3.3 & 19.2 $\pm$ 0.6 & - & $0.20^{+0.05}_{-0.06}$ &  - &$4.8^{+1.4}_{-1.4}$ & SB2:\\ 
277 & O9 V + B1.5:~V & 240.42$\pm$0.13 & 54875.2$\pm$0.6 & 0.928 $\pm$ 0.014 & 302.7 $\pm$  4.5 & 63.6 $\pm$ 7.9 & - & $0.36^{+0.11}_{-0.11}$ &  - &$7.2^{+2.9}_{-2.4}$ & SB2\\ 
314 & O9.7 V(n)  + B & 2.550786$\pm$0.000005\,\tablefootmark{a} &54891.407$\pm$0.010\,\tablefootmark{a} &0.166 $\pm$ 0.012 & 248.9 $\pm$  4.7 & 110.8 $\pm$ 1.3 & 232 $\pm$ 16 & 0.48$\pm$0.03 &  - &$8.9\pm2.6$ & SB2, ecl.\\ 
318 & O9.5  V + O9.2~V & 14.0043$\pm$0.0029 & 54878.0$\pm$0.5 & 0.083 $\pm$ 0.044 & -7.5 $\pm$  15.7 & 23.3 $\pm$ 1.1 & 99 $\pm$ 11 & 0.24$\pm$0.03 &  - &$4.6\pm1.6$ & SB2\\ 
329 & O9.5 V(n) + B1:~V: & 7.0491$\pm$0.0004 & 54855.94$\pm$0.09 & 0.439 $\pm$ 0.021 & 334.6 $\pm$  2.9 & 93.0 $\pm$ 8.0\,\tablefootmark{a} & 131 $\pm$ 24 & 0.71$\pm$0.14 &  - &$13.9\pm5.3$ & SB2\\ \vspace{0.3cm} 
332 & O9 III + O9.2~V & 1025$\pm$9 & 55064$\pm$7 & 0.813 $\pm$ 0.057 & 185.4 $\pm$  1.4 & 79.0 $\pm$ 10.0\,\tablefootmark{a} & 46 $\pm$ 13 & 0.58$\pm$0.18 &  - &$11.4\pm6.6$ & SB2:\\ 
333 & O9 II((f)) + O6.5~V: & 980.1$\pm$1.5 & 55336.1$\pm$1.4 & 0.746 $\pm$ 0.003 & 115.55 $\pm$  0.49 & 43.0 $\pm$ 2.5\,\tablefootmark{a} & 34 $\pm$ 14 & 0.79$\pm$0.33 &  - &$16.1\pm11.7$ & SB2\\ 
350 & O8.5 V + O9.5~V & 69.570$\pm$0.005 & 54904.26$\pm$0.20 & 0.351 $\pm$ 0.008 & 93.4 $\pm$  1.5 & 70.3 $\pm$ 3.8\,\tablefootmark{a} & 91 $\pm$ 20 & 0.77$\pm$0.17 &  - &$16.3\pm5.4$ & SB2\\ 
386 & O9 V(n) + B1~V: & 20.451$\pm$0.020\,\tablefootmark{a} &54816.5$\pm$2.8\,\tablefootmark{a} &0.150 $\pm$ 0.050\,\tablefootmark{a} & 165.0 $\pm$  26\,\tablefootmark{a} & 30.8 $\pm$ 2.1\,\tablefootmark{a} & 64 $\pm$ 12  & 0.48$\pm$0.1 &  - &$9.6\pm3.7$ & SB2\\ 
390 & O5.5 V:((fc)) + O9.7: V: & 21.90590$\pm$0.00020\,\tablefootmark{a} &54990.43$\pm$0.12\,\tablefootmark{a} &0.495 $\pm$ 0.017 & 274.1 $\pm$  2.5 & 69.7 $\pm$ 1.3 & 137 $\pm$ 34 & 0.51$\pm$0.13 &  - &$16.7\pm6.7$ & SB2, ecl.\\ \vspace{0.3cm} 
404 & O3.5:   V:((fc)) + O5~V: & 145.76$\pm$0.08 & 54993.2$\pm$0.9 & 0.718 $\pm$ 0.016 & 99.6 $\pm$  2.6 & 98.0 $\pm$ 11.0\,\tablefootmark{a} & 106 $\pm$ 11 & 0.92$\pm$0.14 &  - &$41.5\pm18.4$ & SB2\\ 
409 & O3.5: V:((f)) + B: & 22.1909$\pm$0.0012 & 54876.60$\pm$0.16 & 0.294 $\pm$ 0.012 & 105.2 $\pm$  3.5 & 43.2 $\pm$ 0.7 & - & $0.21^{+0.07}_{-0.07}$ &  - &$9.3^{+4.4}_{-3.3}$ & SB2:\\ 
429 & O7 V: + B1:~V: & 30.0450$\pm$0.0003\,\tablefootmark{a} &54874.33$\pm$0.05\,\tablefootmark{a} &0.559 $\pm$ 0.005 & 22.36 $\pm$  0.72 & 92.4 $\pm$ 0.7 & 143 $\pm$ 27 & 0.65$\pm$0.12 &  - &$17.4\pm6.3$ & SB2, ecl.\\ 
440 & O6: V:(f) + O8~V & 1019$\pm$9 & $54900\pm40$ & 0.277 $\pm$ 0.026 & 160.7 $\pm$  14.2 & 11.7 $\pm$ 0.7 & 29 $\pm$ 18 & 0.40$\pm$0.25 &  - &$12.4\pm8.8$ & SB2\\ 
441 & O9.2 V + B0.5~V & 6.86858$\pm$0.00022 & 54861.19$\pm$0.09 & 0.217 $\pm$ 0.020 & 340.1 $\pm$  5.6 & 73.0 $\pm$ 5.0\,\tablefootmark{a} & 108 $\pm$ 8 & 0.68$\pm$0.07 &  - &$13.2\pm4.5$ & SB2\\ \vspace{0.3cm} 
475 & O9.7 V + B0~V & 4.05424$\pm$0.00012 & 54862.30$\pm$0.06 & 0.573 $\pm$ 0.057 & -0.1 $\pm$  2.6 & 135.0 $\pm$ 33.0\,\tablefootmark{a} & 169 $\pm$ 58 & 0.80$\pm$0.34 &  - &$14.8\pm7.5$ & SB2\\ 
479 & O4.5 V((fc))z + B: & 14.7254$\pm$0.0009 & 54872.85$\pm$0.13 & 0.310 $\pm$ 0.016 & 189.4 $\pm$  2.2 & 73.0 $\pm$ 1.1 & - & $0.34^{+0.1}_{-0.09}$ &  - &$12.9^{+5.2}_{-4.2}$ & SB2:\\ 
481 & O8.5 V + O9.7:~V: & 141.823$\pm$0.009 & 54986.20$\pm$0.16 & 0.929 $\pm$ 0.004 & 37.99 $\pm$  0.74 & 128.3 $\pm$ 5.5 & 303 $\pm$ 50 & 0.42$\pm$0.07 &  - &$8.9\pm2.7$ & SB2\\ 
514 & O9.7 V & 184.92$\pm$0.11 & 54842.3$\pm$1.5 & 0.411 $\pm$ 0.019 & 41.2 $\pm$  2.6 & 22.9 $\pm$ 0.4 & - & $0.29^{+0.08}_{-0.07}$ & 98\% & $5.3^{+1.9}_{-1.7}$ & SB1, BH?\\ 
532 & O3.5: V:((f*)) + B~III & 5.796223$\pm$0.000002\,\tablefootmark{a} &54861.49$\pm$0.07\,\tablefootmark{a} &0.460 $\pm$ 0.032 & 159.0 $\pm$  3.1 & 37.5 $\pm$ 2.0\,\tablefootmark{a} & 102  $\pm$ 32  & 0.37$\pm$0.12 &  - &$16.5\pm8.6$ & SB2, ecl.\\ \vspace{0.3cm} 
603 & O4 III:(fc) + OB: & 1.756777$\pm$0.000024 & 54865.06$\pm$0.12 & 0.107 $\pm$ 0.032 & 139.2 $\pm$  27.1 & 11.4 $\pm$ 0.3 & - & $0.01^{+0.01}_{-0.03}$ &  - &$1.1^{+0.5}_{-0.4}$ & SB2:\\ 
613 & O9  V + O7.5~V & 69.16$\pm$0.04 & 54804$\pm$5 & 0.351 $\pm$ 0.061 & 293.9 $\pm$  25.7 & 66.0 $\pm$ 15.0\,\tablefootmark{a} & 96 $\pm$ 39 & 0.69$\pm$0.32 &  - &$13.7\pm7.8$ & SB2\\ 
619 & O8: V & 14.5043$\pm$0.0026 & 54869.1$\pm$1.7 & 0.085 $\pm$ 0.040 & 161.6 $\pm$  45.0 & 36.8 $\pm$ 1.4 & - & $0.20^{+0.05}_{-0.05}$ & (95\%) & $4.4^{+1.3}_{-1.3}$ & SB1:\\ 
631 & O9.7 V & 5.37487$\pm$0.00018 & 54870.37$\pm$0.07 & 0.007 $\pm$ 0.005 & 37.0 $\pm$  6.6 & 48.6 $\pm$ 1.1 & - & $0.20^{+0.05}_{-0.06}$ & 73\% & $3.6^{+1.2}_{-1.1}$ & SB1\\ 
645 & O9.5 V & 12.5458$\pm$0.0016 & 54870.8$\pm$0.4 & 0.235 $\pm$ 0.070 & -2.8 $\pm$  10.7 & 31.6 $\pm$ 2.5 & - & $0.15^{+0.04}_{-0.06}$ & 56\% & $3.2^{+1.2}_{-0.8}$ & SB1\\ \vspace{0.3cm} 
657 & O7 II:(f) + OB: & 63.466$\pm$0.008 & 54858.7$\pm$0.4 & 0.480 $\pm$ 0.021 & 312.1 $\pm$  3.9 & 44.6 $\pm$ 1.2 & 169 $\pm$ 38 & 0.26$\pm$0.06 &  - &$9.3\pm8.1$ & SB2:\\ 
702 & (O8 V(n) + OB) + (OB+OB) & - & - & - & - & - & - & - & - & quad.\\ 
702A & O8 V(n) + OB & $1.981595 \pm 0.000030$ & $54869.00 \pm 0.02$ &0\,\tablefootmark{e} & 90\,\tablefootmark{f} & 105.9 $\pm$ 3.0 & - & $0.30^{+0.08}_{-0.08}$ &  - &$7.1^{+2.4}_{-1.9}$ & ecl.\\ 
702B & OB + OB & $2.932818 \pm 0.000025$ & $4869.84 \pm 0.07$ &0\,\tablefootmark{e} & 90\,\tablefootmark{f} & - & - & $0.36^{+0.15}_{-0.14}$ &  - & - & ecl.\\ 
733 & O7.5 V + B1~II & 5.922078$\pm$0.000005 & 54871.08$\pm$0.16 & 0.002 $\pm$ 0.001 & 196.4 $\pm$  10.2 & 39.4 $\pm$ 1.9\,\tablefootmark{a} & 102.2 $\pm$ 0.9 & 0.39$\pm$0.02 &  - &$9.6\pm2.3$ & SB2\\ \vspace{0.3cm} 
736 & O9.5 V + B: & 68.800$\pm$0.021 & 54922.1$\pm$2.6 & 0.086 $\pm$ 0.020 & 255.3 $\pm$  13.7 & 24.6 $\pm$ 0.5 & - & $0.23^{+0.07}_{-0.08}$ &  - &$4.7^{+1.8}_{-1.5}$ & SB2:\\ 
743 & O9.5 V((n)) & 14.9473$\pm$0.0009 & 54866.91$\pm$0.31 & 0.012 $\pm$ 0.008 & 42.4 $\pm$  7.9 & 23.4 $\pm$ 0.6 & - & $0.13^{+0.03}_{-0.04}$ & 28\% & $2.5^{+1.0}_{-0.8}$ & SB1\\ 
750 & O9.5 V + B: & 417$\pm$8 & 55247$\pm$10 & 0.779 $\pm$ 0.039 & 44.2 $\pm$  6.1 & 29.5 $\pm$ 1.6 & - & $0.35^{+0.1}_{-0.07}$ &  - &$6.5^{+2.2}_{-2.0}$ & SB2:\\ 
769 & O9.7 V & 2.365644$\pm$0.000016 & 54868.340$\pm$0.030 & 0.007 $\pm$ 0.005 & 333.9 $\pm$  3.5 & 40.8 $\pm$ 0.8 & - & $0.11^{+0.03}_{-0.05}$ & 25\% & $2.3^{+0.9}_{-0.6}$ & SB1\\ 
779 & B1 II-III & 59.945$\pm$0.025 & 54903.3$\pm$2.4 & 0.046 $\pm$ 0.011 & 84.3 $\pm$  12.2 & 30.8 $\pm$ 0.2 & - & $0.35^{+0.13}_{-0.11}$ & 87\% & $4.8^{+2.3}_{-1.8}$ & SB1, BH?\\ \vspace{0.3cm} 
802 & O7 V: + O8~Vn & 181.88$\pm$0.04 & 54942.98$\pm$0.25 & 0.602 $\pm$ 0.007 & 47.18 $\pm$  0.78 & 70.4 $\pm$ 1.9\,\tablefootmark{a} & 76 $\pm$ 14 & 0.93$\pm$0.17 &  - &$25.0\pm9.0$ & SB2\\ 
810 & O9.7 V  + B1~V & 15.6886$\pm$0.0006 & 54892.32$\pm$0.06 & 0.678 $\pm$ 0.008 & 359.0 $\pm$  1.3 & 108.0 $\pm$ 9.0\,\tablefootmark{a} & 162 $\pm$ 27 & 0.67$\pm$0.12 &  - &$12.4\pm4.2$ & SB2\\ 
812 & O4 V((fc)) & 17.28443$\pm$0.00035 & 54856.95$\pm$0.04 & 0.624 $\pm$ 0.009 & 339.5 $\pm$  1.3 & 42.7 $\pm$ 0.6 & - & $0.15^{+0.05}_{-0.06}$ &  - &$6.4^{+2.6}_{-2.1}$ & SB2:\\ 
827 & B1.5 III & 43.221$\pm$0.017 & 54870.7$\pm$0.5 & 0.244 $\pm$ 0.011 & 97.6 $\pm$  3.4 & 25.3 $\pm$ 0.3 & - & $0.23^{+0.08}_{-0.08}$ & (56\%) & $3.1^{+1.7}_{-1.2}$ & SB1:\\ 
829 & B1.5 III & 202.9$\pm$0.9 & 55024$\pm$10 & 0.273 $\pm$ 0.043 & 145.8 $\pm$  11.4 & 12.6 $\pm$ 0.7 & - & $0.20^{+0.07}_{-0.05}$ & (34\%) & $2.5^{+1.3}_{-0.9}$ & SB1:\\ \vspace{0.3cm} 
887 & O9.7: V: + O9.5:~V & 2.672807$\pm$0.000035 & 54870.390$\pm$0.030 & 0.056 $\pm$ 0.019 & 53.5 $\pm$  4.8 & 105.0 $\pm$ 7.0\,\tablefootmark{a} & 98 $\pm$ 12 & 0.93$\pm$0.13 &  - &$17.3\pm5.5$ & SB2\\ 
\hline
\end{tabular}}
\tablefoot{
Provided are VFTS identifiers,  new spectral types from our study, orbital parameters, mass ratios, probability of the system to host a black hole assuming the companion is a compact object, estimated secondary mass, and individual comments. See text for details.
\tablefoottext{a}{Derived in our study}.
\tablefoottext{b}{Derived by \citet{Almeida2017}, unless otherwise stated}.
\tablefoottext{c}{Values and errors represent the median and 68\% confidence intervals of the respective probability density function (see text for details)}.
\tablefoottext{d}{The probability the companion is a black hole assuming it is a compact object, computed as the probability that $M_2 > 3\,M_\odot$ (see Sect.\,\ref{subsec:BHcandidatesDisc}). Probabilities for uncertain SB1 systems (SB1:), rounded to nearest integer, are provided in parentheses, and are omitted for uncertain SB2 systems (SB2:). }.
\tablefoottext{e}{Adopted based on symmetry of light curve};
\tablefoottext{f}{fixed}}
\label{tab:SampleFin}\end{table*}

\section{Discussion}
\label{sec:disc}

\subsection{Mass ratio distribution}
\label{subsec:massratio}




\begin{figure}
   \centering
\begin{tabular}{cc}
\includegraphics[width=0.5\textwidth]{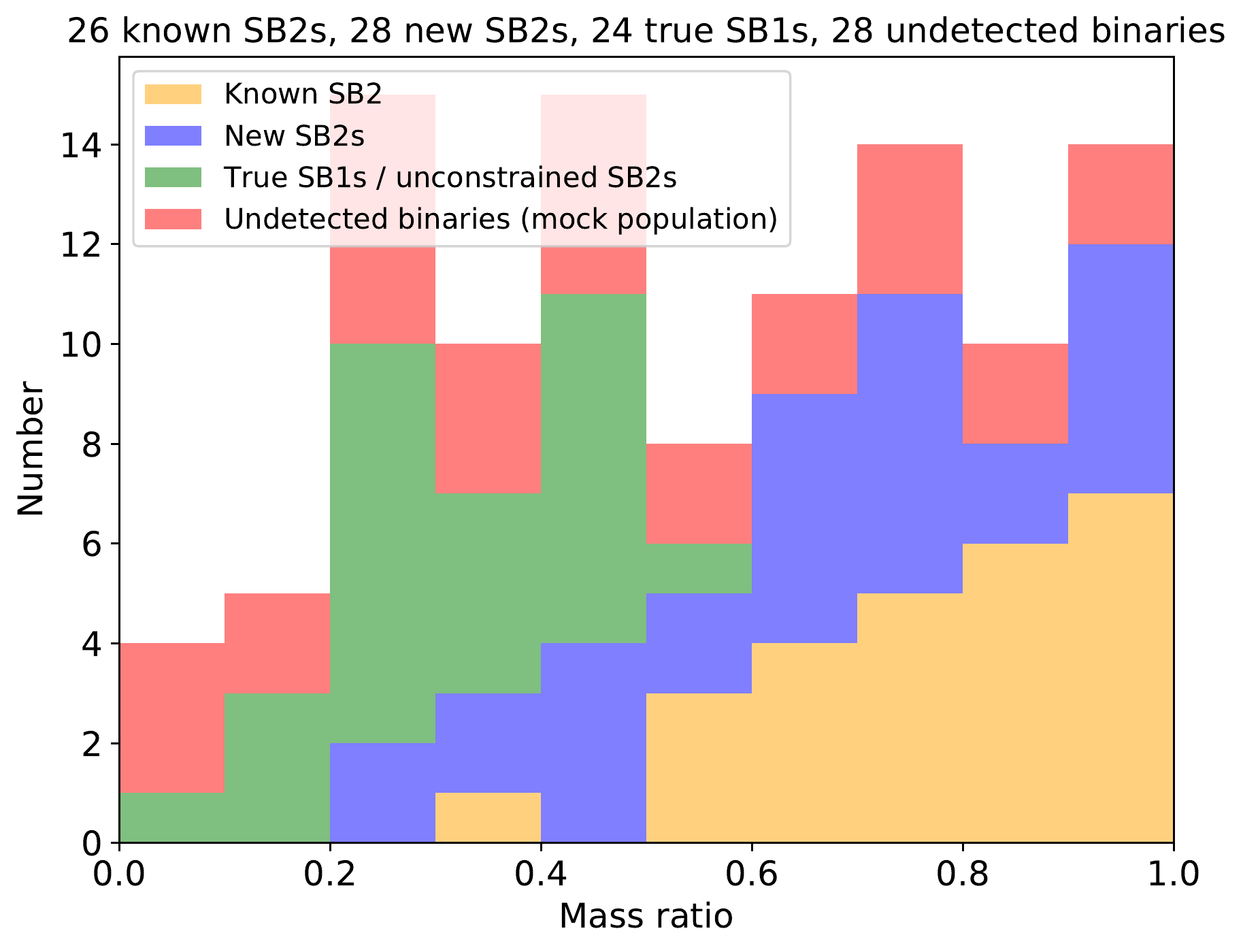}\\
\includegraphics[width=0.47\textwidth]{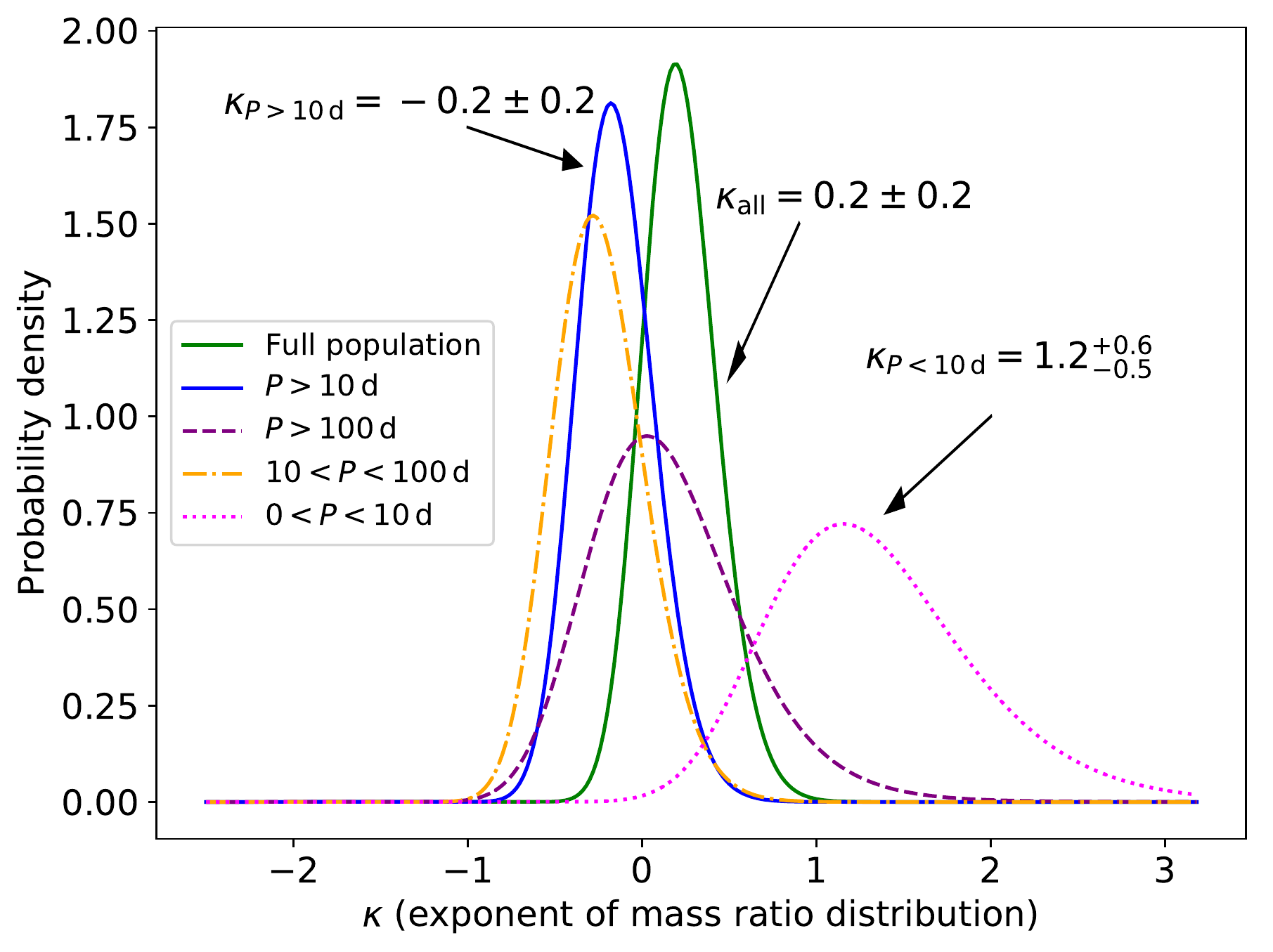}
\end{tabular}
    \caption{Results for the bias-corrected mass-ratio distribution of the O-type binaries in the Tarantula region. Upper panel: histograms of the mass ratio distributions of the known SB2, new SB2, SB1, and hidden binaries (see legend).  Lower panel: corresponding probability densities of the exponent $\kappa$ of the mass ratio distribution ($f(q) \propto q^\kappa$). The analysis is perform for various period cuts (see legend). }
    \label{fig:qhist} 
\end{figure}

In this section, we combine all available constraints on mass ratios coming from the homogeneous VFTS and TMBM surveys to reassess the power index $\kappa$ of the mass ratio distribution, modelled in the form $f(q) \propto q^\kappa$ (with $0 < q \le 1$). 
As discussed in Sect.\,\ref{sec:intro}, 
while SB2 systems allow for direct measurement of $q = K_1/K_2$, they are biased towards short periods, high component masses, and mass ratios close to unity. Disentangling is helpful in this sense, since it extends the sensitivity to wider periods and more extreme mass ratios. Hence, the results become less bias-dominated when disentangling is used. However, even with disentangling, the bias is not fully removed, since a substantial fraction of the SB1 systems for which no companions were retrieved are likely to host  faint non-degenerate companions with even more extreme mass ratios. Finally, given that the binary detection probability becomes low for extreme mass ratios, one can expect a population of binaries with two non-degenerate companions that were classified as presumably single stars in the original VFTS survey \citep{Sana2013}. These undetected binaries need to be accounted for if we are to obtain an unbiased estimate for $\kappa$.

In Fig.\,\ref{fig:qhist}, we show a histogram of the mass ratios measured for SB2 systems in the TMBM sample \citep{Almeida2017} and newly uncovered SB2 systems in this study. We omit VFTS~603 due to its uncertain nature (see Appendix\,\ref{sec:indiv}), and we omit VFTS~243 due to it being identified as an  O+BH system \citep{Shenar2022VFTS243}. We further omit the five SB2 systems classified as post-interaction binaries by \citet{Mahy2020a}, as we are interested in the natal mass ratio distribution. In addition, we show the median $q$ values for the SB1 systems, derived as describe in Sect.\,\ref{subsec:MonteCarlo}. The final population we need to consider  comprises non-degenerate binaries  that were classified as single stars. To visualise this population in the histograms in  Fig.\,\ref{fig:qhist}, we create a mock population of such binaries by using the  binary detection probability function provided by \citet{Sana2013} as a function of $q$, $f_{\rm det}(q)$. We digitise the function, extend it down to $q=0$, where per definition $f_{\rm det}(0) = 0$, and fit a polynomial through the data points (see Appendix\,\ref{sec:DetProb}). For every $q$ bin, the number of undetected binaries is given by $1/f_{\rm det}(q) \times M(q)$, where $M(q)$ is the number of detected binaries in the respective bin.

The distributions match the expected behaviour of the different samples. The known SB2 systems are preferentially found in the range $0.6 < q < 1$, while the new SB2 systems, obtained from disentangling of the SB1 systems in this study, probe more uniformly the range $0.2 < q < 1$. The "true SB1" binaries (which also include the potential SB2 systems for which $K_2$ could not be constrained) cluster around more extreme mass ratios of the order $0 < q < 0.5$, which is reasonable since a companion was not detected in them. Finally, the undetected binaries have the highest relative contribution for the lowest $q$ values, as could be anticipated.



Assuming a power-law in the form $f(q) \propto q^\kappa$, we now estimate $\kappa$ in a Bayesian approach by computing the posterior of $\kappa$ given the set of measurements $\{q_i\}$ as

\begin{equation}
    P(\kappa\,|\,\{q_i \} ) = \frac{P(\{q_i \}\,|\,\kappa ) \cdot P(\kappa)}{P(\{q_i \} )}.
    \label{eq:Bayes}
\end{equation}

We assume a flat prior on $\kappa$, such that $P(\kappa)$ amounts to a normalisation constant,  as does the probability for observing the data ${P(\{q_i \} )}$. We account for the errors on $q_i$ by multiplying the mass-ratio distribution $q^\kappa$ with the probability distributions of each $q_i$, $P(q_i)$. The population of undetected binaries is accounted for via multiplication with the binary detection probability, $f_{\rm det}(q)$. The posterior $P(\kappa\,|\,\{q_i \} )$ is thus given by:

\begin{equation}
    P(\kappa\,|\,\{q_i \} ) = \prod_{i = 1}^{N} \frac{1}{\int_0^1 f_{\rm det}(q)\,q^\kappa \mathrm{d}q} \int_{0}^{1} f_{\rm det}(q)\,q^\kappa\,{\rm P}(q_i)\,\mathrm{d}q,
    \label{eq:kapbias}
\end{equation}
where $i$ runs over the populations of known SB2, new SB2, and SB1 systems. For SB2 binaries, $P(q_i)$ is modelled as a Gaussian using the values given in Table\,\ref{tab:SampleFin}. For SB1 binaries, $P(q_i)$ is directly obtained from our Monte Carlo simulations described above. 

The lower panel of Fig.\,\ref{fig:qhist} shows the results obtained for $P(\kappa | {q_i } )$. The result obtained for the entire sample is $\kappa_{\rm all} = 0.2\pm0.2$, consistent with a uniform distribution ($\kappa=0$). However, we also perform the analysis for various period cuts between 1 and 1\,000\,d to investigate a possible dependence of $\kappa$ on the period interval, as advocated by \citet{Moe2017}. While most period cuts yield consistent results in the range $-0.2 < \kappa < 0.2$, the group of short-period binaries with $P< 10\,$d exhibits a distinctively different distribution, with $\kappa = 1.2^{+0.6}_{-0.5}$, implying that very short period binaries tend to have mass ratios closer to unity. A possible explanation or this result could be that many such short-period binaries at more extreme mass ratios ($q_{\rm ini} \lesssim 0.5$) have already interacted and merged \citep{Soberman1997, Langer2020}, and are therefore not included in our sample. However, it is also possible that the star-formation process at very short orbital separations tends to form equal-mass binaries. This fundamental question should be explored in detail in the future.

\begin{figure}
   \centering   
    \includegraphics[width=0.5\textwidth]{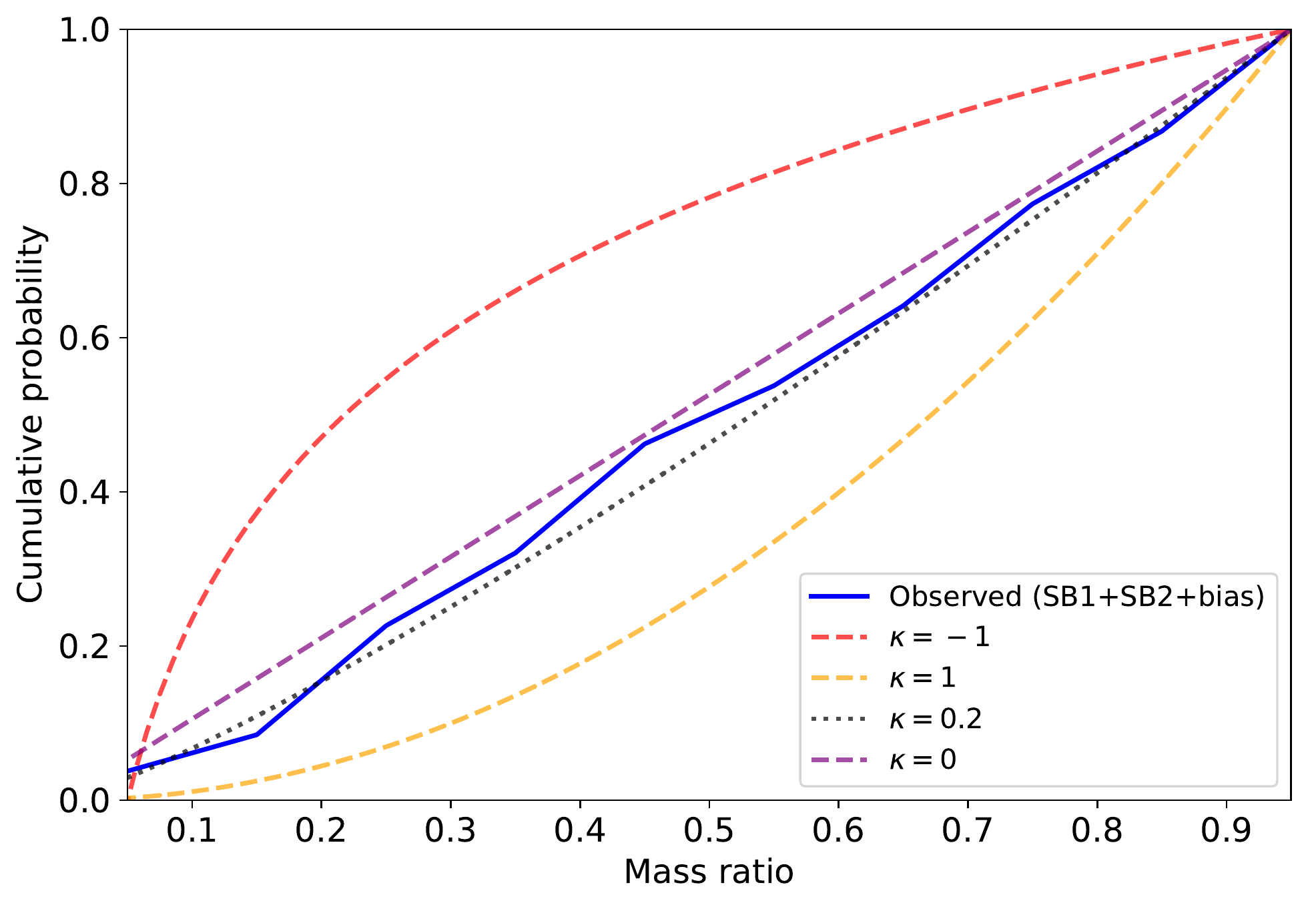}
    \caption{Cumulative distribution of the mass ratios measured for the SB2 and SB1 population, including bias-correction due to the finite binary detection probability (see text for details), defined in the bins 0.05, 0.15, ..., 0.95 (each representing an interval of $\Delta q = 0.1$). The distribution is compared to cumulative functions obtained for $\kappa = -1, 0, 0.2,$ and $1$, defined  in the interval $0.0 \le q \le 0.95$, except for the $\kappa=-1$ case, which is defined for $0.05 \le q \le 0.95$ to avoid a singularity. Evidently, the observed distribution closely follows a flat mass-ratio $\kappa=0$ (uniform).  }
    \label{fig:cumdist} 
\end{figure}

The unbiased $\kappa$ values derived here are consistent with a uniform mass ratio distribution ($\kappa = 0$). This is also evident when comparing the observed cumulative distribution of the sample with that of power laws for different $\kappa$ values (Fig.\,\ref{fig:cumdist}). 
A flat mass-ratio distribution is in contrast with the value of $\kappa=-1.0\pm0.4$ provided by \citet{Sana2013} using the same sample. However, \citet{Sana2013} only had access to RV variability information. A glance at their figure 6 reveals that RV diagnostics are poorly sensitive to $\kappa$. Hence, the discrepancy could be easily masked by statistical and systematic biases. On the other hand, a flat mass ratio distribution is more in line with the findings of \citet{Sana2012} for Galactic O-type stars ($\kappa = -0.1\pm0.6$), and also better matches results for Galactic O-types stars by \citet{Kobulnicky2007}. 

It is possible that a few of the SB1 binaries considered here host neutron stars or black holes, and should have been omitted from the sample of SB1 binaries.  In Sect.\,\ref{subsec:BHcandidatesDisc}, we estimate that $\approx 4-5$ OB+BH are expected to be present in our sample. Such systems have typical mass ratios of the order of 0.5-1 \citep{Langer2020}. Having omitted the confirmed O+BH binary VFTS~243 \citep{Shenar2022VFTS243}, and given our sample size, the impact of a few  O+BH binaries is likely negligible.  While OB+NS binaries would strongly skew the distribution towards extreme mass ratios, such binaries are unlikely to be part of our sample. The RV amplitudes of OB+NS binaries are typically of the order of 10-20\,\kms~or less. Hence, such binaries are very unlikely to have passed the threshold of $\Delta {\rm RV} > 20$\,\kms~imposed by the VFTS survey to be identified as spectroscopic binaries.

\subsection{The OB+BH population}
\label{subsec:BHcandidatesDisc}

Among our sample, only one system could be robustly shown to be  an O+BH binary: VFTS~243 \citep{Shenar2022VFTS243}. In principle, all systems for which a non-degenerate companion could not be found are potential OB+BH candidates. However, not at the same likelihood. The likelihood that a binary in the sample hosts a BH can be written as \mbox{$P({\rm BH}) = P({\rm BH|co}) \cdot P({\rm co})$}, where $P({\rm co})$ is the probability that the companion is a compact object, and $P({\rm BH|co})$ is the probability that the companion is a BH assuming it is a compact object. The latter is relatively easy to compute. The mass of the most massive neutron star known slightly exceeds $2\,M_\odot$ \citep{Fonseca2021}, and neutron stars are not expected to exist above $3\,M_\odot$ given constraints from their equation of state \citep{Oppenheimer1939, Akmal1998, Lattimer2004, Oertel2017} . While the bulk of known BHs in the Galactic neighbourhood exceed $5\,M_\odot$ \citep{Oezel2010, Farr2011, Corral-Santana2016}, the existence of lower-mass BHs  is still debated \citep{Kreidberg2012, Wyrzykowski2020, Lam2022}. Hence, to reasonable approximation, $ P({\rm BH|co}) \approx P3 \coloneqq P(M_2 > 3\,M_\odot)$. 
In Sect.\,\ref{subsec:MonteCarlo}, we outlined our Monte Carlo method for establishing probability distributions for $M_2$ and $q$. We can use the  probability densities $P(M_2)$ to estimate $P3$.

Less trivial is the computation of $P({\rm co})$, which is the complementary probability of the companion being a non-degenerate companion (assuming it is a single star). In principle, one can perform simulations to narrow down the possible types of companions (e.g. faint main sequence stars, helium stars; see \citealt{Shenar2022VFTS243} and \citealt{Mahy2022BH}). One would then need to combine this with assumptions regarding the likelihood of these companions, which involves the initial mass function and binary evolution. Given the non-trivial uncertainties and complications involved, we simply consider $P3$ as a proxy for $P({\rm BH})$. These probabilities are given in Table\,\ref{tab:SampleFin}. For uncertain SB2 systems (SB2:), we provide the values in parentheses, since $P({\rm co})$ is systematically lower in this case.

 Aside from the confirmed O+BH system VFTS~243, two SB1 systems approach or exceed a probability of $ 90$\%: VFTS~514 and 779. We mark these systems as prime OB+BH candidates, noting that non-degenerate companions could only be fully excluded  in VFTS~243. It is also possible that BHs hide among the other five SB1 binaries, but at a lower probability. 
 The 15 uncertain SB2 or SB1 binaries (SB1:, SB2:) could also host BHs -- especially those with high $P3$ probabilities. More data will be needed to clarify this question (see below).

We now estimate the number of OB+BH binaries expected to lurk among the 51 SB1 systems considered here.  According to \citet{Langer2020}, about 2\% of an unbiased population of massive OB-type binaries are expected to host BH companions\footnote{\citet{Langer2020} predicted 3\% for an assumed initial binary fraction of 100\%. Assuming  50-70\% instead \citep{Sana2012, Sana2013} reduces this fraction to $\approx 2\%$.}. There are 360 O-type stars\footnote{The few evolved B-type stars included in the sample are statistically negligible here.} included in the VFTS survey, and hence we expect roughly seven O+BH binaries among them. With typical RV semi-amplitudes in excess of $50\,$\kms, these O+BH binaries are all expected to have been flagged as binaries in the VFTS survey. The TMBM follow-up targeted 69\% of the flagged binaries \citep{Almeida2017}, hence there should be 4-5 O+BH binaries in the TMBM sample. Naturally, none of those would have been classified as SB2 by \citet{Almeida2017}, and so we can conclude that these 4-5 O+BH binaries would be present in our SB1 sample (8-10\%). 
With three promising OB+BH candidates (one of which is unambiguous) and 20 less likely candidates,  our results  seem qualitatively in line with  predictions. However, we refrain from over-interpreting this result until  the OB+BH nature of the remaining candidates can be verified.

It is interesting to note that the orbit of the confirmed O+BH binary VFTS~243 is near-circular, implying negligible kick or ejecta for the BH progenitor \citep{Shenar2022VFTS243}. This is in contrast to the recent report of a BH in the  eccentric ($e\approx 0.5$) binary \object{HD~130298}, which otherwise shares a similar orbital period and spectral type to VFTS~243 \citep{Mahy2022BH}.
The orbit of the OB+BH candidate VFTS~779 is near-circular, while VFTS~514 is eccentric. If all of these systems are truly O+BH binaries, this may give evidence that some BHs are born with strong kicks, while others are not, in line with recent predictions \citep[e.g.][]{Fryer2012}.

Future high-resolution spectroscopy should significantly improve the detection thresholds, since it should allow for a better removal of the nebular lines and thus make the Balmer lines usable. The Balmer lines are key diagnostics to reject the presence of companions contributing as little as $\approx 0.5\%$ to the flux.  Moreover, high resolution is vital for binaries with RV amplitudes comparable or smaller to the resolution element of FLAMES ($K_1 \lesssim 40\,$\kms). Additionally, UV spectroscopy should help reject the presence of non-degenerate companions through dilution of P-Cygni lines \cite[e.g.][]{Georgiev2011, Shenar2016}.





\section{Conclusions}
\label{sec:conclusions}

In this study, we characterised the nature of the hidden companions in 51 SB1 O-type and evolved B-type binaries identified in the framework of the Tarantula Massive Binary Monitoring (TMBM) campaign \citep{Almeida2017}. We implemented the shift-and-add disentangling algorithm to separate the component spectra and establish the RV amplitudes of the two components, whenever possible.  We also investigated OGLE lightcurves in search for eclipses or other orbital modulations. Below, we summarise our main results:

\begin{enumerate}
    \item Out of the 51 SB1 binaries, 43 are found to host non-degenerate companions: 28 are considered certain, and 15 are considered less certain. We also find one triple (VFTS~64) and two quadruple (VFTS~120 and VFTS~702) systems,  and eight eclipsing binaries. The remaining eight targets  retain their SB1 classifications. 
    
    \item Combining constraints from our sample with previously known SB2 binaries \citep{Almeida2017}, and accounting for detection bias, we derive $\kappa =0.2\pm0.2$  for the exponent of the mass-ratio distribution $f(q) \propto q^\kappa$ in the range $0.05 < q < 1$ and $1 < P < 1000\,$d. This drops to $\kappa = -0.2\pm0.2$ if binaries with periods shorter than 10\,d (potential post-interaction products) are removed, and increases to $\kappa = 1.2\pm0.5$ for binaries with $P<10\,$d, which we propose is a result of binary interaction among the tight binaries. Our results are therefore consistent with  a flat natal mass-ratio distribution for O-type stars at LMC metallicity. 
    
    \item Through a probabilistic approach, we identify three O+BH candidates: VFTS~243, 514, 779, of which VFTS~243 is confirmed as an O+BH binary \citep{Shenar2022VFTS243}. The other five SB1 binaries (VFTS~225, 631, 645, 743, 769) and  15 uncertain SB1/SB2 binaries (VFTS~73, 171, 184, 256, 332, 409, 479, 603, 619, 657, 736, 750, 812, 827, 829) could also be O+BH binaries, but at a lower probability.
    
\end{enumerate}

We strongly encourage further investigations of the various important sub-samples uncovered in this study: the OB+BH candidates, the remaining SB1 binaries (including uncertain SB2), the eclipsing binaries, and the higher-order multiples.
The VFTS and TMBM samples offer unparalleled laboratories to homogeneously study binary interaction and evolution at subsolar metallicity. Step by step, the nature of each and every star in this  population is illuminated, providing indispensable constraints on evolution models of massive single, binary, and multiple stellar systems at subsolar metallicity.

\section*{Acknowledgments}
This research has received funding from the European Research Council (ERC) under the European Union's Horizon 2020 research and innovation programme (grant agreement number 772225: MULTIPLES). TS acknowledges support from the European Union's Horizon 2020 under the Marie Sk\l{}odowska-Curie grant agreement No 101024605.  This work is based on observations collected at the European Southern Observatory under programme IDs 090.D-0323 and 092.D-0136. 
L.M.\ thanks the European Space Agency (ESA) and the Belgian Federal Science Policy Office (BELSPO) for their support in the framework of the PRODEX Programme. PAC acknowledges support from the UK Science and Technology Facilities Council research grant ST/V000853/1.  AH acknowledges support by the Spanish MCI through grant PGC-2018-0913741-B-C22 and the Severo Ochoa programme through CEX2019-000920-S.  JMA acknowledges support from the Spanish Government Ministerio de Ciencia, Innovaci\'on y Universidades through grant PGC2018-095\,049-B-C22. PM acknowledges support from the FWO junior postdoctoral fellowship No. 12ZY520N. This work has received funding from the European Research Council (ERC) under the European Union's Horizon 2020 research and innovation programme (Grant agreement No.\ 945806).  ST acknowledges support from the Netherlands Research Council NWO (VENI 639.041.645, VIDI 203.061 grants). MG is supported by the EU Horizon 2020 research and innovation programme under grant agreement No 101004719.

\bibliographystyle{aa}
\bibliography{papers}

\begin{appendix}

\section{The applicability regime of the shift-and-add method}\label{sec:appregg}

In the context of our study, disentangling comprises two main steps: the retrieval of the RV amplitudes ($K_1, K_2$), and the retrieval of the disentangled spectra. 
The robustness of the disentangling procedure depends on many factors, including 
the individual S/N, Doppler-space coverage and number of observations, line profiles, light ratios, and RV amplitudes. 

The underlying assumption of spectral disentangling, as implemented here, is that each observed spectrum is the sum of stellar spectra (two in the case of a binary) that are Doppler-shifted with the respective RVs of the stars, contaminated only by Gaussian noise and static nebular lines that can vary in strength from epoch to epoch (Sect.\,\ref{subsec:neblines}). 
Hence, intrinsic non-Doppler variability can impact the results, especially if it is (quasi-)periodic. In the case of O-type stars, the most important source of contamination is likely stellar pulsations, which can result in apparent peak-to-peak RV variability of up to $\approx 20\,$\kms~\citep{Ritchie2009, Aerts2009, Sana2013}. 
Since all binaries in our sample have $K_1 > 10\,$\kms~per definition, we expect intrinsic variability to play a secondary role  for most targets. Moreover, we note that the procedure of disentangling tends to smear-out features that do not follow the orbital period. We therefore do not expect pulsations to impact the qualitative or quantitative results significantly. An exception is when the dominant pulsation period is mistaken for the orbital period, or when they are otherwise comparable.  A notable example is VFTS~603, which has a derived RV semi-amplitude of $K_1 = 11.4\,$\kms~and a period of $P=1.76\,$d.  The low-amplitude motion, combined with the short period, raises doubts that pulsations cause the underlying variability. Disentangling   does suggest the presence of two non-degenerate stars, but it is possible that the second source is a spurious result of non-Doppler variability (see Appendix\,\ref{sec:indiv}). 

Fully mapping the impact of all parameters and line-profile variability on the output of spectral disentangling is beyond the scope of our study, and would require a dedicated study with prespecified models of line variability. Critical for our study is to establish threshold light ratios below which the spectrum of a non-degenerate secondary would be discerned from the noise.  To estimate it, and to test the applicability regime of our method, we run a set of simulations of synthetic noisy composite spectra that mimic the resolving power, S/N, and timing of the TMBM programme. Specifically, we consider a hypothetical binary with $P=100\,$d, a circular orbit, and a mass ratio of 0.5 (i.e. $K_2 = 2\,K_1$). We use synthetic TLUSTY spectra \citep{Lanz2003, Lanz2007, Hubeny1995} for the primary and secondary. For the primary, we adopt $T_1 = 35\,$kK, $\log g = 4.0\,$[cgs], and $\varv \sin i = 200\,$\kms. For the secondary, we assume $T_2 = 26\,$kK, $\log g = 4.0\,$[cgs], and $\varv \sin i = 100$\,\kms. 
We mimic the data quality of VFTS 231, which has varying S/N for the individual spectra between $\approx$ 10 and 40, with an average of $\approx 30$. VFTS 231 is among the 10\% worst-quality targets in terms of S/N, and is chosen to ensure that our estimates are conservative.

\begin{figure*}[!h]
\centering
\includegraphics[width=.99\textwidth]{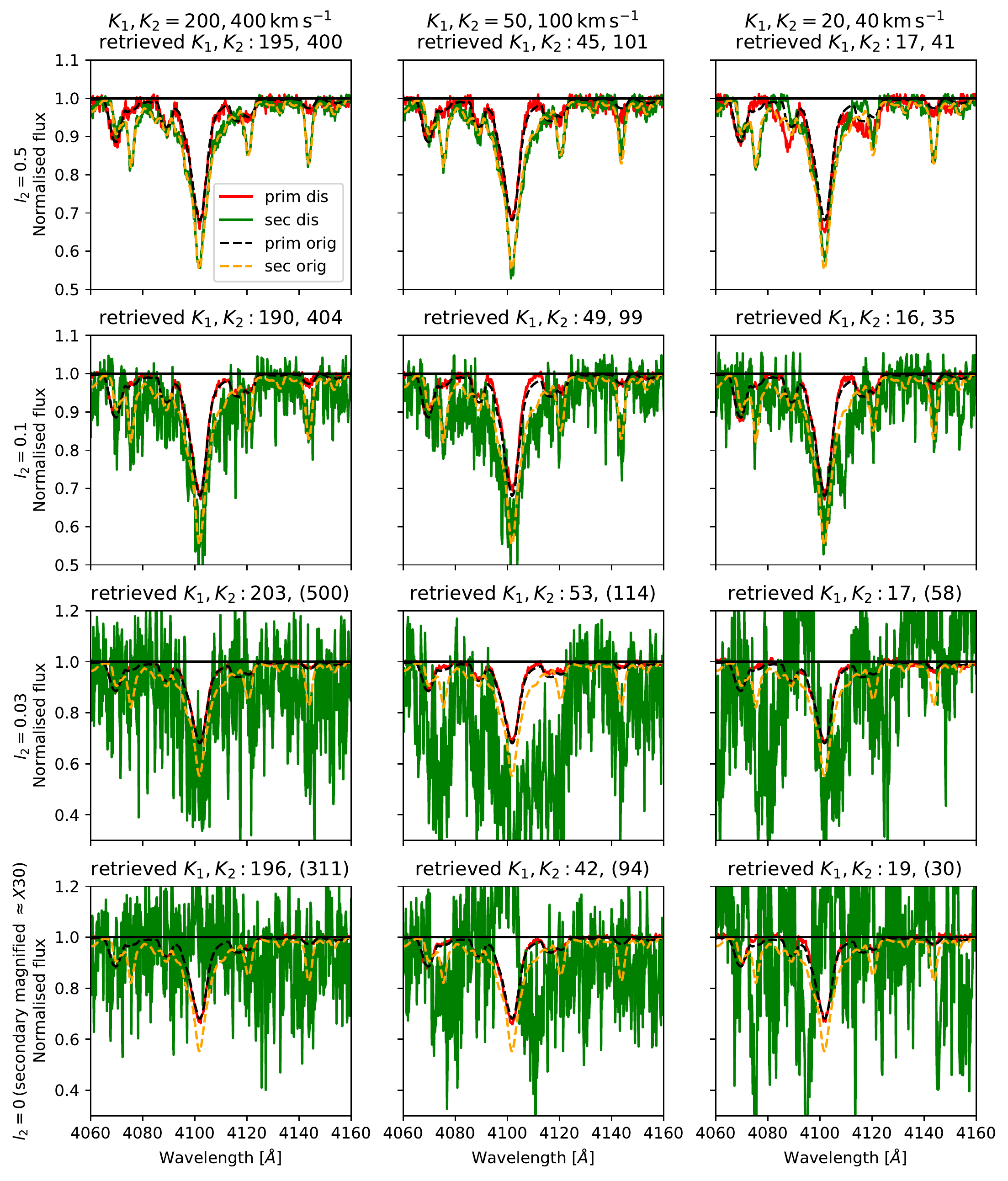}
\caption{Disentangled spectra obtained from the 12 mock binaries, spanning three $K_1, K_2$ (left to right, input values given in uppermost row) sets and four secondary light contributions (top to bottom, input values given in leftmost column). The bottom row ($l_2 = 0$) corresponds to BH companions. The $K_1, K_2$ values derived from grid disentangling are shown in the header of each panel, with values in parentheses denoting results from poorly constrained solutions. The disentangled spectra are scaled by their respective light ratios, except for the bottom row, where the scaling assumes $l_2 = 3\%$ (factor 33 amplification).  } \label{fig:Sims}
\end{figure*}

We constructed 12 mock binaries in total, spanning the values $K_1 = 200, 50, $ and 20\,\kms~(and hence $K_2 = 400, 100, $ and 40\,\kms), and secondary light contributions of $l_2 = f_2/f_{\rm tot}(V)= 0.5, 0.1, 0.03$, and 0 (the latter corresponding to a black hole secondary).
We then apply the 2D shift-and-add disentangling technique to derive $K_1, K_2$ and disentangle the spectra using those values. The results are shown in Fig.\,\ref{fig:Sims}. Evidently, the retrieval procedure works well across the parameter space, although discrepancies become apparent for lower RV amplitudes and fainter secondaries. Figure.\,\ref{fig:Sims} illustrates that, at the low data quality of the simulations, companions contributing as little as $\approx 5\%$ can be retrieved. This improves to $1-3\%$ for higher S/N data available for some of our targets.

\section{Comments on individual targets}\label{sec:indiv}

This appendix provides a detailed commentary on the individual targets and the nature of the unidentified secondaries. We also discuss the light curves of the targets (when available). Unless stated otherwise, cited periods and eccentricities  are taken from \citet{Almeida2017}.

\begin{figure}
\centering
\includegraphics[width=.5\textwidth]{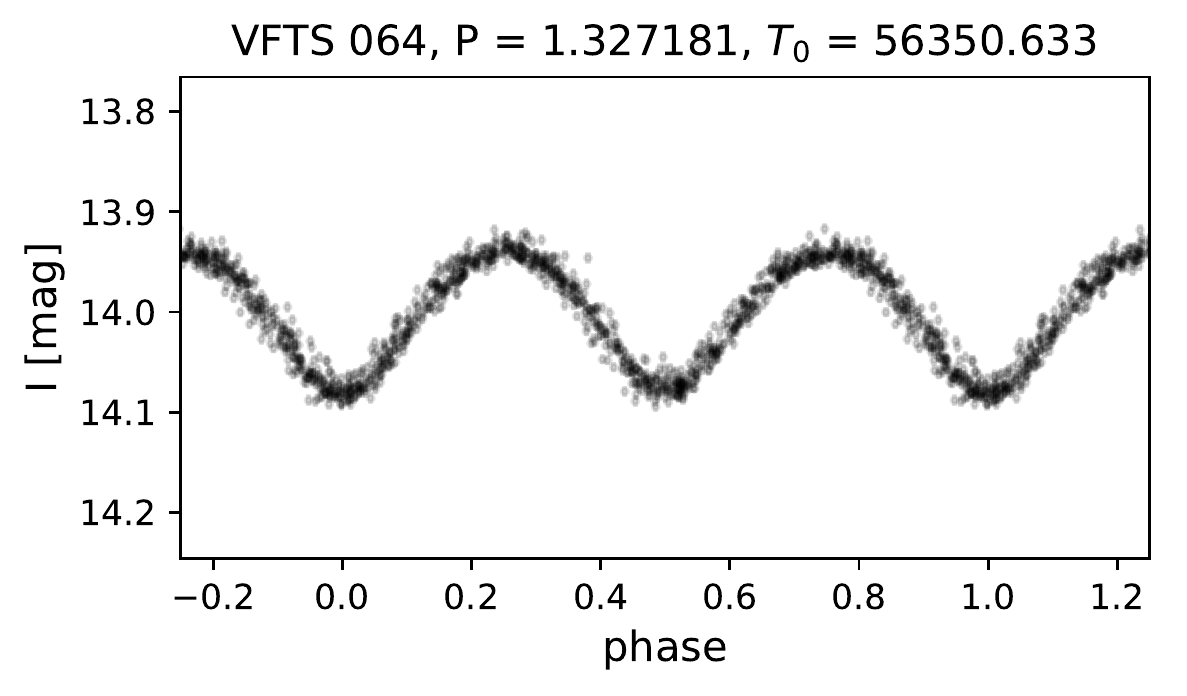}
\caption{OGLE I-band light curve of  the triple system \object{VFTS~64}, phased with the derived period of the inner binary.} \label{fig:VFTS64OGLE}
\end{figure}

\begin{figure}
\centering
\includegraphics[width=.5\textwidth]{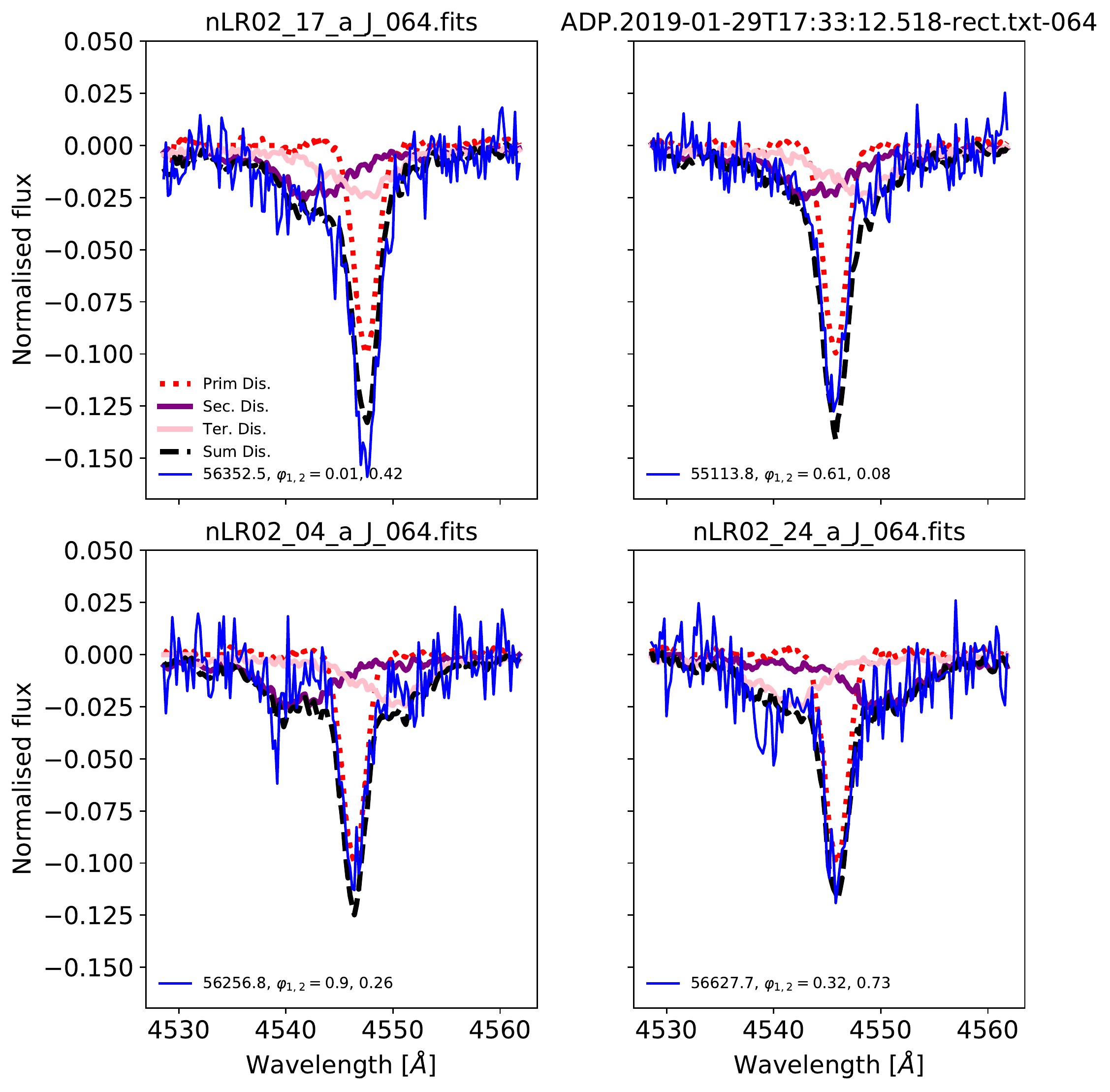}
\caption{Comparison of \HeI4471 spectra of the triple \object{VFTS~64} at RV extremes, along with the disentangled spectra and their sum, as derived for  $K_1 = 57, K_3 = 261, K_4 = 27\,$\kms, where $K_2$ is the RV amplitude of the eclipsing binary as a whole, and equals to $K_2 = 35$\,\kms.} \label{fig:VFTS064EXT}
\end{figure}

{\bf  \object{VFTS~64}, O8~II:(f) + (O7~V + O7~V), } was reported to be a long-period, eccentric binary with $P= 903\,$d and $e = 0.53$. Among our sample, it has the largest mass function, with $f(m) = 10.7 \pm 0.4\,M_\odot$. Spectral disentangling unambiguously shows the presence of a companion. However, disentangling of various He lines results in unrealistically small $K_2$ values, and the disentangled spectra of the secondary appear to centre  around a different systemic velocity than that derived by \citet{Almeida2017}. This suggests that this system may be a multiple.

Available OGLE photometry confirms the triple nature of the system.  Figure\,\ref{fig:VFTS64OGLE} shows the phased I-band light-curve of VFTS~64. A period of
$P=1.327181 \pm 0.000003$ is found from the light-curve analysis. This period is clearly different than the one derived from RVs ($903\,$d).  The light curve suggests that the secondary is in fact an eclipsing contact system consisting of two nearly-identical stars, such that the system is a triple with an outer period of $P_{\rm out} = 903\,$d and an inner period of $P_{\rm in} = 1.33\,$d. 

We extended the 2D disentangling technique to account for three components. Denoting with the subscript '2' the contact binary as a whole, we now need to constrain the RV amplitudes $K_3$ and $K_4$ of the components of the contact binary. We fix the orbital parameters derived by \citet{Almeida2017} for the primary, including its RV amplitude of $K_1 = 57.2\,$\kms. The period and time of primary eclipse $P_{\rm in}$, $T_{\rm 0, in}$ are fixed to those found from the light curve analysis (Fig.\,\ref{fig:VFTS64OGLE}). Given the properties of the light curve, we adopt $e_{\rm in} = 0$ and $\omega_{\rm in} = 90^\circ$, such that $T_{\rm 0, in}$ corresponds to zero RV shift.  Assuming that the inclinations of the inner and outer orbits are aligned given the relatively low separation between the binary and the tertiary,  it can be shown from  Newtonian relations that the RV amplitude of the contact binary itself, $K_2$, is obtained via

\begin{equation}
    K_2 = \left( \frac{P_2}{P_1}\,\left(\frac{1-e_2^2}{1-e_1^2}\right)^{3/2}\,\frac{ (K_3 + K_4)^3}{K_1} \right)^{1/2} - K_1.
\label{eq:K2const}
\end{equation}

We note that by requiring $K_2 > 0\,$\kms, Eq.\,(\ref{eq:K2const}) imposes the condition $K_3 + K_4 \gtrsim 420\,$\kms. We then measured $K_3$ and $K_4$ by disentangling  the strong He\,{\sc ii} and \HeI4471 lines (e.g. Fig.\,\ref{fig:VFTS064EXT}), whose weighted mean yields $K_3 = 282 \pm 36\,$\kms, $K_4 = 323\pm 33$\,\kms, yielding in turn $K_2 = 41 \pm 13$. This unique system could be a prototypical progenitor of the triple WR system \object{BAT99~126}, which was also found to host a contact binary \citep{Janssens2021}. 


\begin{figure}
\centering
\includegraphics[width=.5\textwidth]{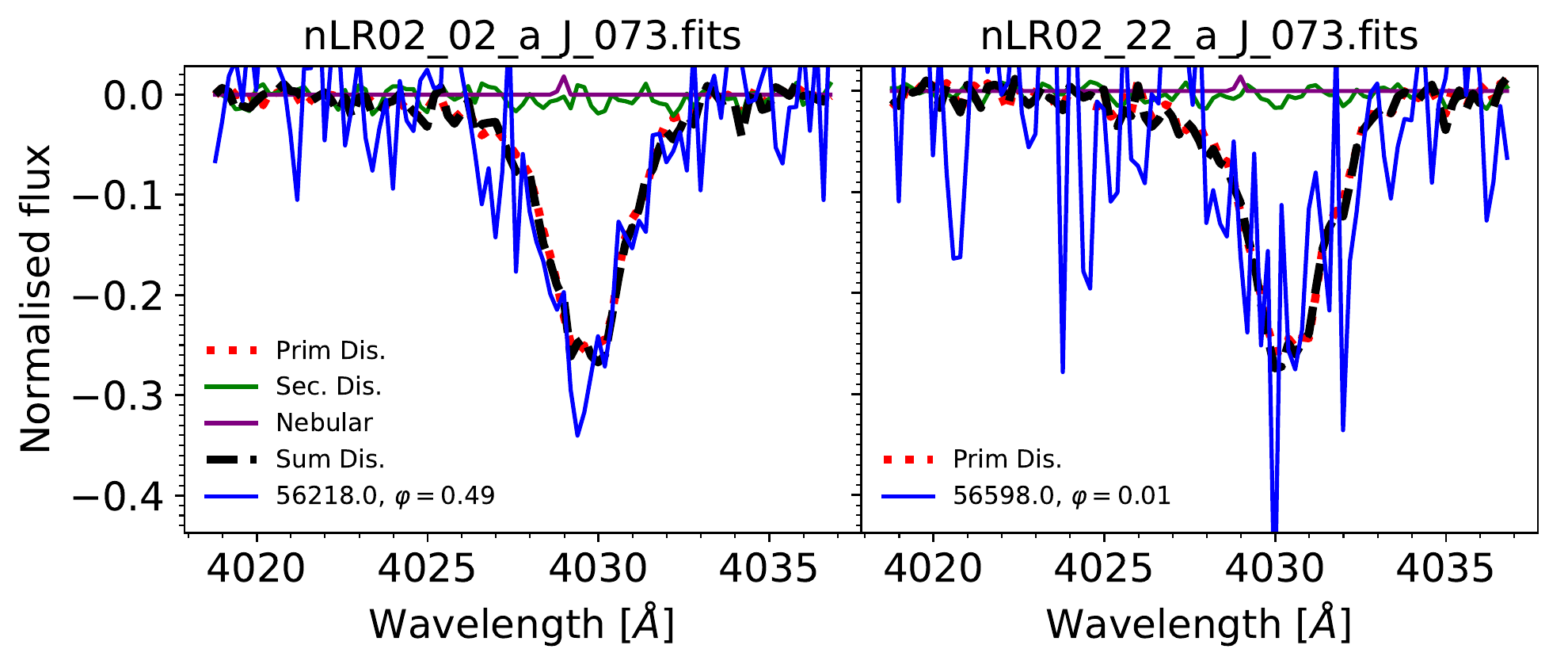}
\caption{Comparison of \HeI4026 spectra of \object{VFTS~73} at RV extremes, along with the disentangled spectra and their sum, as derived for  $K_1 = 27\,$\kms~and $K_2  = 3 \times K_1 = 81\,$\kms. The disentangled spectra are not scaled by the light ratio in this plot.} \label{fig:VFTS73EXT}
\end{figure}

{\bf  \object{VFTS~73}, O9.5~IV + B:} has a reported period of $P=151\,$d and an eccentricity of $e=0.20$. The spectral variability does not readily suggest the presence of a non-degenerate companion (Fig.\,\ref{fig:VFTS73EXT}).  With a mass function of $f(m) = 0.29\,M_\odot$ and an estimated primary mass of $19\,M_\odot$, the companion weighs at least $5.6\pm0.8\,M_\odot$. Disentangling of the spectra does not yield a unique $K_2$ value. When adopting plausible values of the order of 70\,\kms, very faint spectral features appear in the secondary's disentangled spectra, primarily in Balmer lines and strong He\,{\sc i} lines. However, they could be the result of nebular contamination.  We tentatively classify this binary as SB2: with a B-type companion and an assumed a 5\% light contribution, but we cannot rule out a BH. Analysis of the OGLE light curve did not reveal significant periods. 

\begin{figure}
\centering
\includegraphics[width=.5\textwidth]{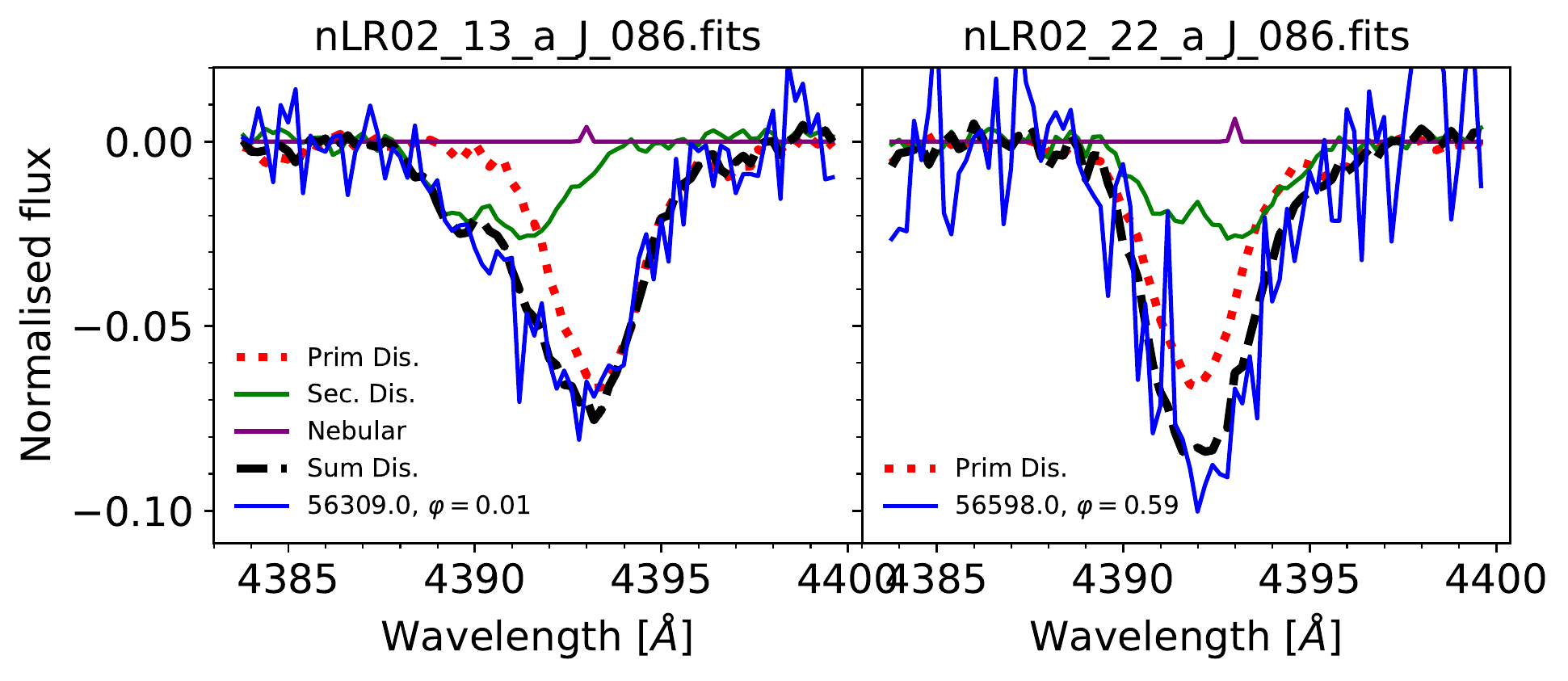}
\caption{Comparison of \HeI4388 spectra of \object{VFTS~86} at RV extremes, along with the disentangled spectra and their sum, as derived for  $K_1 = 47\,$\kms~and $K_2  = 57$\,\kms. The disentangled spectra are not scaled by the light ratio in this plot.} \label{fig:VFTS86EXT}
\end{figure}

{\bf  \object{VFTS~86}, O9.5~III + O8 IV:n} has a reported period of $P=183\,$d and an eccentricity of $e=0.51$. Inspection of the line profiles is suggestive of the presence of two non-degenerate stars in the system, both of which appear to contribute a significant amount to the visual flux (Fig.\,\ref{fig:VFTS86EXT}). For this reason,  we resort to 2D disentangling. We used the strong He\,{\sc i} lines (which suffer very little nebular contamination) and the two He\,{\sc ii} lines. The lines yield broadly consistent results, albeit with large errors. The smallest errors are obtained from the \HeI4388~line, yielding $K_1 = 43\pm8\,$\kms~and $K_2 = 48\pm10$\,\kms. With a derived light contribution of 44\%,   the secondary appears to be an O8 star, that is, of earlier spectral type than the primary. Analysis of the OGLE light curve did not reveal significant frequencies.


\begin{figure}
\centering
\includegraphics[width=.5\textwidth]{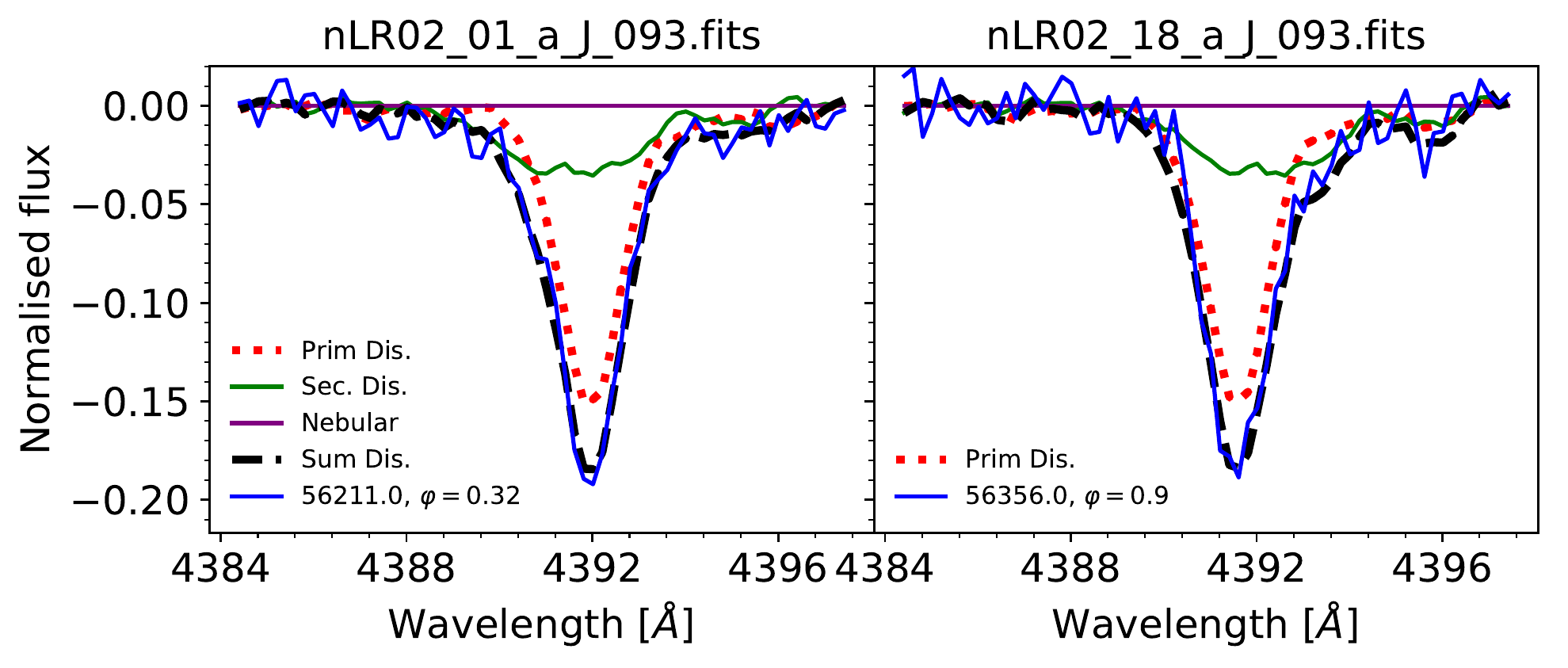}
\caption{Comparison of \HeI4388 spectra of \object{VFTS~93} at RV extremes, along with the disentangled spectra and their sum, as derived for  $K_1 = 13\,$\kms~and $K_2  = 20$\,\kms. The disentangled spectra are not scaled by the light ratio in this plot.} \label{fig:VFTS93EXT}
\end{figure}

{\bf  \object{VFTS~93}, O8.5~V + B0.2:~V} has a reported period of $P=250\,$d and an eccentricity of $e=0.20$. The primary moves with a very low RV amplitude of $K_1 = 11\,$\kms, making it difficult to properly disentangle the spectra. However, careful inspection of the line-profile variability of strong He\,{\sc i} lines at extremes suggests the presence of a non-degenerate companion (Fig.\,\ref{fig:VFTS93EXT}). We re-derived $K_1$ by measuring the RVs of the He\,{\sc ii} lines, which are not contaminated by the secondary, and obtain $K_1 = 12.8 \pm 1.1\,$\kms. Fixing this value, 
the strong He\,{\sc i}~lines all result in minima of the order of $K_2 = 10-30\,$\kms. A weighted mean of all measurements yields $K_2 = 20\pm9$\,\kms. The spectrum is dominated by He\,{\sc i} absorption, and best corresponds to a B0.2 spectral type, with an estimated light contribution of 21\%.  
Analysis of the OGLE light curve did not reveal significant frequencies. 



\begin{figure}
\centering
\includegraphics[width=.5\textwidth]{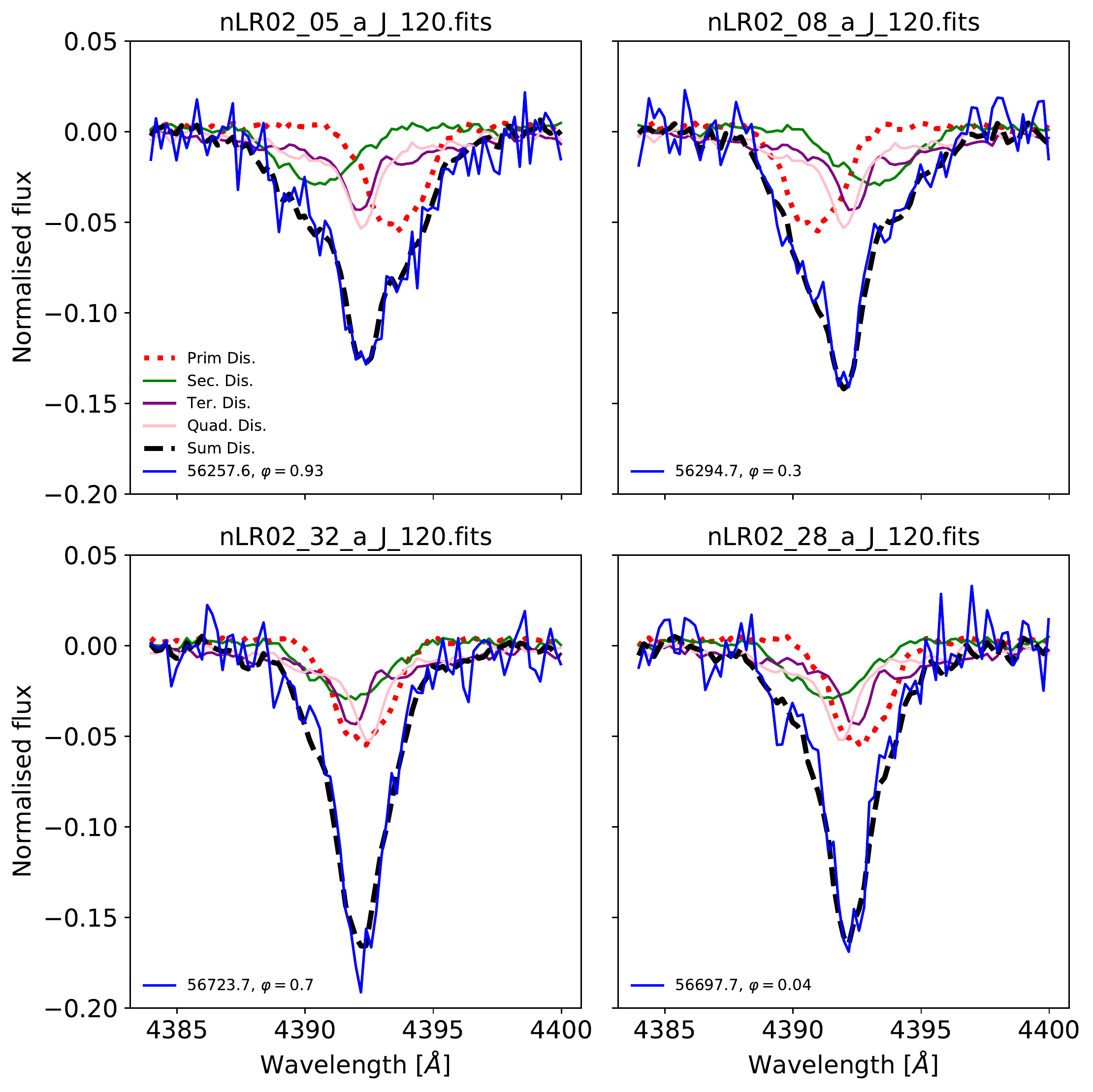}
\caption{Comparison of \HeI4338 spectra of the quadruple \object{VFTS~120} at RV extremes of both binaries, along with the disentangled spectra and their sum, as derived for  $K_1 = 92\,$\kms~and $K_2  = 93$, $K_3 = 20, K_4 = 21$\,\kms. The disentangled spectra are not scaled by the light ratio in this plot.} \label{fig:VFTS120EXT}
\end{figure}

\begin{figure}
\centering
\begin{tabular}{c}
\includegraphics[width=0.5\textwidth]{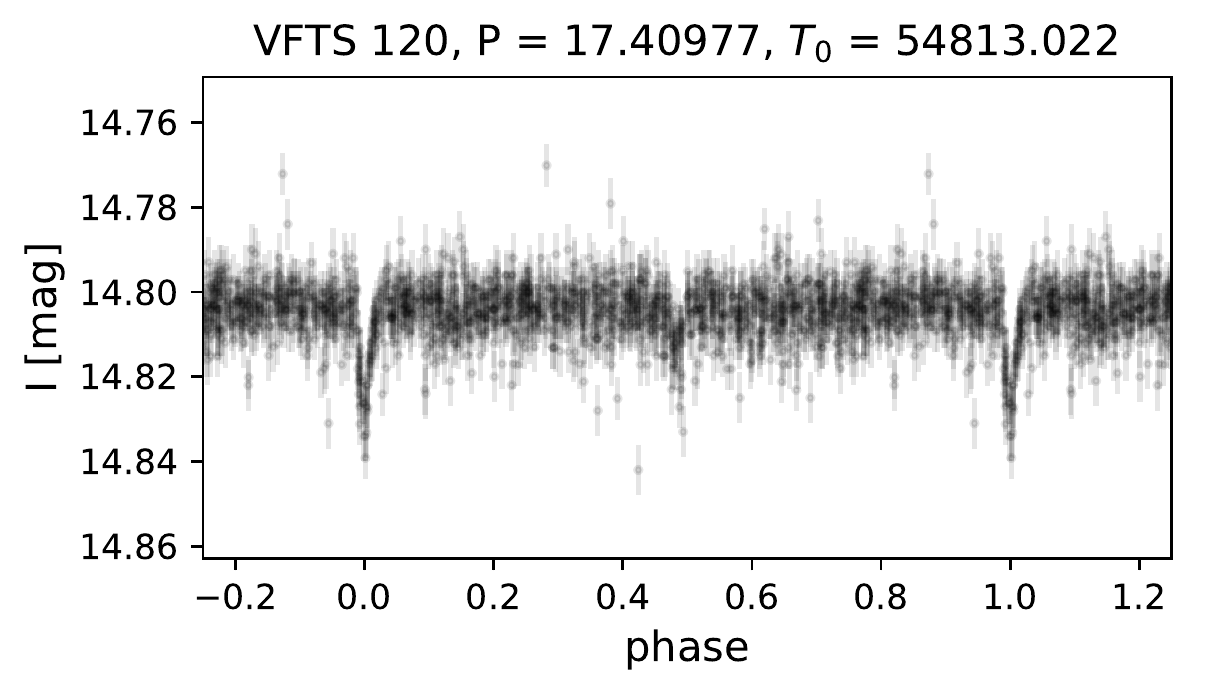}\\ 
\includegraphics[width=0.5\textwidth]{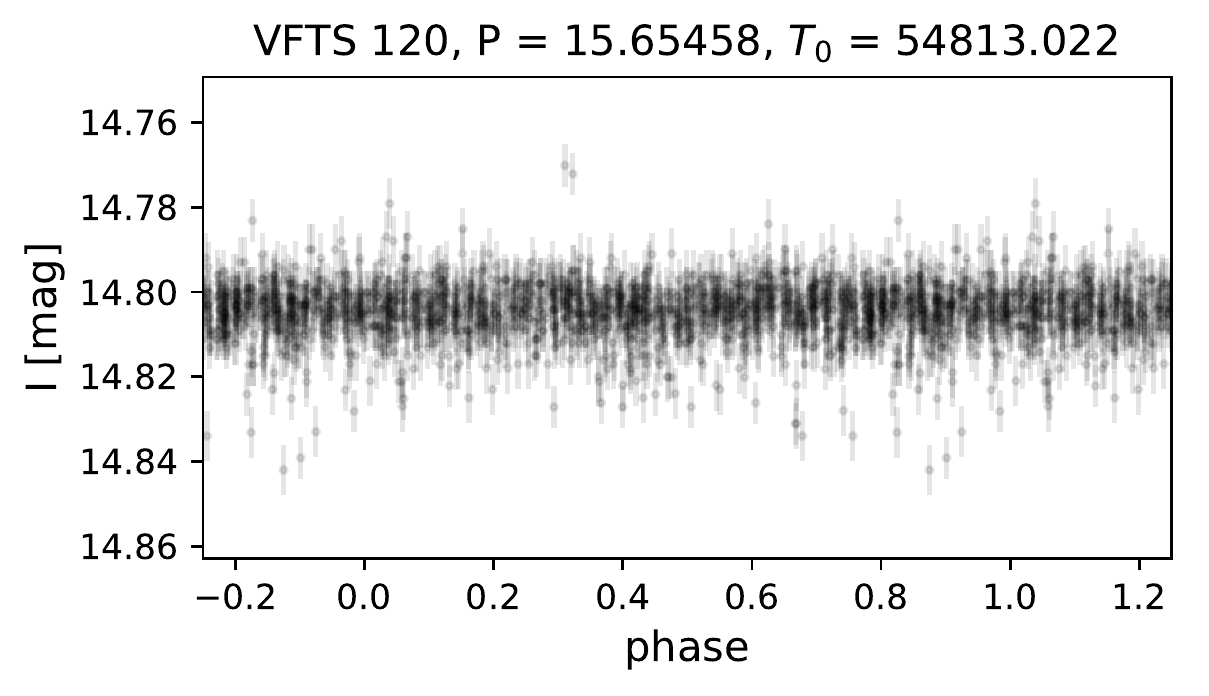} 
\end{tabular}
\caption{OGLE I-band light curve of  the binary \object{VFTS~120}, phased with $P = 17.41\,$d (the peak found in the Fourier transform, top panel) and $15.65\,$d (the spectroscopic period, bottom panel). In both cases, $T_0 = 54813.02$ [JD-2400000].
}
\label{fig:VFTS120OGLE}
\end{figure}



{\bf  \object{VFTS~120}, (O9.2~IV + B) + (O9.5~V + O9.5~V)} was reported 
to be have a period of $P=15.7$\,d and an eccentricity of $e=0.28$. Inspection of the He\,{\sc i} lines reveals line-profile variations that are suggestive of a second non-degenerate star in the spectrum (Fig.\,\ref{fig:VFTS120EXT}). 
For this reason, we implemented the 2D disentangling algorithm. However, the solution appeared to be strongly dependent on the lines used. Moreover, the \HeI4388 and \HeI4471 lines, which are the strongest and appear to be hardly contaminated by nebular emission, result in $K_2$ amplitudes of the order of 0\,\kms. As in the case of VFTS~64, this could imply the presence of additional stellar components in the system.

Interestingly, Fourier analysis of the OGLE light curve shows a peak at $8.7\,$d and its harmonics. Folding the light curve with twice this value, that is, $P = 17.41\,$d, reveals a clear signature of eclipses in the system (Fig.\,\ref{fig:VFTS120OGLE}). It can be fully ruled out that this period corresponds to the spectroscopic period. Given the similarity of the periods, an eclipsing tertiary can be ruled out. The only conceivable way to explain this is that VFTS~120 is a quadruple system. For this reason, we extended the shift-and-add algorithm to account for four components. To reduce the parameter space, we adopt for the eclipsing binary the photometric period, an eccentricity of 0 (as implied by the $\approx 0.5$ phase difference in eclipses), and the time of periastron (defined here as time of primary eclipse). The four RV amplitudes $K_1, K_2, K_3, K_4$ are then scanned through a $\chi^2$ minimisation. We find $K_1 = 92 \pm 13\,$\kms, $K_2 = 93 \pm 17\,$\kms, $K_3 = 20 \pm 7$\kms, and $K_4 = 21 \pm 7\,$\kms. However, the results should be taken with caution, given the assumptions, limited resolution, and data quality available here. The outer period connecting the two binaries cannot be established here.

\begin{figure}
\centering
\includegraphics[width=.5\textwidth]{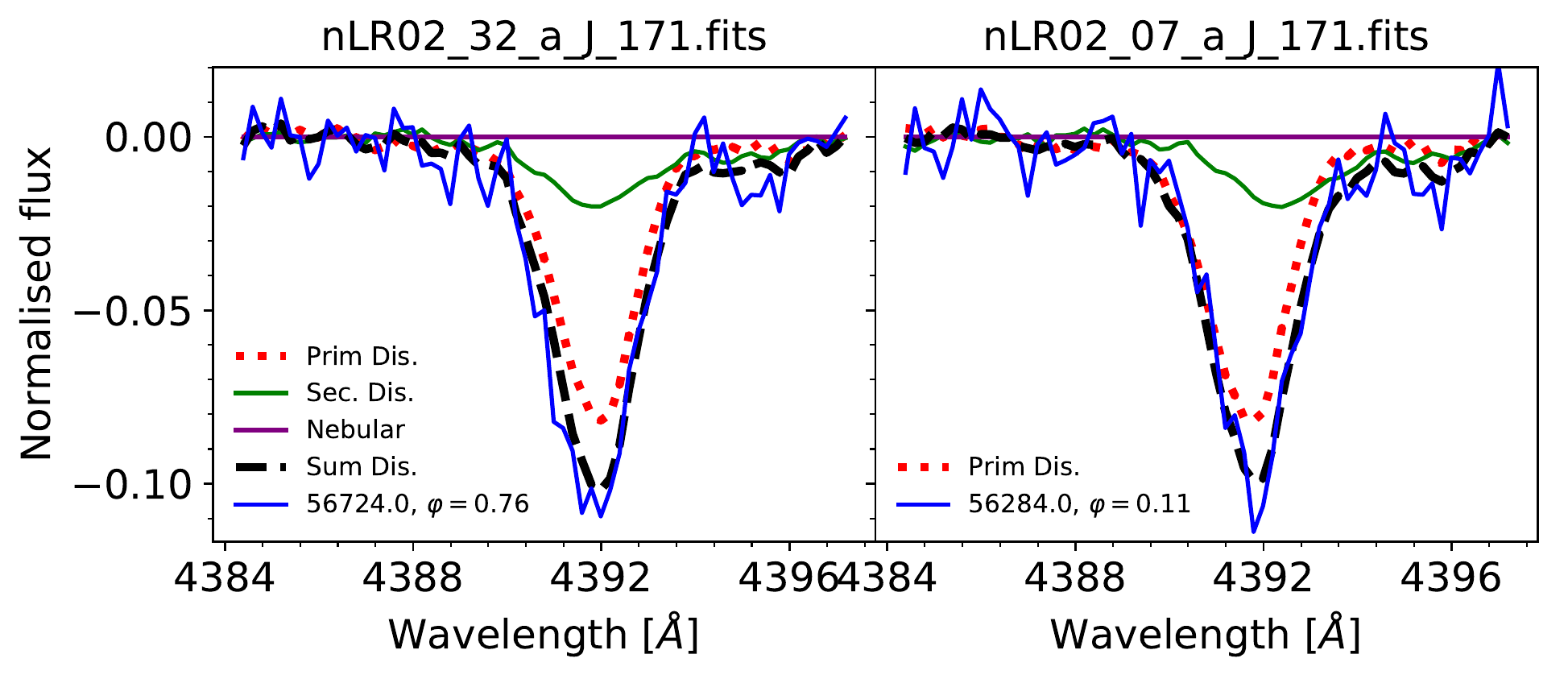}
\caption{Comparison of \HeI4388 spectra of \object{VFTS~171} at RV extremes, along with the disentangled spectra and their sum, as derived for  $K_1 = 12\,$\kms~and $K_2  = 3\times K_1 = 24$\,\kms. The disentangled spectra are not scaled by the light ratio in this plot.} \label{fig:VFTS171EXT}
\end{figure}

\begin{figure}
\centering
\includegraphics[width=.5\textwidth]{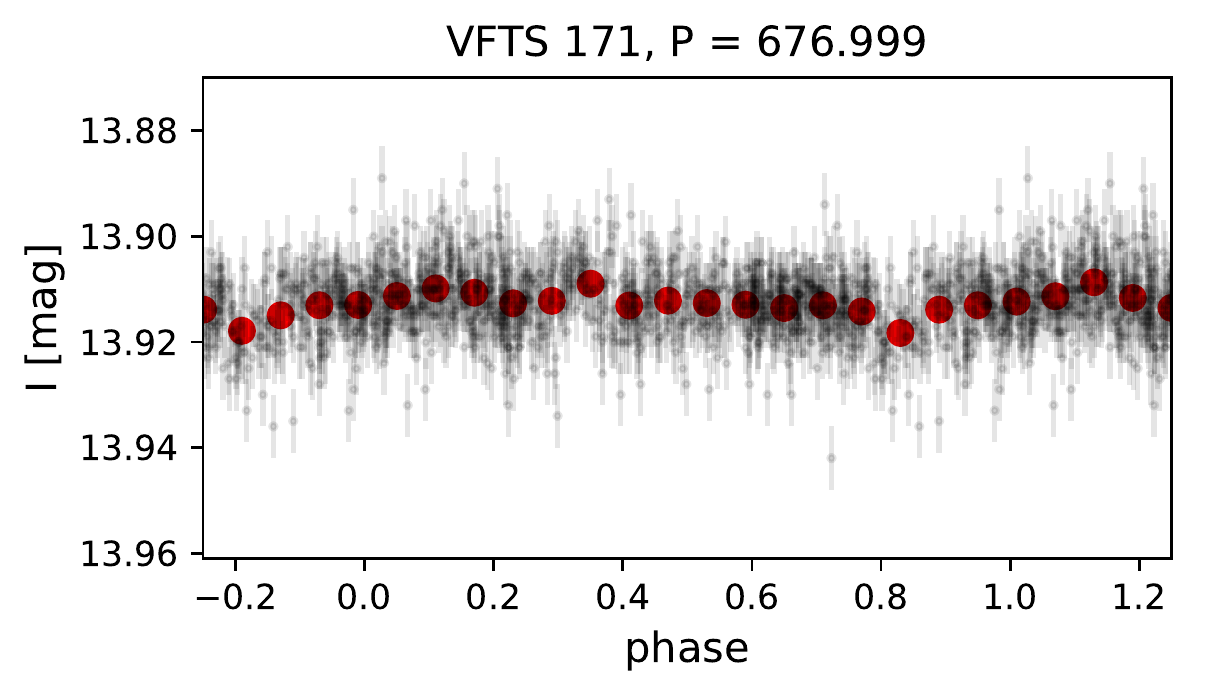}
\caption{OGLE I-band light curve of  the binary \object{VFTS~171}, phased at the orbital period of $P=677\,$d.} \label{fig:VFTS171OGLE}
\end{figure}

{\bf  \object{VFTS~171}, O8.5~III:(f)) + B1.5:~V, } was reported to be a long-period binary with $P=677\,$d and  $e=0.56$. The low-amplitude motion of the primary ($K_1 = 12\,$\kms) prevents us from obtaining unambiguous results for this system, and inspection of the line-profile variability is not readily suggestive of a non-degenerate secondary (Fig.\,\ref{fig:VFTS171EXT}).
Disentangling the spectra results in faint signatures in Balmer lines, He\,{\sc i} lines, and He\,{\sc ii} lines. However, we cannot derive $K_2$ from the relatively flat $\chi^2$ map. We tentatively classify the companion as B1.5:~V and estimate its light contribution at 8\%, but given the low RV amplitude, we consider the SB2 nature uncertain. The OGLE light curve shows a variability pattern  when phased at the orbital period (Fig.\,\ref{fig:VFTS171OGLE}), though it is not clear that this pattern is of really astrophysical origin. 

\begin{figure}
\centering
\includegraphics[width=.5\textwidth]{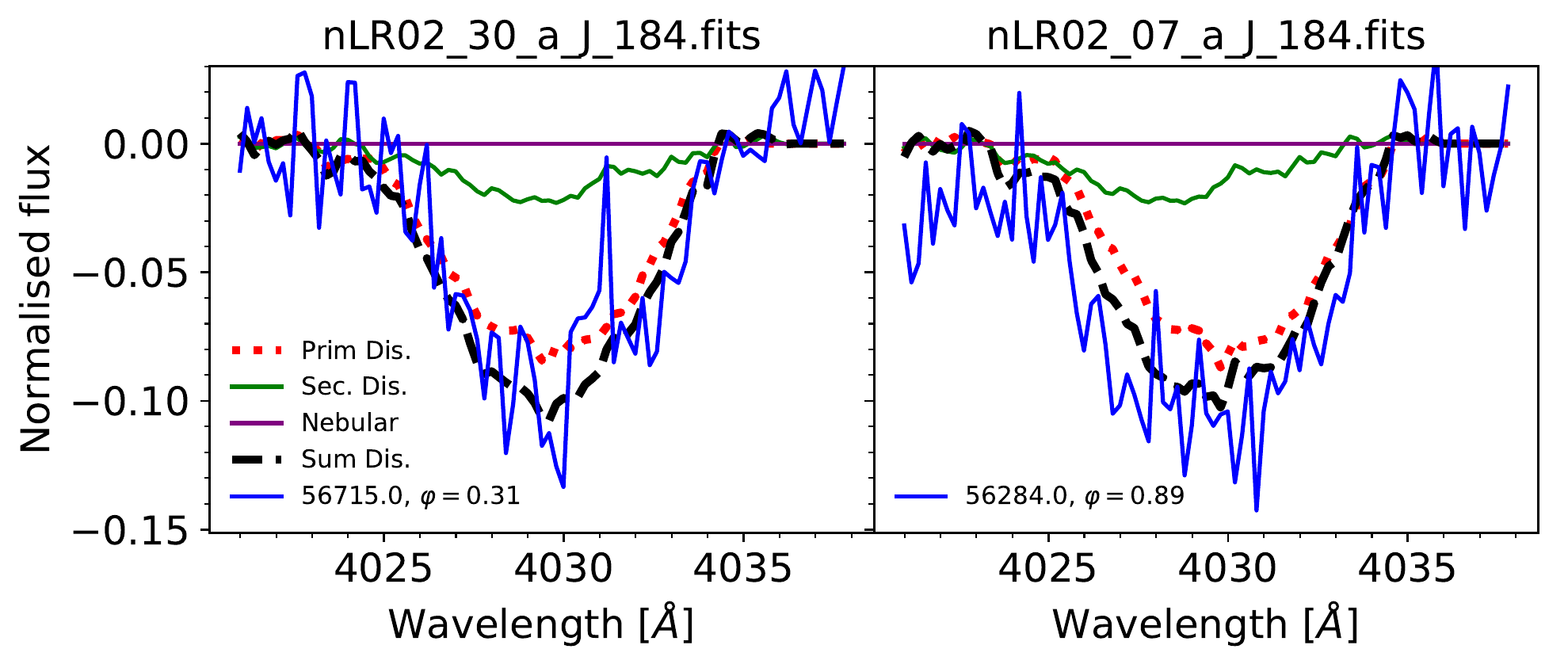}
\caption{Comparison of \HeI4026 spectra of \object{VFTS~184} at RV extremes, along with the disentangled spectra and their sum, as derived for  $K_1 = 12\,$\kms~and $K_2  = 3\times K_1 = 36$\,\kms. The disentangled spectra are not scaled by the light ratio in this plot.} \label{fig:VFTS184EXT}
\end{figure}

{\bf  \object{VFTS~184}, O6.5~Vn +OB:} has a reported period of $P=32\,$d and eccentricity of $e=0.20$. However, the orbital solution derived by \citet{Almeida2017} shows substantial scatter.   The lack of line-profile variability at RV extremes does not readily suggest the presence of a non-degenerate companion (Fig.\,\ref{fig:VFTS184EXT}). Given the scatter in the orbital solution, we attempted 2D disentangling across the $K_1, K_2$ axes, and probing all He\,{\sc i} and He\,{\sc ii} lines, as well as H$\delta$, which is weakly contaminated by nebular lines. The results of the different lines are consistent, but the errors are very large, with final mean averages of $K_1 = 29\pm11$\kms~and $K_2 = 61\pm29\,$\kms. Disentangling the spectrum for these RV amplitudes results in similar spectra for both components, which sheds doubt on the validity of the results, as this is often the case when the adopted $K_1$ is significantly different than the true one. When disentangling for $K_1 = 12.1\,$\kms, the features in the spectrum of the secondary weaken, though they do not fully disappear. Our results are therefore uncertain. We tentatively classify the companion as OB:, and refrain from using our measurements for any further analysis. With a minimum mass of $M_2 > 1.7\pm0.4\,M_\odot$, the nature of the secondary is virtually unconstrained. Higher-resolution spectra will be necessary to disentangle this system.  Analysis of the OGLE light curve did not reveal significant frequencies. 

\begin{figure}
\centering
\includegraphics[width=.5\textwidth]{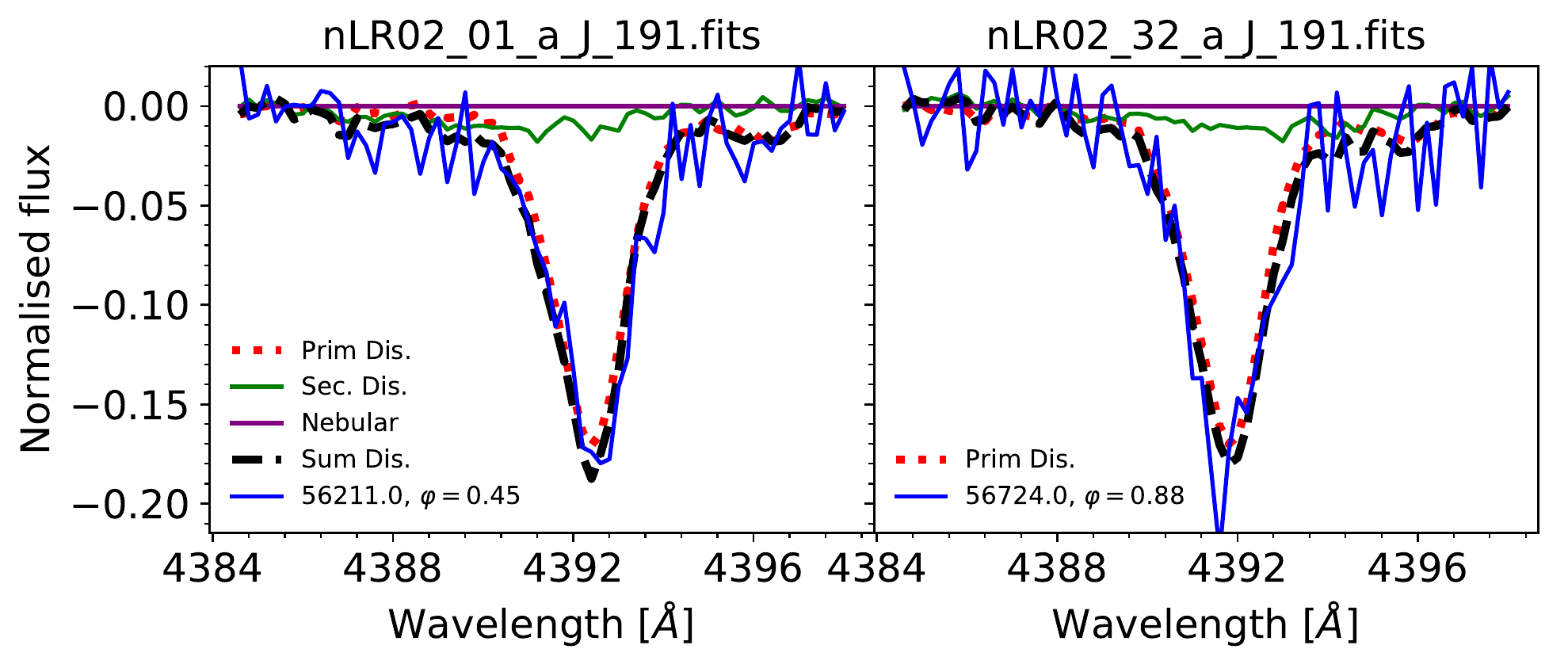}
\caption{Comparison of \HeI4388 spectra of \object{VFTS~191} at RV extremes, along with the disentangled spectra and their sum, as derived for  $K_1 = 23\,$\kms~and $K_2  = 3\times K_1 = 69$\,\kms. The disentangled spectra are not scaled by the light ratio in this plot.} \label{fig:VFTS191EXT}
\end{figure}

{\bf  \object{VFTS~191}, O9.2~V + O9.7:~V:} has a reported period of $P=359\,$d and an eccentricity of $e=0.22$, although the orbital solution (especially the eccentricity) are uncertain. The spectral variability does not readily suggest the presence of a non-degenerate companion (Fig.\,\ref{fig:VFTS191EXT}). However, grid disentangling consistently points towards the presence of a relatively bright companion entangled in the spectrum.
We are not able to constrain a $K_2$ value using disentangling of the strong He\,{\sc i} lines. However, adopting plausible $K_2$ values of the order of $2-3\,K_1 \approx 50-70\,$\kms~results in a secondary spectrum that appears to belong to a  fainter, rapidly rotating O-type star, contributing roughly 13\% to the total flux.  Analysis of the OGLE light curve did not reveal significant frequencies. 

\begin{figure}
\centering
\includegraphics[width=.5\textwidth]{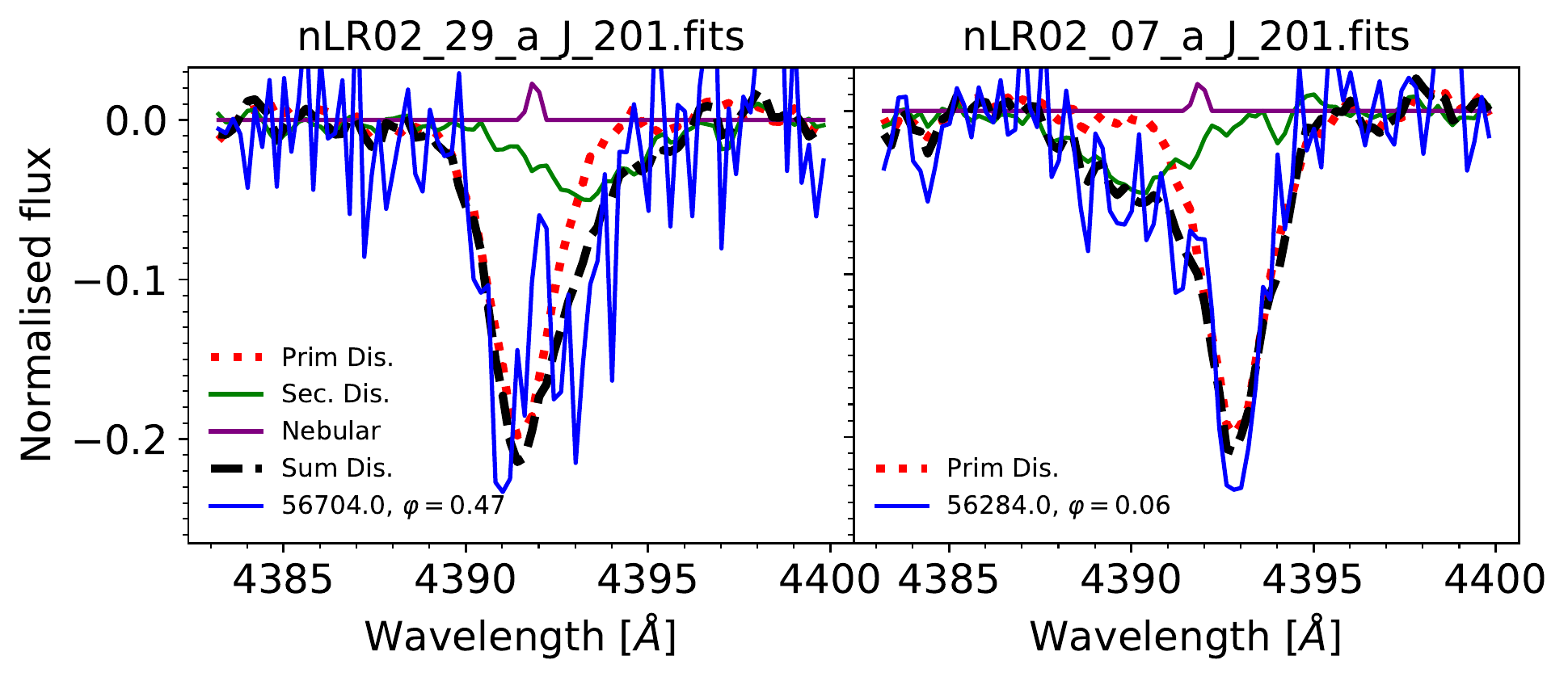}
\caption{Comparison of \HeI4388 spectra of \object{VFTS~201} at RV extremes, along with the disentangled spectra and their sum, as derived for  $K_1 = 79\,$\kms~and $K_2  = 105\,$\kms. The disentangled spectra are not scaled by the light ratio in this plot.} \label{fig:VFTS201EXT}
\end{figure}

{\bf  \object{VFTS~201}, O9.7~V + B1.5:~V} has a reported period of $P=15.3\,$d and eccentricity of $e=0.46$. The variability seen in the He\,{\sc i} lines (Fig.\,\ref{fig:VFTS201EXT}) is suggestive of a second non-degenerate component in the system. Due to the lack of strong He\,{\sc ii} lines and the apparent contribution of the secondary, we resort to 2D disentangling.  All He\,{\sc i} lines are in broad agreement with each other, but show substantial errors on the RV amplitudes. Disentangling all strong He\,{\sc i} simultaneously, we find $K_1 =  79 \pm 9$\,\kms~and $K_2 = 105 \pm 26$\,\kms.  The disentangled spectrum of the secondary, which has an estimated light contribution of roughly $30\%$, shows strong He\,{\sc i} lines and no He\,{\sc ii} lines, with only faint evidence for the Mg\,{\sc ii}\,$\lambda \lambda 4481.1, 4481.3$ doublet, and it is classified as B1.5:~V. Analysis of the OGLE light curve did not reveal significant frequencies. 

\begin{figure}
\centering
\includegraphics[width=.5\textwidth]{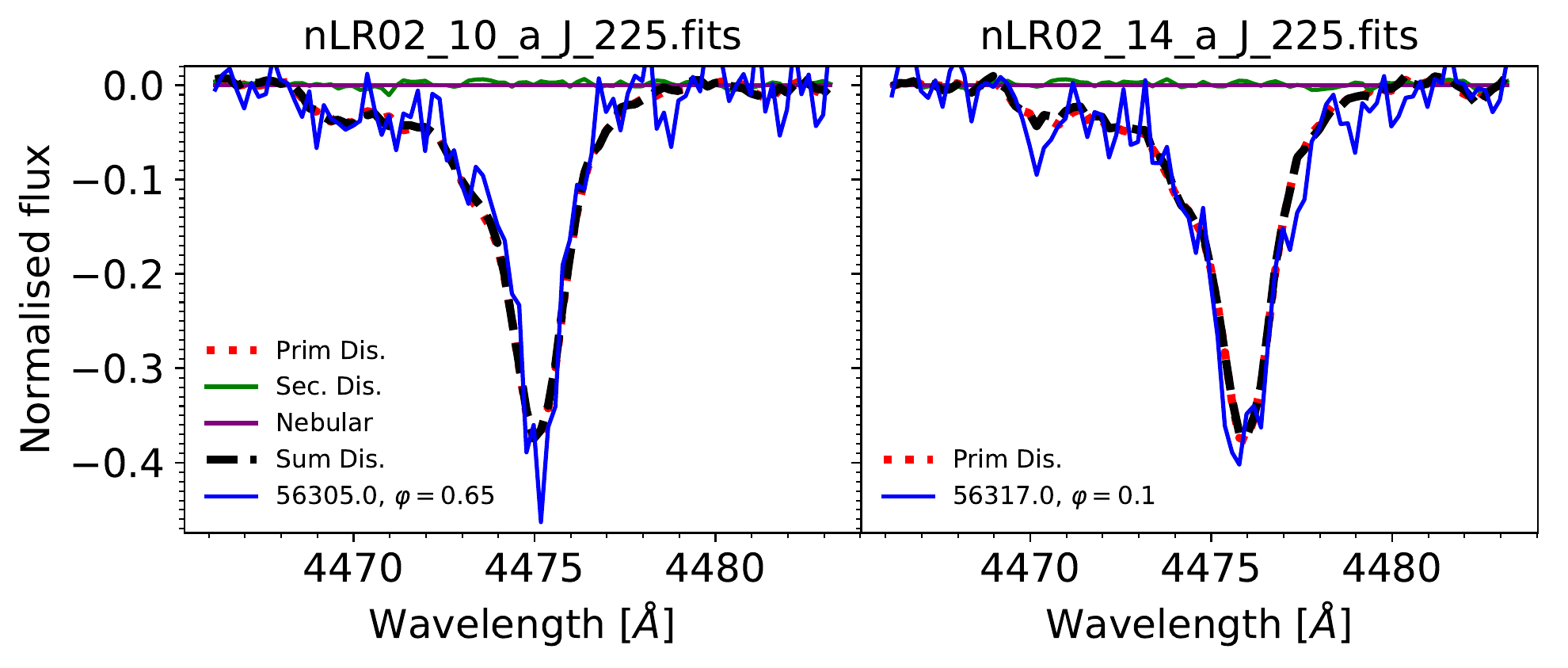}
\caption{Comparison of \HeI4388 spectra of \object{VFTS~225} at RV extremes, along with the disentangled spectra and their sum, as derived for  $K_1 = 29\,$\kms~and $K_2  = 3\times K_2 = 3 \times K_1 = 87$\,\kms. The disentangled spectra are not scaled by the light ratio in this plot.} \label{fig:VFTS225EXT}
\end{figure}

\begin{figure}
\centering
\includegraphics[width=.5\textwidth]{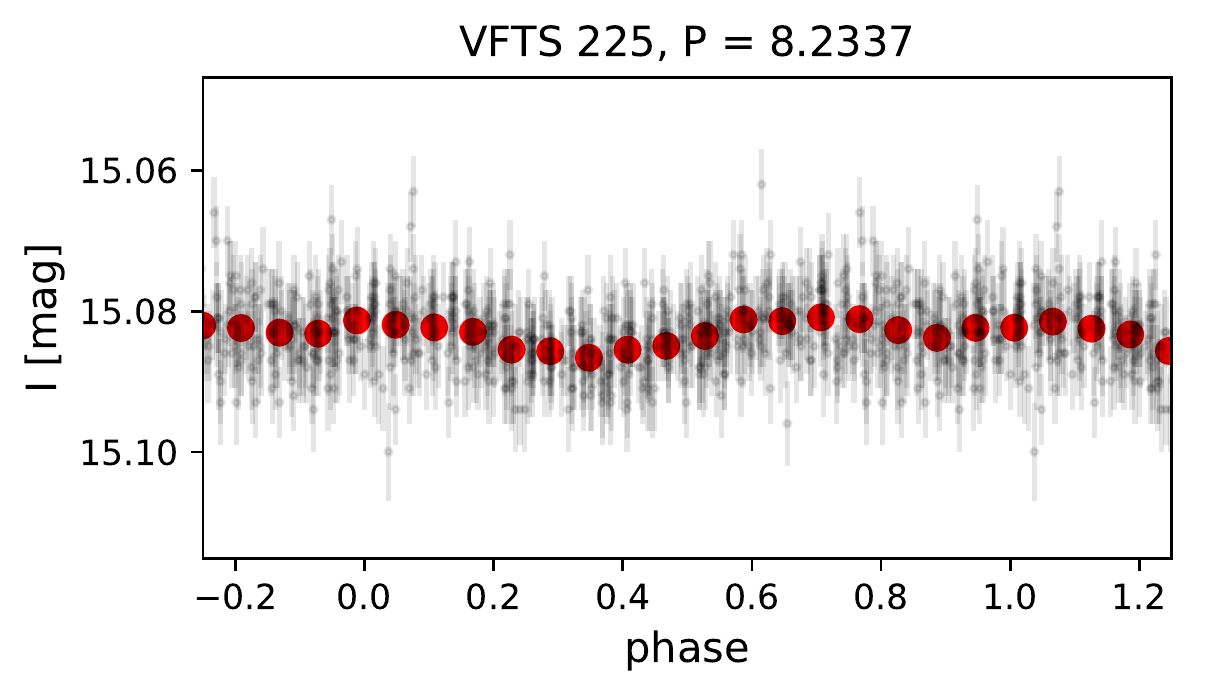}
\caption{OGLE I-band light curve of  the binary \object{VFTS~225}, phased at the orbital period of $P=8.2\,$d.} \label{fig:VFTS225OGLE}
\end{figure}

{\bf  \object{VFTS~225}, B0.7~III} has a reported period of  $P= 8.2\,$d and a virtually circular orbit.  Disentangling of the strong He\,{\sc i} lines (Fig.\,\ref{fig:VFTS225EXT}) does not reveal a significant minimum for $K_2$, and the disentangled spectrum of the secondary is virtually featureless in He\,{\sc i} lines, and shows very little signatures in Balmer lines that appear to originate from inaccurate subtractions of the weak nebular lines. The system is thus a bona fide SB1 system. However,  the minimum mass imposed by the orbital parameters and the estimated mass of the primary yield $M_2 > 1.8\pm0.3\,M_\odot$, such that the nature of the secondary remains unconstrained. 

The OGLE~III data show substantial scatter and a long-term magnitude increase that is not supported by the OGLE~IV data. We therefore remove them from the photometric analysis. Fourier analysis of the OGLE~IV data reveals a peak at the orbital period. The phased light curve is shown in Fig.\,\ref{fig:VFTS225OGLE}.

\begin{figure}
\centering
\includegraphics[width=.5\textwidth]{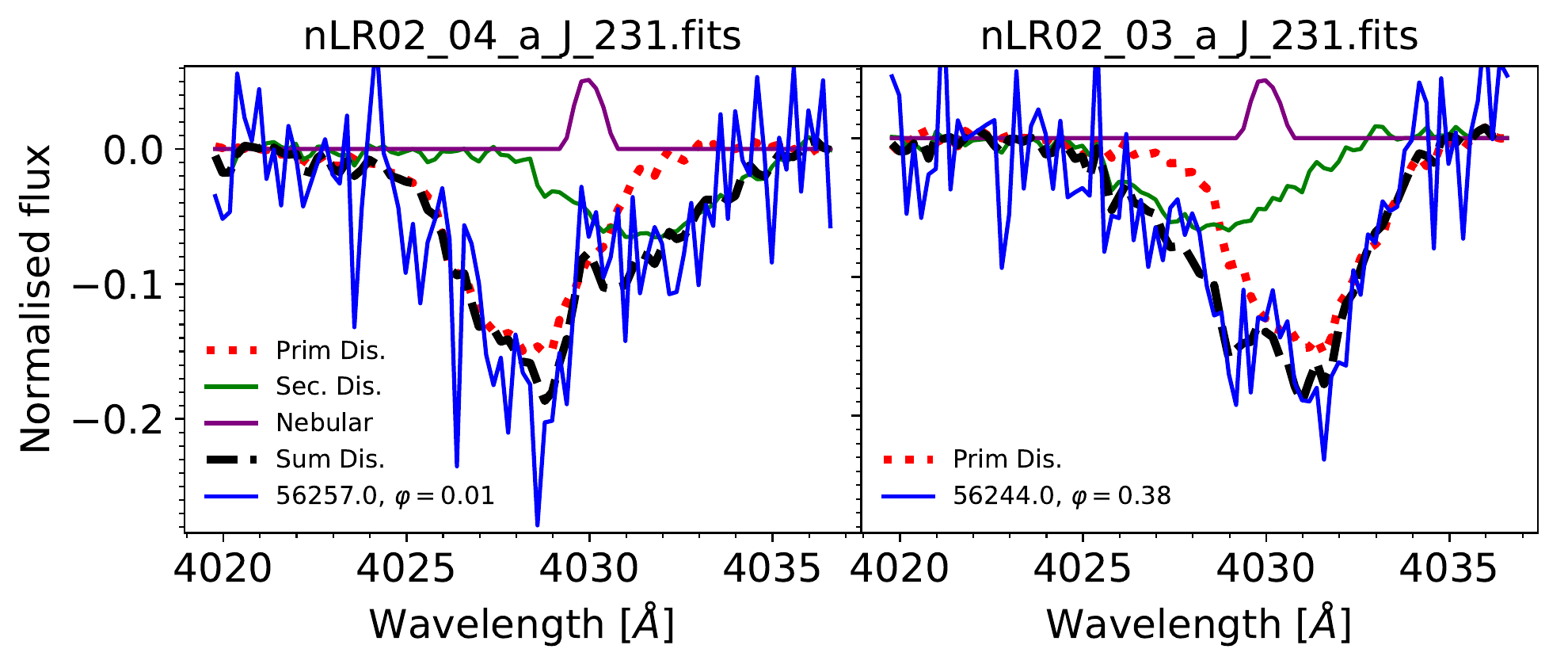}
\caption{Comparison of \HeI4026 spectra of \object{VFTS~231} at RV extremes, along with the disentangled spectra and their sum, as derived for the best-fitting $K_1, K_2$ values of  $109, 117\,$\kms, respectively. The disentangled spectra are not scaled by the light ratio in this plot.} \label{fig:VFTS231EXT}
\end{figure}

{\bf  \object{VFTS~231}, O9.7~V + B1.5:~V} has a reported period of $P= 7.9\,$d and eccentricity of $e=0.41$. \citet{Almeida2017} noticed the presence of a secondary star in the system, but could not derive its RVs. Indeed, comparing line profiles belonging to strong He\,{\sc i} lines such as \HeI4026 (Fig.\,\ref{fig:VFTS231EXT}) clearly shows the presence of a non-degenerate companion. We implemented 2D disentangling and find consistent results for all He\,{\sc i} lines. A mean average of the measurements obtained for the strong He\,{\sc i} lines (with exception of the \HeI4471 line, which suffers strong nebular contamination) yields $K_1 = 99 \pm 12$ and $K_2 = 130 \pm 25$\,\kms. From the strengths of the spectral lines, we estimate a light contribution of 33\% for the secondary. We classify it as B1.5:. Despite the short period of the system, we cannot find significant periodicities in the OGLE light curve.


{\bf  \object{VFTS~243}, O7~V(n)((f)) + BH} was shown to be a bona fide O+BH binary with a $P=10.4\,$d period and a near-circular orbit in \citet{Shenar2022VFTS243}, to which we refer for a detailed analysis of the spectra and light curve of this unique binary.

\begin{figure}
\centering
\includegraphics[width=.5\textwidth]{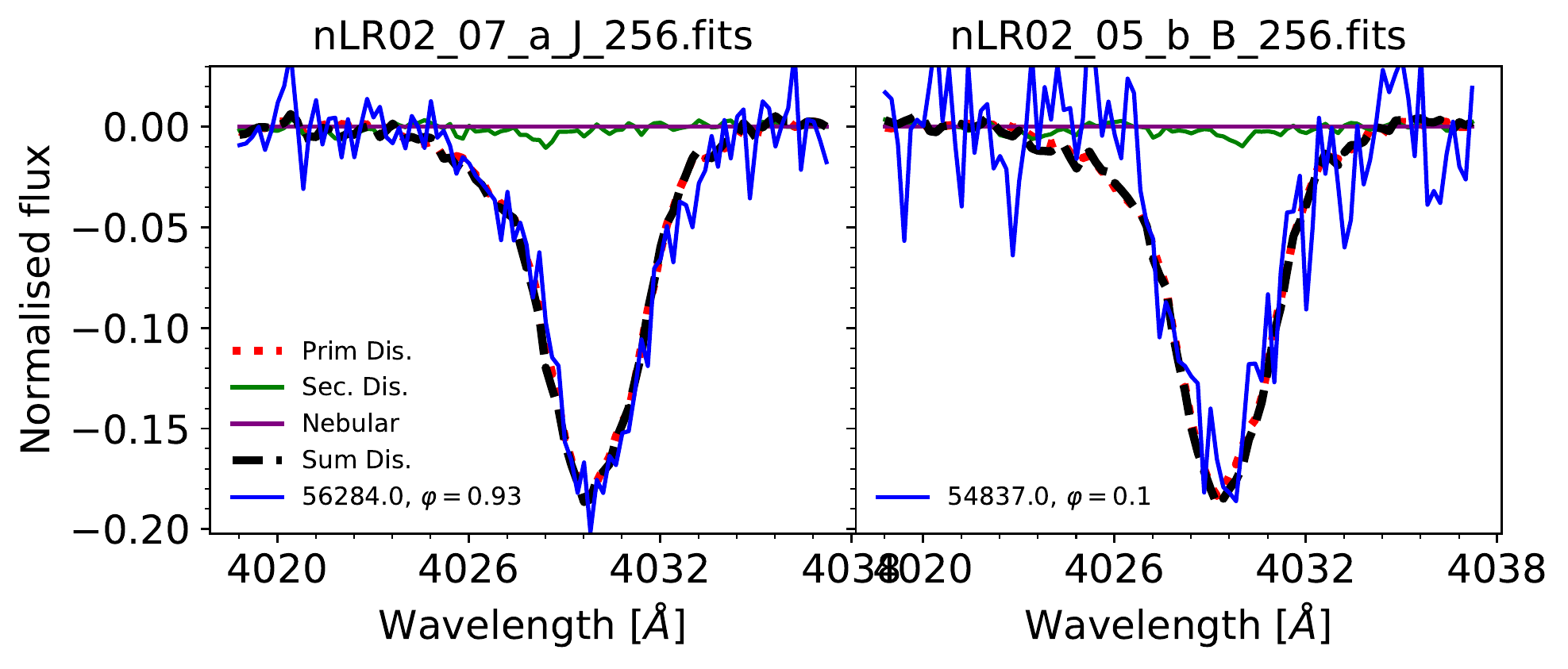}
\caption{Comparison of \HeI4471 spectra of \object{VFTS~256} at RV extremes, along with the disentangled spectra and their sum, as derived for $K_1 = 18\,$\kms and $K_2 = 3\times K_1 = 57\,$\kms. The disentangled spectra are not scaled by the light ratio in this plot.} \label{fig:VFTS256EXT}
\end{figure}

{\bf  \object{VFTS~256}, O7.5~V: +OB:} has  a  period of $P=246\,$d and an eccentricity of $e=0.63$. Inspection of the line-profile variability at RV extremes does not readily reveal the presence of a non-degenerate companion (Fig.\,\ref{fig:VFTS256EXT}). The low-amplitude motion ($K_1 =19.2\,$\kms) and nebular contamination make disentangling of the system challenging.  Given the potential contamination of the secondary, we performed 2D disentangling, but find a consistent $K_1$ value to that derived by \citet{Almeida2017}, and therefore adopt their value. We cannot find a minimum in the $\chi^2(K_2)$ map, and the disentangled spectra provide marginal evidence for the presence of a companion at faint He\,{\sc i}~lines features and Balmer features, which may however be impacted by nebular contamination. Comparably weak He\,{\sc ii}~features are seen, thought these features may be spurious due to an inaccurate $K_1$ value.  We tentatively classify this as an SB2:. With a minimum mass of $M_2 > 4.1\pm 0.5\,M_\odot$, the secondary could be a faint non-degenerate star or a BH. Analysis of the OGLE light curve does not reveal any periodicity at the orbital period, but does show periodicity at $P = 341\,$d. This periodicity may well be related to the yearly cycle, and hence cannot be confirmed to represent an intrinsic period of the system.

\begin{figure}
\centering
\includegraphics[width=.5\textwidth]{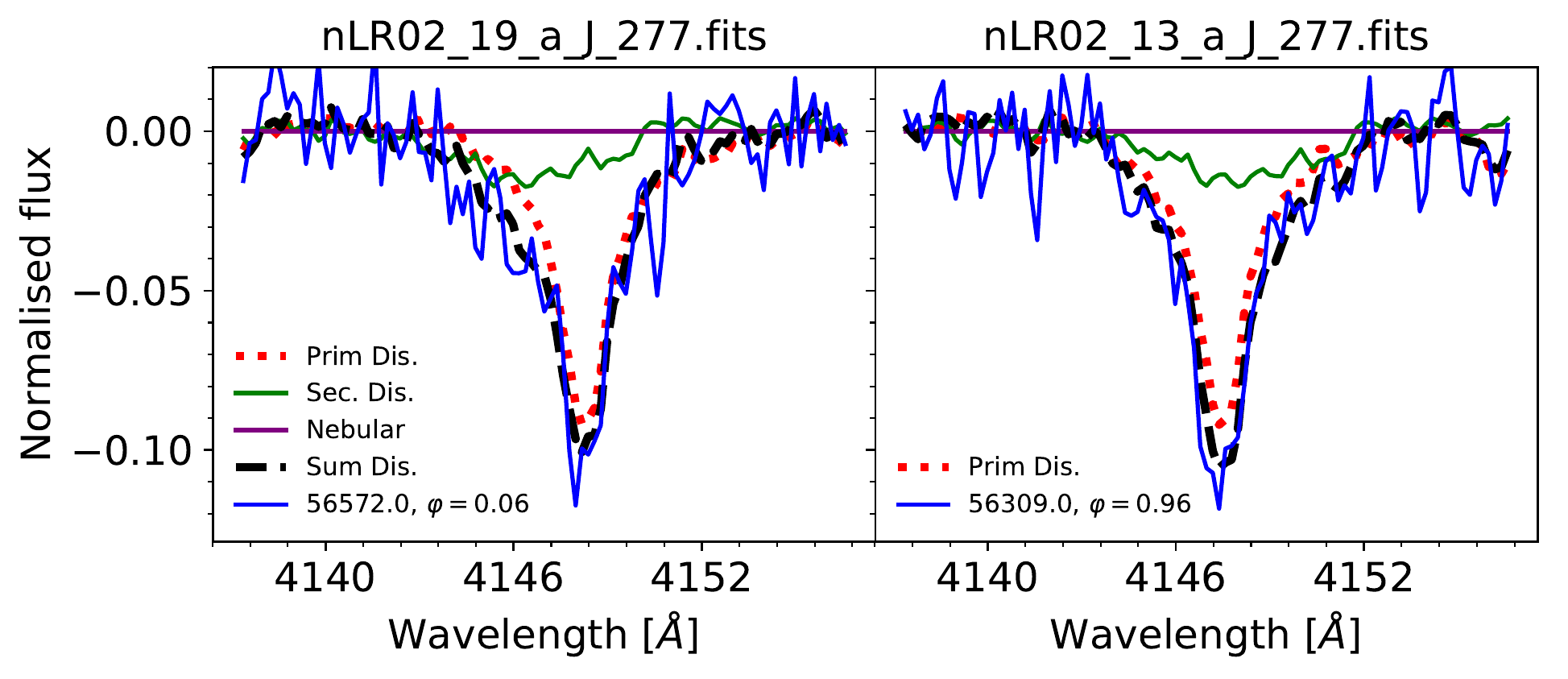}
\caption{Comparison of \HeI4144 spectra of \object{VFTS~277} at RV extremes, along with the disentangled spectra and their sum, as derived for $K_1 = 64$ and $K_2 = 2\times K_1 = 128\,$\kms. The disentangled spectra are not scaled by the light ratio in this plot.} \label{fig:VFTS277EXT}
\end{figure}

{\bf  \object{VFTS~277}, O9~V + B1.5:~V} has  a  period of $P=240\,$d and a  high eccentricity of $e=0.93$. The mass function and estimated mass of the primary imply a companion of at least $\approx 3\,M_\odot$. The spectral variability of He\,{\sc i} lines  weakly suggests the presence of a non-degenerate companion (Fig.\,\ref{fig:VFTS277EXT}), though not conclusively. Disentangling reveals significant He\,{\sc i} absorption for the  secondary but no He\,{\sc ii} absorption. However, $K_2$ cannot be constrained with the current data quality. Based on the appearance of the H$\delta$ region and the Mg\,{\sc ii}\,$\lambda \lambda 4481.1, 4481.3$ doublet, we tentatively classify the companion as B1.5. However,  as the spectral appearance of the secondary depends on the adopted $K_2$ value (here, we use $K_1 = 2\times K_2 = 128\,$\kms), the classification is not certain. Moreover, the presence of a faint, blue-shifted \HeII{4200} absorption in the spectrum of the secondary implies that this system may be a higher-order multiple. Analysis of the OGLE light curve did not reveal significant frequencies.

\begin{figure}
\centering
\includegraphics[width=.5\textwidth]{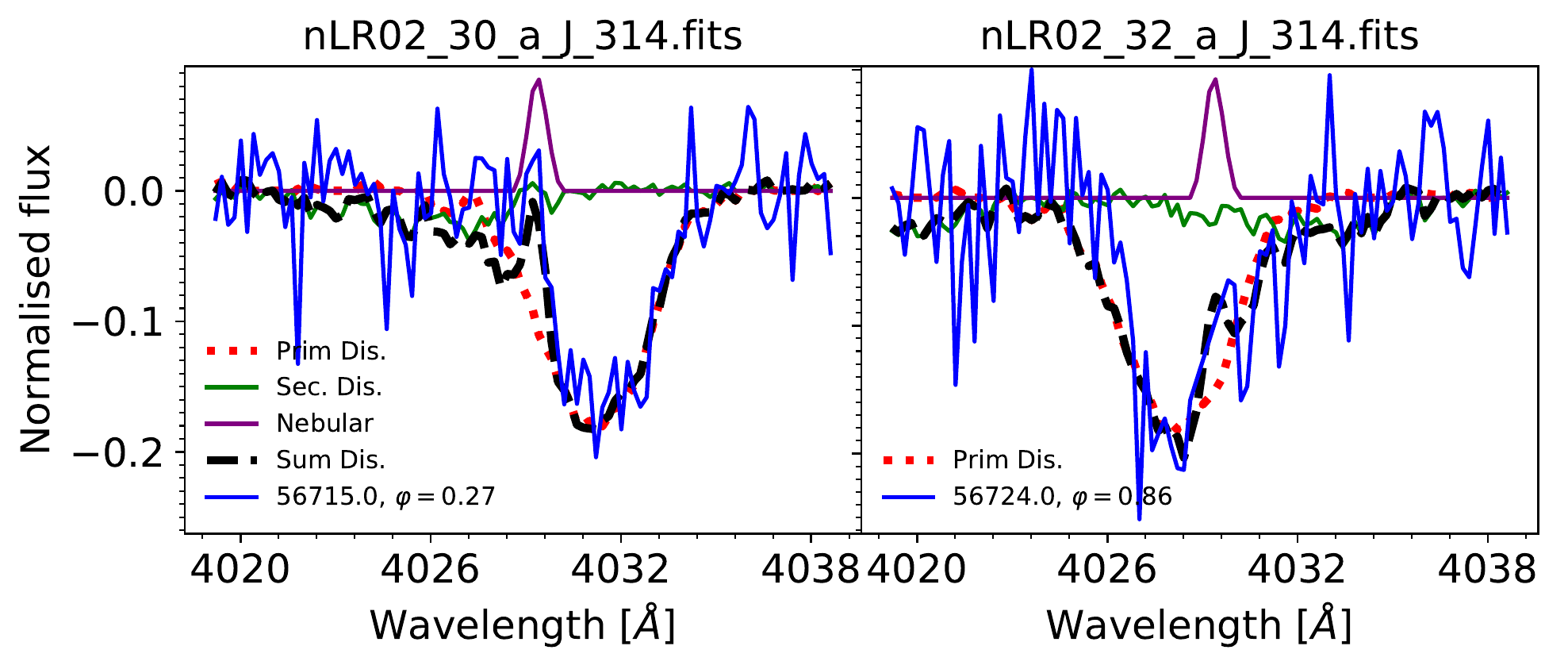}
\caption{Comparison of \HeI4026 spectra of \object{VFTS~314} at RV extremes, along with the disentangled spectra and their sum, as derived for the best-fitting RV amplitudes $K_1 = 111$ and $K_2 = 232$\kms. The disentangled spectra are not scaled by the light ratio in this plot.} \label{fig:VFTS314EXT}
\end{figure}

{\bf  \object{VFTS~314}, O9.7~V(n) + B} has a short period of $P=2.5\,$d and an eccentricity of $e=0.17$. \citet{Almeida2017} noted the presence of a secondary in the individual spectra, but could not measure its RVs. Indeed, disentangling of the He\,{\sc i} lines (Fig.\,\ref{fig:VFTS314EXT}) results in a relatively well defined minimum around 200\,\kms; the most robust result is found from disentangling of the \HeI4026 line, which yields $K_2 = 232\pm16$\,\kms. The corresponding disentangled spectrum shows a spectrum with clear He\,{\sc i} absorption and no He\,{\sc ii} absorption, but a precise classification of the secondary is difficult due to the lack of metal lines. We estimate that the companion contributes 5\% or less to the visual flux.

\begin{figure}
\centering
\includegraphics[width=.5\textwidth]{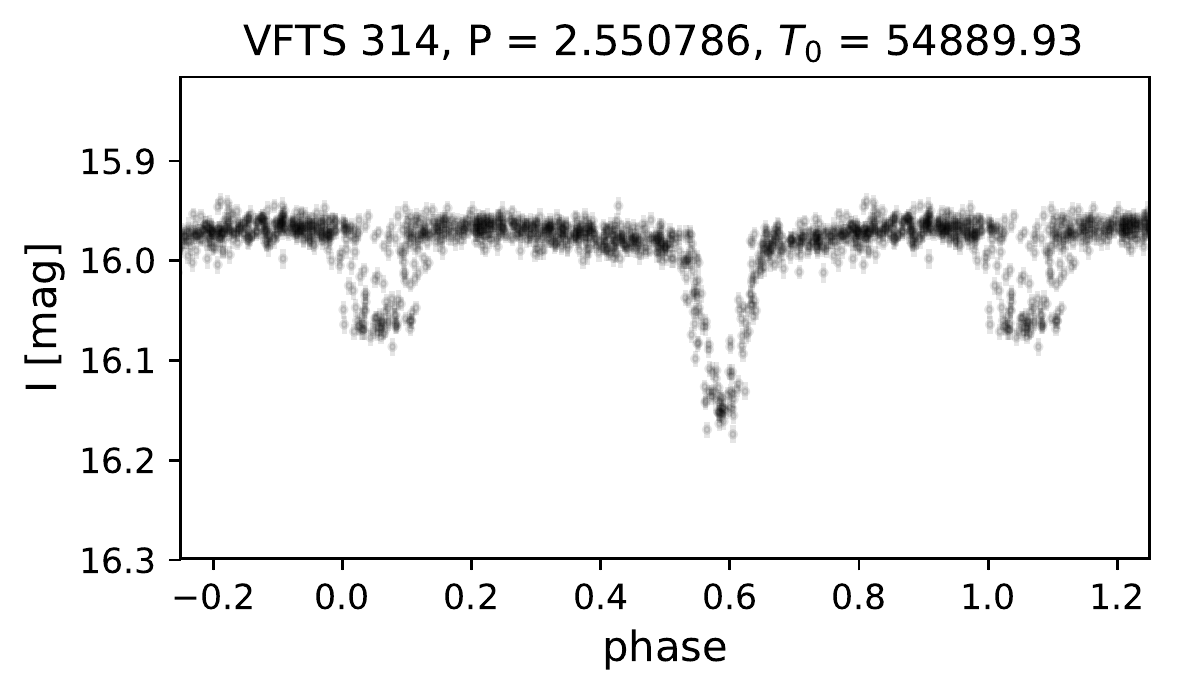}
\caption{OGLE I-band light curve of  the binary \object{VFTS~314}, phased with the derived period. We note that the system experiences apsidal motion, causing the eclipse epochs to shift with time.} \label{fig:VFTS314OGLE}
\end{figure}

As could be anticipated from the short period, VFTS~314 is an eclipsing system. The light curve of VFTS~314 is shown in Fig.\,\ref{fig:VFTS314OGLE}. From analysis of the light curve, the period can be refined to $P = 2.550786 \pm 0.000005$. Interestingly, the primary and secondary minima do not yield an identical period, suggesting that the system experiences apsidal motion. 

\begin{figure}
\centering
\includegraphics[width=.5\textwidth]{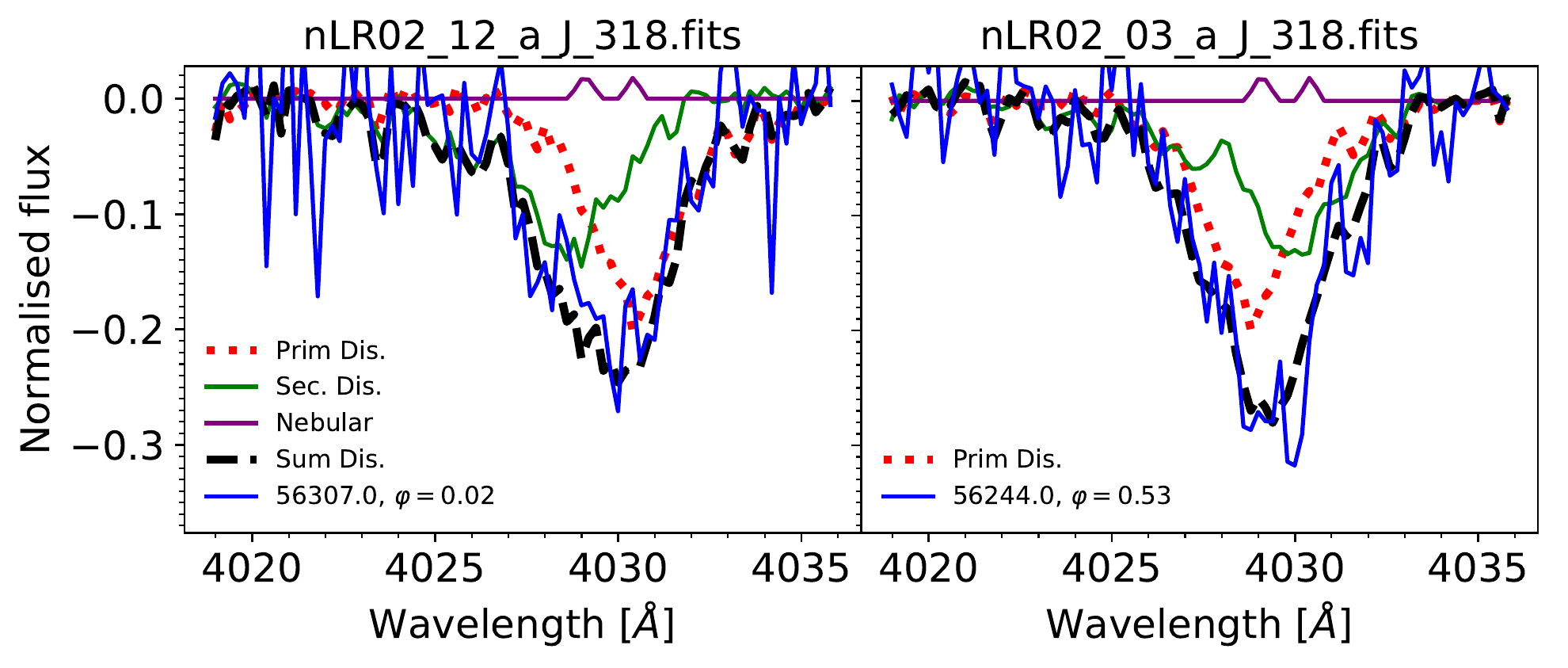}
\caption{Comparison of \HeI4026 spectra of \object{VFTS~318} at RV extremes, along with the disentangled spectra and their sum, as derived for the best-fitting $K_1, K_2$ values of  $53, 93\,$\kms, respectively. The disentangled spectra are not scaled by the light ratio in this plot.} \label{fig:VFTS318EXT}
\end{figure}

{\bf  \object{VFTS~318}, O9.5~V + O9.2~V} is a nearly circular binary with a reported period of $P=14\,$d. Careful inspection of the line variability at RV extremes of He\,{\sc i} lines such as \HeI4026 or \HeI4388 (Fig.\,\ref{fig:VFTS318EXT}) suggests the presence of a second non-degenerate star in the binary. For this reason, we implemented 2D disentangling on various He\,{\sc i} lines. A weighted mean of the RVs measured from the He\,{\sc i} lines yields $K_1 = 53 \pm 12$\,\kms~and $K_2 = 93 \pm 24\,$\kms. The He\,{\sc ii} lines do not yield well-constrained measurements. The value of $K_1$ is more than twice the amplitude derived by \citet{Almeida2017}, a consequence of accounting for the secondary.  The two stars a very similar  spectral type, and the secondary is estimated to contribute 33\% to the visual flux. It is therefore somewhat surprising that the secondary is found to have roughly half the mass of the primary, suggesting that the primary may be an evolved star. Analysis of the OGLE light curve did not yield significant periods.

\begin{figure}
\centering
\includegraphics[width=.5\textwidth]{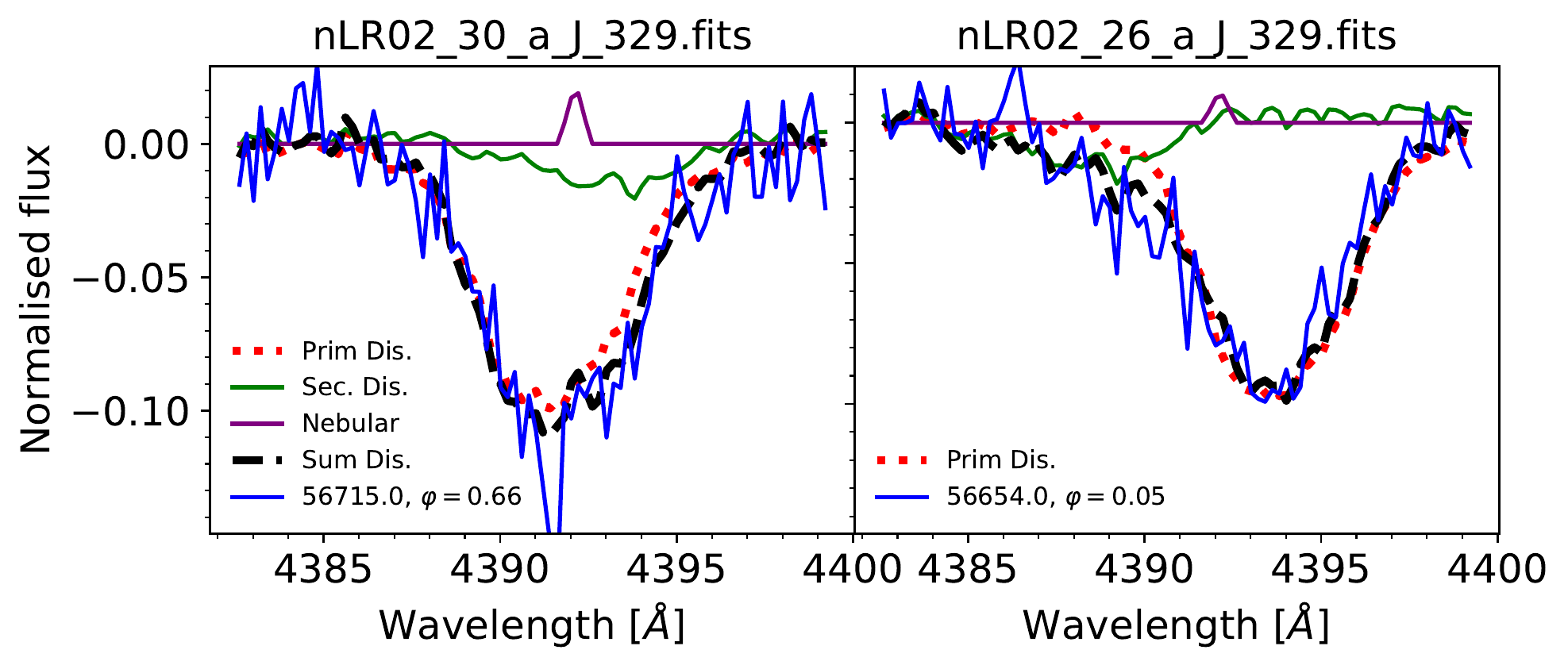}
\caption{Comparison of \HeI4388 spectra of \object{VFTS~329} at RV extremes, along with the disentangled spectra and their sum, as derived for the best-fitting $K_1, K_2$ values of  $93, 131\,$\kms, respectively. The disentangled spectra are not scaled by the light ratio in this plot.} \label{fig:VFTS329EXT}
\end{figure}

{\bf  \object{VFTS~329}, O9.5 V(n) + B1:~V:} has a  reported period of $P=7.0\,$d and eccentricity of $e = 0.44$. The spectral-line variability of He\,{\sc i} lines suggests the presence of a fainter non-degenerate secondary in the spectrum, as is shown for the \HeI4338 line in Fig.\,\ref{fig:VFTS329EXT}. Given the non-negligible contribution, we perform a 2D disentangling for the system.
A mean average of the  of strong He\,{\sc i}, aside from the contaminated \HeI4471 line, yields $K_1 = 93 \pm 8$, $K_2 = 131 \pm 24$\,\kms.  The disentangled spectrum of the secondary, which has an estimated light contribution of  22\%,  matches a B1~V spectral type. Despite the relatively short period, analysis of the OGLE light curve did not yield significant periods. 
\begin{figure}
\centering
\includegraphics[width=.5\textwidth]{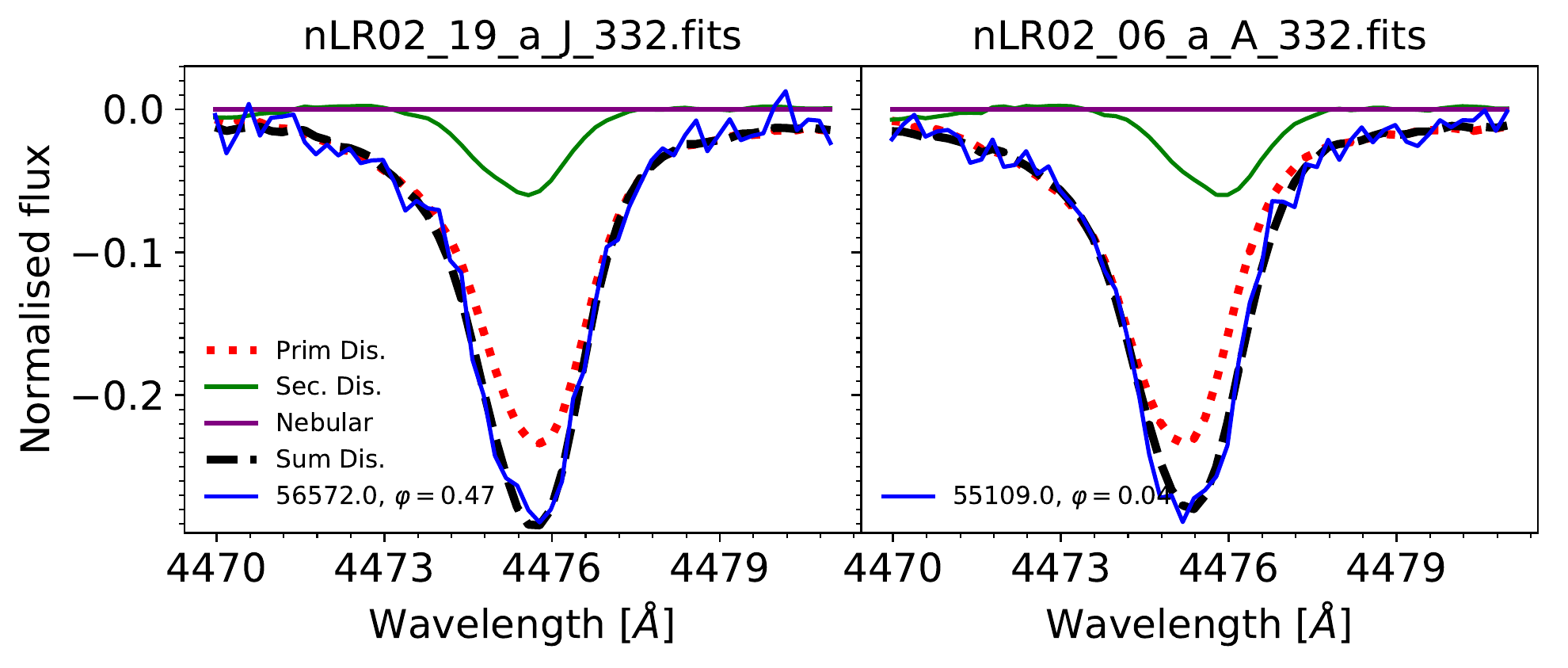}
\caption{Comparison of \HeI4388 spectra of \object{VFTS~332} at RV extremes, along with the disentangled spectra and their sum, as derived for the best-fitting $K_1, K_2$ values of  $79, 46\,$\kms, respectively. The disentangled spectra are not scaled by the light ratio in this plot.} \label{fig:VFTS332EXT}
\end{figure}

{\bf  \object{VFTS~332}, O9~III + O9.2~V} is an eccentric long-period  binary with $P=1025$\,d and  $e=0.81$. The coverage of the spectra in Doppler space is very poor, with only one epoch showing significant RV shifts compared to the others. Therefore, results from disentangling of this system should be taken with caution. The results point to the presence of a second, non-degenerate object in the system, although its presence is not readily seen from the spectral variability (Fig.\,\ref{fig:VFTS332EXT}). Disentangling of all He lines results in a weighted mean of $K_1 = 79 \pm 10\,$\kms~and $K_2 = 46 \pm 13\,$\kms. As in the case of VFTS~184, the similarity of the disentangled spectra suggests that the secondary spectrum may be contaminated by the primary due to a mismatch of the $K_1$ value. However, regardless of the $K_1$ value used for disentangling, a stellar spectrum is always obtained for the secondary, suggesting that it is likely real.

Taking our results at face value, it appears that the hidden secondary is the more massive star, with a spectral class of roughly O9. Despite of this, the disentangled spectra imply that it contributes only 24\% to the total light, suggesting that the primary is perhaps more evolved. On the other hand, the spectral appearance of the secondary also corresponds to an evolved object. These results combined are hard to reconcile with each other, suggesting that the disentangled spectra suffer from cross-contamination between the components. High resolution spectroscopy, obtained during periastron passage, will be required to disentangle this object unambiguously. Analysis of the OGLE light curve did not yield significant periods. 

\begin{figure}
\centering
\includegraphics[width=.5\textwidth]{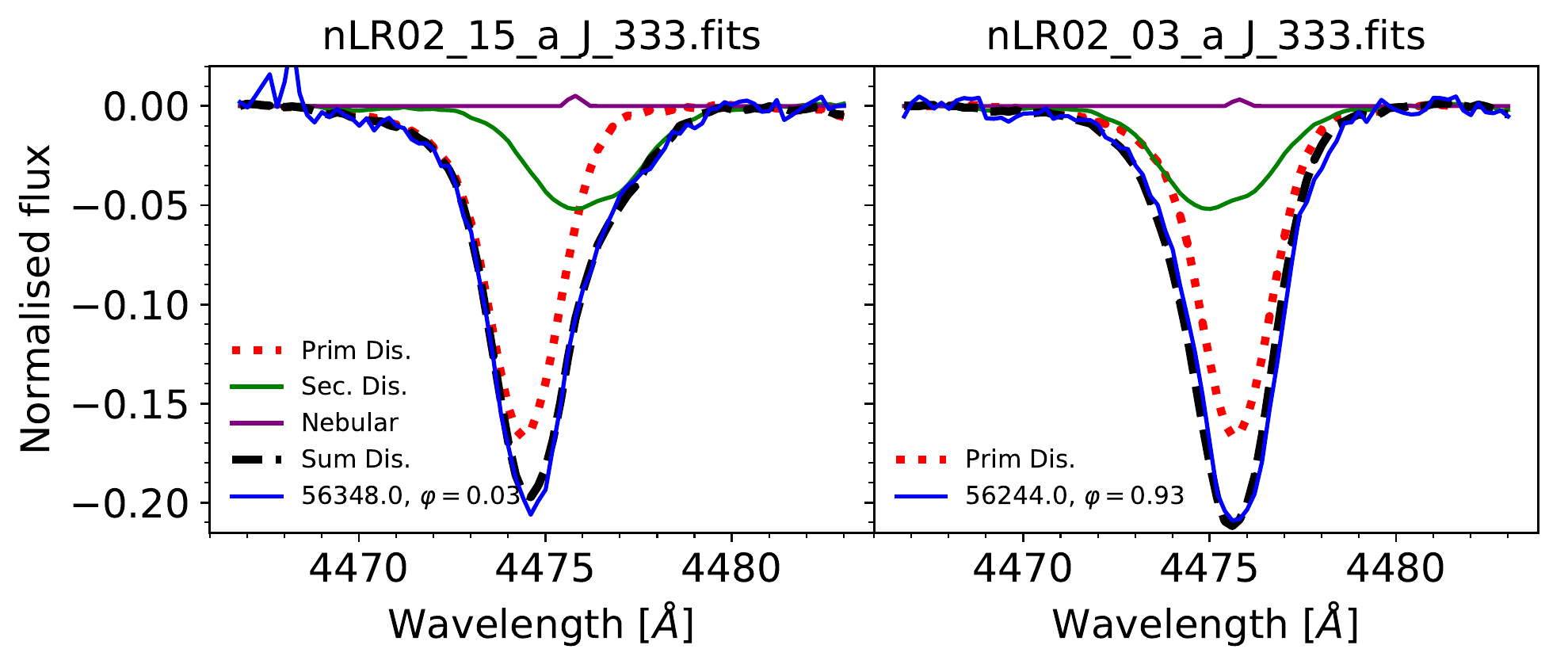}
\caption{Comparison of \HeI4471 spectra of \object{VFTS~333} at RV extremes, along with the disentangled spectra and their sum, as derived for the best-fitting $K_1, K_2$ values of  $43, 34\,$\kms, respectively. The disentangled spectra are not scaled by the light ratio in this plot.} \label{fig:VFTS333EXT}
\end{figure}

{\bf  \object{VFTS~333}, O9~II((f)) + O6.5~V:} is an eccentric long-period binary with $P=980\,$d and $e=0.75$,. Unlike  \object{VFTS~332}, the spectra offer a good coverage in Doppler space. A non-degenerate companion can be seen at the RV extreme of strong He\,{\sc i} lines (Fig.\,\ref{fig:VFTS333EXT}). Disentangling 
of the strong He\,{\sc ii} and He\,{\sc i} lines yields values between 20-40\,\kms~for $K_2$, and suggest that the secondary is relatively bright. We therefore resort to 2D disentangling. In all cases, $K_1$ is found to be of the order of 40\,\kms, with a weighted mean of  $K_1 = 43.0 \pm 2.5$ and $K_2 = 34\pm14\,$\kms. All measurements suggests that the object previously identified as the primary is in fact the less massive companion. We estimate it to contribute  42\% to the flux based on the strengths of the spectral lines.   Analysis of the OGLE light curve did not reveal significant frequencies. 


{\bf  \object{VFTS~350}, O8.5~V + O9.5~V} was thoroughly discussed in Sect.\,\ref{subsec:examples}. Analysis of the OGLE light curve does not reveal any significant frequencies.

\begin{figure}
\centering
\includegraphics[width=.5\textwidth]{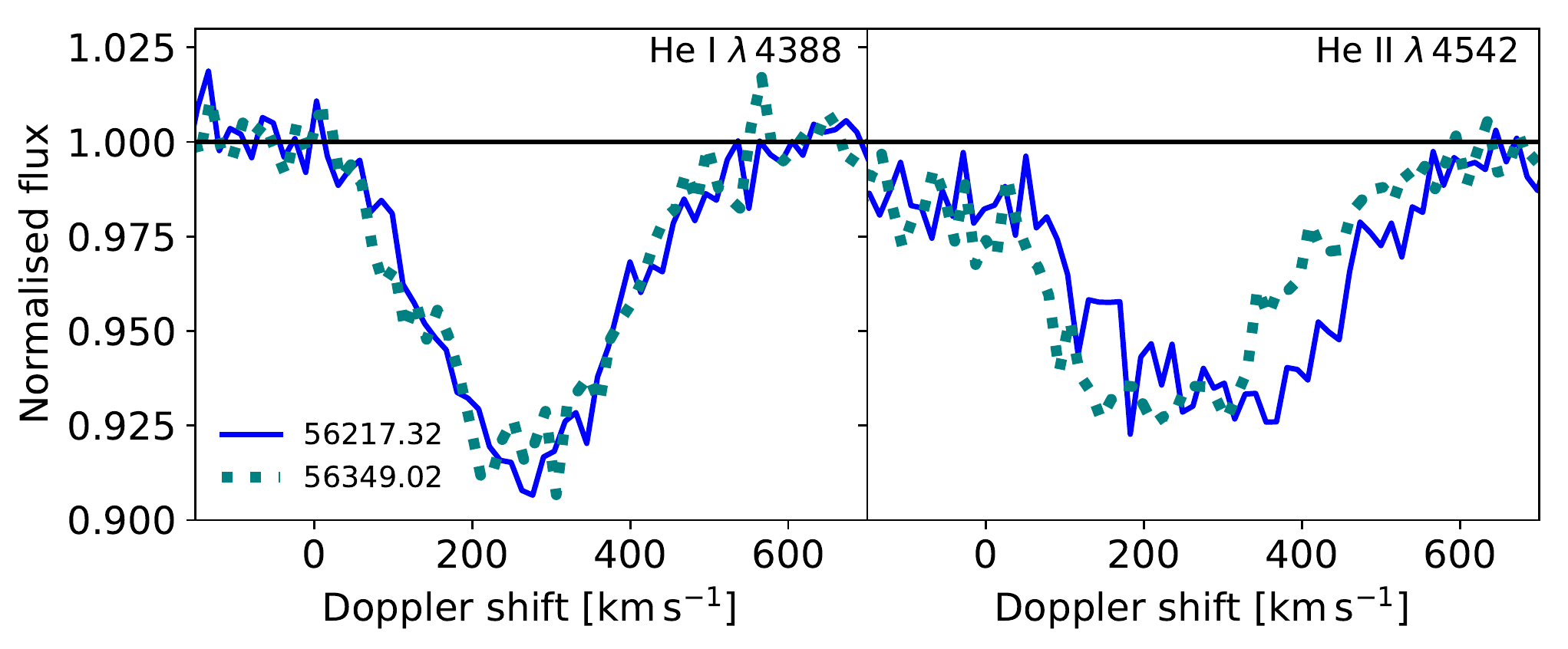}
\caption{Comparison between two spectra of VFTS~386 taken close to RV extremes (legend gives JD - 2400000). The motion of the He\,{\sc ii} lines show that they originate mainly in one star (the primary), while the apparent static behaviour of the He\,{\sc i} lines is the results of the anti-phase motion of both components, indicating the SB2 nature of the system.
} \label{fig:VFTS386Hecomp}
\end{figure}

\begin{figure}
\centering
\includegraphics[width=.5\textwidth]{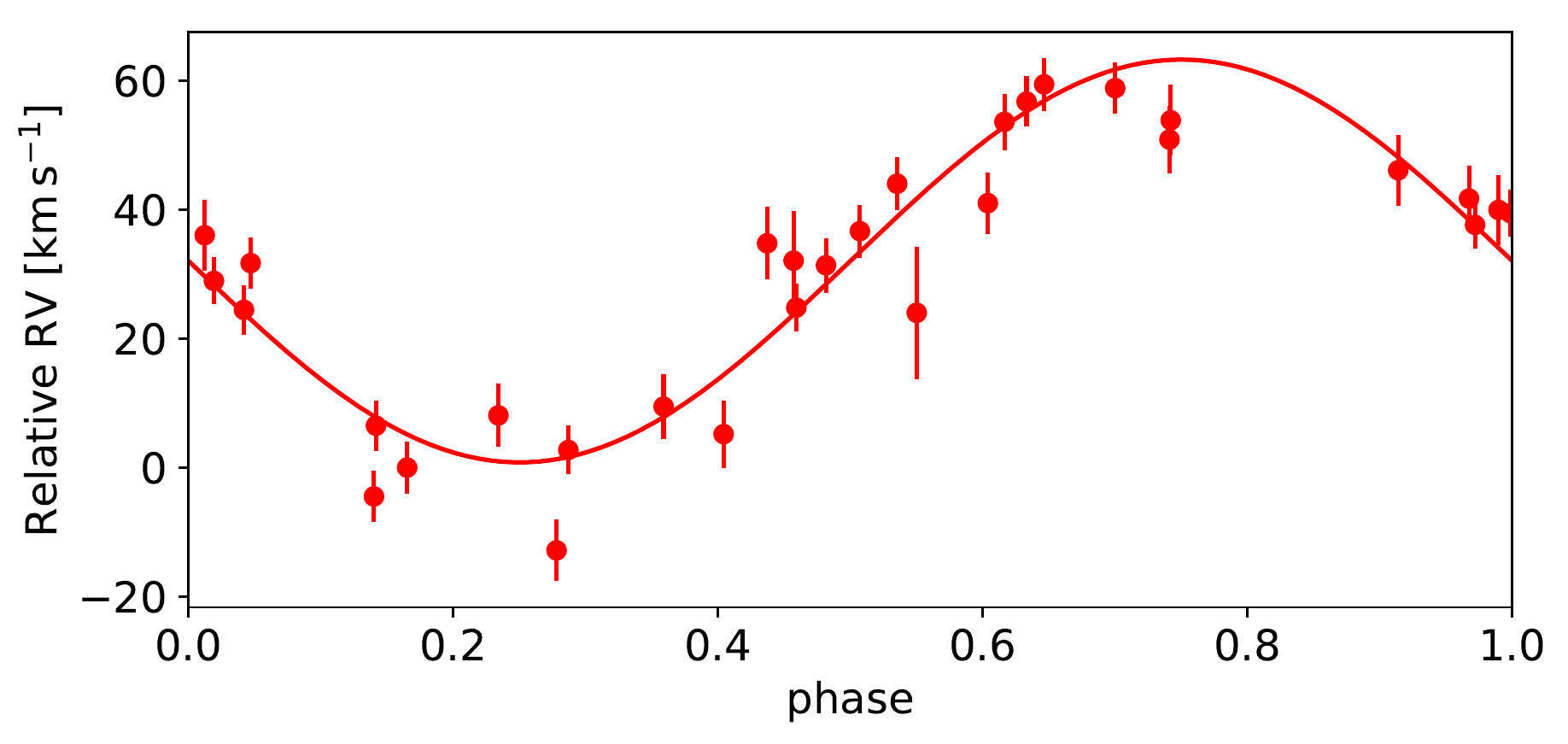}
\caption{Revised orbital solution for VFTS~386 using the He\,{\sc ii} lines only, with a reduced $\chi^2$ of 2.4. While the period is similar to that derived by \citet{Almeida2017}, the orbit is found to be circular, and the RV amplitude is more than doubled ($K_1 = 31\,$\kms~versus $K_1 = 14.3$\,\kms.)
} \label{fig:VFTS386Orbit}
\end{figure}

\begin{figure}
\centering
\includegraphics[width=.5\textwidth]{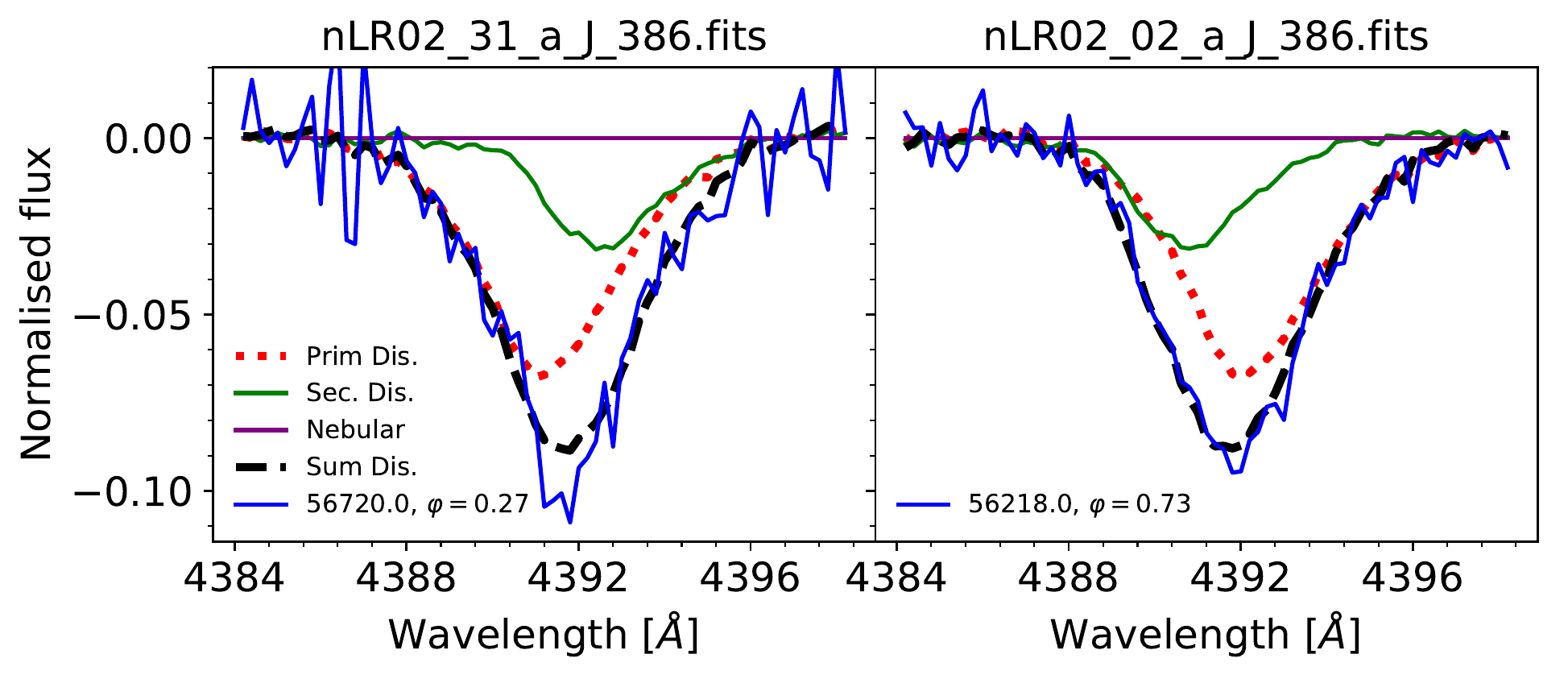}
\caption{Comparison of \HeI4388 spectra of \object{VFTS~386} at RV extremes, along with the disentangled spectra and their sum, as derived for the best-fitting $K_1, K_2$ values of  $31, 60\,$\kms, respectively. The disentangled spectra are not scaled by the light ratio in this plot.} \label{fig:VFTS386EXT}
\end{figure}

{\bf  \object{VFTS~386}, O9~V(n) + B1~V:} has a reported period of $P= 20\,$d, an eccentricity of $e= 0.25$, and a low-amplitude motion of $K_1 = 14.3\,$\kms~for the primary. However, the orbital solution established by \citet{Almeida2017} exhibits substantial scatter, with a reduced $\chi^2$ of 9.5. While the presence of a non-degenerate companion is not readily clear from the line profile variability, comparison of the variability in He\,{\sc i} and He\,{\sc ii} lines reveals that this binary hosts two non-degenerate companions (Fig\,\ref{fig:VFTS386Hecomp}). Given the low amplitude motion and the lack of significant variability in He\,{\sc i} lines, we do not proceed with 2D disentangling. Instead, we re-derive the orbital solution of the system by focusing only on the He\,{\sc ii} lines. The revised orbital solution yields a reduced $\chi^2$ of 2.4 (Fig.\,\ref{fig:VFTS386Orbit}). The orbital parameters are then $P = 20.45 \pm 0.02\,$d, $T_0 = 54816.5 \pm 2.8$ [JD - 2400000], $K_1 = 30.8 \pm 2.1$\,\kms, and $e=0.15 \pm 0.05$.
We then proceed with 1D disentangling of the He\,{\sc i} lines, fixing the orbital parameters derived above. The best constraints are obtained from the \HeI4388 (Fig.\,\ref{fig:VFTS386EXT}) and \HeI4471 lines, which yield $K_2 = 64 \pm 12\,$\kms. The spectral type of the secondary corresponds to a B1 star, and we estimate its light contribution to be $16$\%. 
 Analysis of the OGLE light curve did not yield significant periodicities.

\begin{figure}
\centering
\includegraphics[width=.5\textwidth]{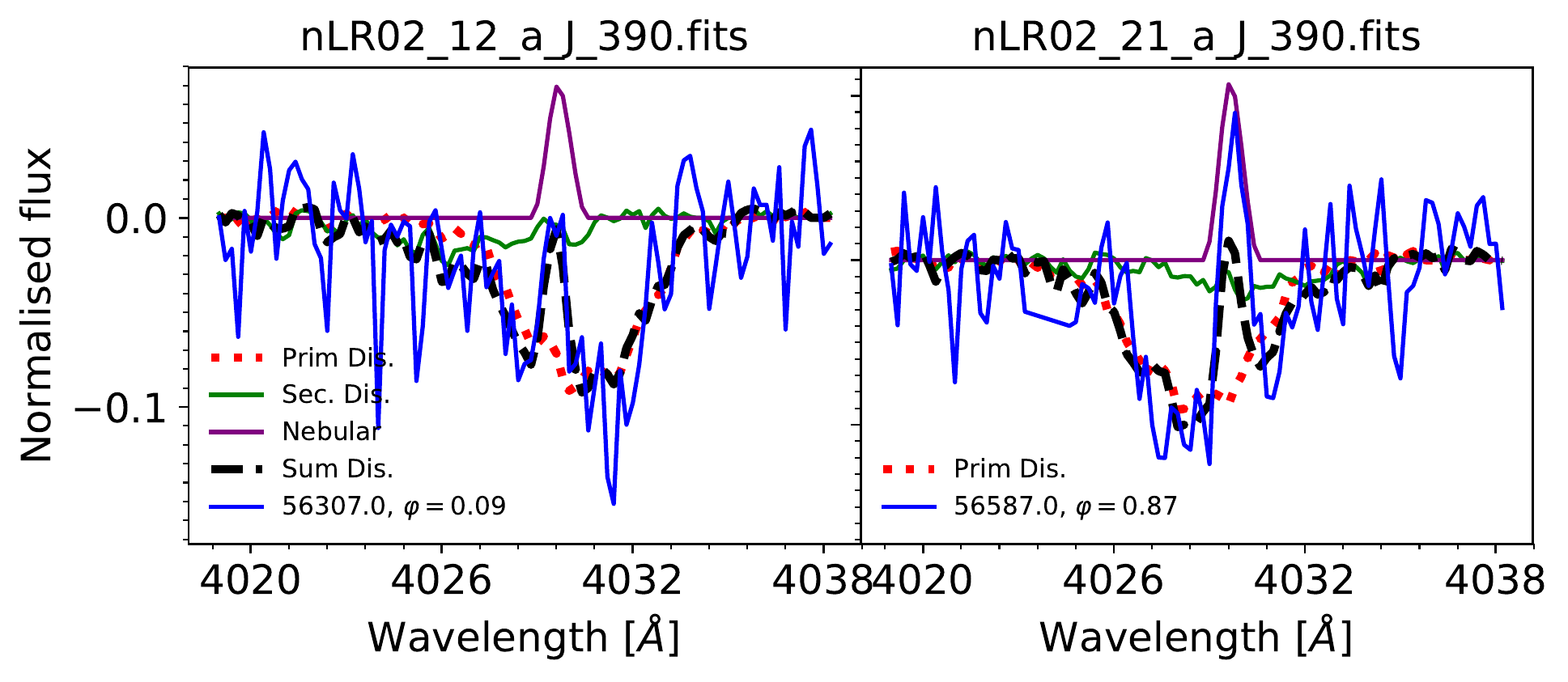}
\caption{Comparison of \HeI4388 spectra of \object{VFTS~386} at RV extremes, along with the disentangled spectra and their sum, as derived for the best-fitting $K_1, K_2$ values of  $70, 137\,$\kms, respectively. The disentangled spectra are not scaled by the light ratio in this plot.} \label{fig:VFTS390EXT}
\end{figure}

\begin{figure}
\centering
\includegraphics[width=.5\textwidth]{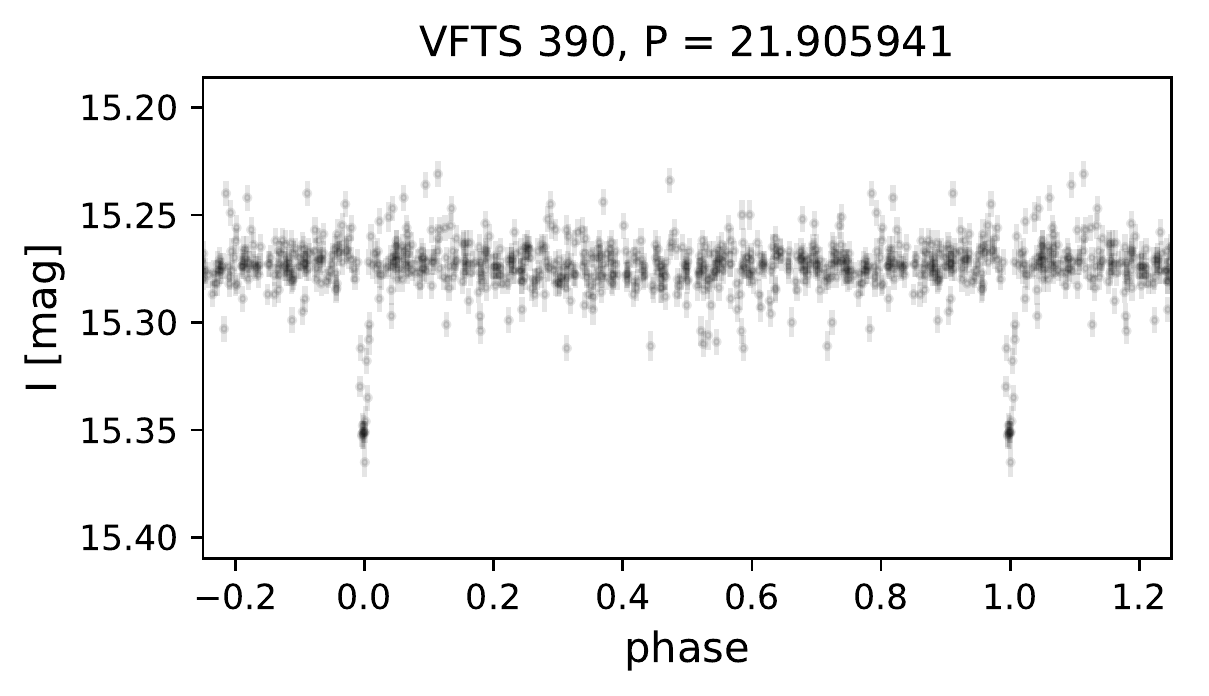}
\caption{OGLE I-band light curve of  the binary \object{VFTS~390}, phased with the derived period. The single eclipse occurs when the hotter primary eclipses the cooler secondary. } \label{fig:VFTS390OGLE}
\end{figure}

{\bf  \object{VFTS~390}, O5.5~V:((fc)) + O9.7:~V:} has a reported period of 22\,d and an  eccentricity of $e=0.49$. Disentangling of the spectra proved to be very challenging due to very strong nebular-line contamination and the fact that the secondary appears to possess very weak lines (Fig.\,\ref{fig:VFTS390EXT}). While disentangling of various He\,{\sc i} lines supports $K_2$ values in the range $100-250\,$\kms, the errors are very large. A weighted mean of the RVs obtained for \HeI4026 and \HeI4388 yields $K_2 = 137 \pm 34\,$\kms, but the nebular emission and very low S/N may entail systemic errors. The secondary's spectrum matches a O9.7~V star, and has an estimated light contribution of $15\%$. 

Interestingly, the system shows a single eclipse (Fig.\,\ref{fig:VFTS390OGLE}), which substantiates beyond doubt the presence of a non-degenerate companion in the system, and allows us to slightly refine the ephemeris. From the orbital configuration, the observed eclipse occurs when the hotter, brighter  primary eclipses the cooler, fainter secondary. Due to the orbital configuration, a second eclipse is not seen.

\begin{figure}
\centering
\includegraphics[width=.5\textwidth]{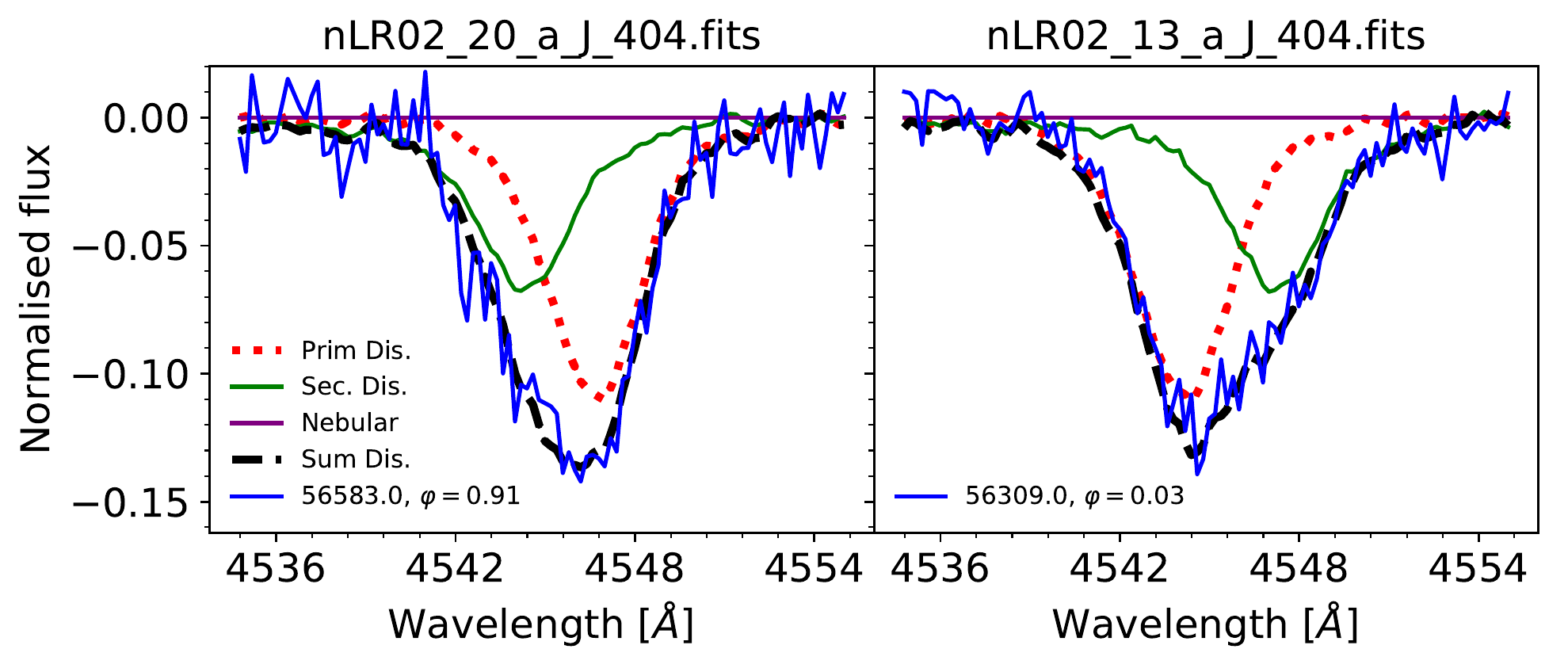}
\caption{Comparison of \HeII4542 spectra of \object{VFTS~404} at RV extremes, along with the disentangled spectra and their sum, as derived for the best-fitting $K_1, K_2$ values of  $98, 106\,$\kms, respectively. The disentangled spectra are not scaled by the light ratio in this plot.} \label{fig:VFTS404EXT}
\end{figure}

{\bf  \object{VFTS~404}, O3.5: V:((fc)) + O5~V:} has a reported period of $P=146\,$d and an eccentricity of $e=0.72$.
\citet{Almeida2017} noted the presence of a non-degenerate companion, but could not derive its RVs. Indeed, the secondary is clearly seen in both He\,{\sc i} and He\,{\sc ii} lines (Fig.\,\ref{fig:VFTS404EXT}). Given the significant contribution of the secondary to these lines and the fact that the components are constantly blended, it is very likely that the RVs measured by \citet{Almeida2017} for the primary were strongly impacted by the secondary. Therefore, we utilise 2D disentangling for this system. For this, we use the He\,{\sc ii} lines, which are free of nebular contamination and are stronger than the He\,{\sc i} lines. A weighted mean yields $K_1 = 98 \pm 11\,$\kms~and $K_2 = 106 \pm 11\,$\kms. We note that the $K_1$ value derive here is roughly three times larger than derived by \citet{Almeida2017}, which is a result of the secondary's anti-phase motion that leads to an apparent lower amplitude of motion. The companion is estimated to contribute 42\% to the visual flux, and is classified O5~V. Analysis of the OGLE light curve did not reveal significant frequencies. We note that VFTS~404 is one of the four targets for which X-rays are detected, and exhibits the highest X-ray luminosity in our sample ($\log L_X = 32.84\,$[\ergs]). This luminosity lies slightly above the canonical value of $L_X / L \approx -7$, perhaps due to the presence of colliding winds in the system.

\begin{figure}
\centering
\includegraphics[width=.5\textwidth]{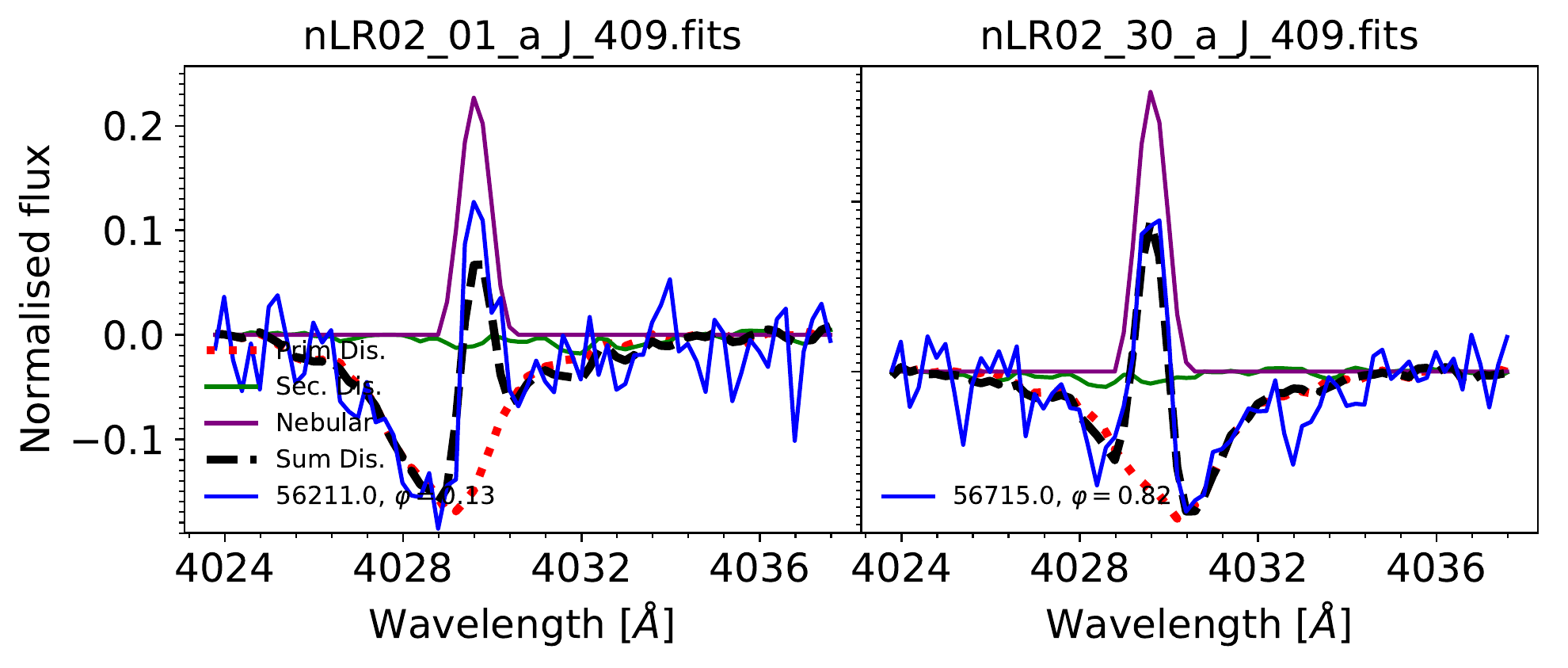}
\caption{Comparison of \HeII4542 spectra of \object{VFTS~409} at RV extremes, along with the disentangled spectra and their sum, as derived for the $K_1 = 43\,$\kms~and the adopted value of $K_2 = 3\times K_1 = 130\,$\kms. The disentangled spectra are not scaled by the light ratio in this plot.} \label{fig:VFTS409EXT}
\end{figure}

{\bf  \object{VFTS~409}, O3.5:~V:((f)) + B:} has a reported period of 22\,d and an eccentricity of $e=0.29$. Unfortunately, the spectra are extremely contaminated by nebular lines that vary in strength, making only the He\,{\sc ii} lines available for a reliable analysis. Disentangling of the He\,{\sc ii} lines reveals no features. Disentangling of the He\,{\sc i} lines reveals no clear minimum for $K_2$ is retrieved, with values in the range $20-300\,$\kms~all yielding acceptable an $\chi^2$. The disentangled spectrum shows faint signatures in He\,{\sc i} lines (Fig.\,\ref{fig:VFTS409EXT}), though nebular contamination questions the validity of these results. If these signatures are real, then the light contribution of the secondary is $\approx 0.05$. We tentatively classify the secondary as B:, and the binary as SB2:. With a minimum mass of $M_2 > 7.6\pm1.1\,M_\odot$, the secondary could be an early B-type star, a helium star, or a BH. Analysis of the OGLE light curve did not reveal significant frequencies.

\begin{figure}
\centering
\includegraphics[width=.5\textwidth]{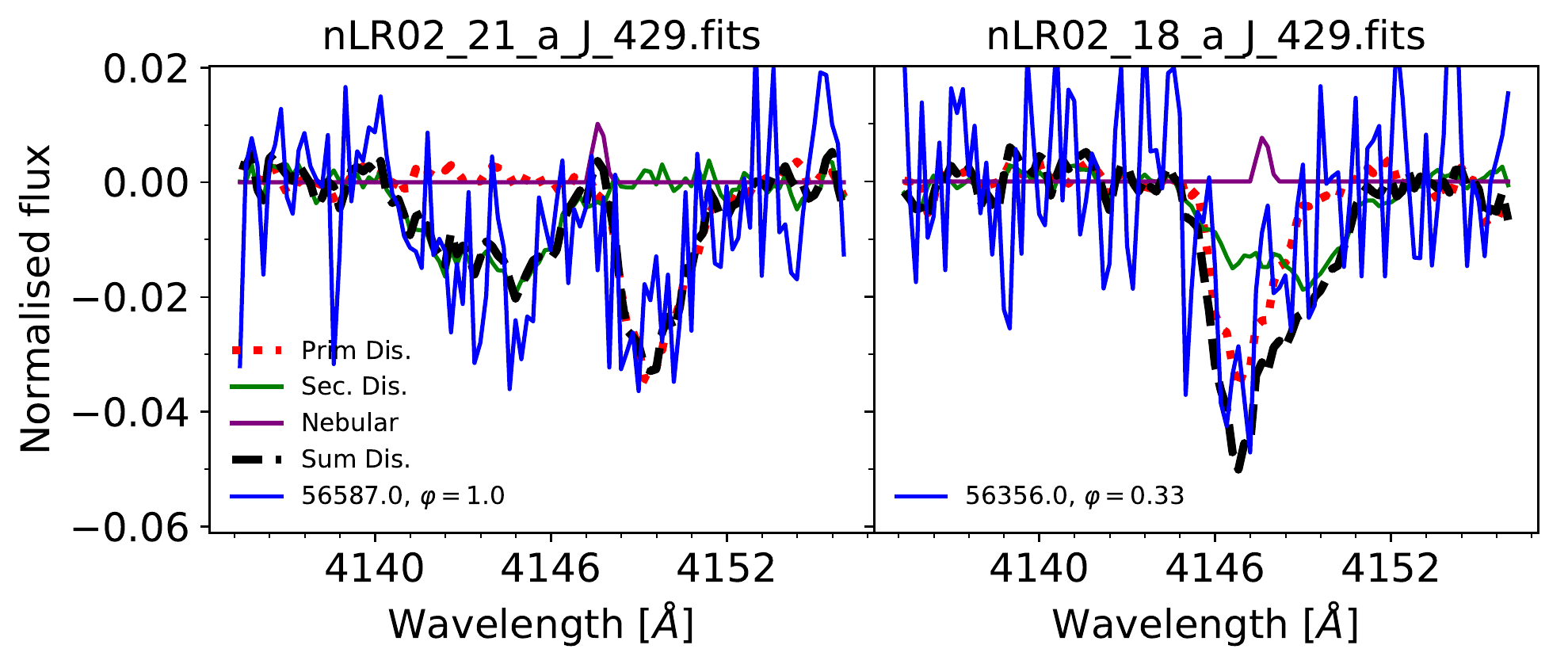}
\caption{Comparison of \HeII4144 spectra of \object{VFTS~429} at RV extremes, along with the disentangled spectra and their sum, as derived for the best-fitting $K_1, K_2$ values of  $92, 143\,$\kms, respectively. The disentangled spectra are not scaled by the light ratio in this plot.} \label{fig:VFTS429EXT}
\end{figure}

\begin{figure}
\centering
\includegraphics[width=.5\textwidth]{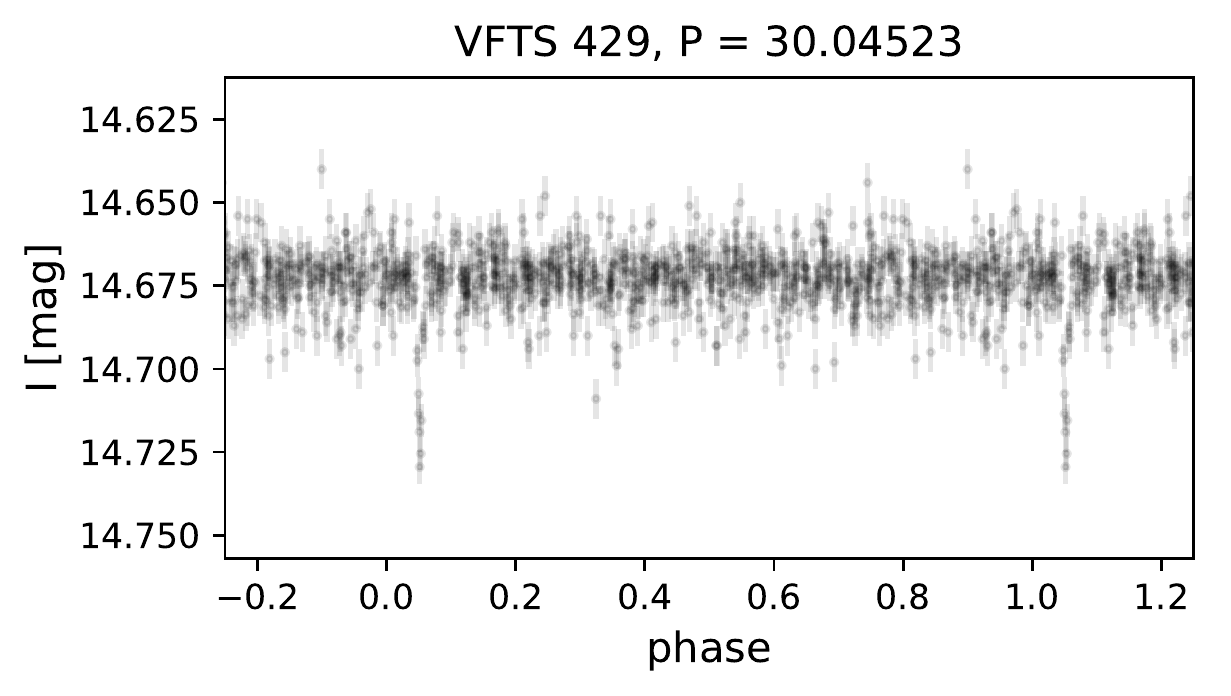}
\caption{OGLE I-band light curve of  the binary \object{VFTS~429}, phased with the derived period. The single eclipse occurs when the cooler secondary eclipses the hotter primary. } \label{fig:VFTS429OGLE}
\end{figure}

{\bf  \object{VFTS~429}, O7~V: + B1:~V:} has a reported period of $P=30\,$d and an eccentricity of $e=0.56$. This system is clearly SB2, as is readily seen from the spectral variability of strong He\,{\sc i} lines. However, the spectra are strongly contaminated by nebular lines, making the derivation of $K_2$ challenging. The least contaminated line clearly showing both stars isolated is the \HeI4144 line (Fig.\,\ref{fig:VFTS429EXT}). A weighted mean of the strong He\,{\sc i} lines (with exception of \HeI4471 due to very strong nebular contamination) yields $K_2 = 143 \pm 27$\,\kms.  Judging by its spectral appearance, we classify the secondary as B1:~V:, and estimate its light contribution in the visual at 10\%.

Similar to VFTS~409, phasing the OGLE light curve of VFTS~429 with its spectroscopic ephemeris reveals a single eclipse (Fig.\,\ref{fig:VFTS429OGLE}). We used it to slightly refine the orbital period. Unlike VFTS~409 however, the eclipse in VFTS~429 occurs when the cooler B0~V secondary eclipses the hotter O7.5~V primary.

\begin{figure}
\centering
\includegraphics[width=.5\textwidth]{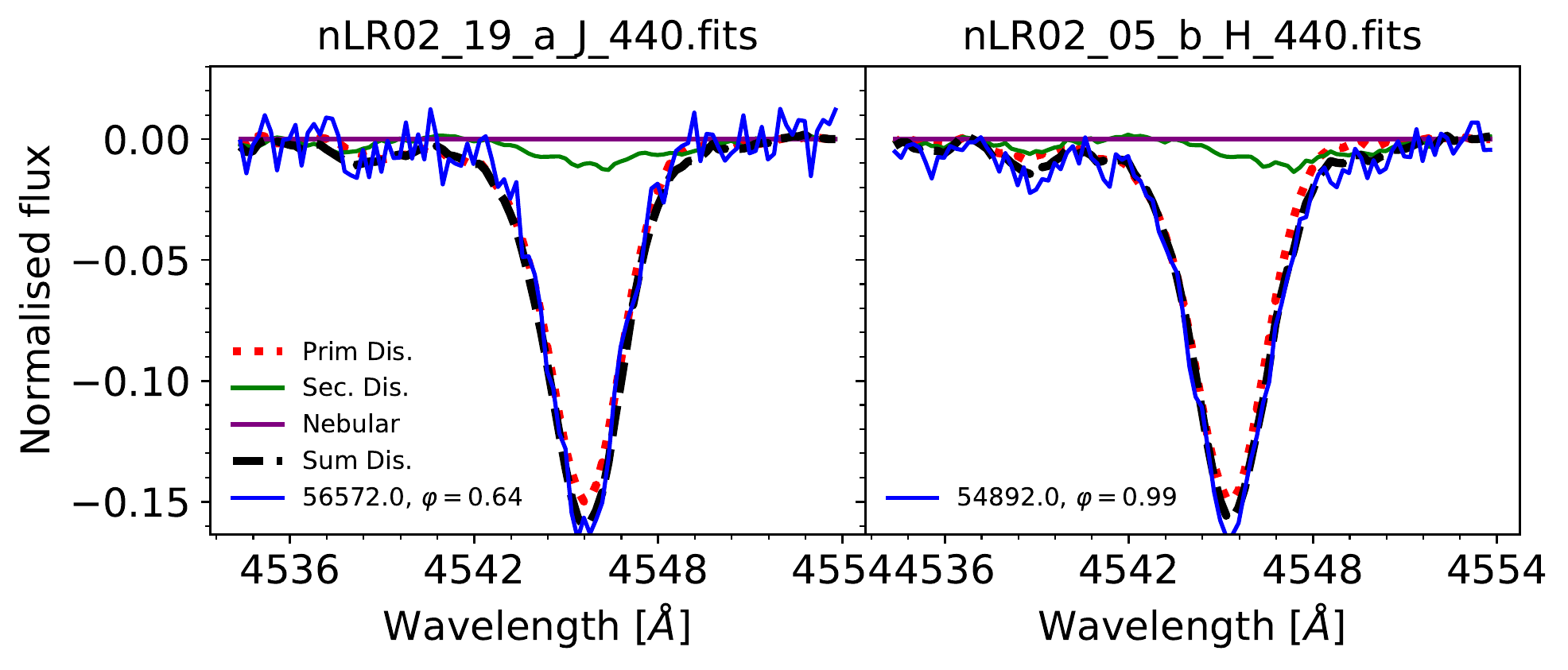}
\caption{Comparison of \HeII4542 spectra of \object{VFTS~440} at RV extremes, along with the disentangled spectra and their sum, as derived for the best-fitting $K_1, K_2$ values of  $11.6, 29\,$\kms, respectively. The disentangled spectra are not scaled by the light ratio in this plot.} \label{fig:VFTS440EXT}
\end{figure}

{\bf  \object{VFTS~440}, O6: V:(f) + O8~V}, is a long-period binary with $P=  1019\,$d and $e = 0.28$. The low RV amplitude of the primary ($K_1 = 11.6\,$\kms) and the strong nebular contamination make the disentangling of this binary challenging. We first implemented the 2D disentangling of the He\,{\sc ii} lines to ensure that the low RV amplitude is not impacted by a secondary contributing significantly to the flux. We find $K_1$ that agree well with the value derived by \citet{Almeida2017}, and we therefore fix it and continue with 1D disentangling across the $K_2$ axis. While not readily apparent, careful inspection of the \HeII4542 line at RV extremes suggests the presence of a faint secondary (Fig.\,\ref{fig:VFTS440EXT}). Disentangling of the He\,{\sc ii} lines results in poor constraints on $K_2$, since all values in the range [0,100]\,\kms result in acceptable $\chi^2$ values (within $1\sigma$). However, disentangling suggests that the \HeI4009 line is present only in the secondary. We therefore use this line for disentangling as well. A weighted mean of the poorly constrained He\,{\sc ii} measurements and the \HeI4009 line yields $K_2 =  29 \pm 18\,$\kms. The RV amplitude of the secondary is thus poorly constrained.  We  classify the companion as O8~V and estimate it contributes 16\% to the total flux. Analysis of the OGLE light curve did not reveal significant frequencies.

\begin{figure}
\centering
\includegraphics[width=.5\textwidth]{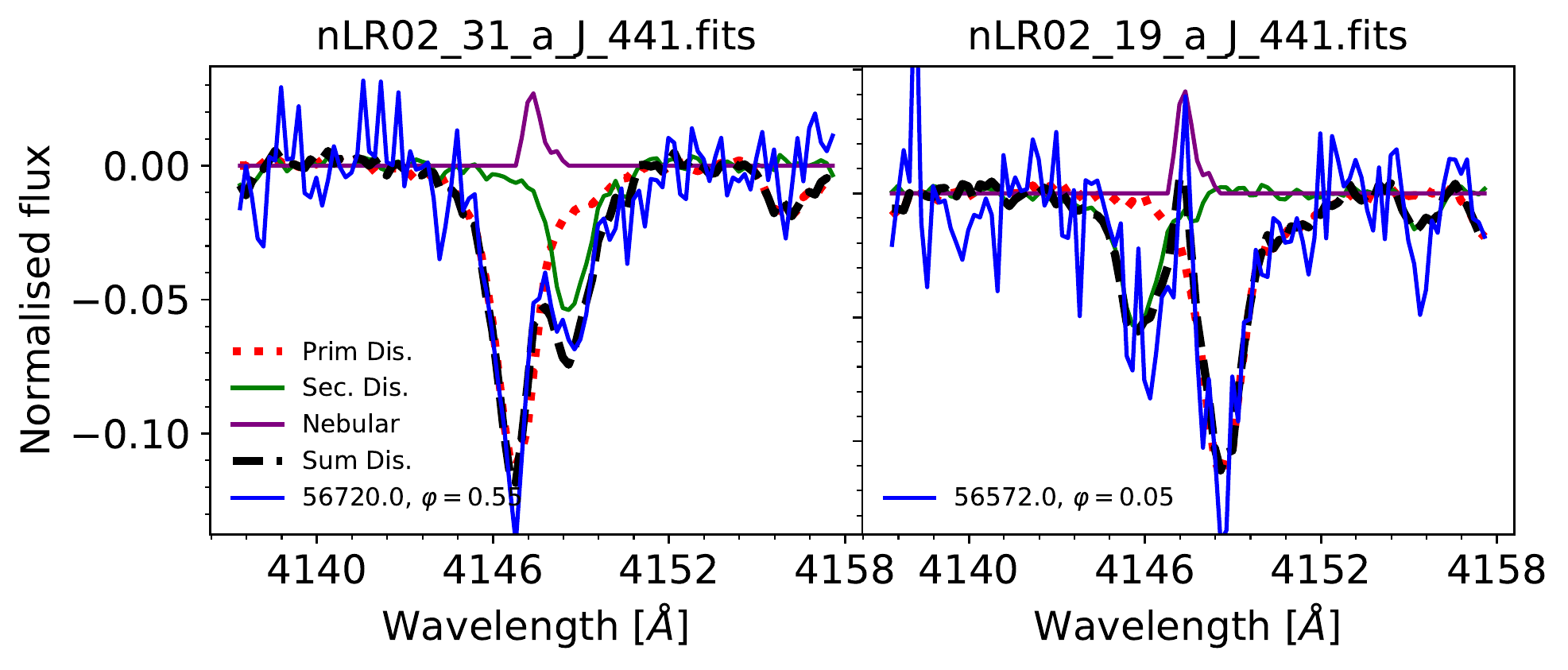}
\caption{Comparison of \HeI4144 spectra of \object{VFTS~441} at RV extremes, along with the disentangled spectra and their sum, as derived for the best-fitting $K_1, K_2$ values of  $73, 108\,$\kms, respectively. The disentangled spectra are not scaled by the light ratio in this plot.} \label{fig:VFTS441EXT}
\end{figure}

{\bf  \object{VFTS~441}, O9.2~V + B0.5~V} has a reported period and eccentricity of $P=6.9\,$d and $e=0.22$.  Inspection of the He\,{\sc i} lines unambiguously reveals this system as an SB2 binary, albeit the strong nebular contamination of the He\,{\sc i} lines makes the disentangling challenging. Because of the presence of a non-negligible secondary, we implement the 2D disentangling technique. The line least contaminated by nebular lines is \HeI4144, shown in Fig.\,\ref{fig:VFTS441EXT} at RV extremes. A weighted mean of all He\,{\sc i} lines but \HeI4471 (due to dominating nebular contamination) yields $K_1 = 73 \pm 5$ and $K_2 = 108\pm 8\,$\kms.  The disentangled spectrum of the secondary, estimated to contribute 21\% to the visual flux, shows no evidence for He\,{\sc ii} absorption. Its spectral appearance matches a B0.5 spectral type. Despite of the short orbital period, analysis of the OGLE light curve of VFTS~441 did not reveal significant periods.

\begin{figure}
\centering
\includegraphics[width=.5\textwidth]{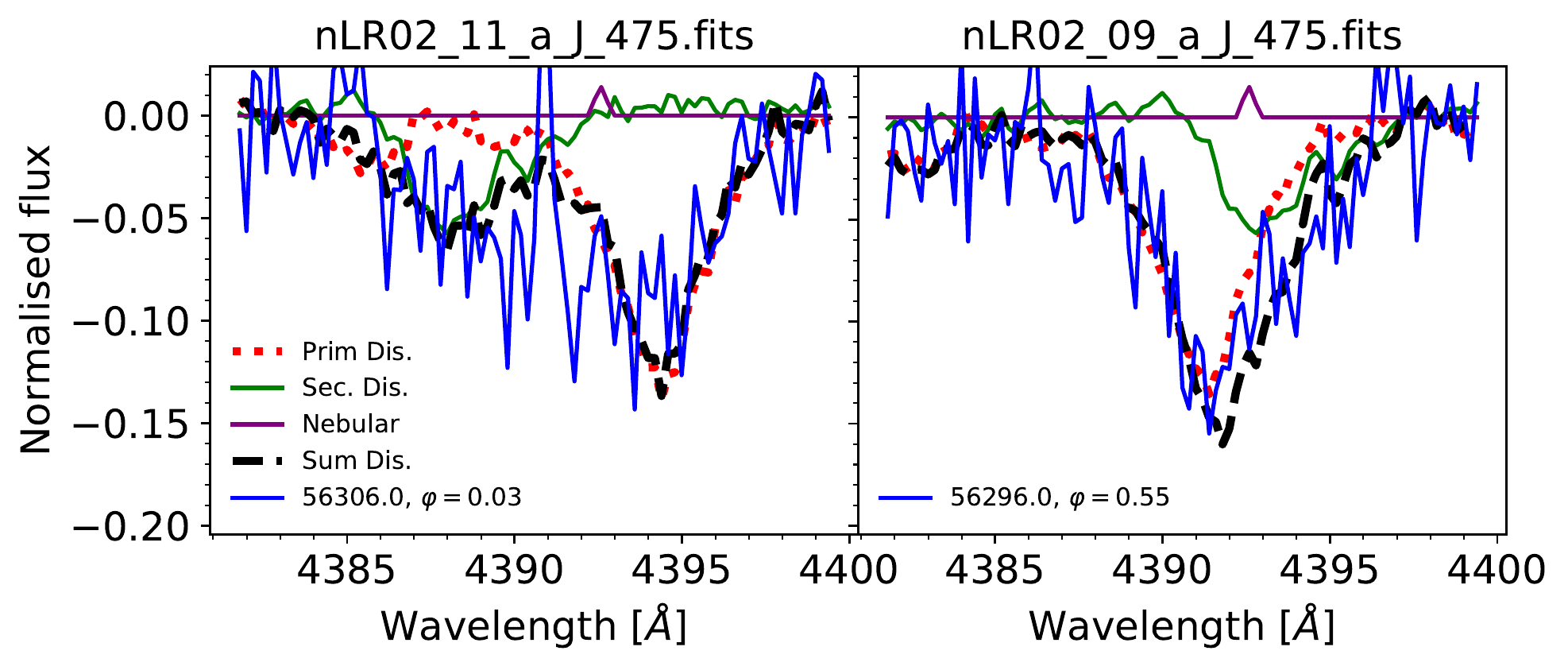}
\caption{Comparison of \HeI4388 spectra of \object{VFTS~475} at RV extremes, along with the disentangled spectra and their sum, as derived for the best-fitting $K_1, K_2$ values of  $92, 207\,$\kms, respectively. The disentangled spectra are not scaled by the light ratio in this plot.} \label{fig:VFTS475EXT}
\end{figure}

\begin{figure}
\centering
\includegraphics[width=.5\textwidth]{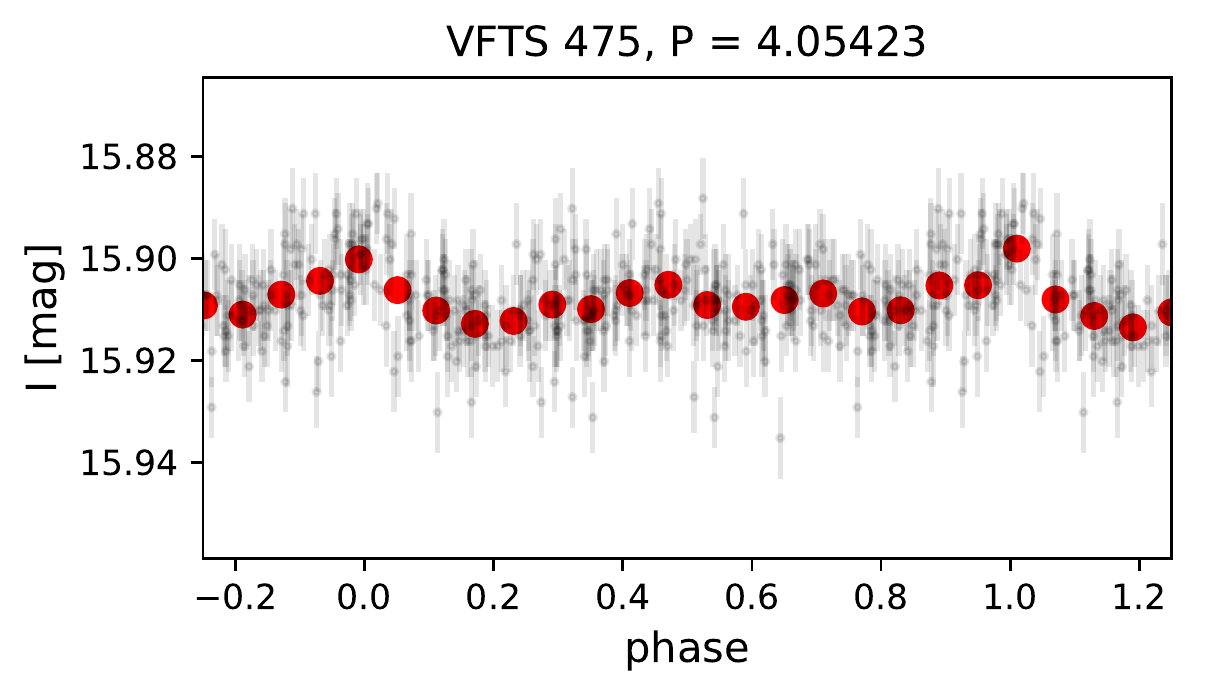}
\caption{OGLE I-band light curve of  the binary \object{VFTS~475}, phased with the orbital period of $P=4.0542\,$d. Only OGLE IV data are shown, as OGLE III data appear to be contaminated.} \label{fig:VFTS475OGLE}
\end{figure}

{\bf  \object{VFTS~475}, O9.7~V + B0~V} has a reported period of $P=4.1\,$d and an eccentricity of $e=0.57$. The spectral variability of He\,{\sc i} lines implies the presence  of a non-degenerate secondary in the system (Fig.\,\ref{fig:VFTS475EXT}), and we therefore implement 2D disentangling. Plausible constraints are obtained with the \HeI4388 and \HeI4475 lines, albeit with substantial errors due to the low S/N of the data. A weighted mean of the measurements yields $K_1 = 135 \pm 33$, $K_2 =169 \pm 58$\,\kms.  The \HeI4026 line results in low $K_2$ amplitudes, which do not agree with the other lines. The reason for this discrepancy is not clear, and could imply that the system is a higher-order multiple. The spectral appearance of the disentangled spectrum off the secondary matches a B0~V spectral type, and we estimate its light contribution at 35\%. The OGLE light curve, phased with the orbital period, shows a clear periodic signature, potentially originating in ellipsoidal variations and irradiation effects (Fig.\,\ref{fig:VFTS475OGLE}).

\begin{figure}
\centering
\includegraphics[width=.5\textwidth]{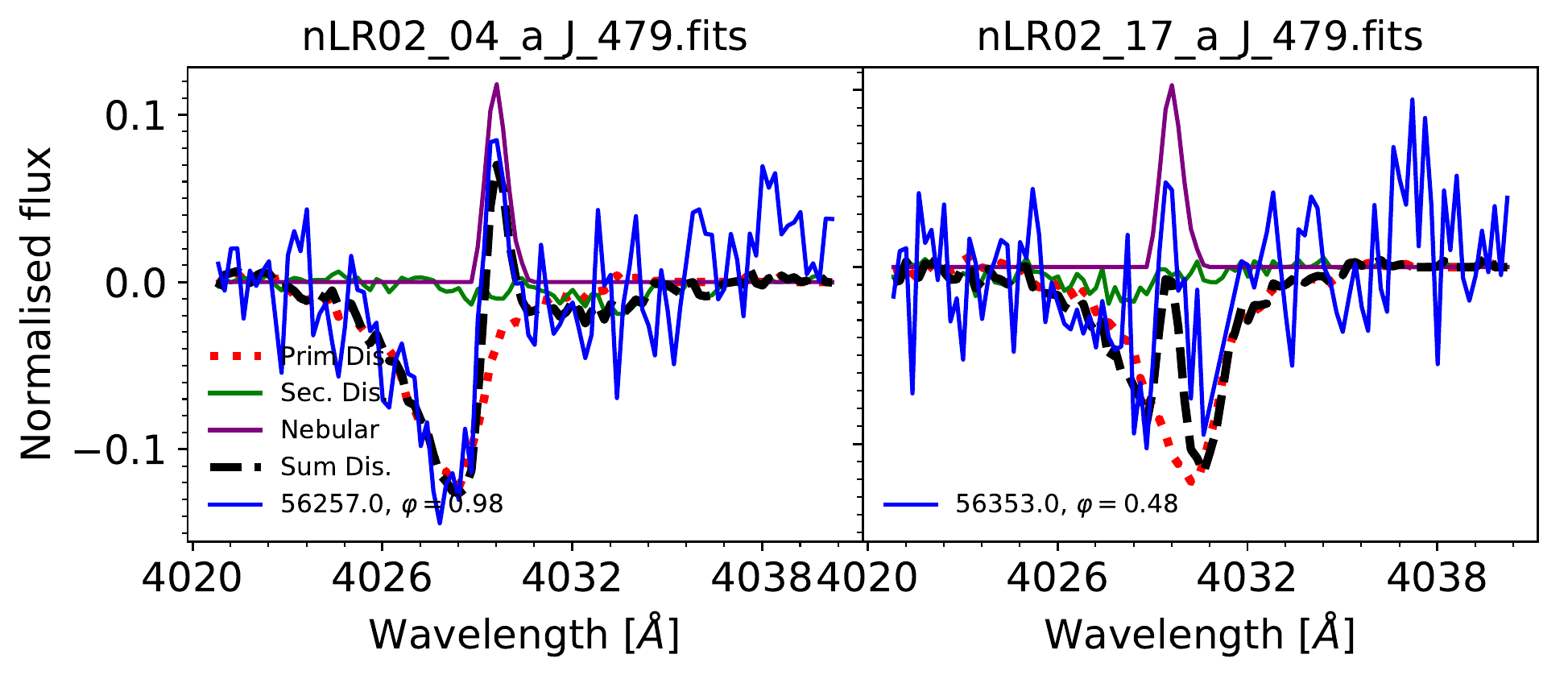}
\caption{Comparison of \HeI4026 spectra of \object{VFTS~479} at RV extremes, along with the disentangled spectra and their sum, as derived for the best-fitting $K_1, K_2$ values of  $73, 238\,$\kms, respectively. The disentangled spectra are not scaled by the light ratio in this plot.} \label{fig:VFTS479EXT}
\end{figure}

\begin{figure}
\centering
\includegraphics[width=.5\textwidth]{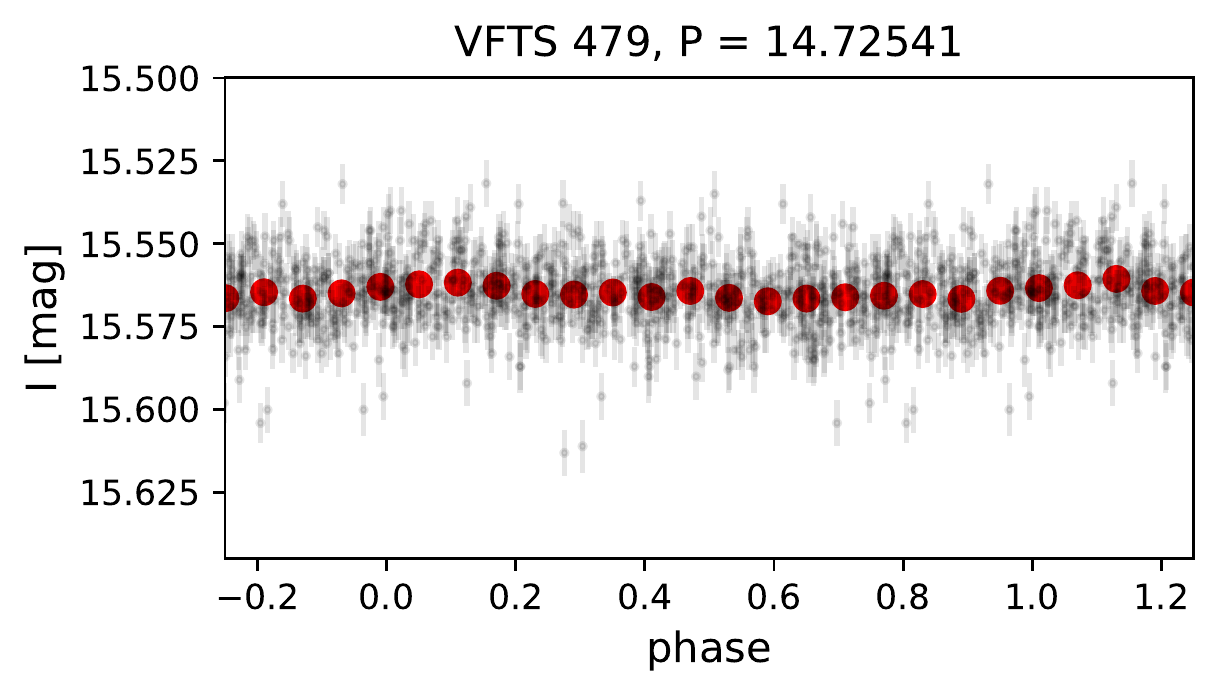}
\caption{OGLE I-band light curve of  the binary \object{VFTS~479}, phased with the orbital period. } \label{fig:VFTS479OGLE}
\end{figure}

{\bf  \object{VFTS~479}, O4.5~V((fc)) + B:} has a reported period of $P=14.7\,$d and an eccentricity of $e=0.31$. The spectra are very noisy and contaminated by nebular emission, such that disentangling of the system is challenging (Fig.\,\ref{fig:VFTS479EXT}). Disentangling of He\,{\sc i} lines results in $K_2$ values of the order of 200\,\kms~with very large errors. A weighted mean of the measurements obtained for the \HeI4026, \HeI4144, and \HeI4388 lines yields $K_2 = 238 \pm 96$\,\kms, but given the large error, we refrain from providing this value. The disentangled spectrum of the secondary exhibits He\,{\sc i} features, but no He\,{\sc ii} features. We therefore tentatively classify this system as SB2:, and the secondary as B:, although we cannot exclude that the nebular lines may bias our interpretation.  We estimate the light contribution of the secondary at $\approx 5\%$. With a minimum mass of $M_2 > 12.9^{+4.7}_{-3.8}\,M_\odot$, the secondary is likely a B0 star, though we cannot fully rule out a BH due to the nebular contamination. The OGLE light curve, phased with the orbital period, is shown in Fig.\,\ref{fig:VFTS479OGLE}, showing a very faint periodic signature. 

\begin{figure}
\centering
\includegraphics[width=.5\textwidth]{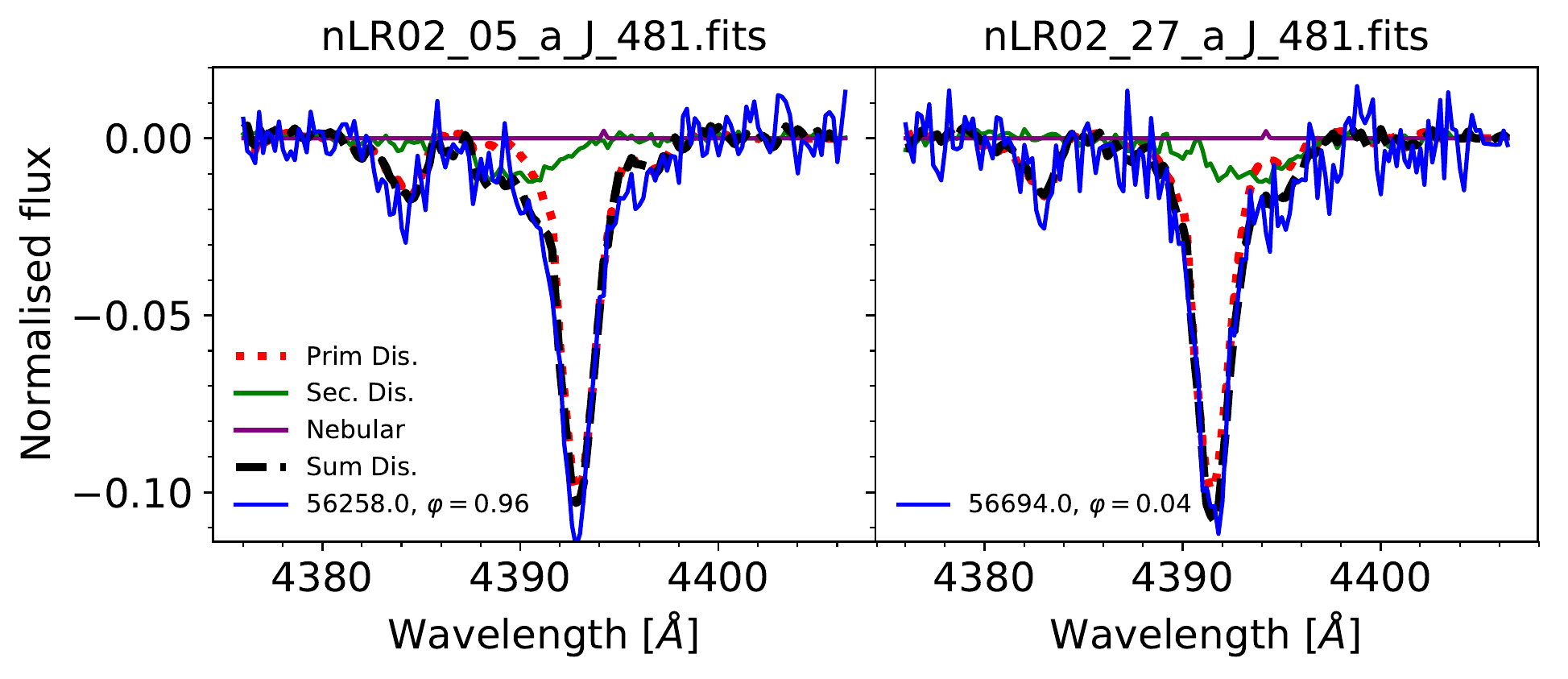}
\caption{Comparison of \HeI4026 spectra of \object{VFTS~481} at RV extremes, along with the disentangled spectra and their sum, as derived for the best-fitting $K_1, K_2$ values of  $73, 303\,$\kms, respectively. The disentangled spectra are not scaled by the light ratio in this plot.} \label{fig:VFTS481EXT}
\end{figure}

{\bf  \object{VFTS~481}, O8.5~V + O9.7:~V:} has a reported period of $P=142\,$d and a  high eccentricity of $e=0.93$. The presence of a non-degenerate companion is barely seen in the wings of strong He\,{\sc i} lines such as \HeI4388 (Fig.\,\ref{fig:VFTS481EXT}). Disentangling supports the presence of a non-degenerate companion. The $K_2$ value is difficult to constrain given the faintness of the companion. A weighted mean of the strong He\,{\sc i} lines yields $K_2=303\pm50\,$\kms. The corresponding disentangled spectrum approx.\ matches an O9.7~V star, with an estimated light contribution of 10\%.  The OGLE light curve does not reveal any significant periods.

\begin{figure}
\centering
\includegraphics[width=.5\textwidth]{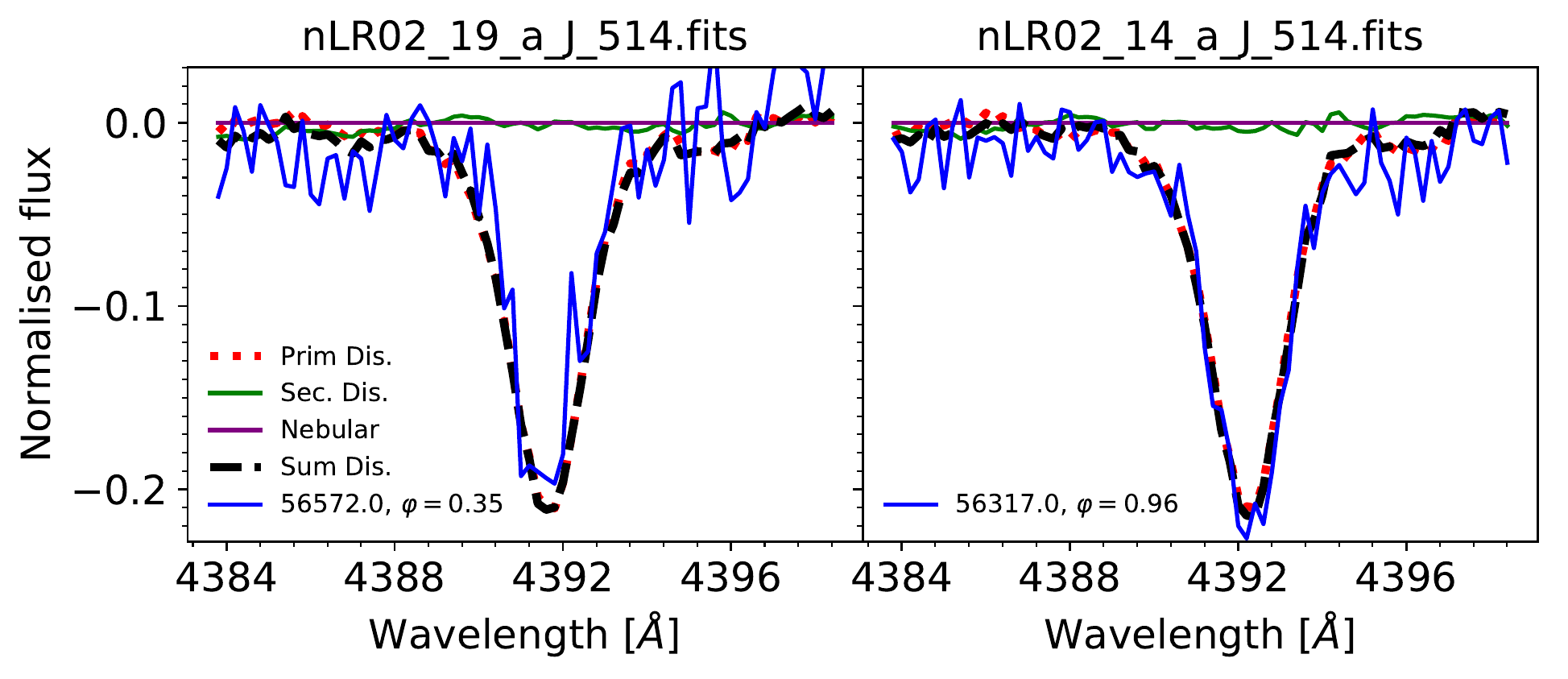}
\caption{Comparison of \HeI4388 spectra of \object{VFTS~514} at RV extremes, along with the disentangled spectra and their sum, as derived for $K_1 = 23\,$\kms~\citep{Almeida2017} and $K_2 = 2\,K_1 = 46\,$\kms. } \label{fig:VFTS514EXT}
\end{figure}

{\bf  \object{VFTS~514}, O9.7~V,} has a reported period of $P=185\,$d and eccentricity of  $e=0.41$. No clear minimum for $K_2$ is obtained, regardless of the disentangled line. The spectrum of the secondary appears to be featureless (Fig.\,\ref{fig:VFTS514EXT}), with exception of the Balmer lines, which are strongly contaminated by nebular lines and are therefore not considered. We can rule out a companion contributing more than $\approx 5\%$ to the visual light, corresponding roughly to B3~V. With a minimum mass of $M_2 > 5.3\pm1.8\,M_\odot$, we cannot rule out that the secondary is a faint non-degenerate star, but given the high BH probability (Table\,\ref{tab:SampleFin}), we consider VFTS~514 an O+BH candidate. The OGLE light curve did not reveal any significant frequencies.

\begin{figure}
\centering
\includegraphics[width=.5\textwidth]{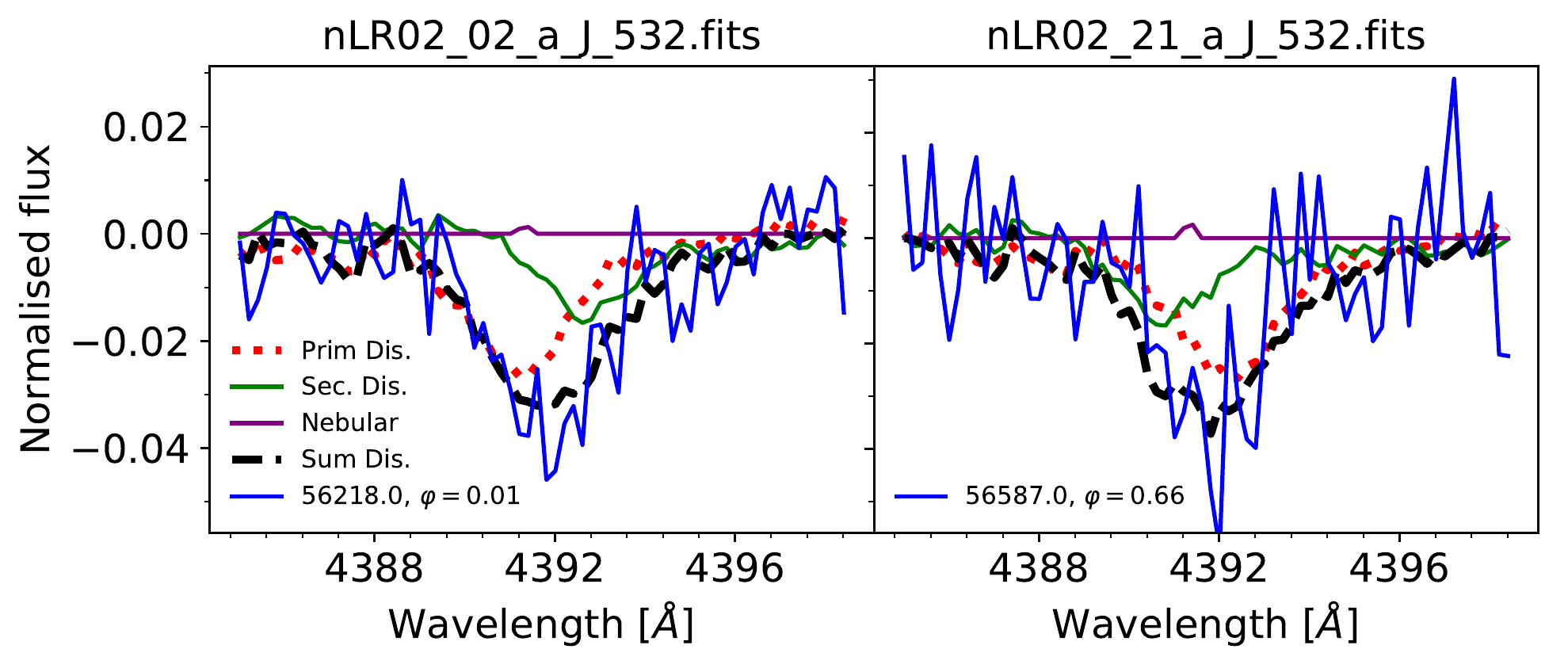}
\caption{Comparison of \HeI4388 spectra of \object{VFTS~532} at RV extremes, along with the disentangled spectra and their sum, as derived for the best-fitting $K_1, K_2$ values of  $34, 81\,$\kms, respectively. The disentangled spectra are not scaled by the light ratio in this plot.} \label{fig:VFTS532EXT}
\end{figure}

\begin{figure}
\centering
\includegraphics[width=.5\textwidth]{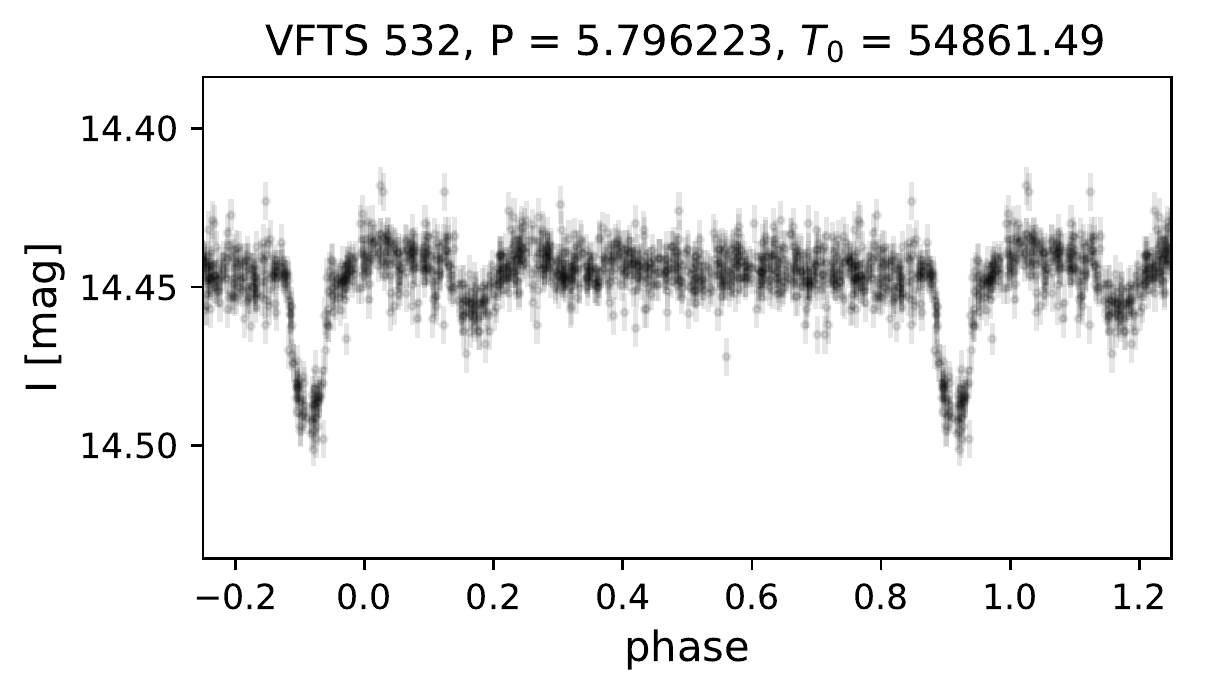}
\caption{OGLE I-band light curve of  the eclipsing binary system \object{VFTS~532}, phased with the derived period of $P = 5.796223\,$d, which matches the spectroscopic period.} \label{fig:VFTS532GLE}
\end{figure}

{\bf  \object{VFTS~532}, O3.5:~V:((f*)) + B~III,} has a reported period of $P=5.8\,$d and an eccentricity of  $e=0.46$. Given the very early spectral class of the primary, its orbital solution and RV amplitude of $K_1 = 34.4\,$\kms~can be well constrained using the He\,{\sc ii} lines, which are not impacted by nebular contamination. Similarly to VFTS~386, the He\,{\sc ii} lines exhibit clear Doppler motion, while the He\,{\sc i} lines appear to be relatively static, which indicates that the secondary contributes mainly to He\,{\sc i} lines (Fig.\,\ref{fig:VFTS532EXT}). Because of the low S/N, spectral disentangling of various He\,{\sc i} lines does not yield a clear minimum for $K_2$, and the results are generally line-dependent, with the $K_2$ values lying in the range 40 - 100\,\kms. A weighted mean of the measurements of the \HeI4026, \HeI4388, and \HeI4471 lines yields $K_2 = 81\pm32\,$\kms. The disentangled spectrum of the secondary matches best a B-type star, but the exact subtype is difficult to retrieve.

Analysis of the OGLE light curve yields a period of $P = 5.796223 \pm 0.000002\,$d, and reveals the unmistakable presence of two eclipses in the system (Fig.\,\ref{fig:VFTS532GLE}) . This further supports the fact that the system hosts two non-degenerate stars. The eclipse ratio implies that the secondary is much cooler than the primary, which agrees with the estimates spectral type. However, a B2~V star would contribute a mere $1\%$ to the flux, and would exhibit RV amplitudes that are $4-5$ times larger than derived here. Therefore, the secondary is likely a giant star, and we classify it as such. We note that VFTS~532 is one of the four objects in our sample for which X-rays were detected (Table\,\ref{tab:Sample}).

\begin{figure}
\centering
\includegraphics[width=.5\textwidth]{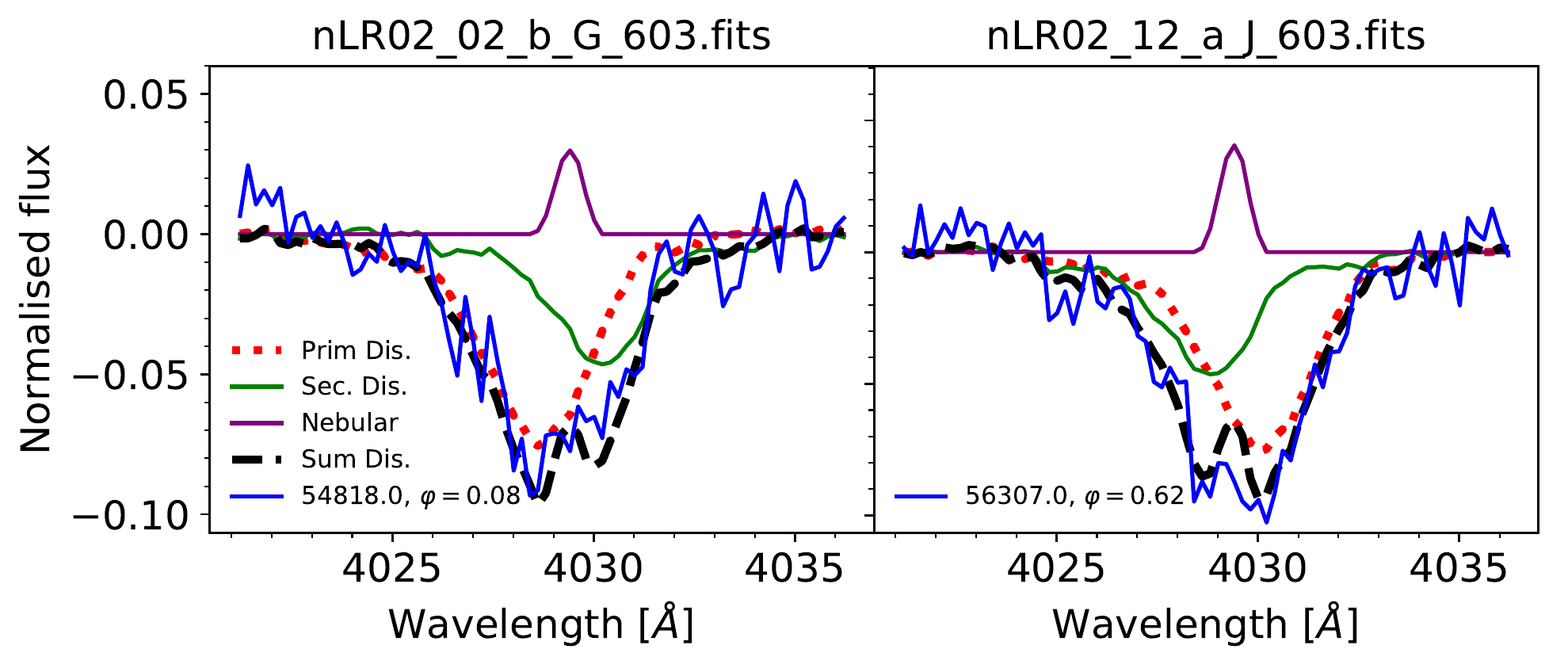}
\caption{Comparison of \HeII4026 spectra of \object{VFTS~603} at RV extremes, along with the disentangled spectra and their sum, as derived for the best-fitting $K_1, K_2$ values of  $45, 49\,$\kms, respectively. The disentangled spectra are not scaled by the light ratio in this plot.} \label{fig:VFTS603EXT}
\end{figure}

{\bf  \object{VFTS~603}, O4~III:(fc) +OB:,} was reported as a short period binary of $P = 1.76$ with an  eccentricity of $e=0.11$. The primary has a very low reported RV amplitude of $11\,$\kms, and the presence of a non-degenerate secondary is not readily evident (Fig.\,\ref{fig:VFTS603EXT}). However, the orbital solution obtained by \citet{Almeida2017} has a fairly high reduced $\chi^2$ value of 6.9.  Indeed, 1D disentangling of the strong He\,{\sc ii} lines results in a clear residual for the companion that appears distorted. We therefore resorted to 2D disentangling. We focus on the He\,{\sc ii} lines, including the \HeII4026 line, as the remaining lines are too weak and are dominated by nebular emission. However, we do not get consistent results from all lines. The $K_1$ amplitude is found consistently to be in the range $30-50\,$\kms, but $K_2$ ranges from vanishingly small values for \HeII4542 to $\approx 50\,$\kms~for \HeI4026. The latter is the only line that gives plausible results for the secondary,  with $K_1 = 45\pm 8$ and $K_2 = 49\pm7\,$\kms. These amplitudes are peculiarly low for such a short period, resulting in very small minimum masses of $M_1 \sin^3 i = 0.08\,M_\odot$ and $M_2 \sin^3 i = 0.07\,M_\odot$, and implying a very small inclination of $i \approx 6^\circ$. While the inclination is not likely (the probability of $i \le 6.3^\circ$ is 0.6\%), there is an appreciable probability for this occurring ones in our sample of 51 stars ($\approx 25\%$). 

Given the short period and low RV amplitudes, it is possible that intrinsic pulsations led to a spurious classification of this object as a binary, and specifically an SB2.  Hence, we do not readily adopt our solution in Tables\,\ref{tab:Sample} \ref{tab:SampleFin}. Higher-resolution data will be needed to verify the nature of this binary system. 

Interestingly, despite of the short period, the OGLE lightcurve does not reveal any clear periodicities beyond integer numbers that are not of astrophysical origin.  Moreover, this target is one of the four targets of our sample for which X-rays were detected, albeit at rather low luminosity (Table\,\ref{tab:Sample}).

\begin{figure}
\centering
\includegraphics[width=.5\textwidth]{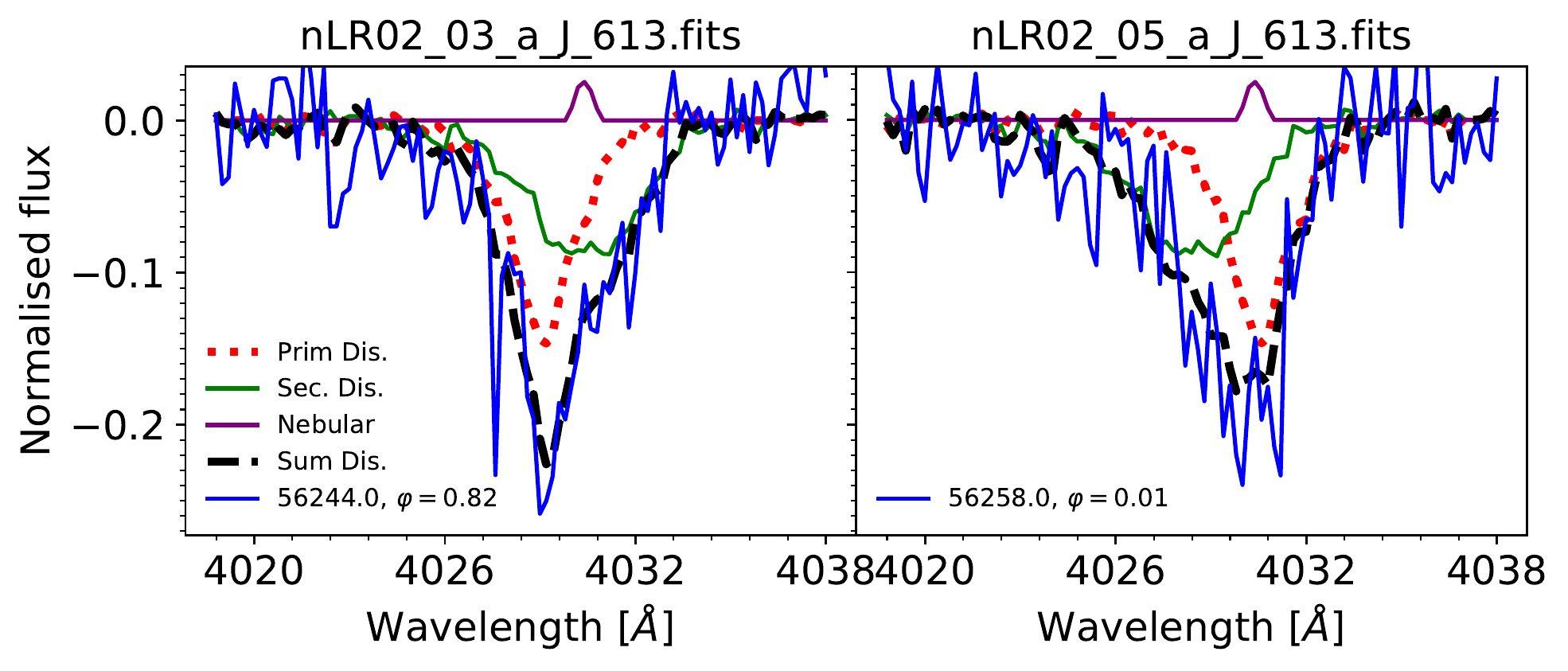}
\caption{Comparison of \HeI4388 spectra of \object{VFTS~613} at RV extremes, along with the disentangled spectra and their sum, as derived for the best-fitting $K_1, K_2$ values of  $66, 96\,$\kms, respectively. The disentangled spectra are not scaled by the light ratio in this plot.} \label{fig:VFTS613EXT}
\end{figure}

{\bf  \object{VFTS~613}, O9~V + O7.5~V,} was reported to have  a period of $P=69\,$d and an eccentricity of $e=0.35$. However, the RV measurements by \citet{Almeida2017} shows substantial scatter with respect to their derived orbital solution. Inspection of the spectral line variability suggests that the system is in fact SB2, consisting of two similar stars (Fig.\,\ref{fig:VFTS613EXT}). We therefore performed 2D grid disentangling of various lines. The only lines to yield sensible constraints are the \HeI4388 and \HeII4542 lines, and a weighted mean of the measured RV amplitudes yields $K_1 = 66\pm15$\kms and $K_2 = 96\pm 39$\kms. The spectral appearance of the secondary suggests that, while less massive, it is hotter than the primary, with an estimated spectral type of O7 and visual light contribution of  49\%. This may imply that the primary is an evolved giant or supergiant, though the spectra quality does not enable us to establish this empirically.   There exists no OGLE data for this object.

\begin{figure}
\centering
\includegraphics[width=.5\textwidth]{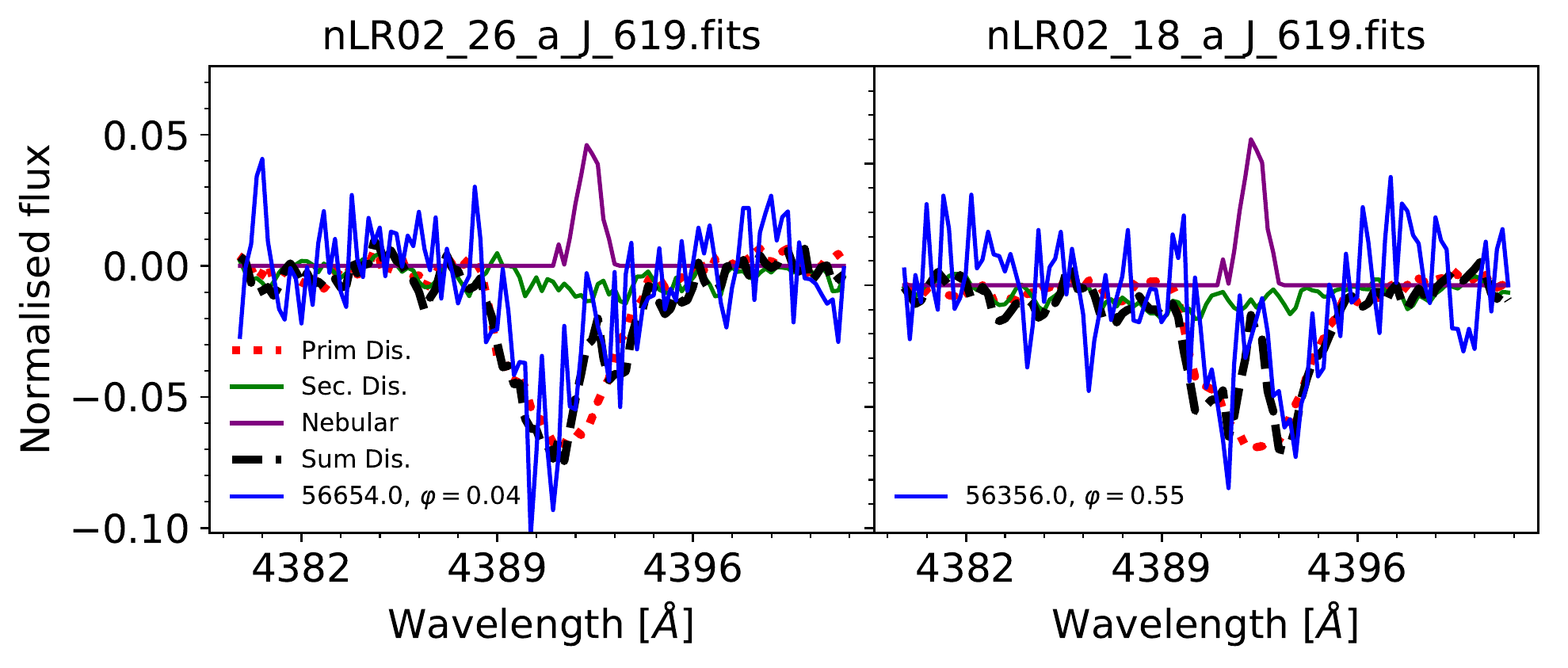}
\caption{Comparison of \HeI4388 spectra of \object{VFTS~619} at RV extremes, along with the disentangled spectra and their sum, as derived for $K_1 = 34$\,\kms~and $K_2 = 3\,K_1 = 102\,$\kms.  The disentangled spectra are not scaled by the light ratio in this plot.} \label{fig:VFTS619EXT}
\end{figure}

{\bf  \object{VFTS~619}, O8:~V,} was reported to have $P=14.5\,$d and $e=0.08$. The spectra are strongly contaminated with nebular lines, which makes the disentangling challenging. We cannot find any clear minima in the $\chi^2$ maps of $K_2$. The disentangled spectra of the secondary plausible $K_2$ values yields faint signatures for the secondary, but they are likely the result of the nebular lines (e.g. Fig.\,\ref{fig:VFTS619EXT}). We therefore tentatively classify this system as SB1.  With a minimum mass of $3.8\pm0.4\,M_\odot$, the companion may be a  B-type star, a helium star, or a BH. The OGLE light curve did not reveal any significant frequencies.

\begin{figure}
\centering
\includegraphics[width=.5\textwidth]{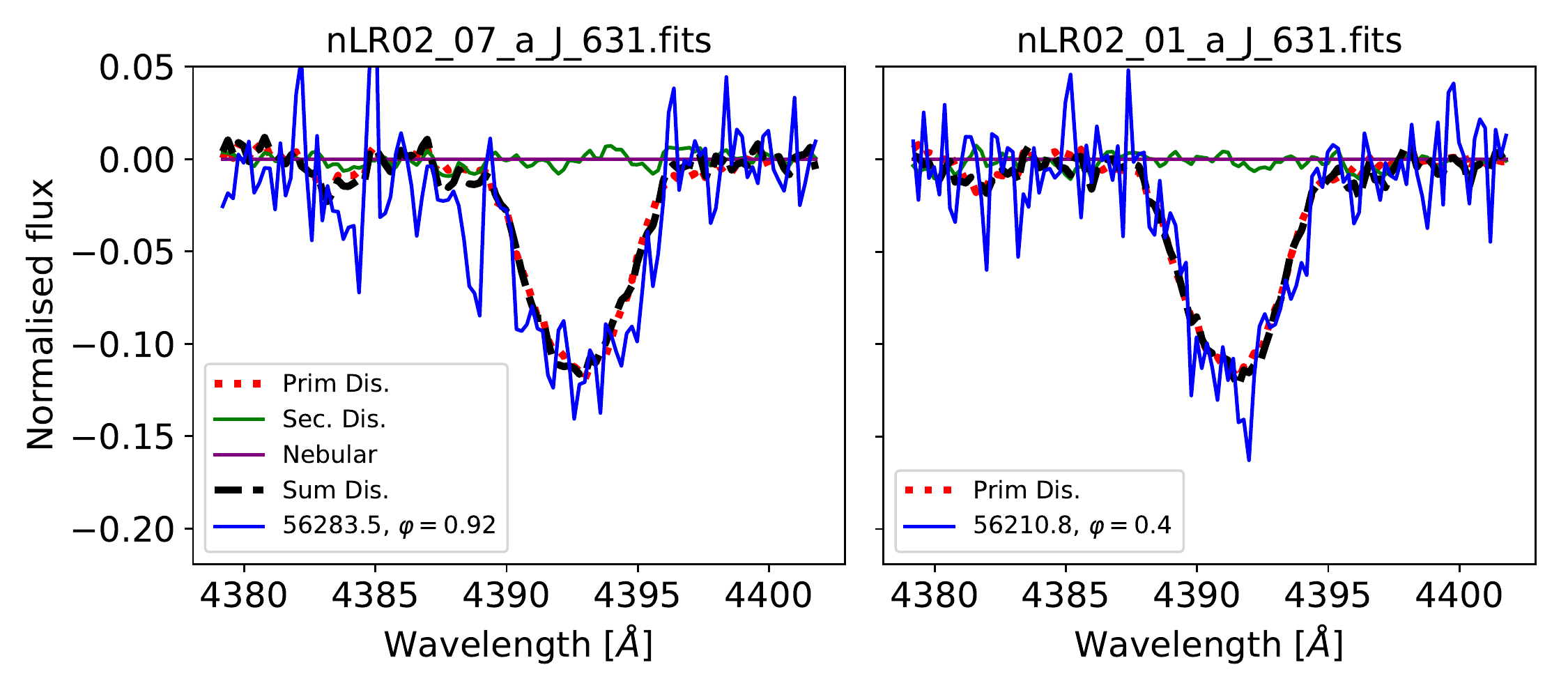}
\caption{Comparison of \HeI4388 spectra of \object{VFTS~631} at RV extremes, along with the disentangled spectra and their sum, as derived for the best-fitting $K_1, K_2$ values of  $49, 285\,$\kms, respectively. The disentangled spectra are not scaled by the light ratio in this plot.} \label{fig:VFTS631EXT}
\end{figure}

\begin{figure}
\centering
\includegraphics[width=.5\textwidth]{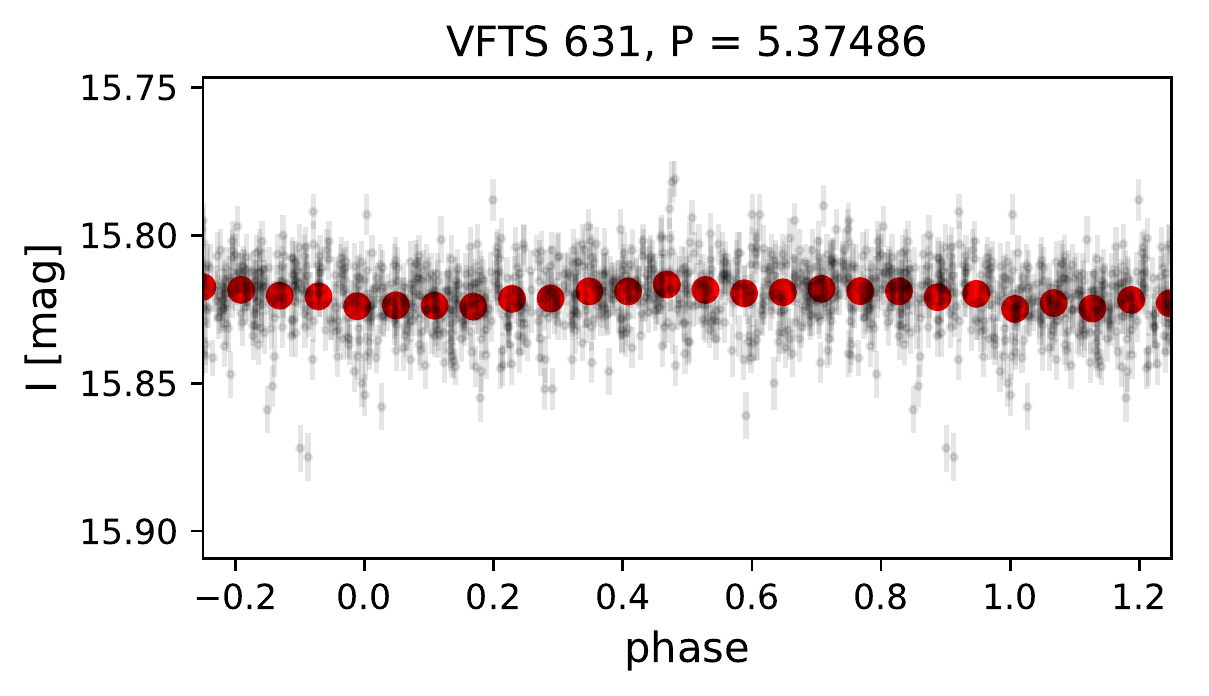}
\caption{OGLE I-band light curve of  the binary system \object{VFTS~631}, phased with the spectroscopic period.} \label{fig:VFTS631OGLE}
\end{figure}

{\bf  \object{VFTS~631}, O9.7 V,} was reported to have a virtually circular orbit and an orbital period of $P=5.4\,$d. The spectral variability is not readily suggestive of the presence of a non-degenerate companion. Disentangling of the \HeI4388 line (Fig.\,\ref{fig:VFTS631EXT}) suggests a $\chi^2$ minimum at $K_2 = 285 \pm 33\,$\kms, but it is barely significant and therefore disregarded. Disentangling of the other He\,{\sc i} lines does not result in a clearly defined minimum. Since the nebular contamination is relatively low, our conservative fiducial simulations suggest that we could probe down companions with brightness ratios down to $\approx 3\%$. \citet{Walborn2014} estimated $M_V = -3.92\,$mag for the system (dominated by the primary), such that our threshold would correspond to a companion brightness of $M_V \approx -0.1\,$mag, which matches roughly B6-8~V, or $\approx 3-5\,M_\odot$.  With a minimum mass of $3.1\pm0.4\,M_\odot$, the companion could therefore be a late type B or an early type A star, a helium star, or a BH. A Fourier analysis shows a peak at the orbital period, and the phased light curve shows a  periodic pattern at a low amplitude when phased at the orbital period (Fig.\,\ref{fig:VFTS631OGLE}).

\begin{figure}
\centering
\includegraphics[width=.5\textwidth]{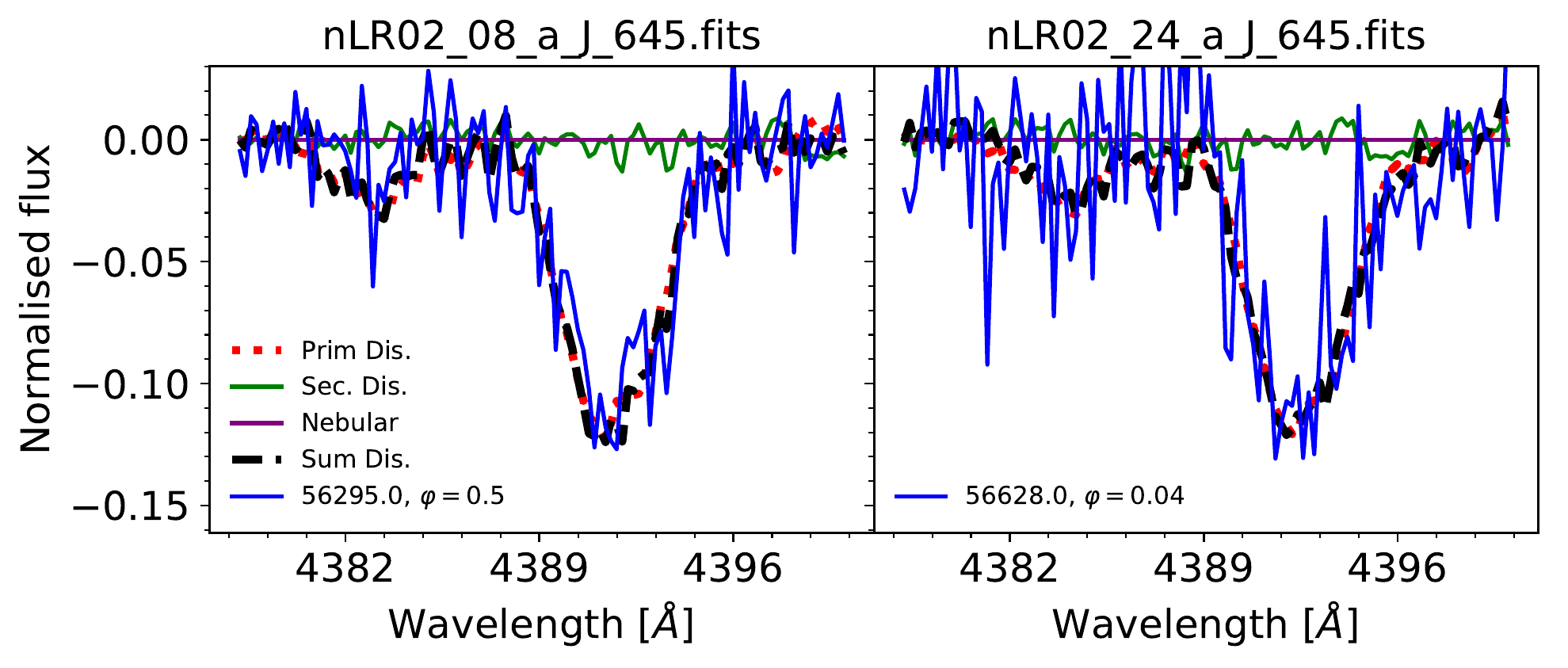}
\caption{Comparison of \HeI4388 spectra of \object{VFTS~631} at RV extremes, along with the disentangled spectra and their sum, as derived for $K_1 = 32\,$\kms and $K_2 = 4 \times K_1 = 128\,$\kms. The disentangled spectra are not scaled by the light ratio in this plot.} \label{fig:VFTS645EXT}
\end{figure}

{\bf  \object{VFTS~645}, O9.5 V,} has a reported period of $12.5\,$d and  eccentricity of $e=0.23$. The disentangling of various strong He\,{\sc i} lines such as \HeI4026 (Fig.\,\ref{fig:VFTS645EXT}) did not result in a well defined minimum for $K_2$, and the disentangled spectra of the secondary for various $K_2$ values does not portray any features beyond Balmer lines, which may be the result of nebular contamination. We therefore classify this system as SB1. \citet{Walborn2014} estimated a visual absolute magnitude of $M_V = -3.86\,$mag for the system, which is dominated by the primary. Adopting our conservative threshold estimate of $3\%$ for the light contribution for an unseen companion, we can rule out dwarfs with spectral types as late as ~B6-8~V. With a minimum mass of $M_2 \gtrsim 2.6\pm0.5\,M_\odot$, the companion could be an A or B-type dwarf, a helium star, a NS or a BH. The OGLE light curve does not show any significant periods.

\begin{figure}
\centering
\includegraphics[width=.5\textwidth]{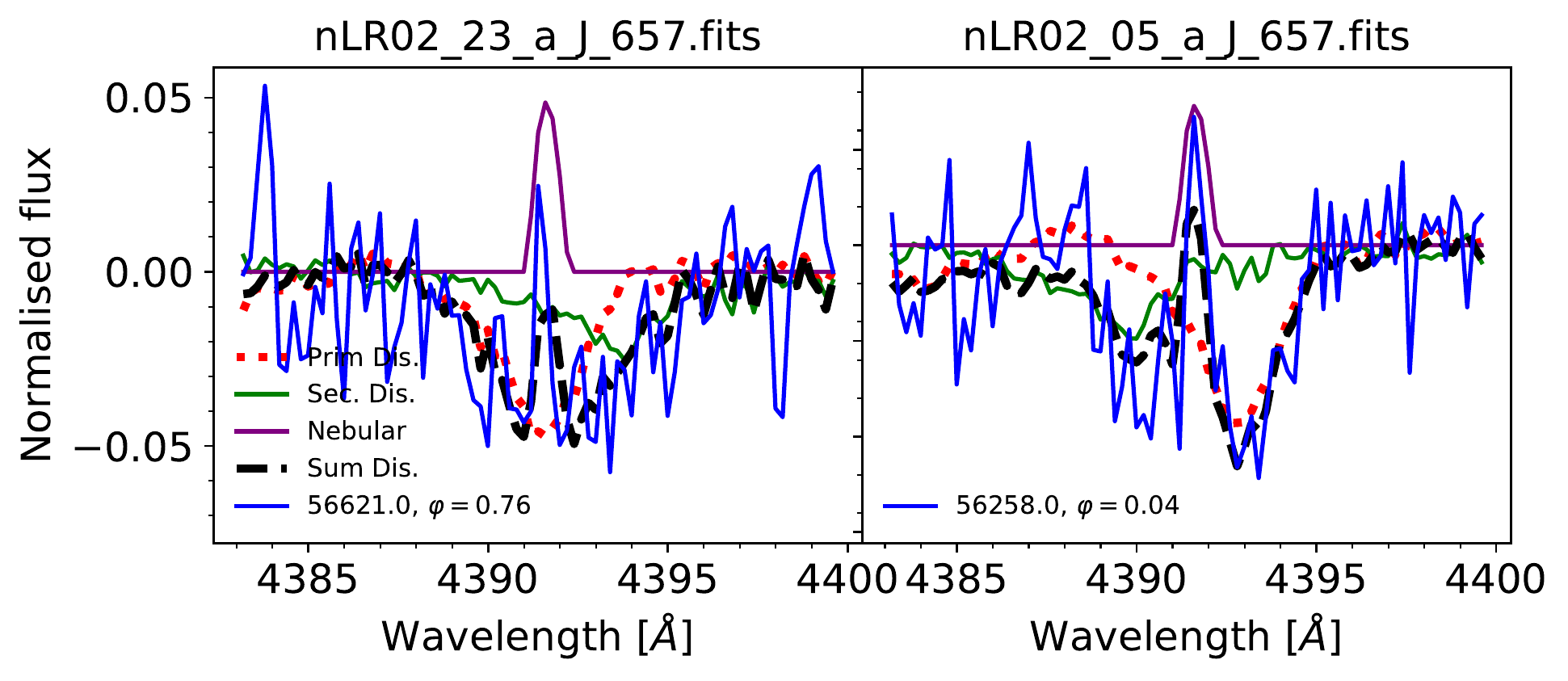}
\caption{Comparison of \HeI4388 spectra of \object{VFTS~657} at RV extremes, along with the disentangled spectra and their sum, as derived for the best-fitting RV amplitudes of $K_1 = 45\,$\kms~and $K_2 = 169\,$\kms. The disentangled spectra are not scaled by the light ratio in this plot.} \label{fig:VFTS657EXT}
\end{figure}

{\bf  \object{VFTS~657}, O7~II:((f)) + OB:,} has a reported period of $P=63\,$d and eccentricity of $e=0.48$. The strong nebular line contamination makes it difficult to determine whether a non-degenerate companion is causing the observed variability (Fig.\,\ref{fig:VFTS657EXT}), and also hampers our ability to disentangle the system robustly. Nonetheless, the \HeI4388 and \HeI4144 lines yield a consistent minimum for $K_2$, and the weighted mean of the measurements yields $K_2 = 169 \pm 38\,$\kms. The spectrum of the companion best matches an early B-type star, but its line coincide with nebular lines, making the SB2 nature of the system questionable. With a minimum mass of $9.2\pm2.7\,M_\odot$, the companion is either a non-degenerate early-type star or a BH.  Analysis of the OGLE light curve did not reveal significant frequencies.

\begin{figure}
\centering
\begin{tabular}{c}
\includegraphics[width=0.5\textwidth]{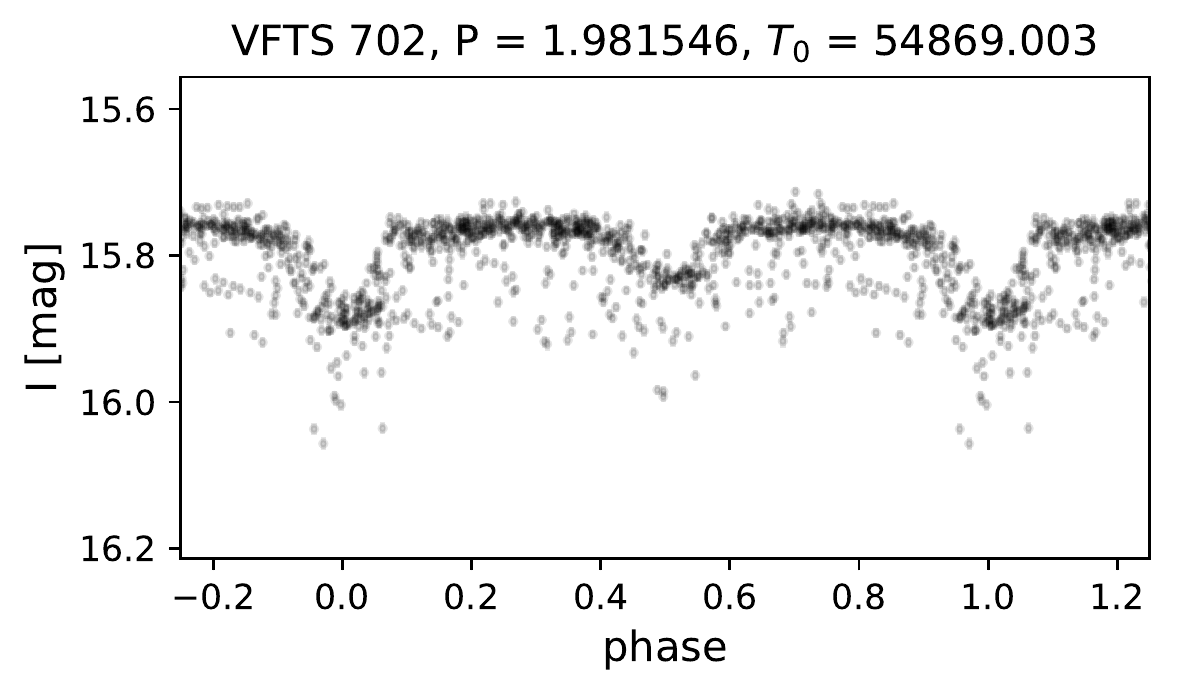}\\ 
\includegraphics[width=0.5\textwidth]{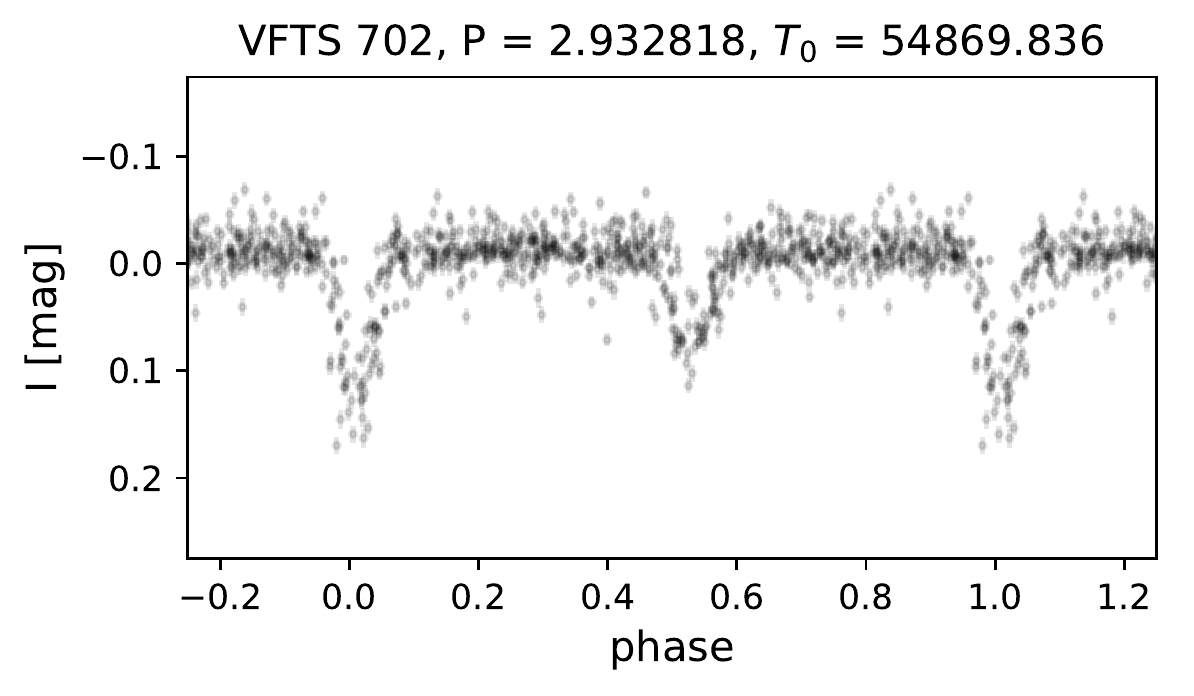} 
\end{tabular}
\caption{{\rm Upper panel:} OGLE I-band light curve of  the eclipsing binary \object{VFTS~702}, phased with the derived ephemeris of $P=1.981546\,$d (the dominant spectroscopic period) and $T_0 = 54869.033$ [JD-2400000].{\rm Lower panel:} Same light curve, phased with the second identified period of $P = 2.932818$\,d and $T_0 = 54869.84$ [JD-2400000]. The 1.98\,d signature was subtracted from the data shown in this panel.
}
\label{fig:VFTS702OGLE}
\end{figure}



{\bf  \object{VFTS~702}, O8~V(n) + OB + (OB + OB),} was reported to be a virtually circular short-period binary with $P=1.98\,$d. The spectra suffer from low S/N and very strong nebular contamination, hampering our ability to disentangle them robustly. However, Fourier analysis of the OGLE light curve reveals the presence of two eclipsing binaries in the system. The period of the first one closely matches the spectroscopic period derived by \citet{Almeida2017}, refined to $P=1.981546 \pm 0.000030\,$d (Fig.\,\ref{fig:VFTS702OGLE}). The second period is $P = 2.932818 \pm 0.000025\,$d. In Fig.\,\ref{fig:VFTS702OGLE}, we also show the light curve phased with this period, after subtracting the 1.98\,d signal. Given the short periods, it is clear that the system must be a doubly-eclipsing quadruple system. From the comparable strengths of the eclipses, it is likely that all components are OB-type stars. Examination of available HST images does not reveal any nearby source within ~0.1'' (roughly $5\,000\,$au), suggesting that the two binaries are gravitationally bound. Unfortunately, the data are not sufficient for disentangling the system, let alone with four components.

\begin{figure}
\centering
\includegraphics[width=.5\textwidth]{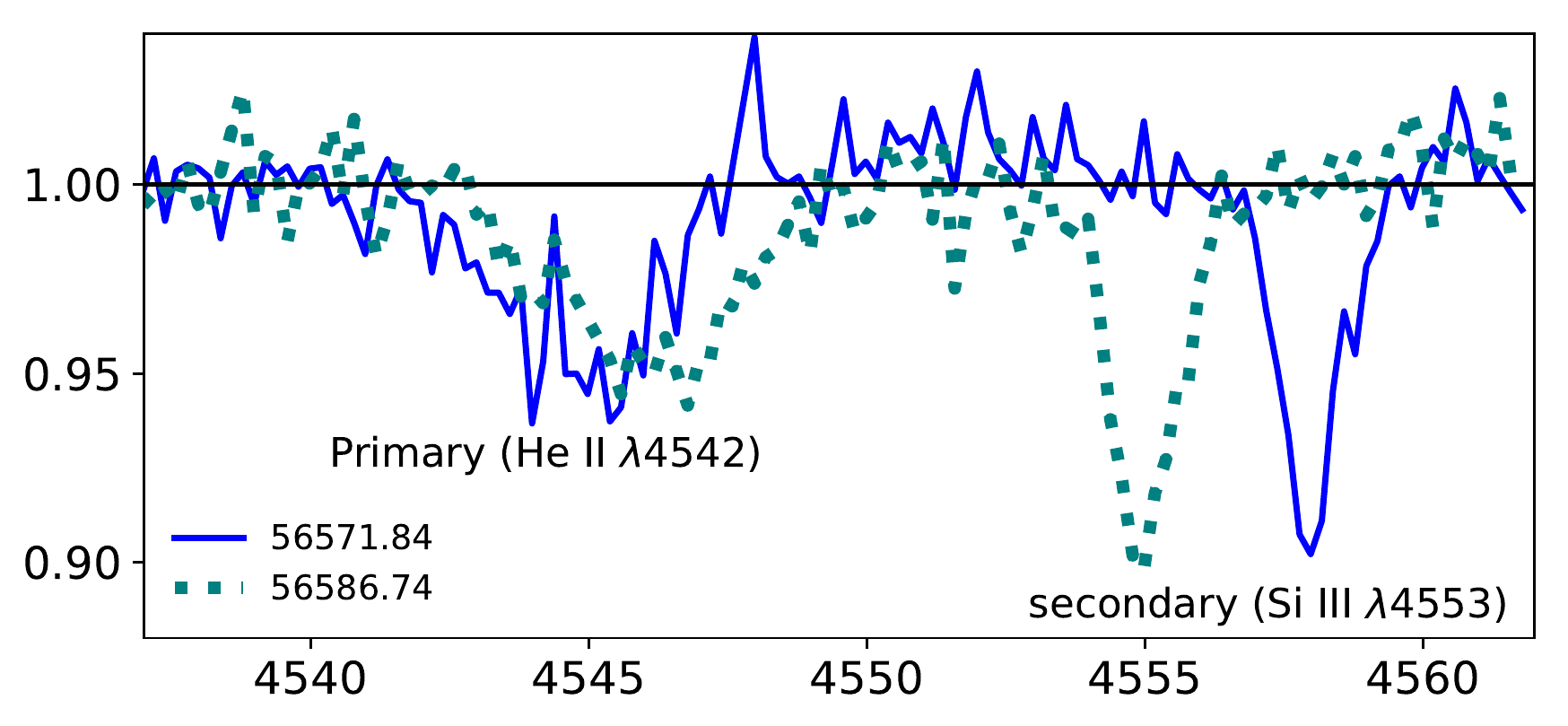}
\caption{Two spectra of VFTS~733 (JD-2400000 given in legend) showing the anti-phase motion of the hotter primary (dominating the \HeII4542 line) and the cooler secondary (dominating the Si\,{\sc iii}\,$\lambda 4553$ line).} \label{fig:VFTS733RVcomp}
\end{figure}

\begin{figure}
\centering
\includegraphics[width=.5\textwidth]{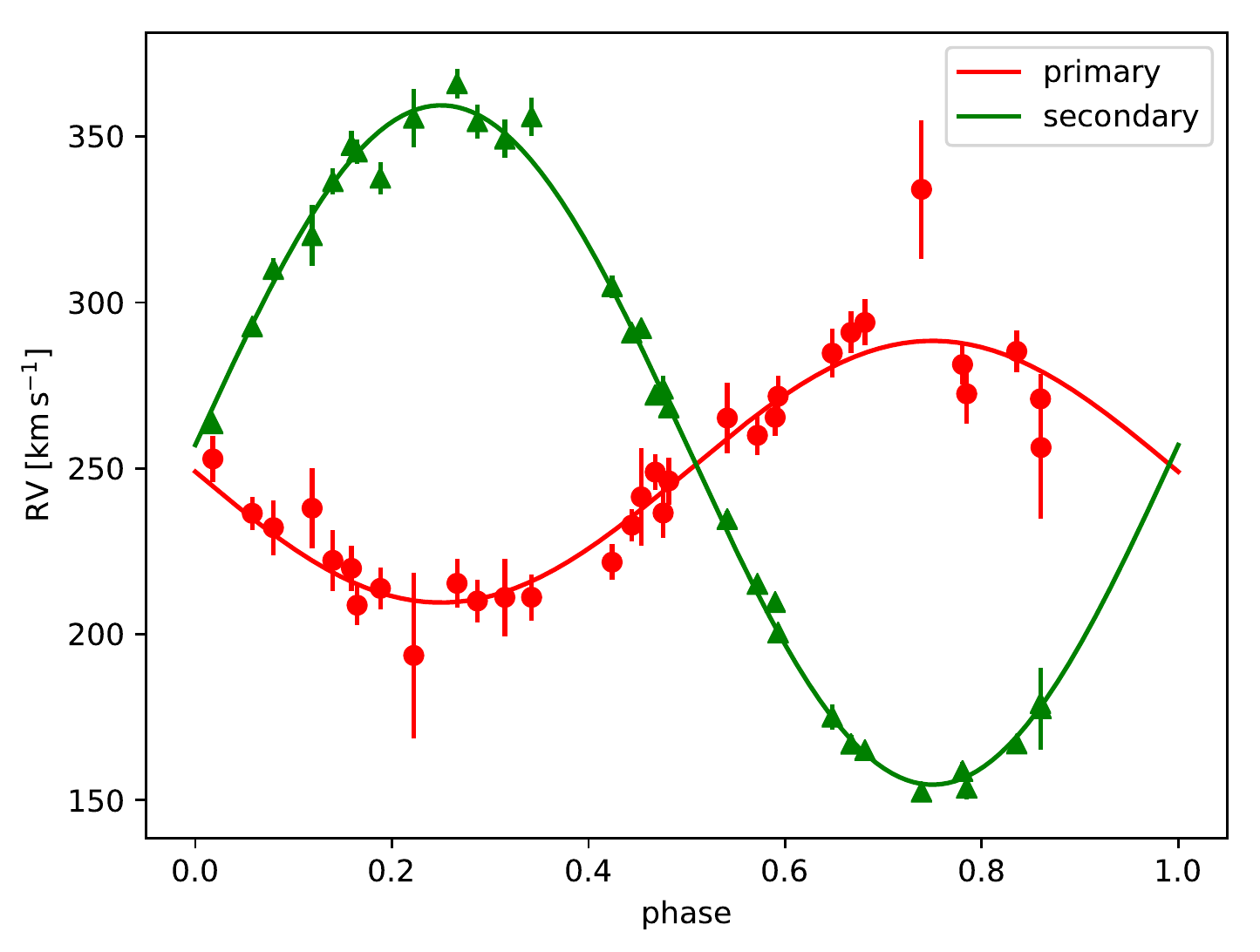}
\caption{SB2 orbital solution for the binary VFTS~733, where the RVs of both components could be measured individually.} \label{fig:VFTS733ORB}
\end{figure}

\begin{figure}
\centering
\includegraphics[width=.5\textwidth]{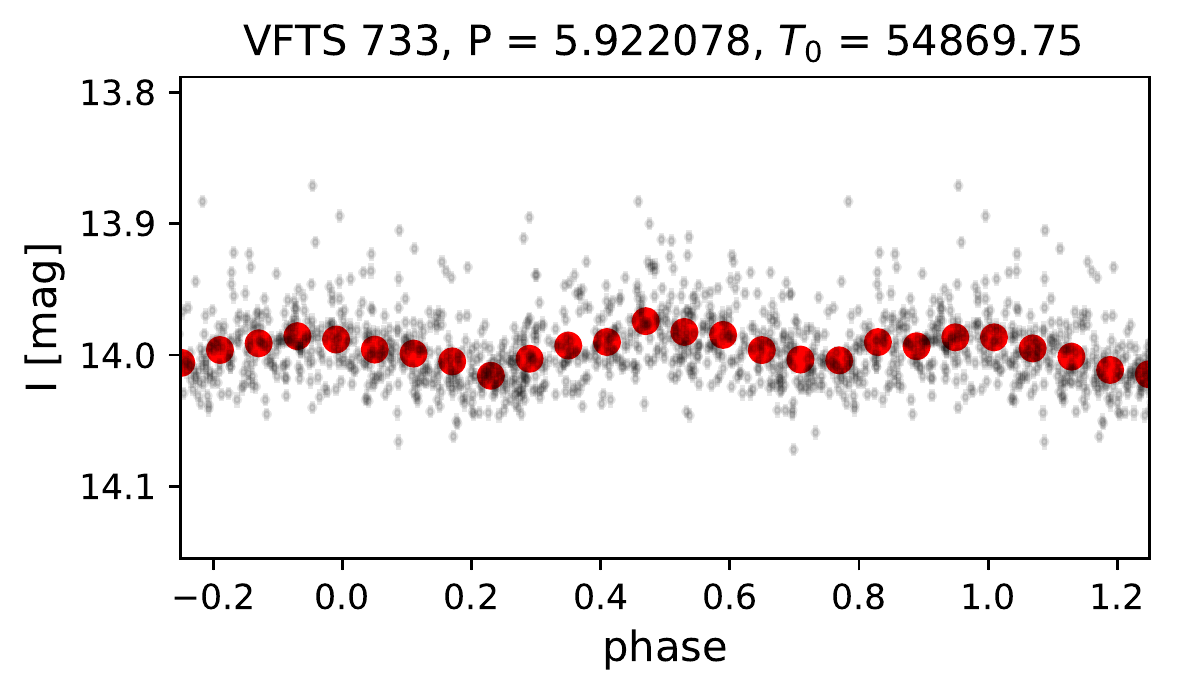}
\caption{OGLE I-band light curve of  the SB2 binary \object{VFTS~733}, phased with the revised orbital period of $P=5.922078\,$d.} \label{fig:VFTS733OGLE}
\end{figure}

{\bf  \object{VFTS~733}, O7.5~V + B1~II,} was classified as an O9.7p binary with a period of $P=5.9\,$d and a virtually-circular orbit. Inspection of the line profiles readily reveals that this system is SB2. In fact, the He\,{\sc ii} lines belong almost solely to hotter primary, while the He\,{\sc i} and a multitude of metal lines belong to the cooler secondary. Therefore, deriving the nature and orbital solution of this system does not require spectral disentangling, as the RVs can be well measured from isolated lines. Here, we use the He\,{\sc ii}\,$\lambda 4542$ and the Si\,{\sc iii}\,$\lambda 4553$ for the hotter primary and cooler secondary, respectively (Fig.\,\ref{fig:VFTS733RVcomp}). The phased RV measurements and orbital solution are shown in Fig.\,\ref{fig:VFTS733ORB}. We find $K_1 = 39.4 \pm 1.9\,$\kms~and $K_2 = 102.2\pm 0.9\,$\kms~for the primary and secondary, respectively. The less massive secondary appears to contribute roughly half of the flux in the visual, suggesting that it is evolved (luminosity class III-I). This is supported by the spectral appearance of the secondary.  The low minimum masses of $M_1 \sin^3 = 1.26\,M_\odot$ and $M_2 \sin^3 i = 0.49\,M_\odot$ imply a low inclination of $i \approx 20^\circ$. Eclipses are therefore not to be expected. The OGLE light curve shows a periodic pattern when phased at the orbital period (Fig.\,\ref{fig:VFTS733OGLE}).

\begin{figure}
\centering
\includegraphics[width=.5\textwidth]{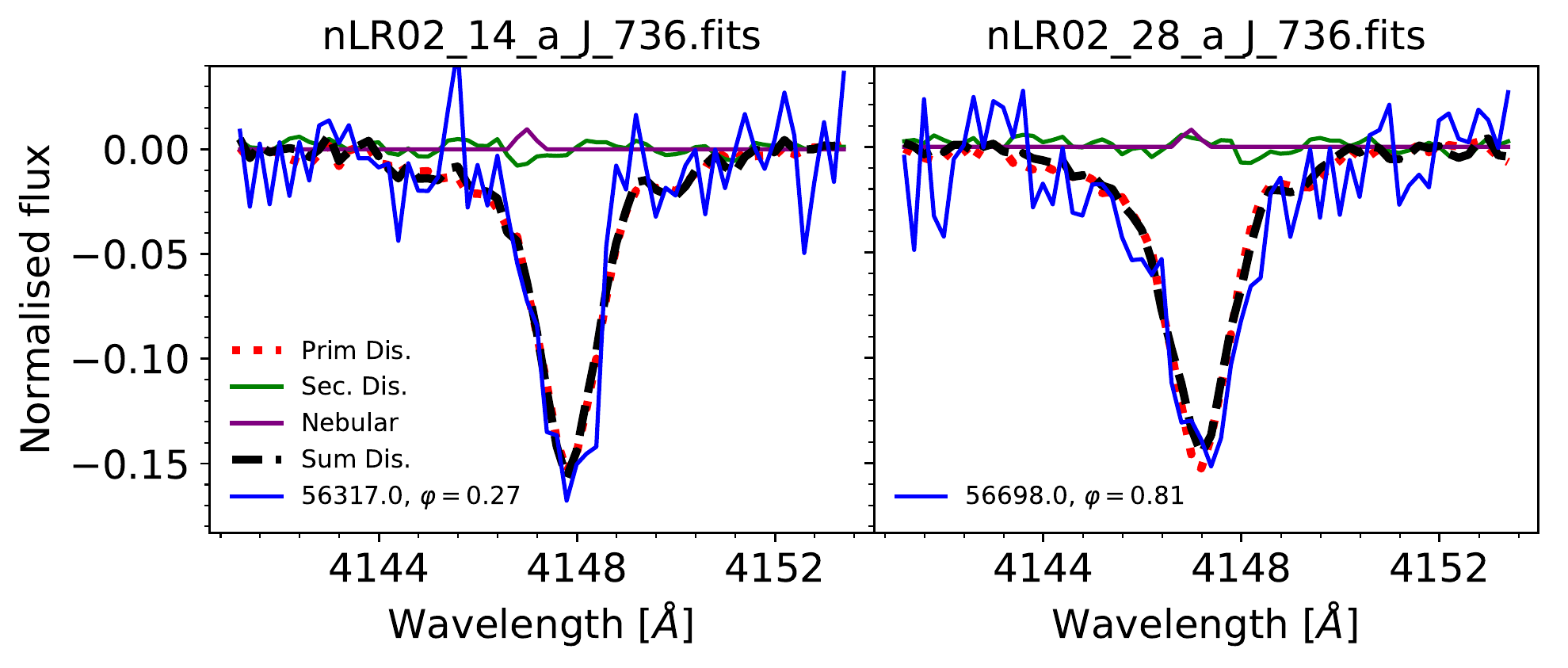}
\caption{Comparison of \HeI4388 spectra of \object{VFTS~736} at RV extremes, along with the disentangled spectra and their sum, as derived for $K_1 = 25\,$\kms and $K_2 = 2 \times 25 = 50\,$\kms. The disentangled spectra are not scaled by the light ratio in this plot.} \label{fig:VFTS736EXT}
\end{figure}

{\bf  \object{VFTS~736}, O9.5~V + B:,} has a reported period of $P=68.8\,$d and  eccentricity of $e=0.09$. Disentangling of various He\,{\sc i} lines does not yield a consistent solution for $K_2$, and the spectral variability is not readily suggestive of a non-degenerate companion (Fig.\,\ref{fig:VFTS736EXT}). The disentangled spectrum of the secondary for plausible $K_2$ shows very faint He\,{\sc i} signatures for strong lines, but these may be the result of nebular contamination.  As $M_2 > 3.9\pm0.5\,M_\odot$, the companion could either be an B- or A-type star, a helium star, or a black hole. Analysis of the OGLE light curve did not reveal significant frequencies. 

\begin{figure}
\centering
\includegraphics[width=.5\textwidth]{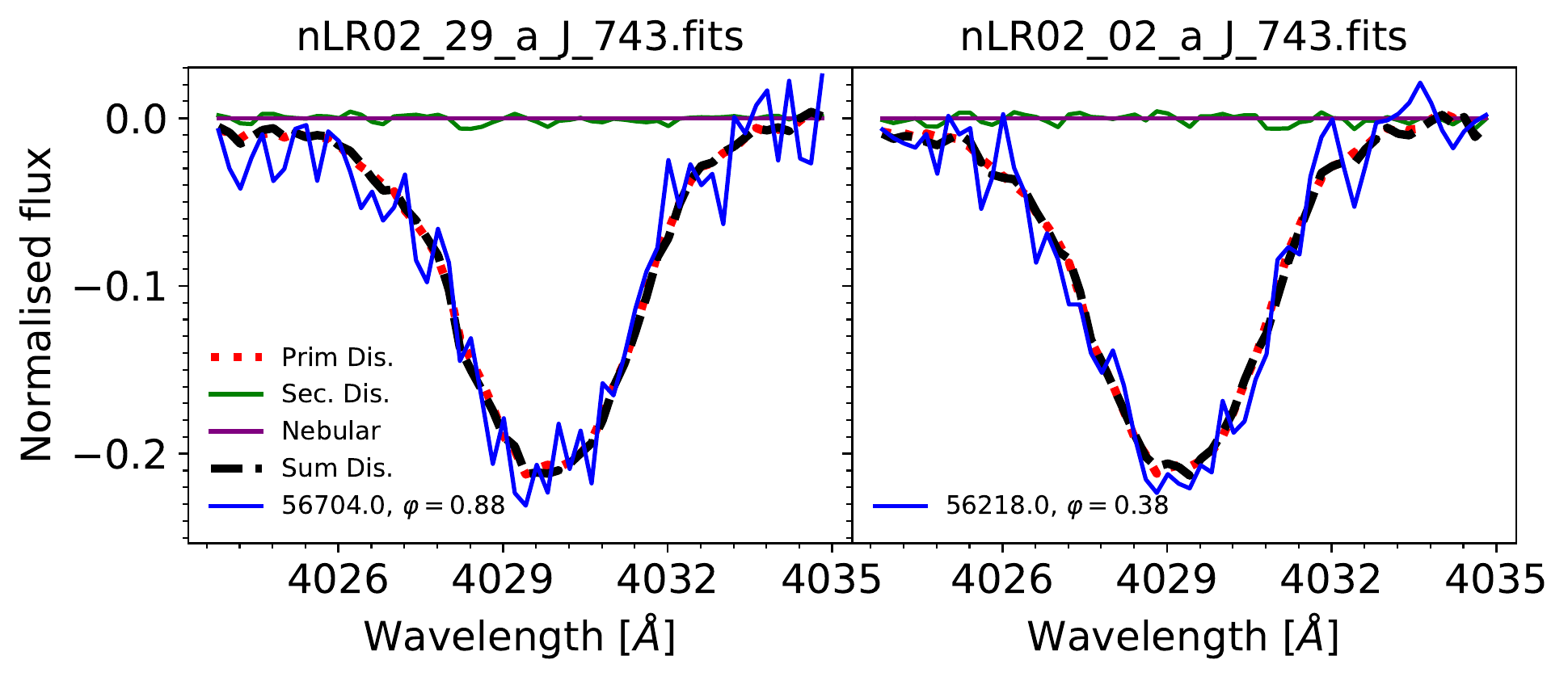}
\caption{Comparison of \HeI4388 spectra of \object{VFTS~743} at RV extremes, along with the disentangled spectra and their sum, as derived for $K_1 = 23\,$\kms and $K_2 = 4 \times K_1 = 92\,$\kms. The disentangled spectra are not scaled by the light ratio in this plot.} \label{fig:VFTS743EXT}
\end{figure}

{\bf  \object{VFTS~743}, O9.5 V((n)),} has a reported period of $P=14.9\,$d and a virtually circular orbit. The spectral variability and disentangled spectrum of the secondary for various $K_2$ values do not support the presence of a non-degenerate companion in the system (Fig.\,\ref{fig:VFTS743EXT}). Some faint He\,{\sc i} features, as well as Balmer features, are seen in the disentangled spectrum of the secondary, but their appearance suggests that they originate primarily  in nebular contamination, and we therefore classify the system as SB1. Given the minimum mass of $M_2 > 2.1\pm0.3\,M_\odot$, the nature of the secondary is virtually unconstrained.  Analysis of the OGLE light curve did not reveal significant frequencies.

\begin{figure}
\centering
\includegraphics[width=.5\textwidth]{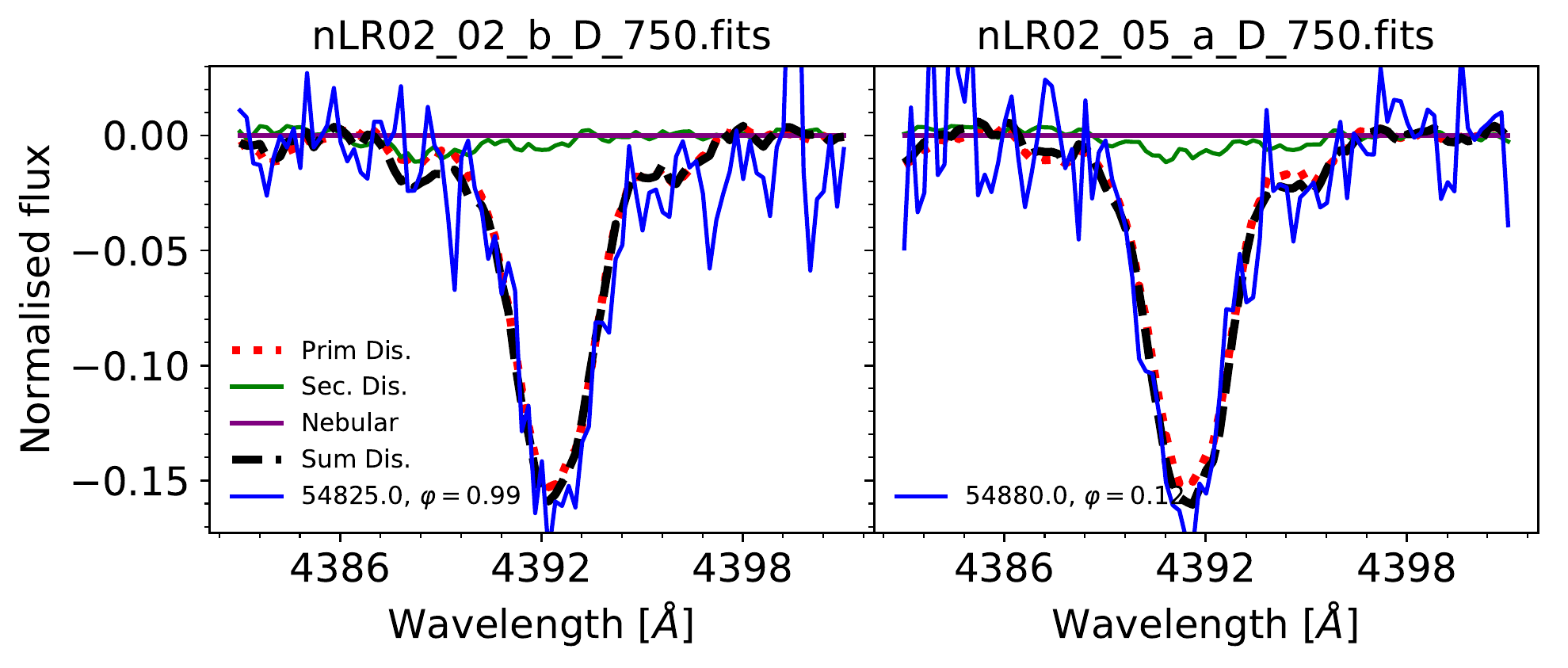}
\caption{Comparison of \HeI4388 spectra of \object{VFTS~750} at RV extremes, along with the disentangled spectra and their sum, as derived for $K_1 = 29\,$\kms and $K_2 = 3 \times K_1 = 87\,$\kms. The disentangled spectra are not scaled by the light ratio in this plot.} \label{fig:VFTS750EXT}
\end{figure}

{\bf  \object{VFTS~750}, O9.5~V + B:,} is reported to be an eccentric long period binary with $P=417\,$d and $e=0.78$ . The spectral variability is not readily suggestive of the presence of a non-degenerate companion (Fig.\,\ref{fig:VFTS750EXT}).
When disentangling for plausible $K_2$ values in the range 2-5\,$K_1$, faint features of He\,{\sc i} can be seen. Since the nebular contamination is modest in this star, it is likely that these features are real. However, due to the low-amplitude motion of the primary, the relatively poor Doppler coverage, and the faintness of the potential companion, we cannot retrieve an unambiguous value for $K_2$.  
We therefore tentatively classify the companion as B: and the system as SB2:. With a minimum mass of $M_2 > 5.6\pm1.2\,M_\odot$, the companion could be a non-degenerate object or a BH. Analysis of the OGLE light curve did not reveal significant frequencies.

\begin{figure}
\centering
\includegraphics[width=.5\textwidth]{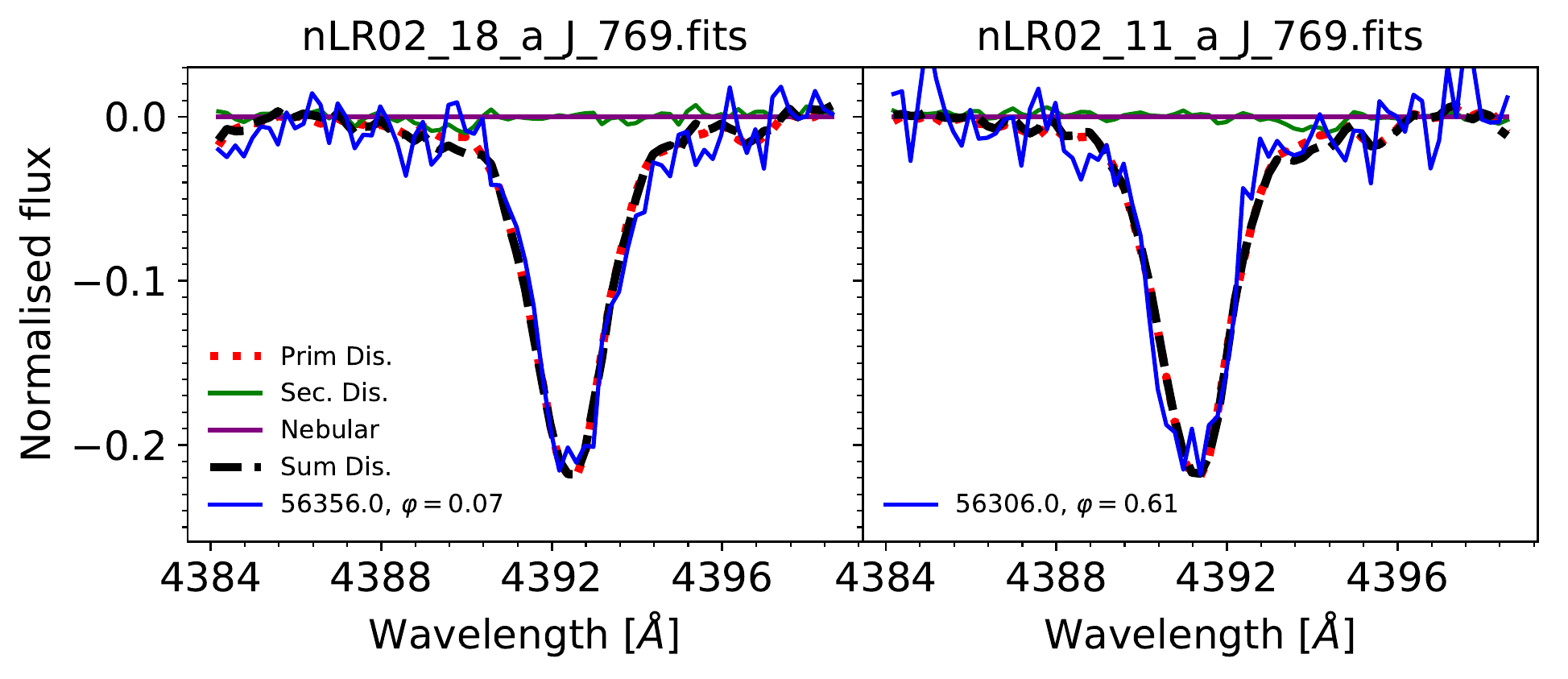}
\caption{Comparison of \HeI4026 spectra of \object{VFTS~631} at RV extremes, along with the disentangled spectra and their sum, as derived for $K_1 = 40\,$\kms and $K_2 = 4 \times K_1 = 160\,$\kms. The disentangled spectra are not scaled by the light ratio in this plot.} \label{fig:VFTS769EXT}
\end{figure}

\begin{figure}
\centering
\includegraphics[width=.5\textwidth]{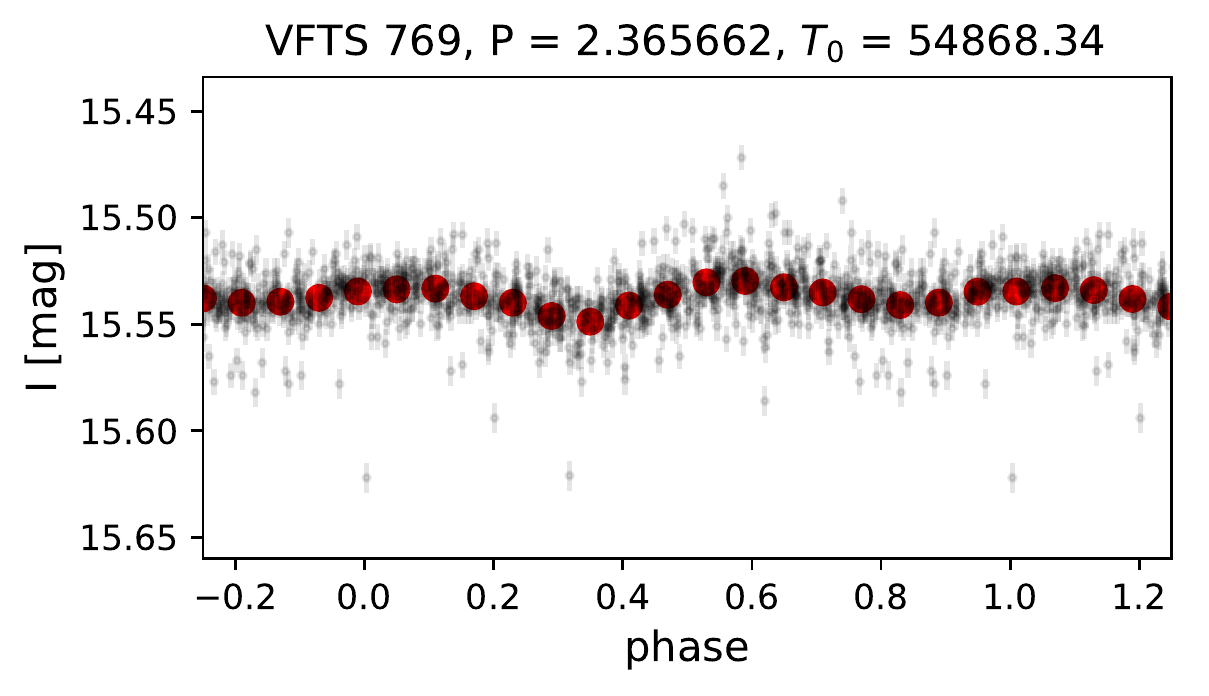}
\caption{OGLE I-band light curve of  the binary \object{VFTS~769}, phased with the spectroscopic period derived by \citet{Almeida2017}.} \label{fig:VFTS769OGLE}
\end{figure}

{\bf  \object{VFTS~769}, O9.7~V,} has a short period of $P=2.4\,$d and a virtually circular orbit. The spectral variability of strong He\,{\sc i} lines such as \HeI4388 (Fig.\,\ref{fig:VFTS769EXT}) is not readily suggestive of the presence of a non-degenerate companion.  The disentangled spectra of the secondary for various $K_2$ values are suggestive of Balmer absorption and hints of He\,{\sc i} lines, but we cannot exclude that they are the result of nebular contamination. With a minimum mass of $M_2 > 1.9\pm0.2\,M_\odot$, the nature of the secondary is unconstrained. Analysis of the OGLE light curve reveals a significant peak at the orbital period of $P=2.365644\,$d and its harmonics. The phased light curve is shown in Fig.\,\ref{fig:VFTS769OGLE}.

\begin{figure}
\centering
\includegraphics[width=.5\textwidth]{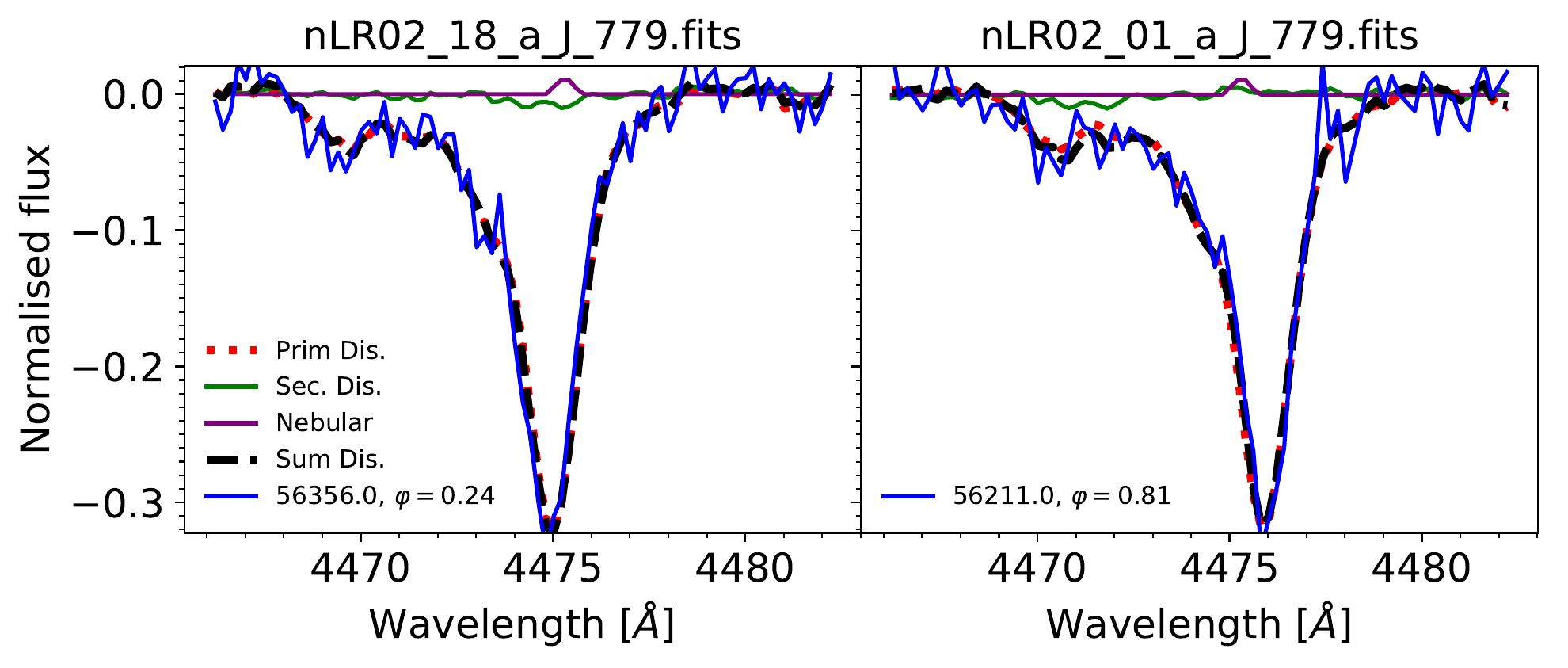}
\caption{Comparison of \HeI4026 spectra of \object{VFTS~631} at RV extremes, along with the disentangled spectra and their sum, as derived for $K_1 = 31\,$\kms and $K_2 = 4 \times K_1 = 124\,$\kms. The disentangled spectra are not scaled by the light ratio in this plot.} \label{fig:VFTS779EXT}
\end{figure}

{\bf  \object{VFTS~779}, B1~III,} has a reported period of $P=59.9\,$d and a virtually circular orbit. The spectral variability of the He\,{\sc i} does not readily suggest the presence of a non-degenerate secondary (e.g. Fig.\,\ref{fig:VFTS779EXT}) . Disentangling of various He\,{\sc i} lines results in nearly-flat $\chi^2(K_2)$ maps, though an insignificant tendency towards values in the range $K_2 \approx 2-4 \times K_2$ is noted. The disentangled spectrum of the secondary for such $K_2$ values shows very faint He\,{\sc i} absorption in some of the lines, though it is not clear whether this is due to low levels of nebular contamination. Given the relatively high quality of the data in this case, we can exclude companions contributing more than $2-3\%$ to the total light, which would correspond roughly to a B3. We classify this system as SB1. As $M_2 >  3.9\pm0.6\,M_\odot$, the companion is either  a mid- to late-type B dwarf, a He star, or a BH. Given the BH probability (Table\,\ref{tab:SampleFin}), we consider this system a good B+BH candidate. Analysis of the OGLE light curve did not reveal significant frequencies. 

\begin{figure}
\centering
\includegraphics[width=.5\textwidth]{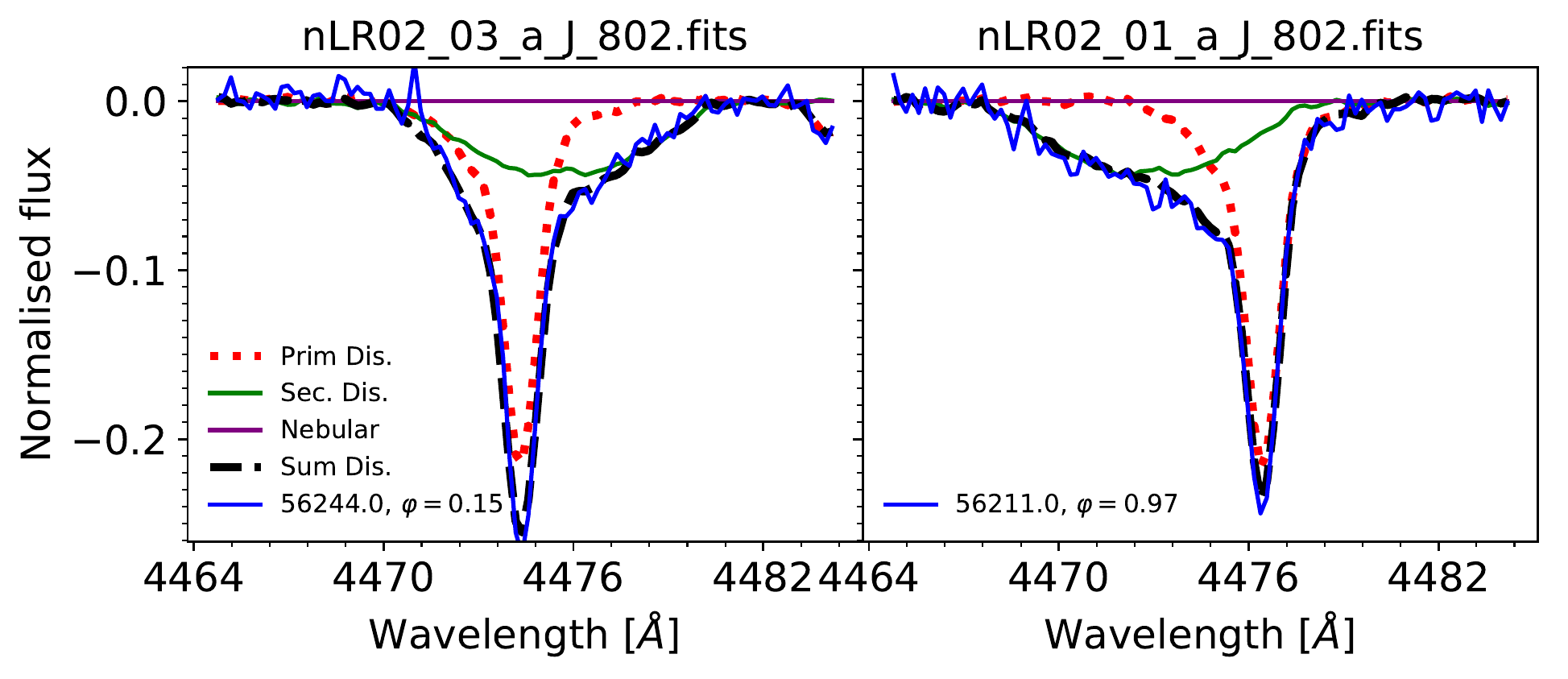}
\caption{Comparison of \HeI4026 spectra of \object{VFTS~802} at RV extremes, along with the disentangled spectra and their sum, as derived for the best-fitting values of $K_1 = 70\,$\kms and $K_2 = 76\,$\kms. The disentangled spectra are not scaled by the light ratio in this plot.} \label{fig:VFTS802EXT}
\end{figure}

{\bf  \object{VFTS~802}, O7~V: + O8~Vn,} has a reported period of  $P=182\,$d and an  eccentricity of $e=0.60$. The presence of a non-degenerate companion is readily seen in a multitude of He lines as a shallow, broad feature (Fig.\,\ref{fig:VFTS802EXT}). Given the strong contribution of the secondary, we perform 2D Disentangling of the strong He\,{\sc i} and He\,{\sc ii} lines. A mean average yields: $K_1 = 70.4\pm 1.9\,$\kms, $K_2 = 76 \pm 14\,$\kms. The RV amplitude of the secondary is more difficult to constrain due to the substantial rotational broadening of its spectrum.   The companion is estimated to contribute 36\% of the visual flux, and is classified as O8. Analysis of the OGLE light curve did not reveal significant frequencies. 

\begin{figure}
\centering
\includegraphics[width=.5\textwidth]{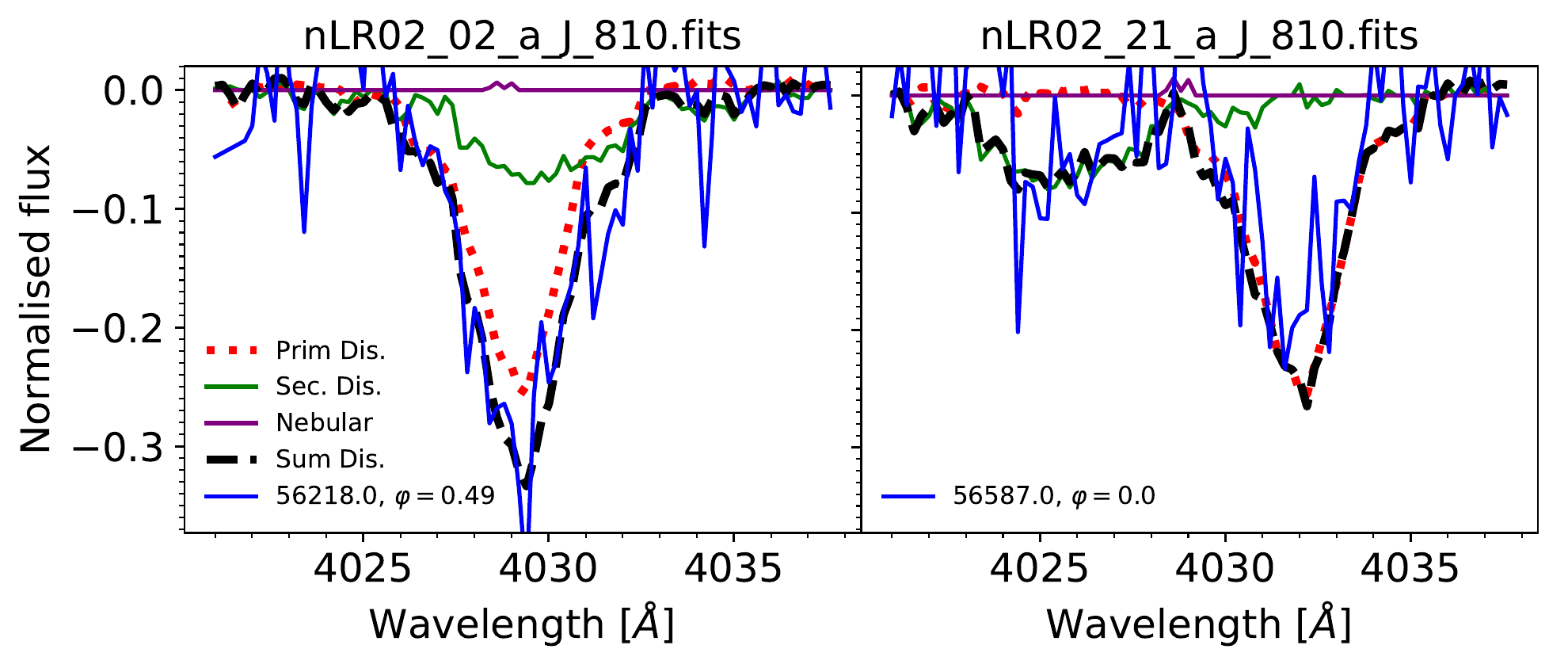}
\caption{Comparison of \HeI4026 spectra of \object{VFTS~810} at RV extremes, along with the disentangled spectra and their sum, as derived for the best-fitting values of $K_1 = 108\,$\kms and $K_2 = 162\,$\kms. The disentangled spectra are not scaled by the light ratio in this plot.} \label{fig:VFTS810EXT}
\end{figure}

{\bf  \object{VFTS~810}, O9.7~V + B1~V,} has a reported period of $P=15.7\,$d with an eccentricity of $e=0.68$. The presence of a non-degenerate companion was already recognised by \citep{Almeida2017}, who however could not measure its RVs. The lines of the secondary become deblended in certain epochs, making its presence readily evident (Fig.\,\ref{fig:VFTS810EXT}). 2D Disentangling of the strong He\,{\sc i} lines yield a weighted mean of  $K_1 = 108 \pm 9$\,\kms~and $K_2 = 162 \pm 27$\,\kms. The companion is estimated to contribute 29\% of the visual flux, and is classified as B1. Analysis of the OGLE light curve did not reveal significant frequencies. 

\begin{figure}
\centering
\includegraphics[width=.5\textwidth]{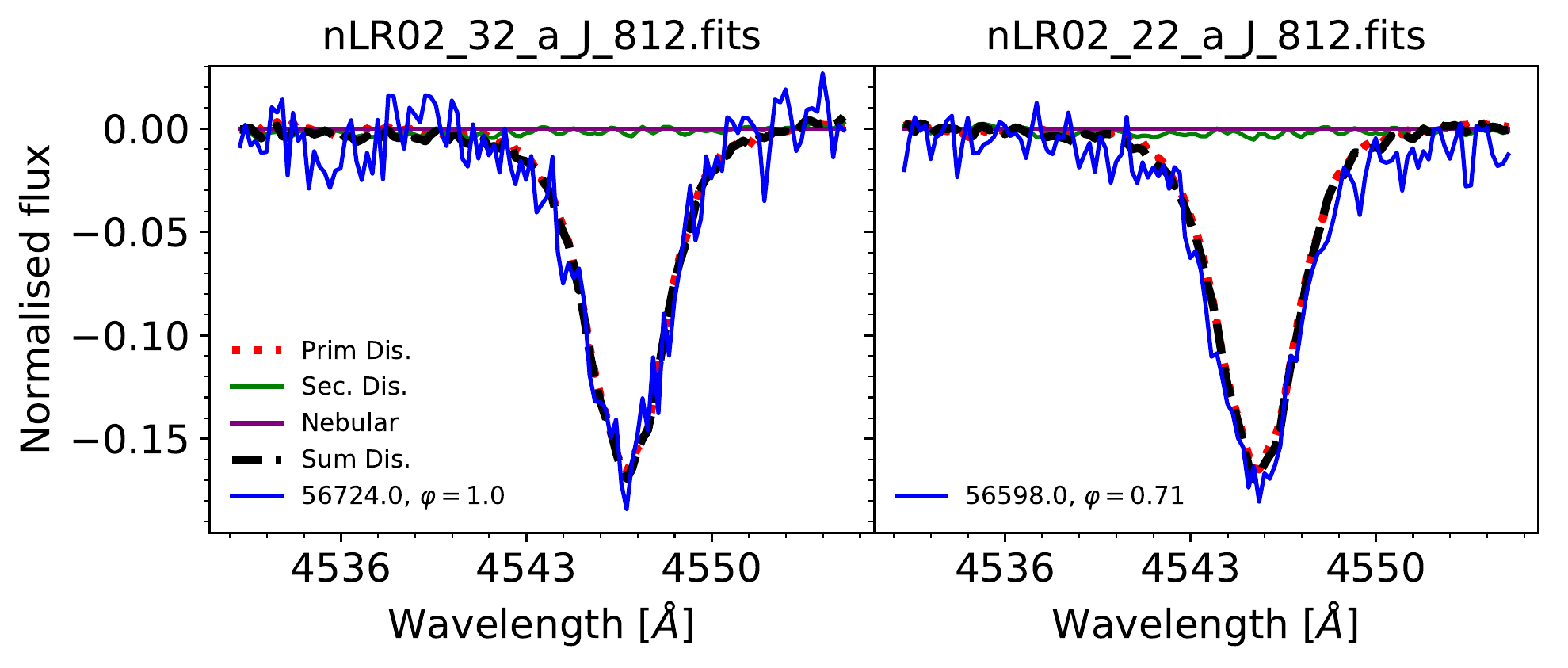}
\caption{Comparison of \HeI4542 spectra of \object{VFTS~812} at RV extremes, along with the disentangled spectra and their sum, as derived for $K_1 = 43\,$\kms and $K_2 = 3 \times K_1 = 129\,$\kms. The disentangled spectra are not scaled by the light ratio in this plot.} \label{fig:VFTS812EXT}
\end{figure}

{\bf  \object{VFTS~812}, O4~V((fc)),} has a reported period of $P=17.3\,$d and eccentricity of $e=0.62$. No sign for a companion is seen in He\,{\sc ii} lines (e.g. Fig.\,\ref{fig:VFTS812EXT}), and He\,{\sc i} lines are contaminated by nebular lines, hampering our ability to use them robustly. No clear value for $K_2$ is retrieved from disentangling. The disentangled spectra of the secondary obtained for plausible $K_2$ values show Balmer lines and some hints of He\,{\sc i} lines, but not sufficiently clear to be able to classify this system SB2 unambiguously. With $M_2 \ge 5.1\pm0.7\,M_\odot$, the companion may be a late-type OB star or a BH.  Analysis of the OGLE light curve did not reveal significant frequencies. 

\begin{figure}
\centering
\includegraphics[width=.5\textwidth]{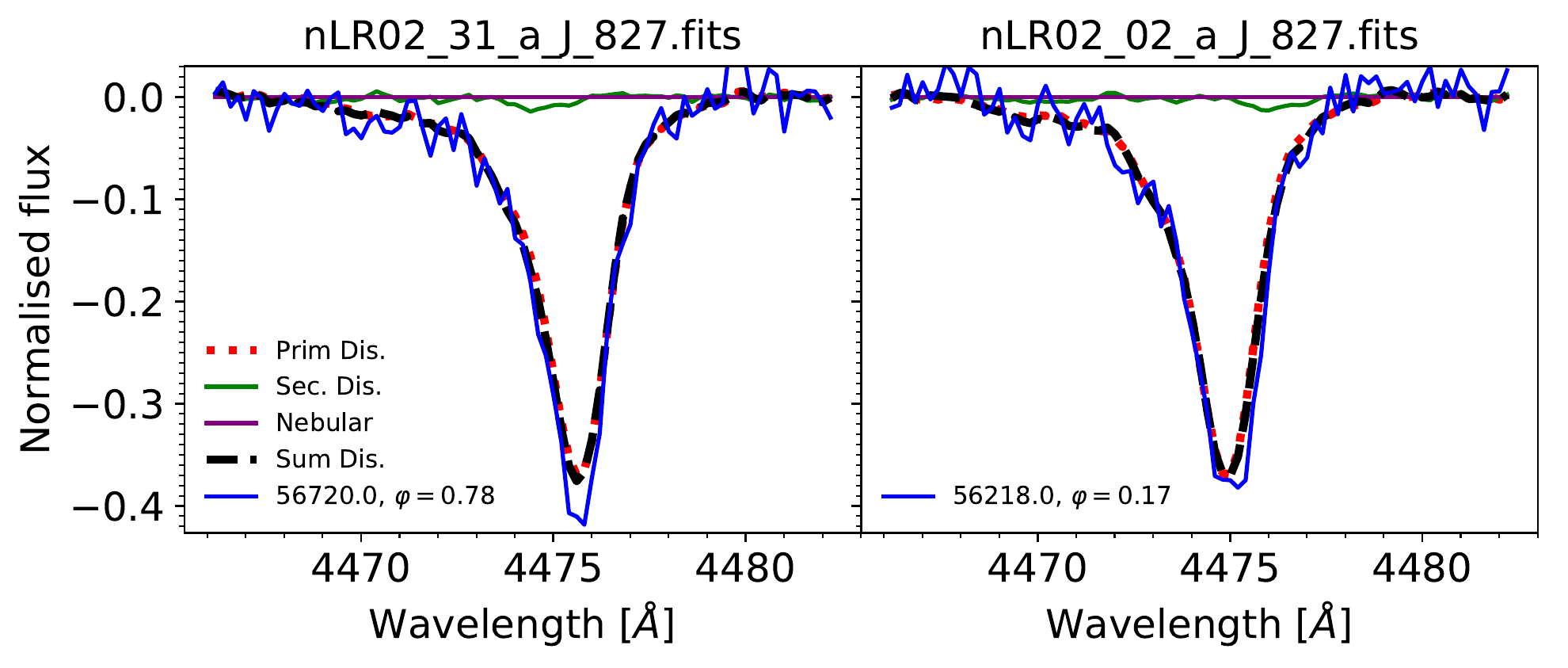}
\caption{Comparison of \HeI4471 spectra of \object{VFTS~827} at RV extremes, along with the disentangled spectra and their sum, as derived for $K_1 = 25\,$\kms and $K_2 = 2 \times K_1 = 50\,$\kms. The disentangled spectra are not scaled by the light ratio in this plot.} \label{fig:VFTS827EXT}
\end{figure}

{\bf  \object{VFTS~827}, B1.5~III, } has a reported period and eccentricity of $P=43$\,d and $e=0.24$. The spectral line variability is not suggestive of the presence of a second non-degenerate companion (e.g. Fig.\,\ref{fig:VFTS827EXT}). However, disentangling of the strong \HeI4471 favours $K_2$ values in the range $K_2 \lesssim 70\,$\kms. Disentangling for various $K_2$ values does not result a clear stellar spectrum for the  secondary. However, strong Balmer lines are seen in the disentangled spectrum of the secondary. While the nebular contamination in this case is very low, the narrow widths of the Balmer lines in the disentangled spectrum of the secondary raise the suspicion that they originate in nebular contamination. We therefore classify the system as SB1:. With a minimum mass of $M_2 > 2.5\pm0.5\,M_\odot$, the companion might also a late B-type star or an A-type star, and we cannot fully rule out that it is a NS or a BH.  No OGLE light curve is available for this object.

\begin{figure}
\centering
\includegraphics[width=.5\textwidth]{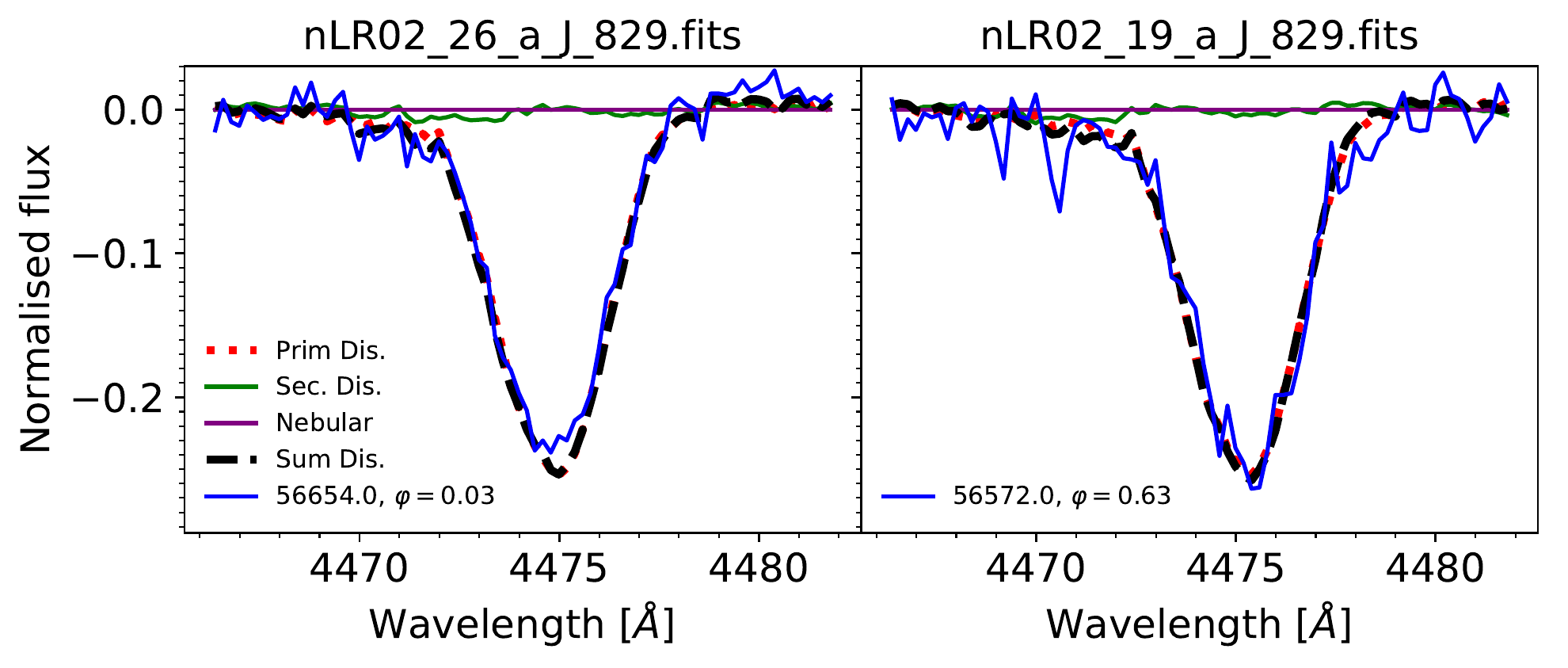}
\caption{Comparison of \HeI4471 spectra of \object{VFTS~829} at RV extremes, along with the disentangled spectra and their sum, as derived for $K_1 = 13\,$\kms and $K_2 = 4 \times K_1 = 52\,$\kms. The disentangled spectra are not scaled by the light ratio in this plot.} \label{fig:VFTS829EXT}
\end{figure}

{\bf  \object{VFTS~829}, B1.5~III,} has a reported period and eccentricity of $P=203$\,d and $e=0.27$. The spectral line variability is not suggestive of the presence of a second non-degenerate companion (e.g. Fig.\,\ref{fig:VFTS829EXT}) , and disentangling of various He\,{\sc i} results in relatively flat $\chi^2(K_2)$ maps. The disentangled spectrum of the secondary for $K_2$ values in the range $2-10 \times K_1$ is fully free of features with the exception of weak Balmer absorption that could be the result of nebular-line contamination. Given the low levels of nebular-line contamination and the relatively high S/N ratios of the spectra, we classify this system tentatively as SB1:. However, the low minimum mass of $M_2 > 2.0\pm0.5\,M_\odot$ leaves the nature of the secondary virtually unconstrained.

\begin{figure}
\centering
\includegraphics[width=.5\textwidth]{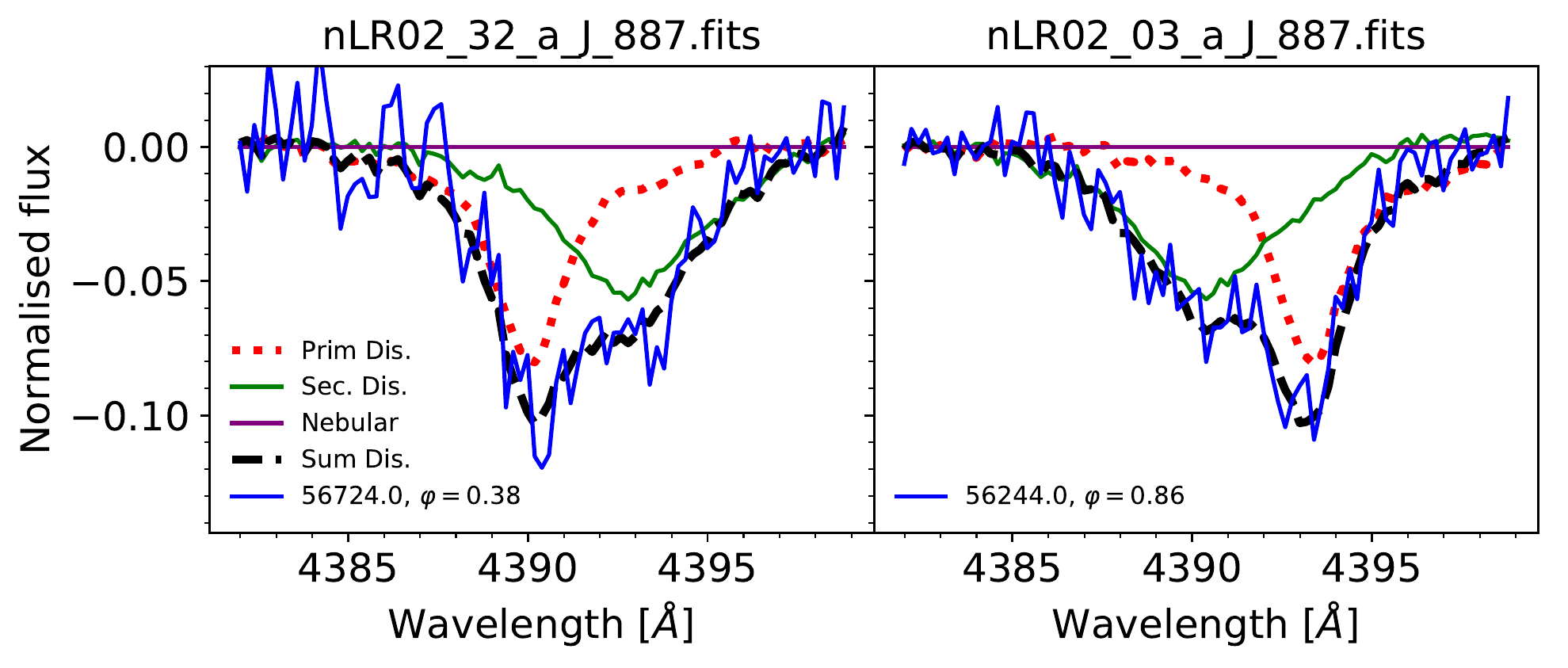}
\caption{Comparison of \HeI4026 spectra of \object{VFTS~887} at RV extremes, along with the disentangled spectra and their sum, as derived for $K_1 = 105\,$\kms~and $K_2 = 98\,$\kms. The disentangled spectra are not scaled by the light ratio in this plot.} \label{fig:VFTS887EXT}
\end{figure}

\begin{figure}
\centering
\includegraphics[width=.5\textwidth]{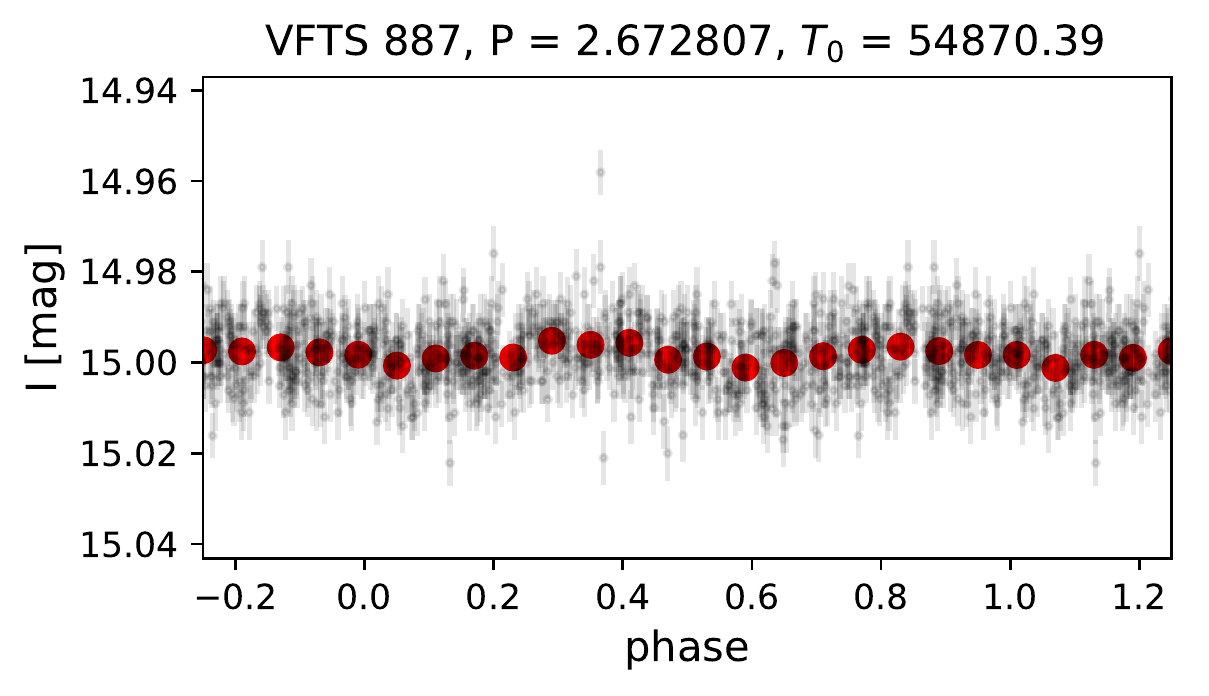}
\caption{OGLE I-band light curve of  the SB2 binary \object{VFTS~887}, phased with the orbital period of $P=2.672807\,$d.} \label{fig:VFTS887OGLE}
\end{figure}

{\bf  \object{VFTS~887}, O9.7:~V: + O9.5:~V,} has a reported period  of $P=2.7$\,d and a virtually circular orbit. The spectral line variability (Fig.\,\ref{fig:VFTS887EXT}) clearly reveals the presence of two components in the spectrum. We therefore utilise 2D disentangling on various He\,{\sc i} and He\,{\sc ii} lines, which yield consistent results. A weighted mean yields $K_1 = 105\pm 7\,$\kms~and $K_2 = 98 \pm 12\,$\kms, implying that the object identified as the primary by \citet{Almeida2017} is the slightly less massive component. The disentangled spectra correspond to O9.7~V and O9.5~V spectral types, respectively, and are estimated to contribute 45\% to the visual flux.  Despite the short period, the OGLE light curve does not show a clear period pattern (Fig.\,\ref{fig:VFTS887OGLE}).

\clearpage

\section{Disentangled spectra}
\label{sec:DisSpec}

\begin{figure}
\centering
\includegraphics[width=.5\textwidth]{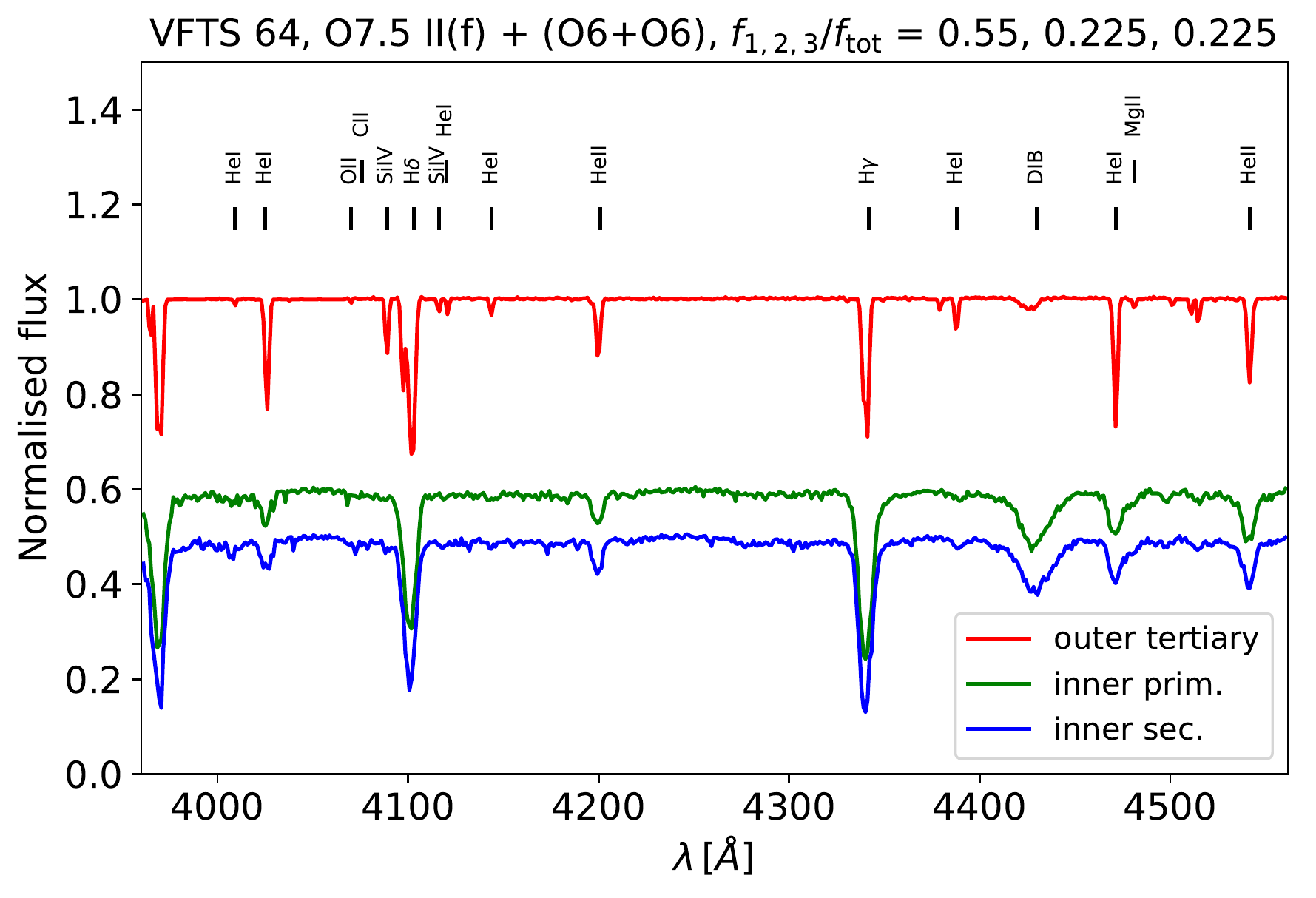}
\caption{Disentangled spectra of the system VFTS~64. Adopted light ratios are given in the plot header, and adopted $K_1, K_2$ values are given in Table\,\ref{tab:SampleFin} or in Appendix\,\ref{sec:indiv}. }
\label{fig:VFFTS64_DISSPEC}
\end{figure}

\begin{figure}
\centering
\includegraphics[width=.5\textwidth]{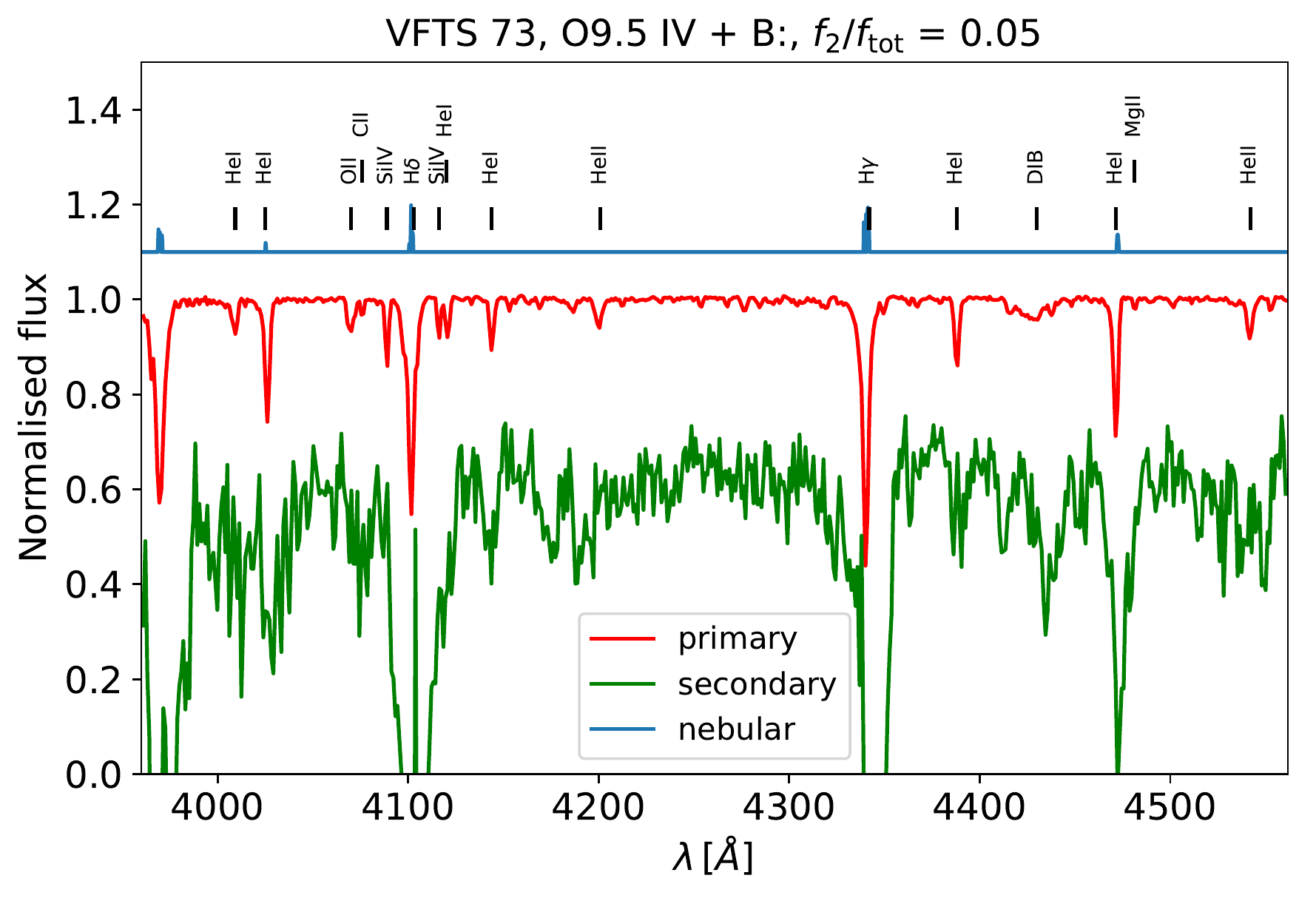}
\caption{As Fig.\,\ref{fig:VFFTS64_DISSPEC}, but for VFTS~73}
\label{fig:DISSPECTRA_VFTS73}
\end{figure}

\begin{figure}
\centering
\includegraphics[width=.5\textwidth]{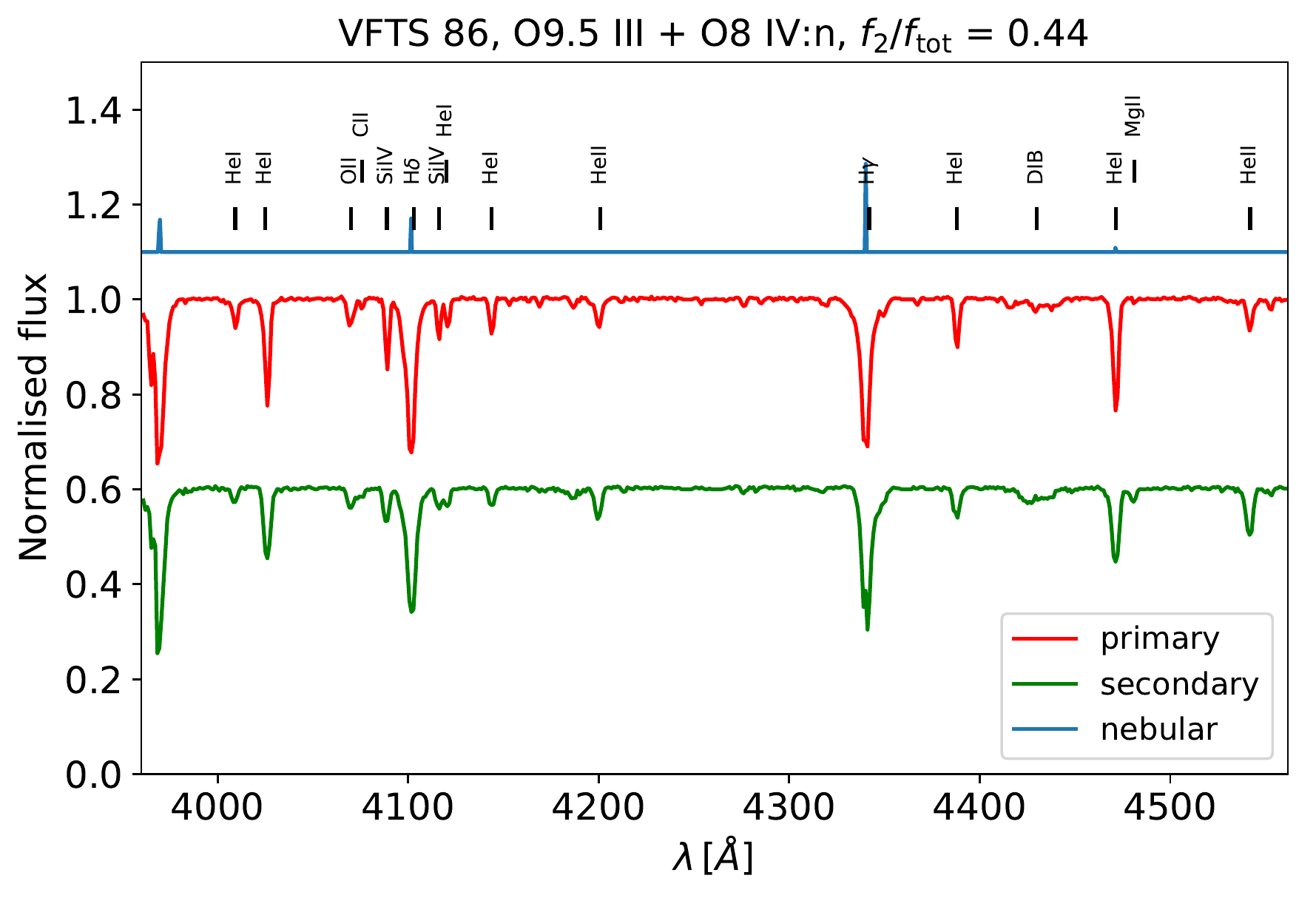}
\caption{As Fig.\,\ref{fig:VFFTS64_DISSPEC}, but for VFTS~86}
\label{fig:DISSPECTRA_VFTS86}
\end{figure}

\begin{figure}
\centering
\includegraphics[width=.5\textwidth]{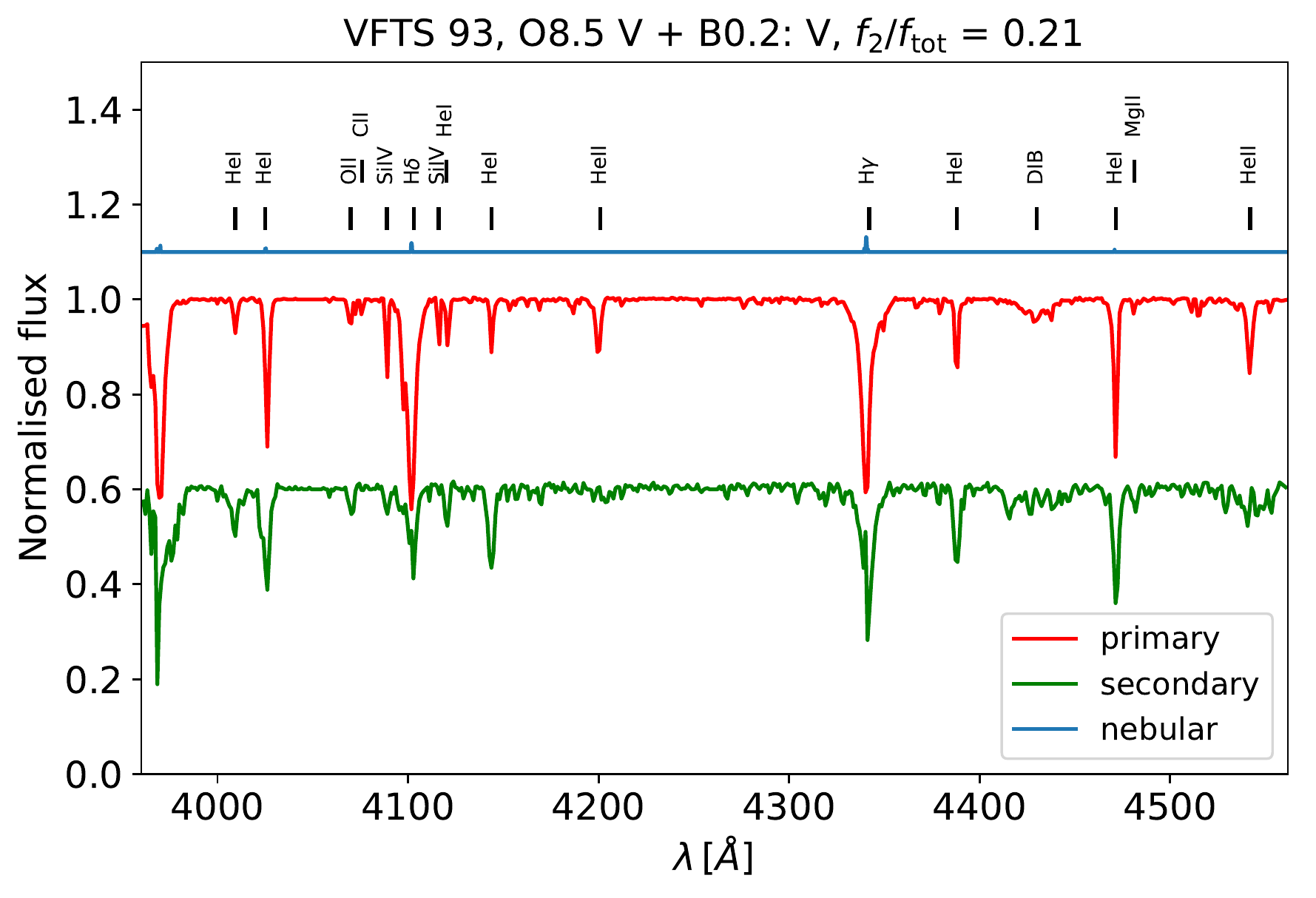}
\caption{As Fig.\,\ref{fig:VFFTS64_DISSPEC}, but for VFTS~93}
\label{fig:DISSPECTRA_VFTS93}
\end{figure}

\begin{figure}
\centering
\includegraphics[width=.5\textwidth]{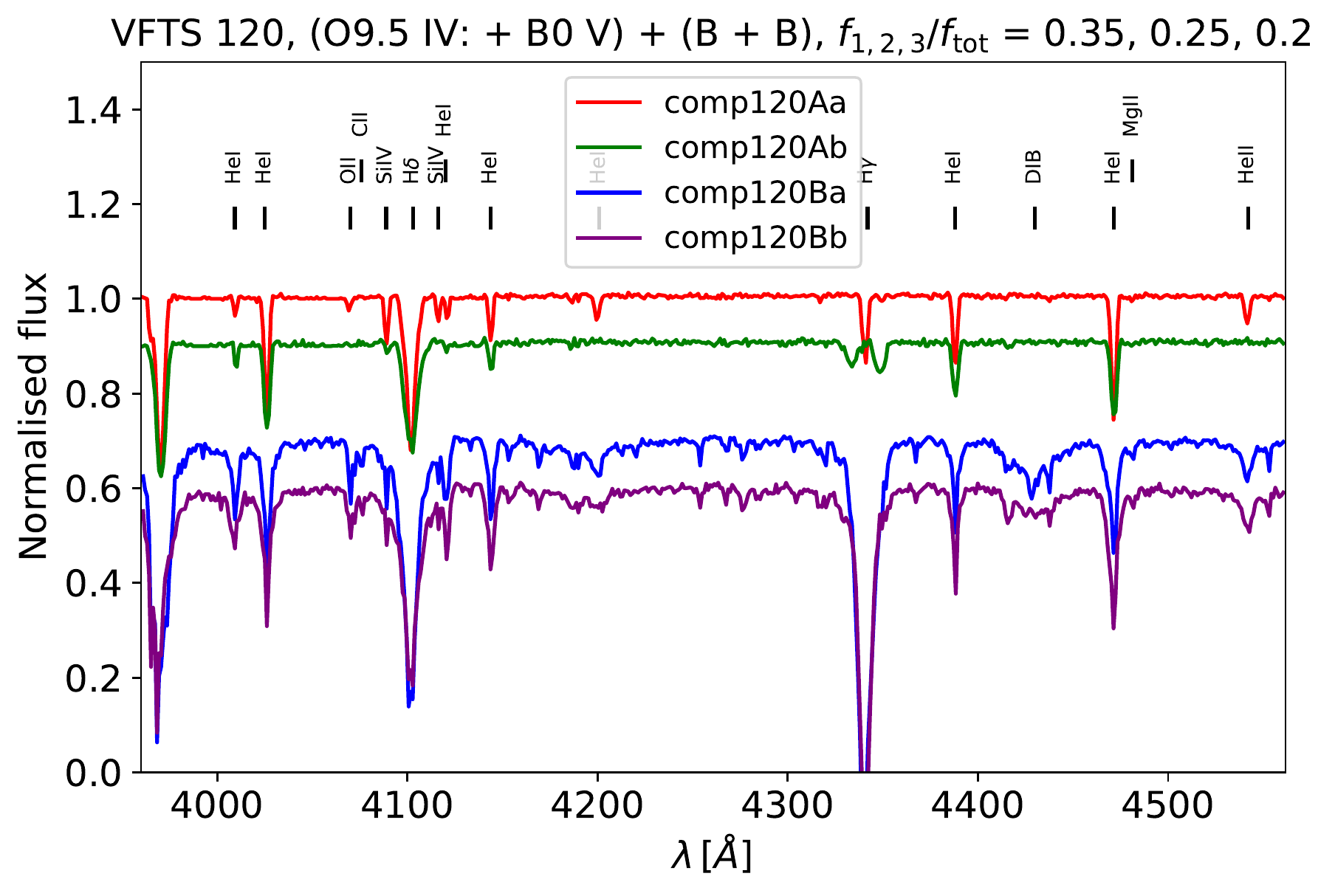}
\caption{As Fig.\,\ref{fig:VFFTS64_DISSPEC}, but for VFTS~120}
\label{fig:DISSPECTRA_VFTS120}
\end{figure}

\begin{figure}
\centering
\includegraphics[width=.5\textwidth]{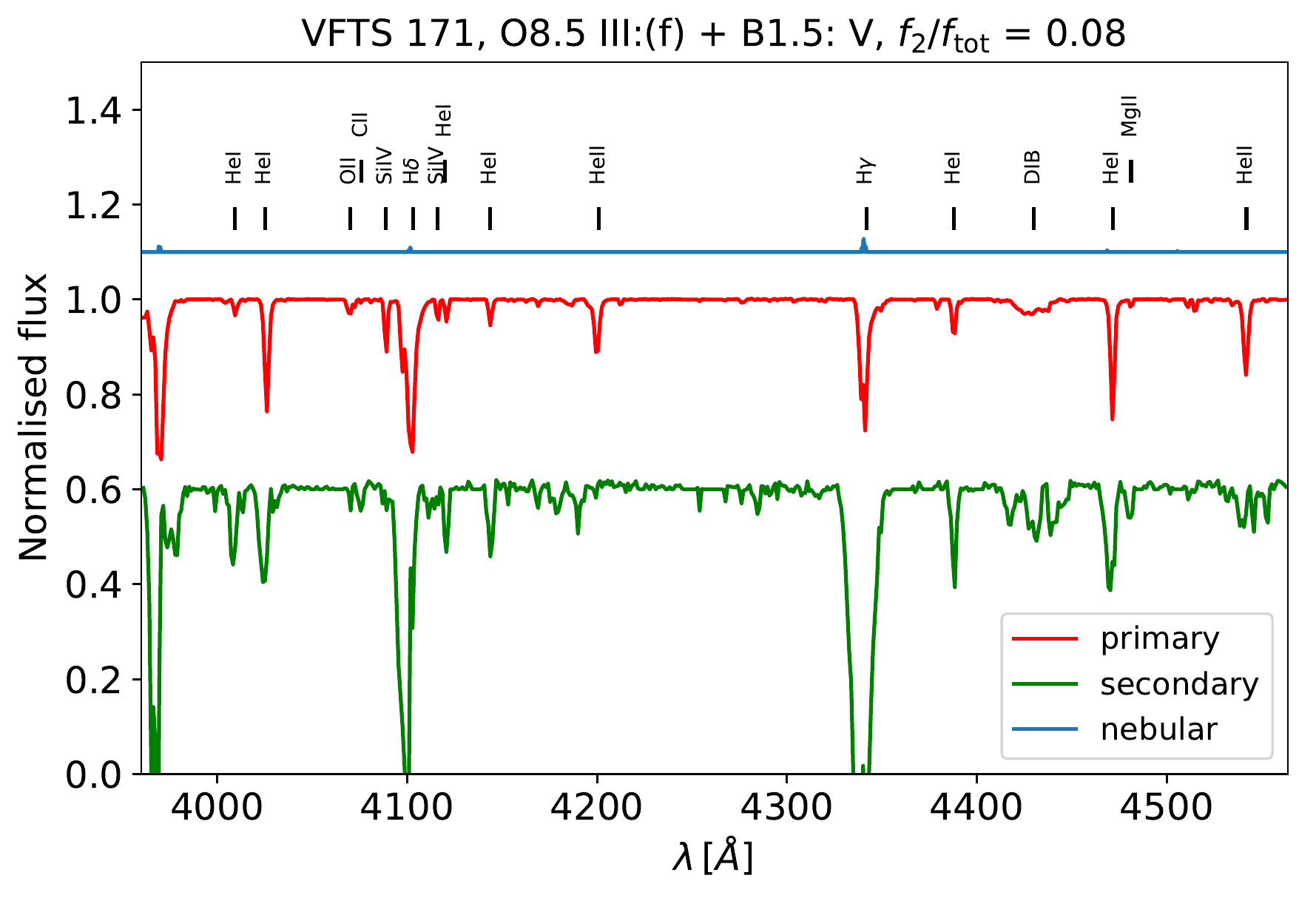}
\caption{As Fig.\,\ref{fig:VFFTS64_DISSPEC}, but for VFTS~171}
\label{fig:DISSPECTRA_VFTS171}
\end{figure}

\begin{figure}
\centering
\includegraphics[width=.5\textwidth]{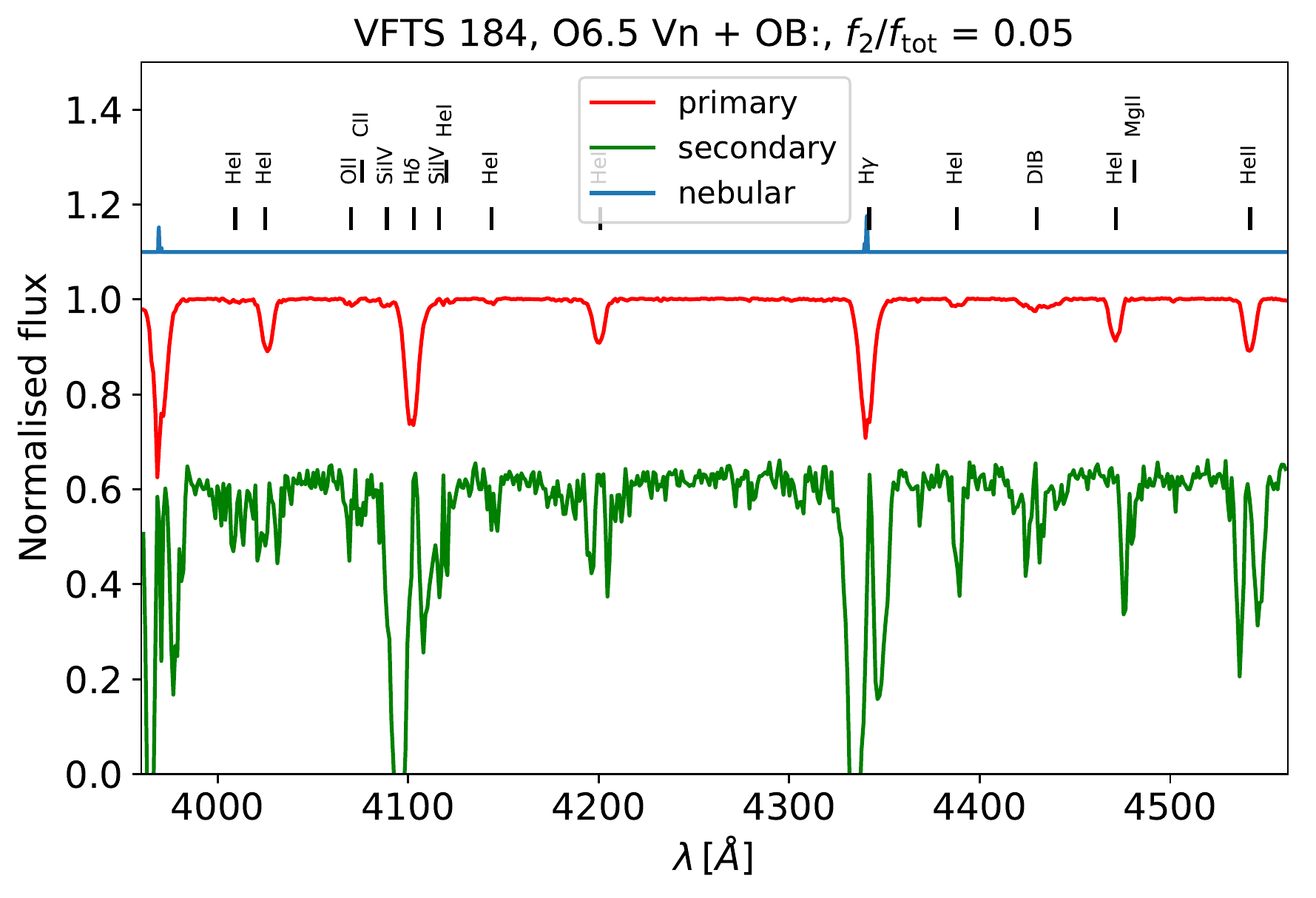}
\caption{As Fig.\,\ref{fig:VFFTS64_DISSPEC}, but for VFTS~184}
\label{fig:DISSPECTRA_VFTS184}
\end{figure}

\begin{figure}
\centering
\includegraphics[width=.5\textwidth]{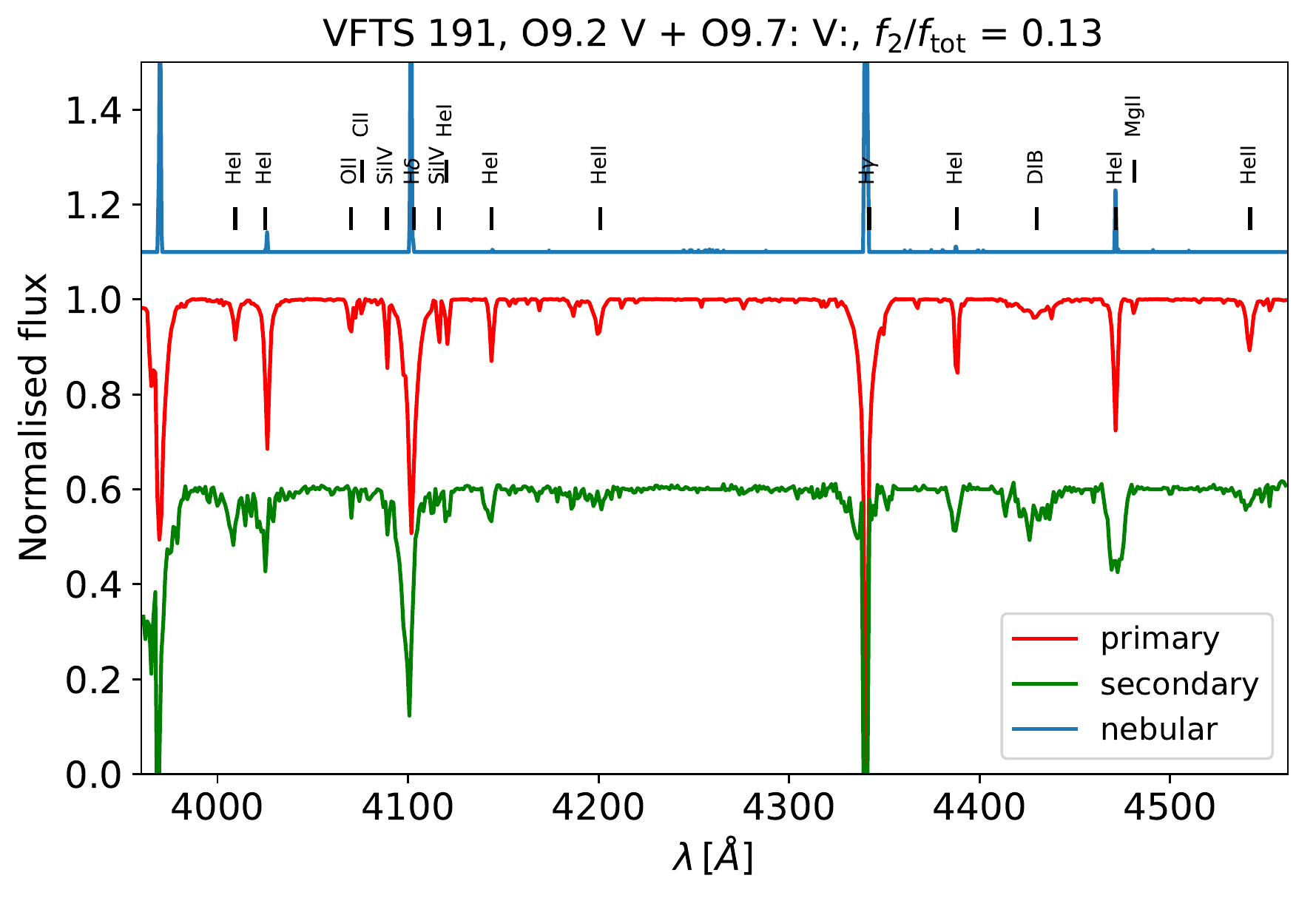}
\caption{As Fig.\,\ref{fig:VFFTS64_DISSPEC}, but for VFTS~191}
\label{fig:DISSPECTRA_VFTS191}
\end{figure}

\begin{figure}
\centering
\includegraphics[width=.5\textwidth]{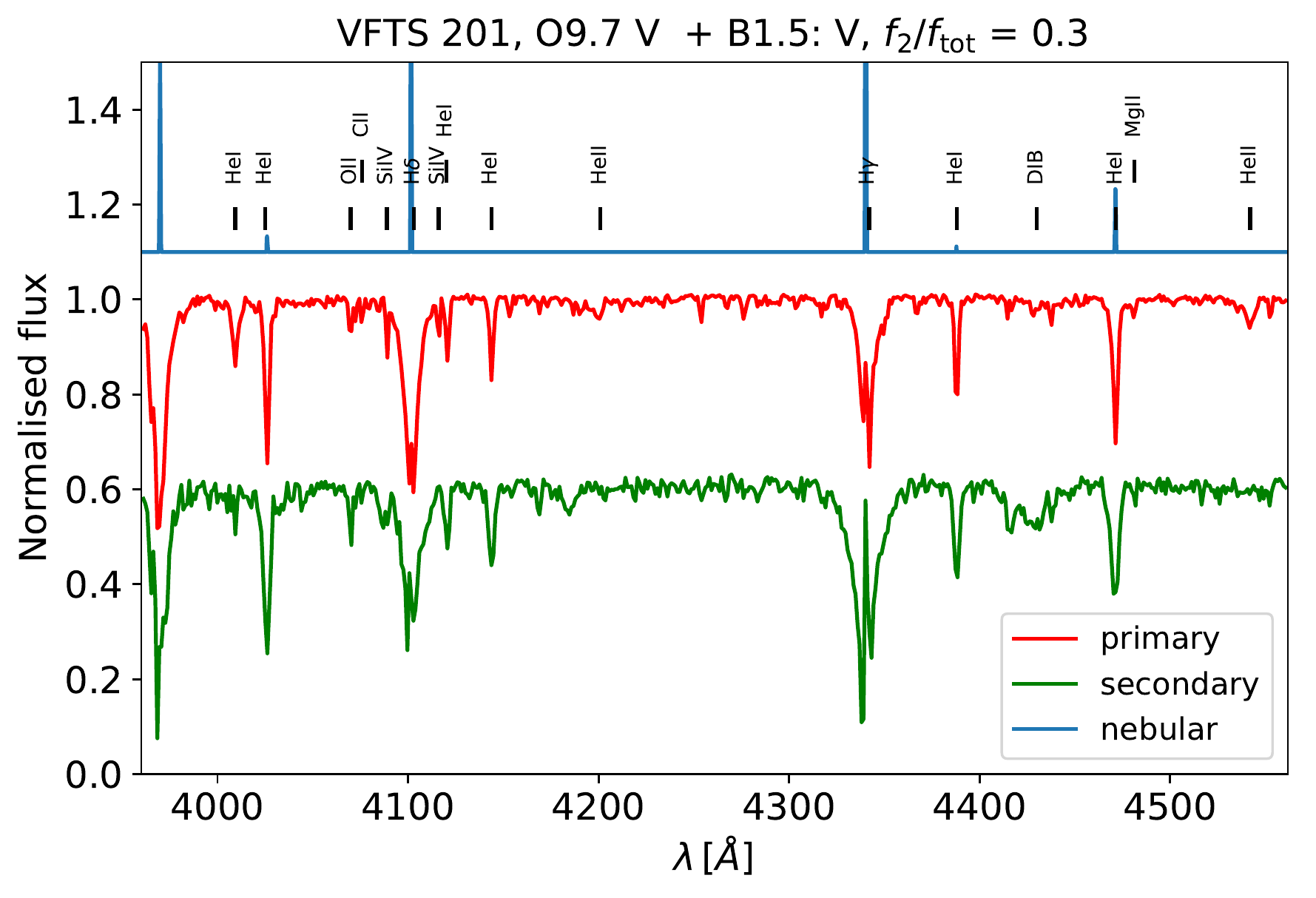}
\caption{As Fig.\,\ref{fig:VFFTS64_DISSPEC}, but for VFTS~201}
\label{fig:DISSPECTRA_VFTS201}
\end{figure}

\begin{figure}
\centering
\includegraphics[width=.5\textwidth]{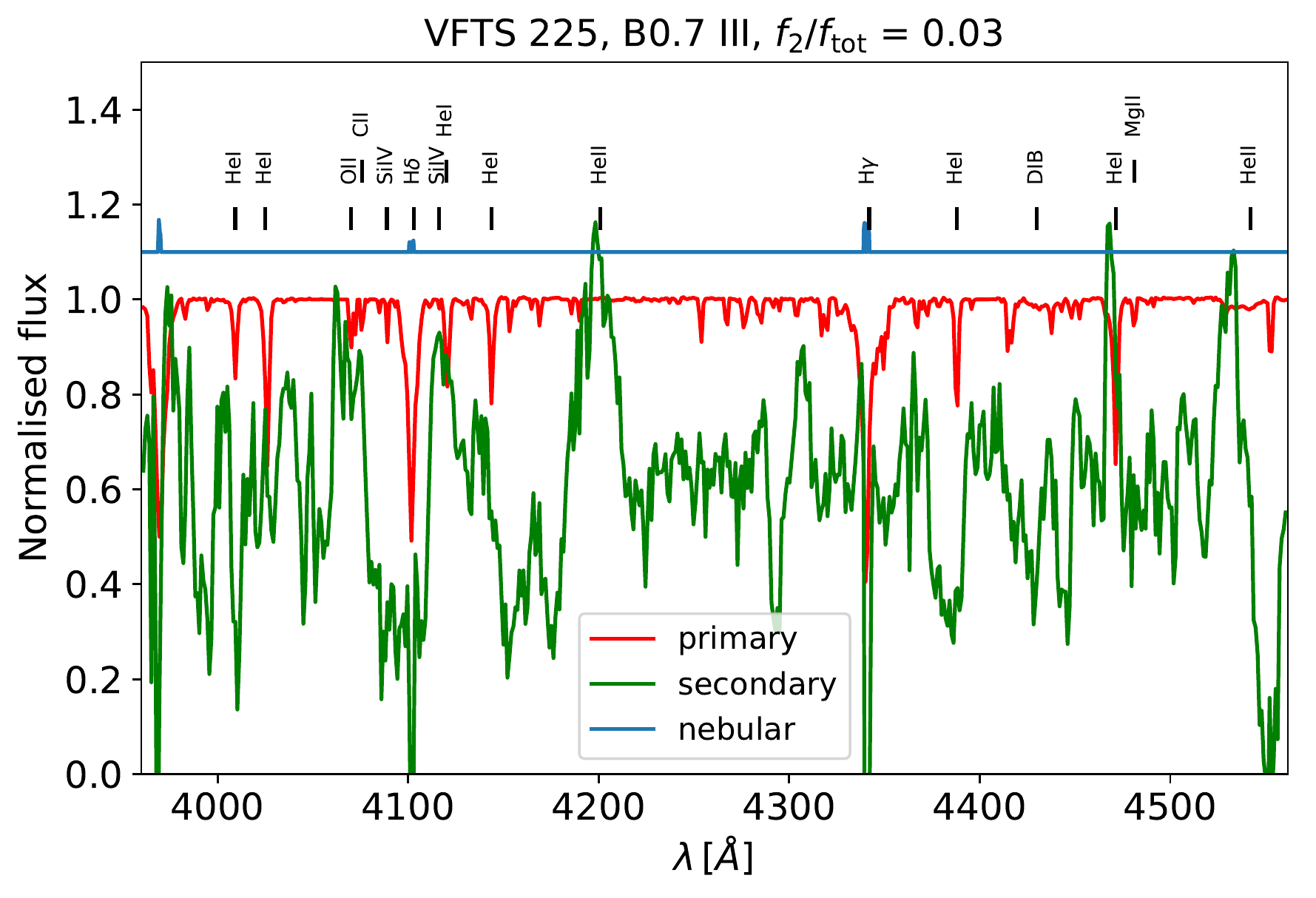}
\caption{As Fig.\,\ref{fig:VFFTS64_DISSPEC}, but for VFTS~225}
\label{fig:DISSPECTRA_VFTS225}
\end{figure}

\begin{figure}
\centering
\includegraphics[width=.5\textwidth]{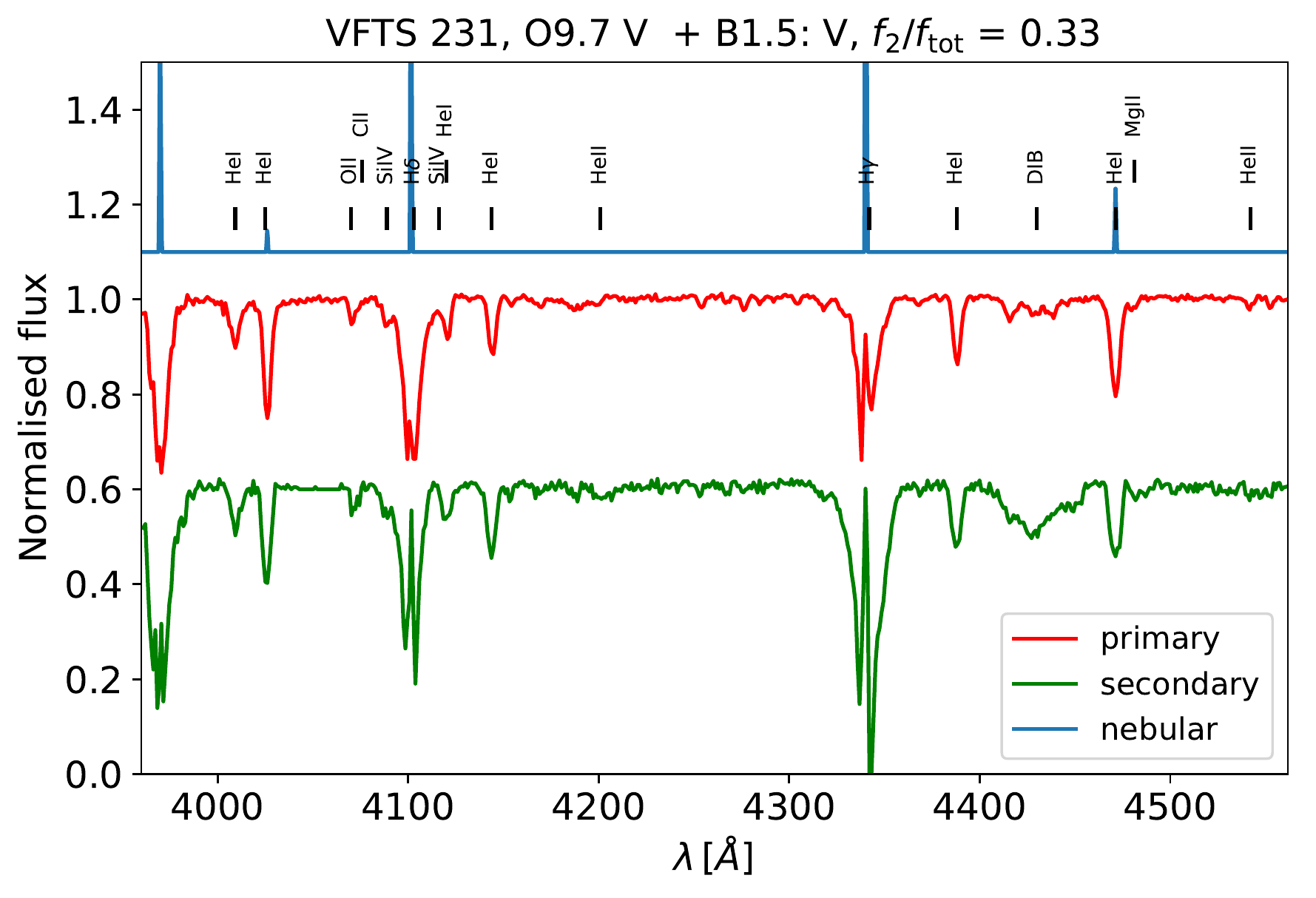}
\caption{As Fig.\,\ref{fig:VFFTS64_DISSPEC}, but for VFTS~231}
\label{fig:DISSPECTRA_VFTS231}
\end{figure}

\begin{figure}
\centering
\includegraphics[width=.5\textwidth]{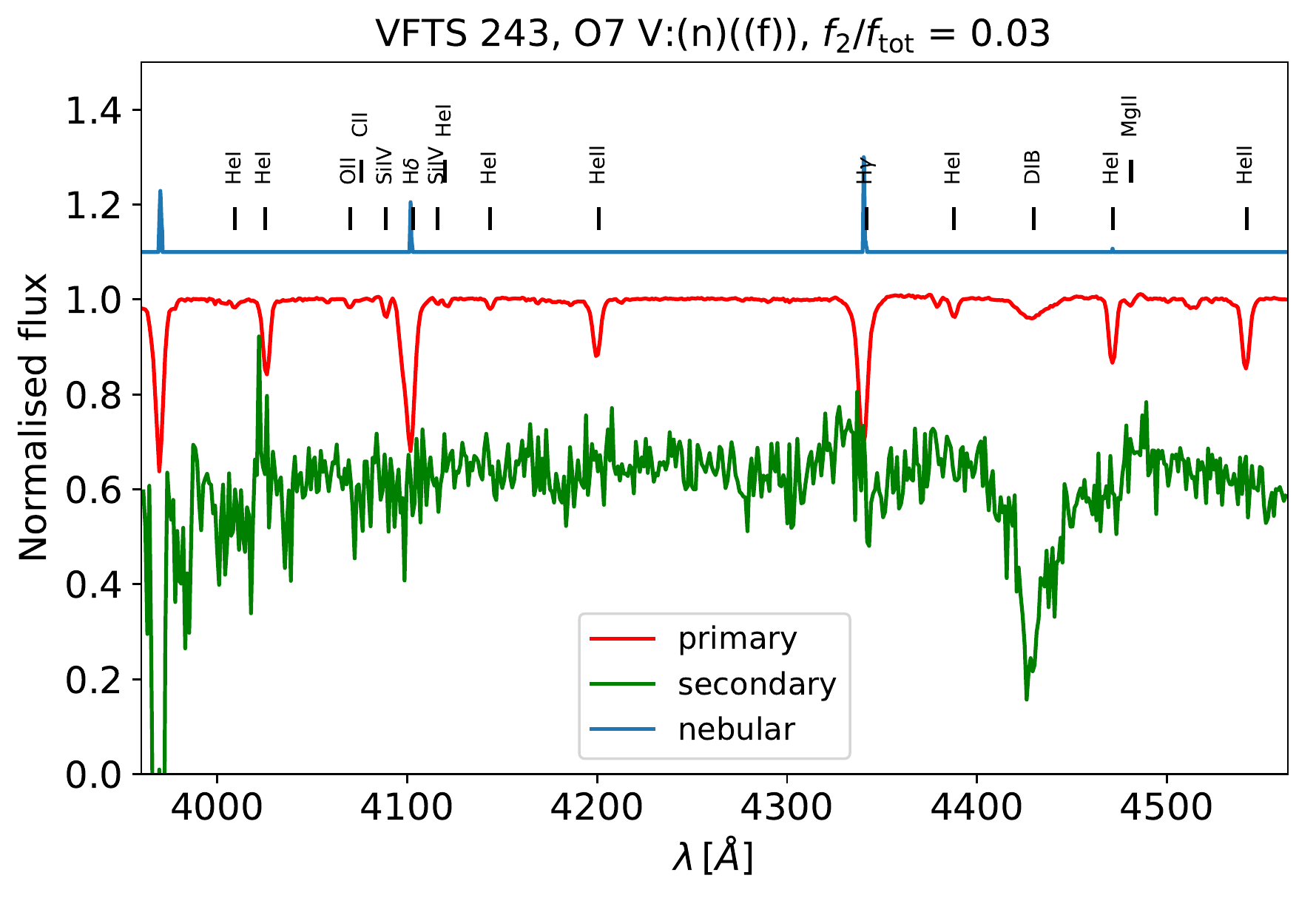}
\caption{As Fig.\,\ref{fig:VFFTS64_DISSPEC}, but for VFTS~243}
\label{fig:DISSPECTRA_VFTS243}
\end{figure}

\begin{figure}
\centering
\includegraphics[width=.5\textwidth]{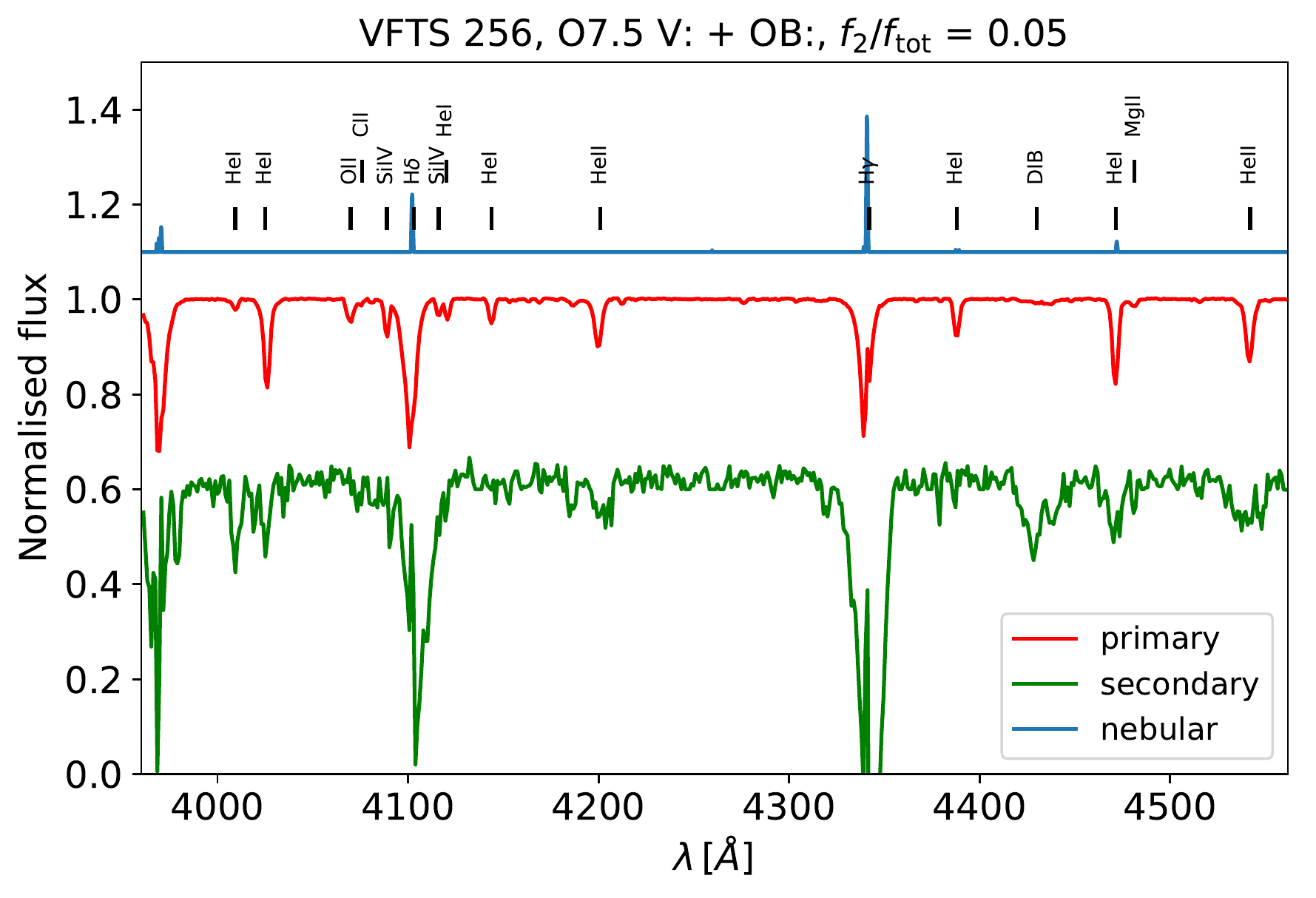}
\caption{As Fig.\,\ref{fig:VFFTS64_DISSPEC}, but for VFTS~256}
\label{fig:DISSPECTRA_VFTS256}
\end{figure}

\begin{figure}
\centering
\includegraphics[width=.5\textwidth]{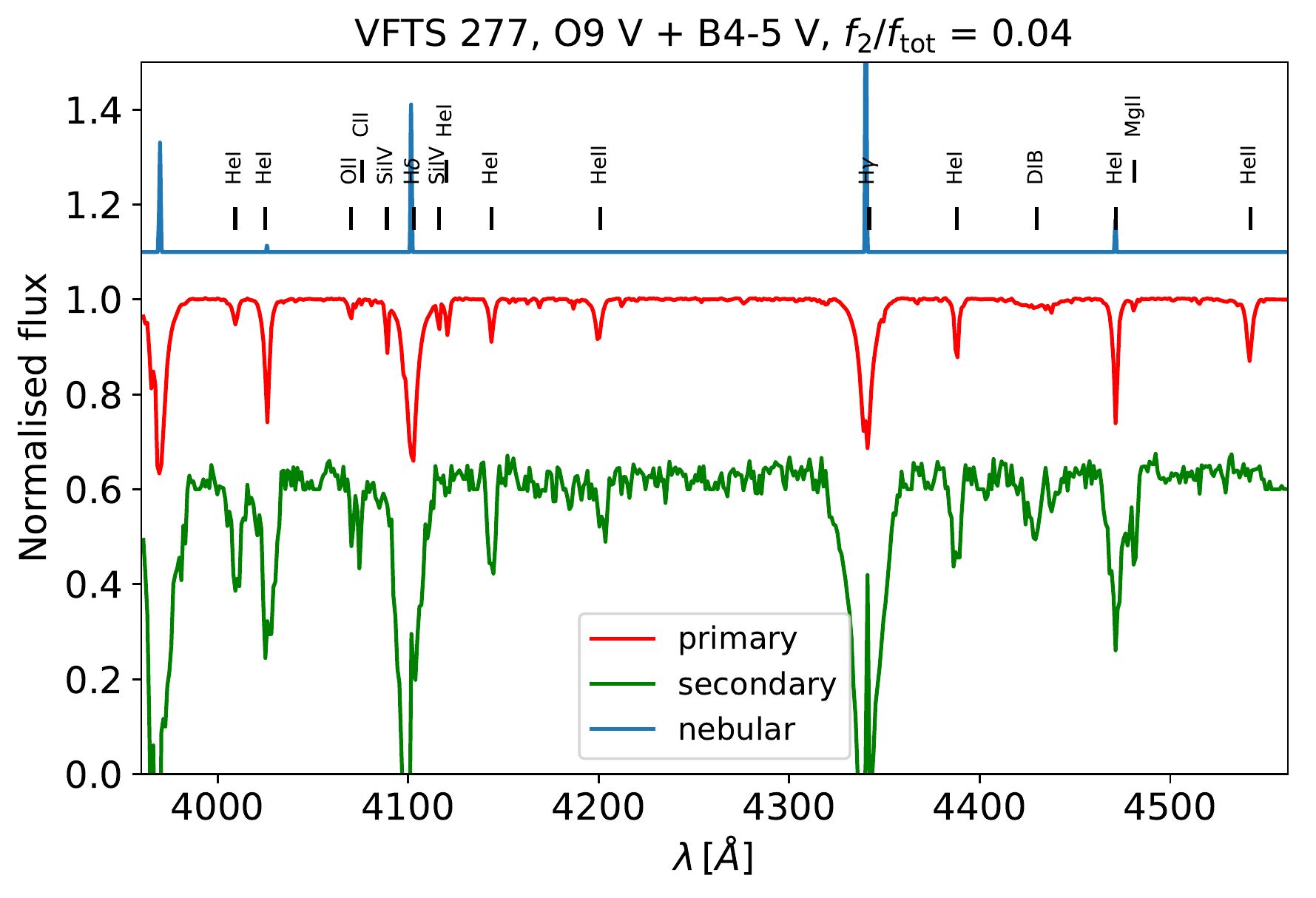}
\caption{As Fig.\,\ref{fig:VFFTS64_DISSPEC}, but for VFTS~277}
\label{fig:DISSPECTRA_VFTS277}
\end{figure}

\begin{figure}
\centering
\includegraphics[width=.5\textwidth]{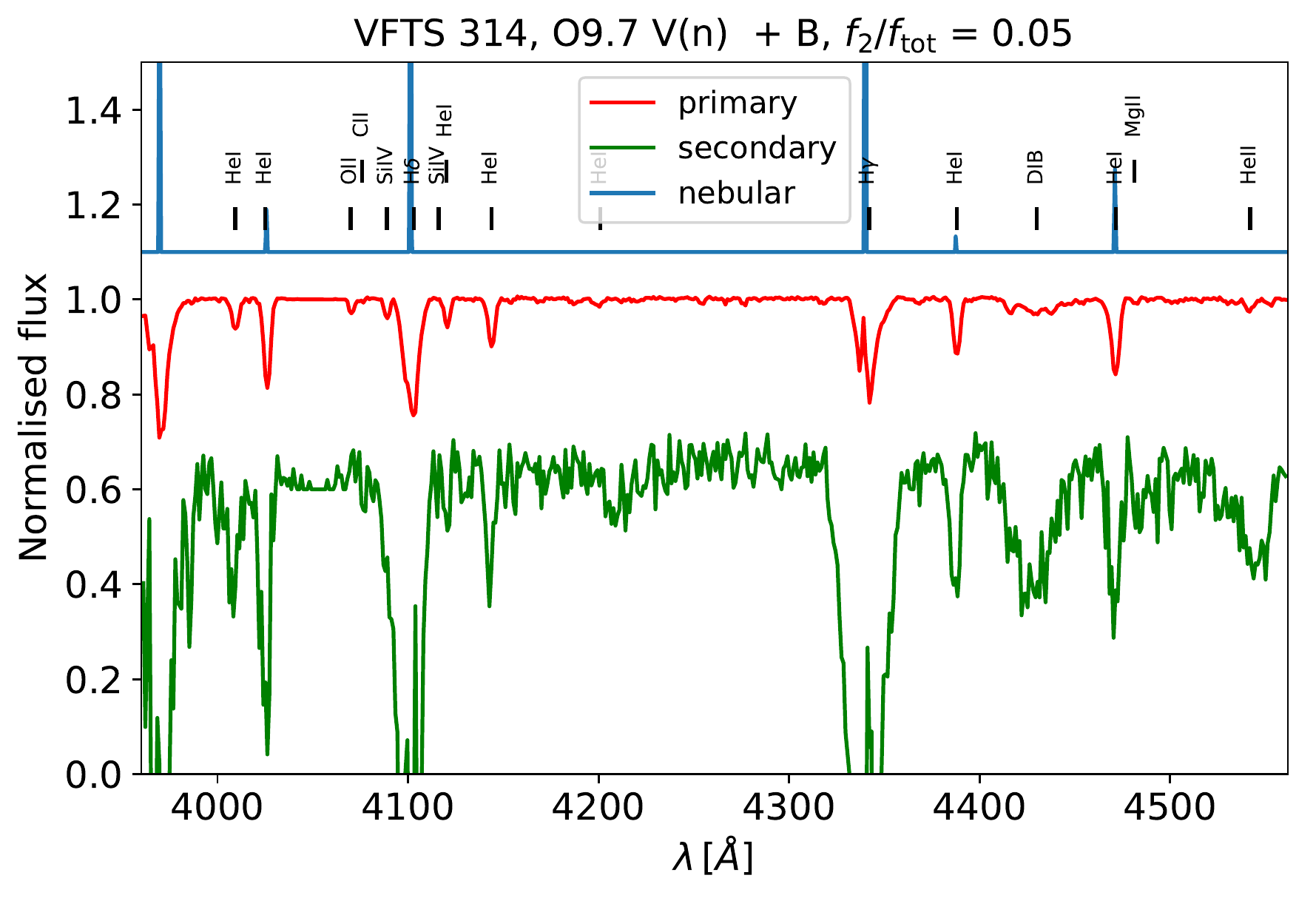}
\caption{As Fig.\,\ref{fig:VFFTS64_DISSPEC}, but for VFTS~314}
\label{fig:DISSPECTRA_VFTS314}
\end{figure}

\begin{figure}
\centering
\includegraphics[width=.5\textwidth]{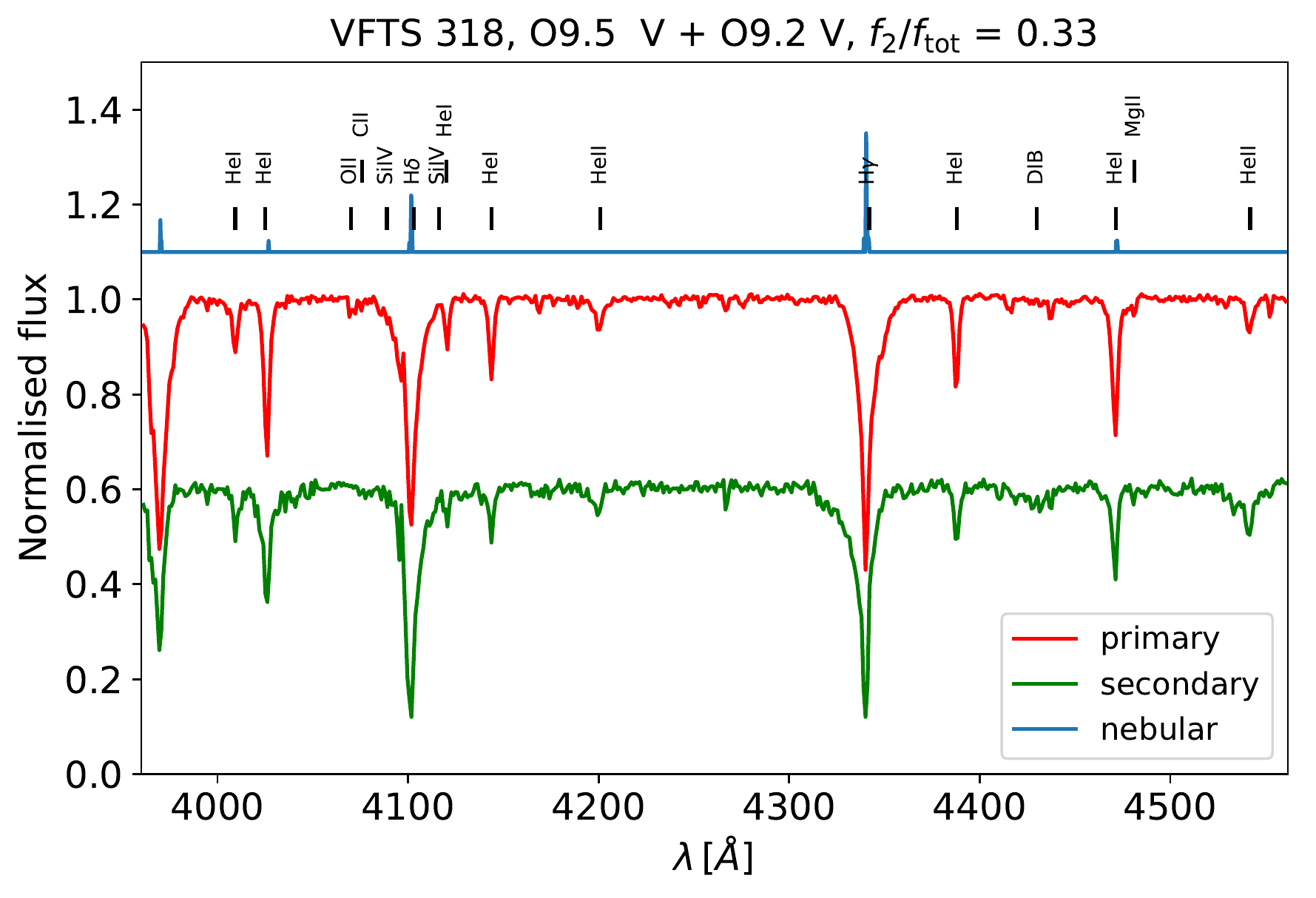}
\caption{As Fig.\,\ref{fig:VFFTS64_DISSPEC}, but for VFTS~318}
\label{fig:DISSPECTRA_VFTS318}
\end{figure}

\begin{figure}
\centering
\includegraphics[width=.5\textwidth]{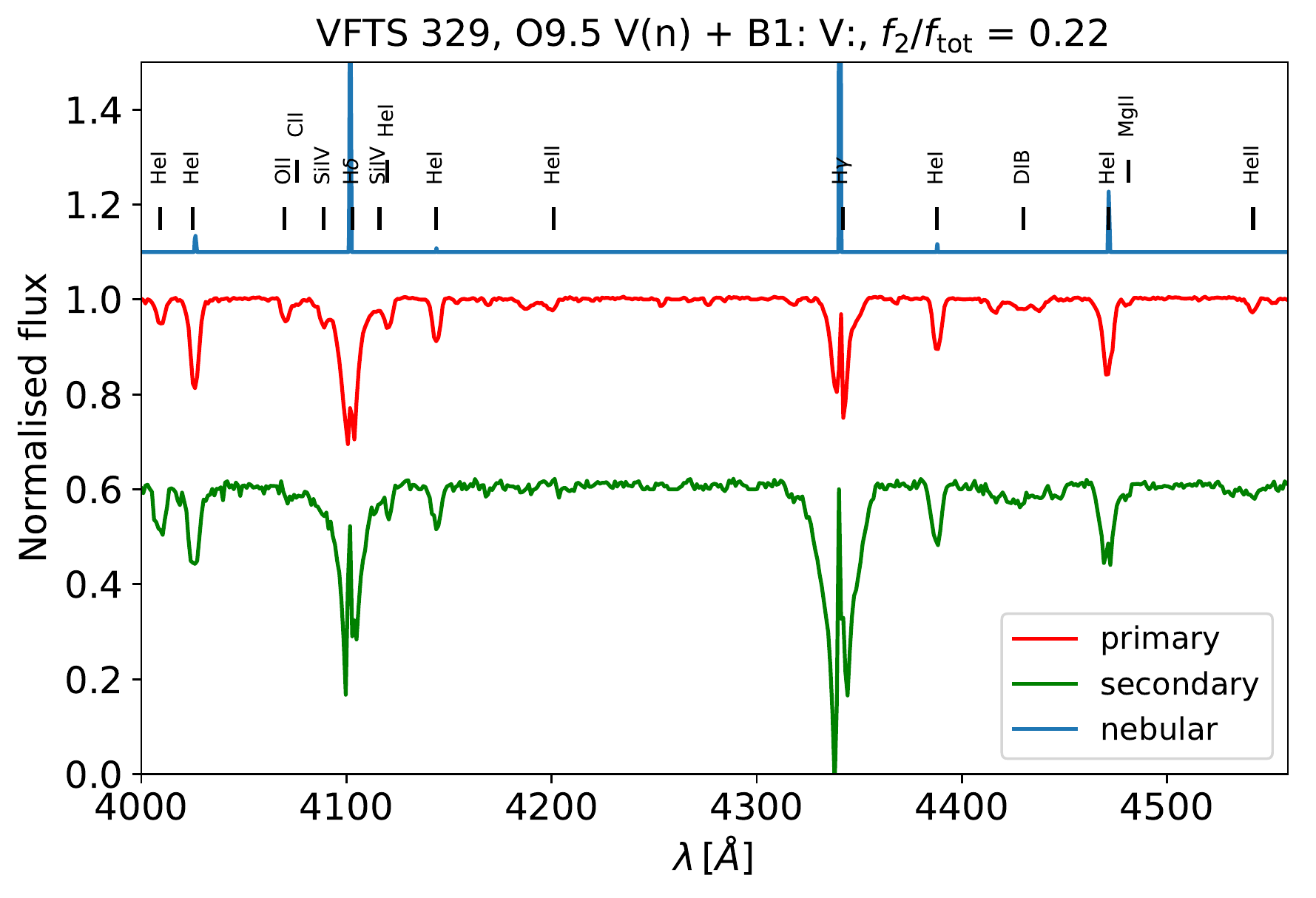}
\caption{As Fig.\,\ref{fig:VFFTS64_DISSPEC}, but for VFTS~329}
\label{fig:DISSPECTRA_VFTS329}
\end{figure}

\begin{figure}
\centering
\includegraphics[width=.5\textwidth]{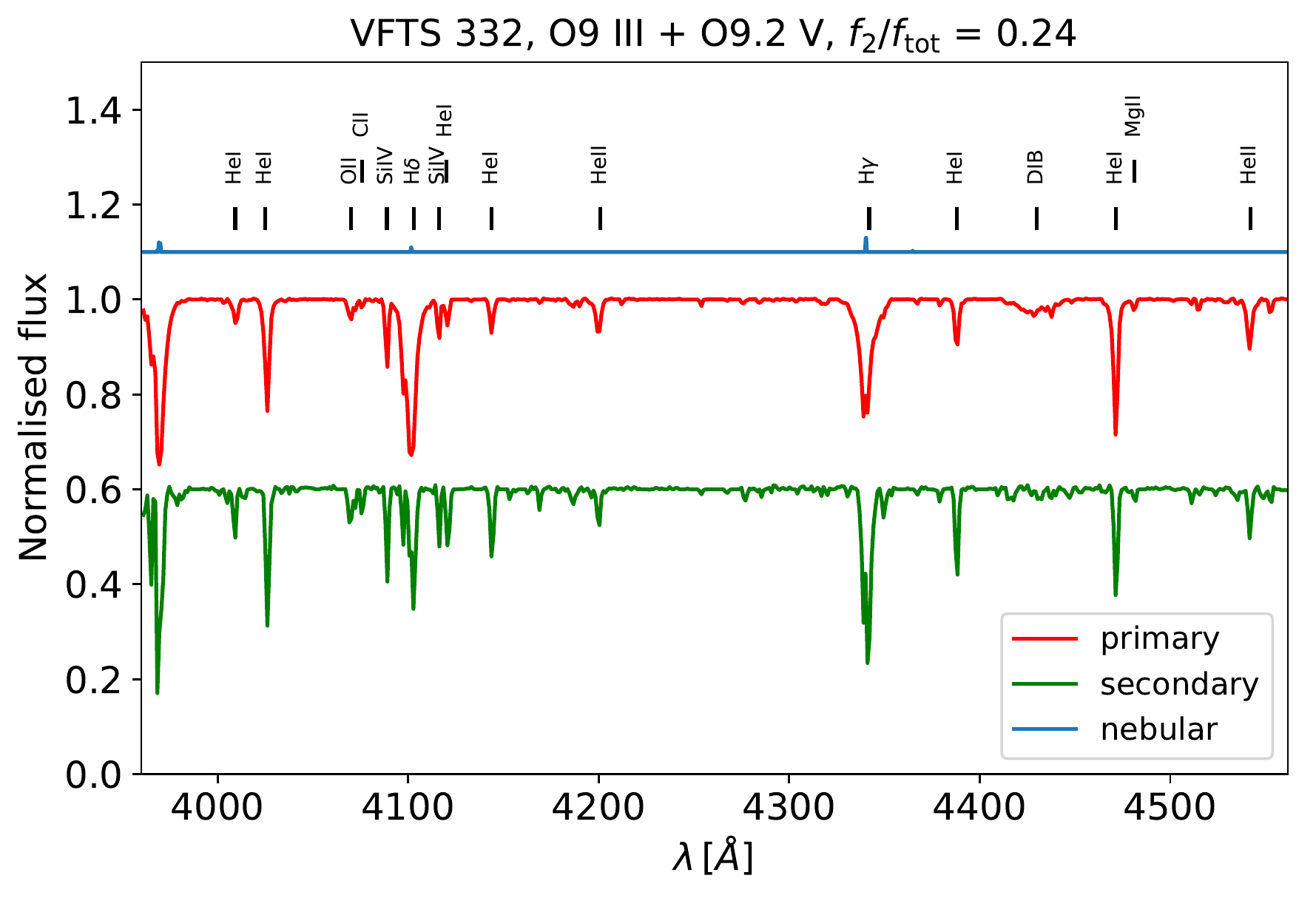}
\caption{As Fig.\,\ref{fig:VFFTS64_DISSPEC}, but for VFTS~332}
\label{fig:DISSPECTRA_VFTS332}
\end{figure}

\begin{figure}
\centering
\includegraphics[width=.5\textwidth]{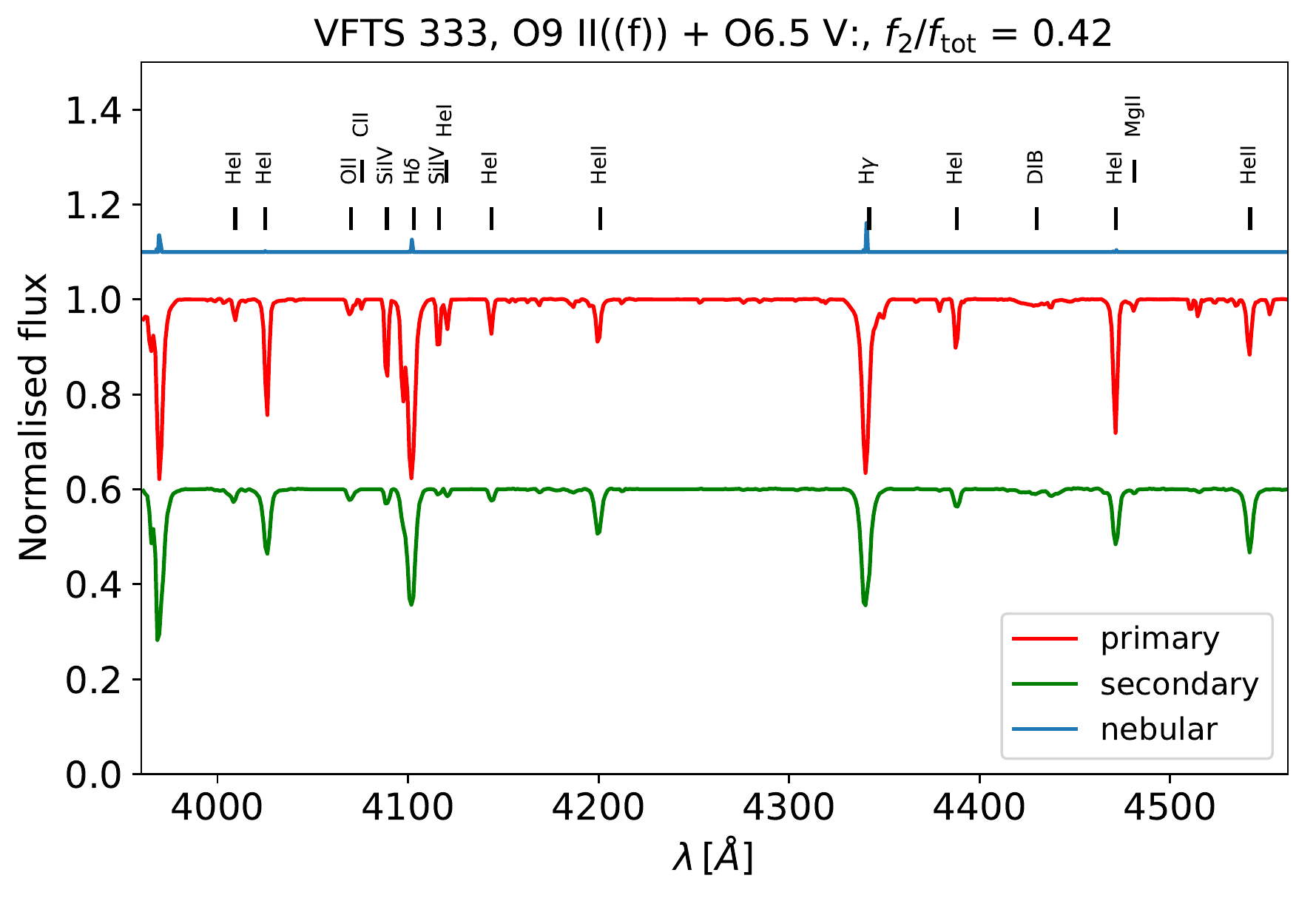}
\caption{As Fig.\,\ref{fig:VFFTS64_DISSPEC}, but for VFTS~333}
\label{fig:DISSPECTRA_VFTS333}
\end{figure}

\begin{figure}
\centering
\includegraphics[width=.5\textwidth]{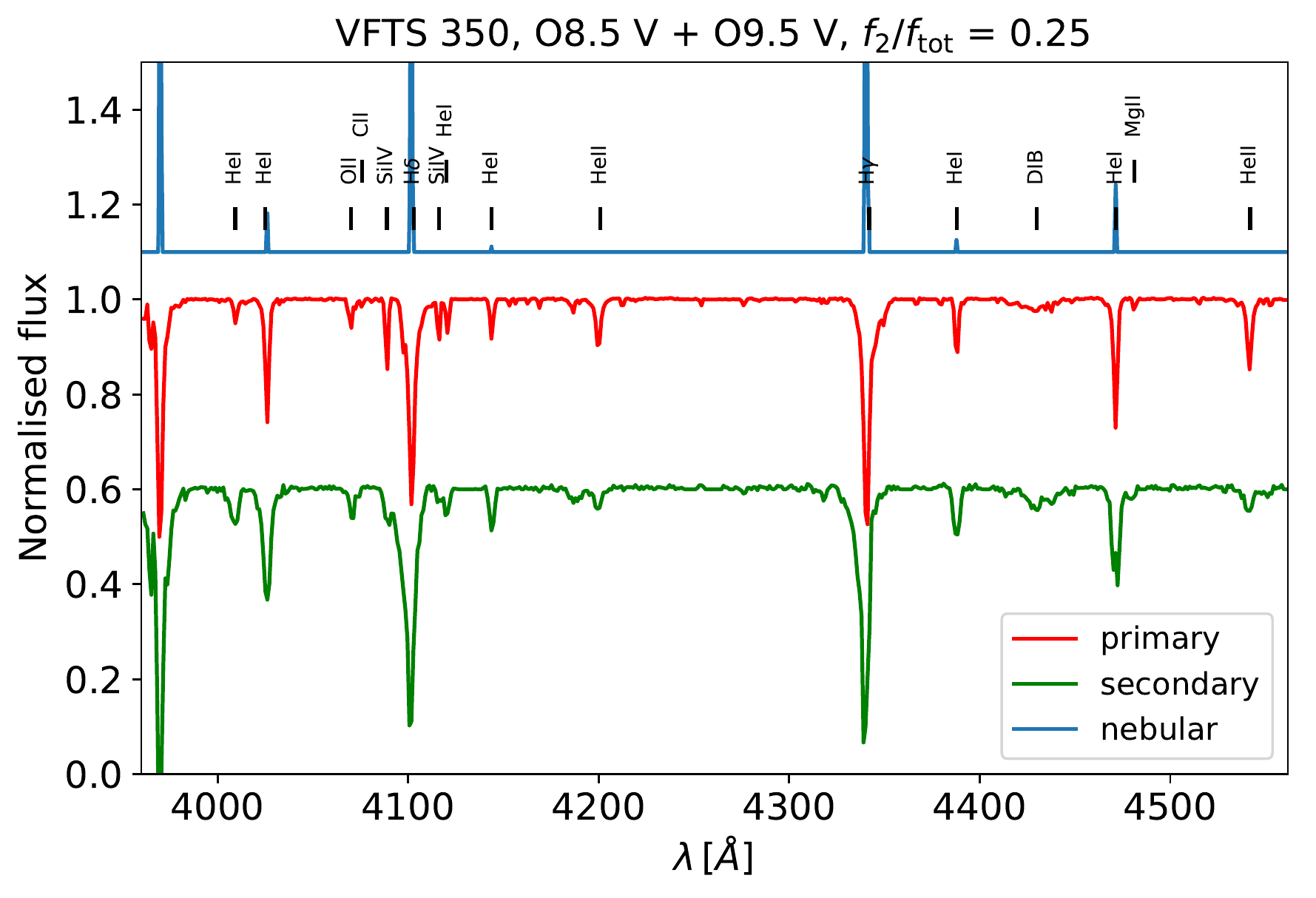}
\caption{As Fig.\,\ref{fig:VFFTS64_DISSPEC}, but for VFTS~350}
\label{fig:DISSPECTRA_VFTS350}
\end{figure}

\begin{figure}
\centering
\includegraphics[width=.5\textwidth]{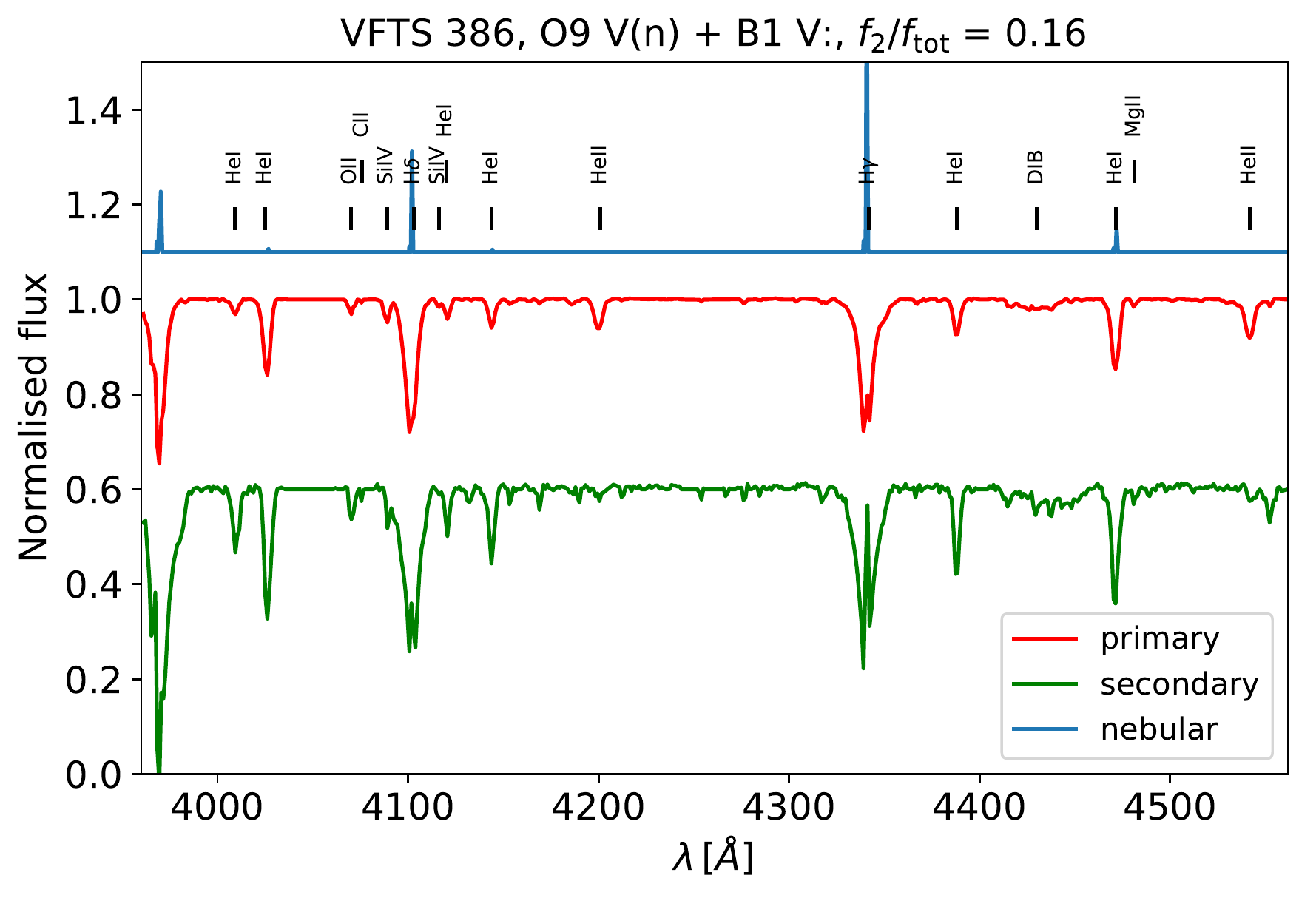}
\caption{As Fig.\,\ref{fig:VFFTS64_DISSPEC}, but for VFTS386}
\label{fig:DISSPECTRA_VFTS386}
\end{figure}

\begin{figure}
\centering
\includegraphics[width=.5\textwidth]{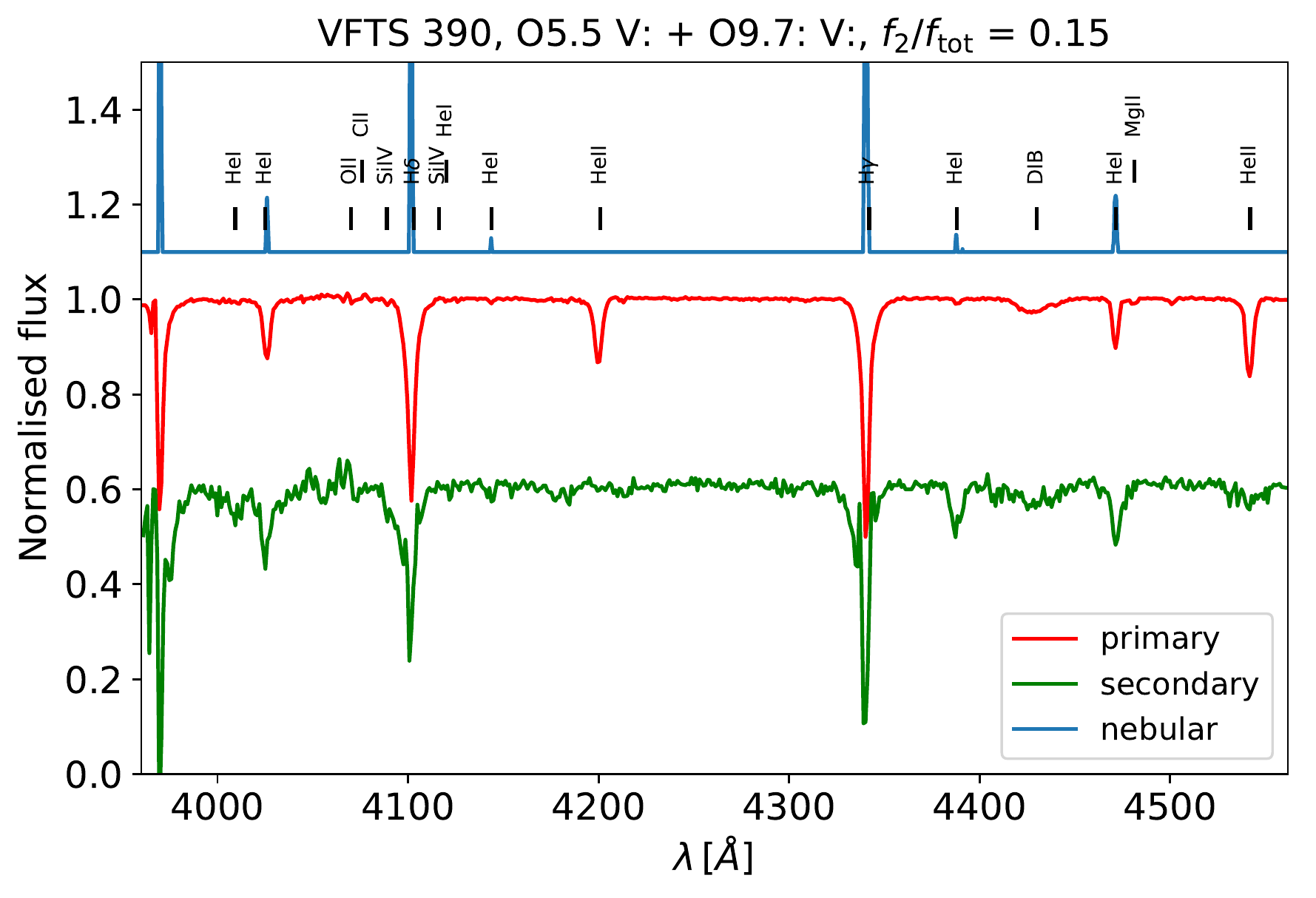}
\caption{As Fig.\,\ref{fig:VFFTS64_DISSPEC}, but for VFTS~390}
\label{fig:DISSPECTRA_VFTS390}
\end{figure}

\begin{figure}
\centering
\includegraphics[width=.5\textwidth]{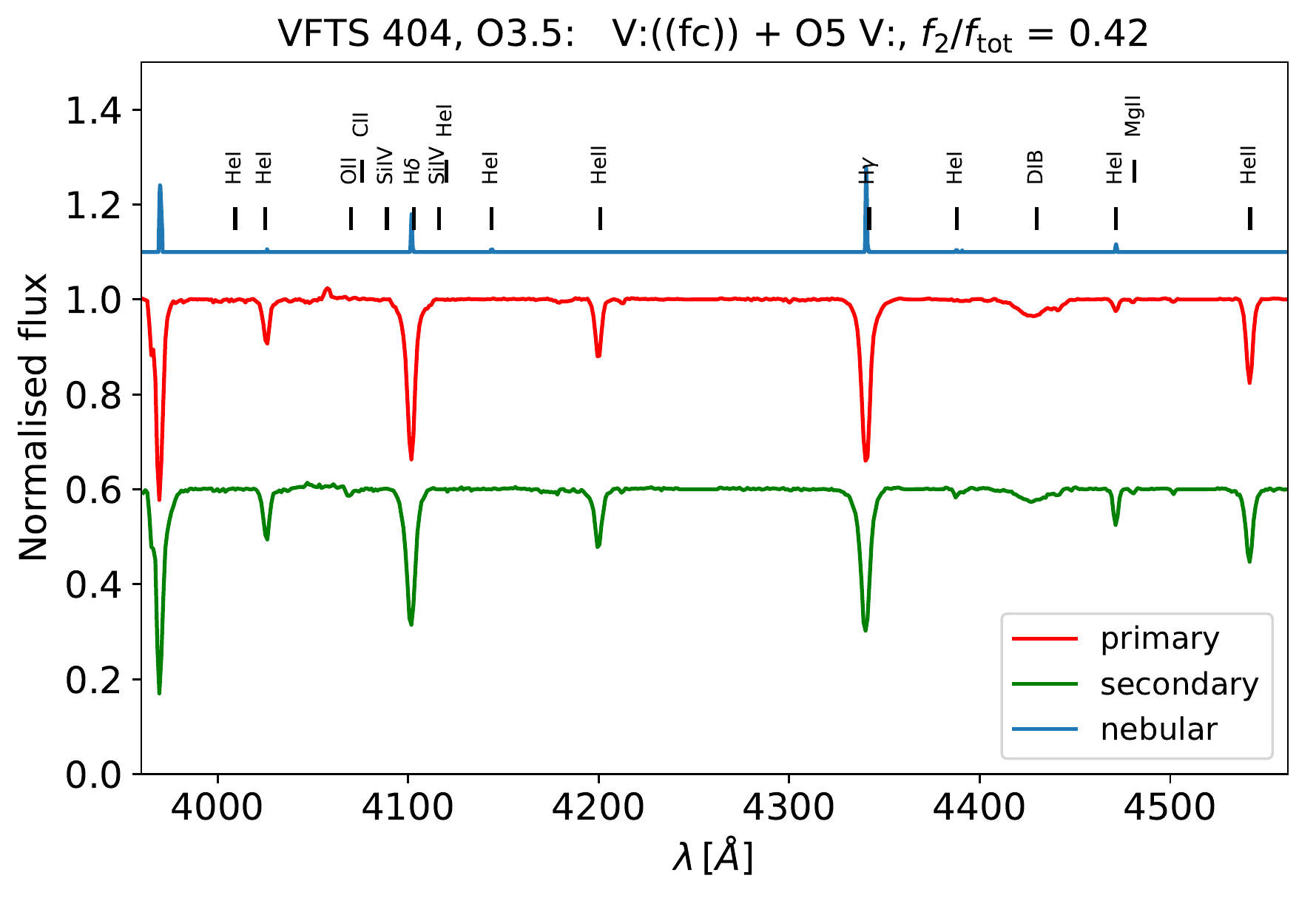}
\caption{As Fig.\,\ref{fig:VFFTS64_DISSPEC}, but for VFTS~404}
\label{fig:DISSPECTRA_VFTS404}
\end{figure}

\begin{figure}
\centering
\includegraphics[width=.5\textwidth]{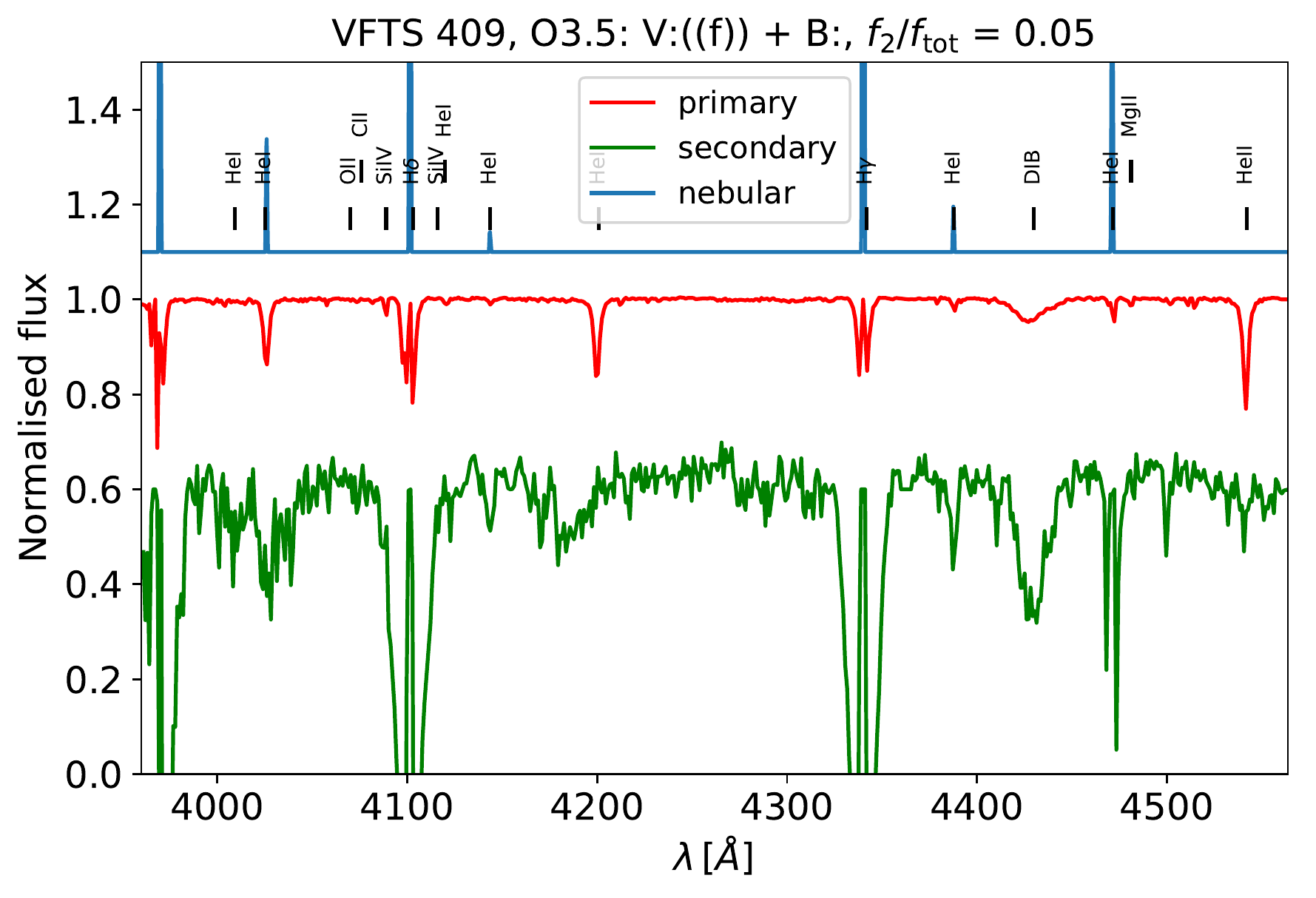}
\caption{As Fig.\,\ref{fig:VFFTS64_DISSPEC}, but for VFTS~409}
\label{fig:DISSPECTRA_VFTS409}
\end{figure}

\begin{figure}
\centering
\includegraphics[width=.5\textwidth]{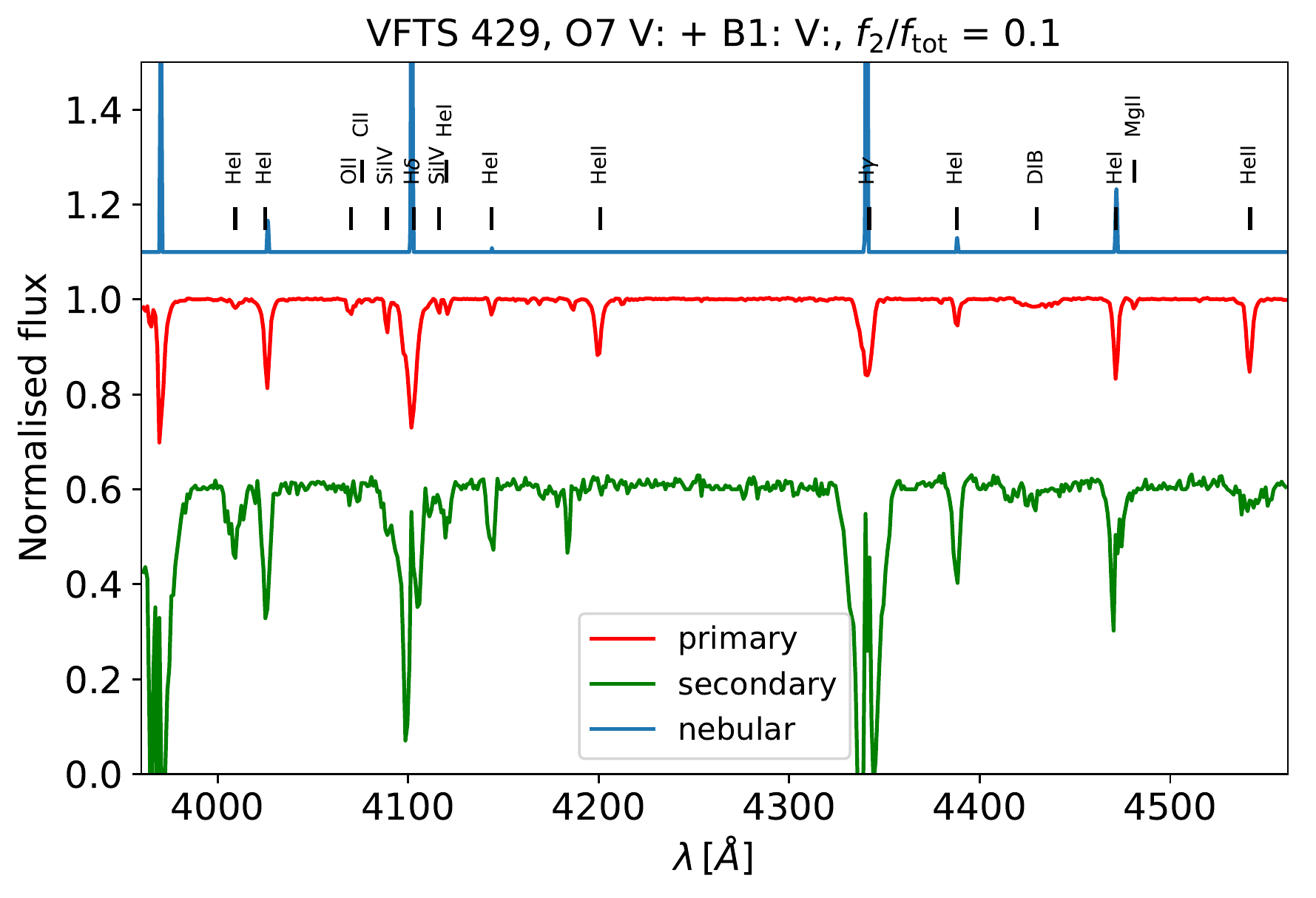}
\caption{As Fig.\,\ref{fig:VFFTS64_DISSPEC}, but for VFTS~429}
\label{fig:DISSPECTRA_VFTS429}
\end{figure}

\begin{figure}
\centering
\includegraphics[width=.5\textwidth]{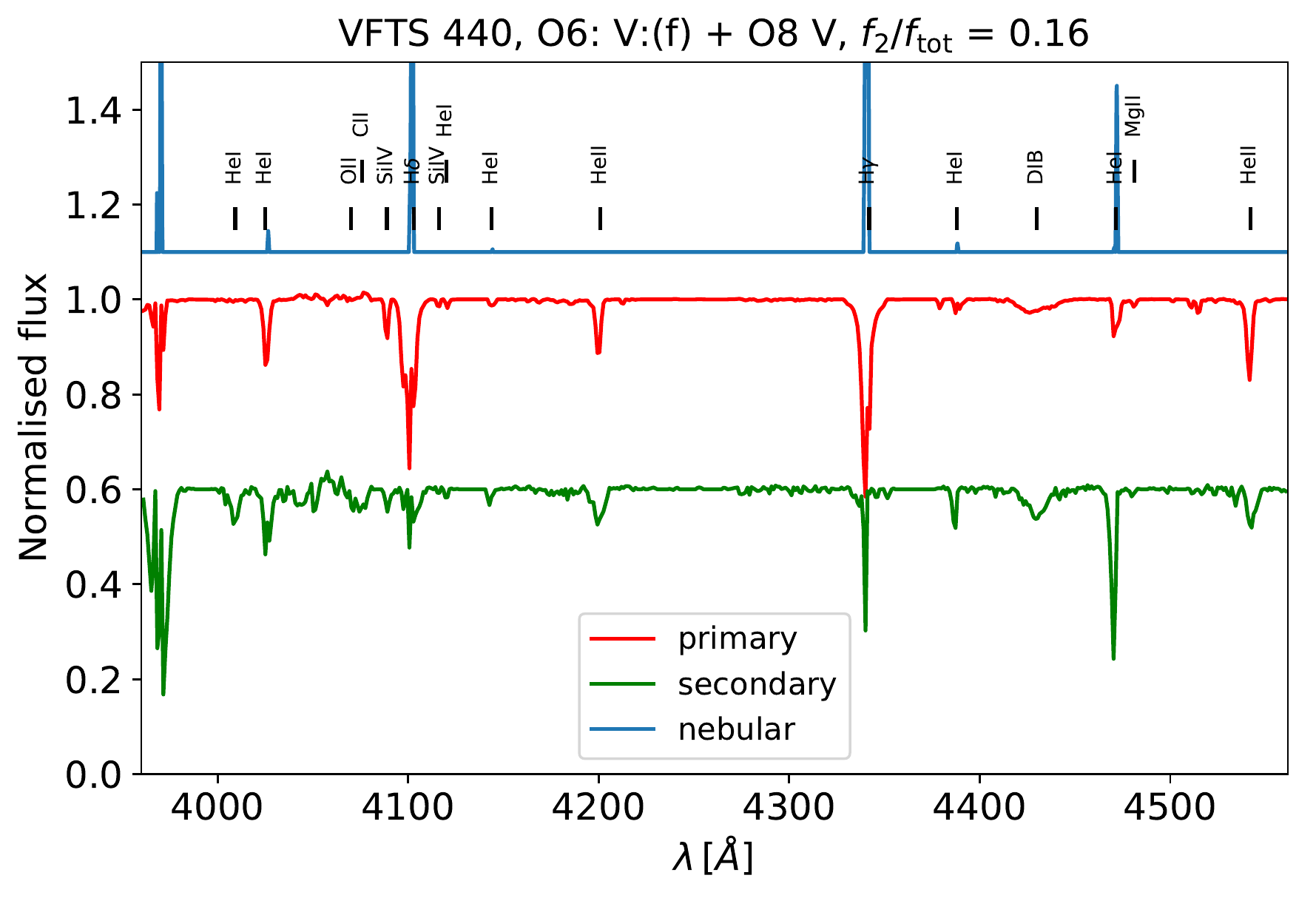}
\caption{As Fig.\,\ref{fig:VFFTS64_DISSPEC}, but for VFTS~440}
\label{fig:DISSPECTRA_VFTS440}
\end{figure}

\begin{figure}
\centering
\includegraphics[width=.5\textwidth]{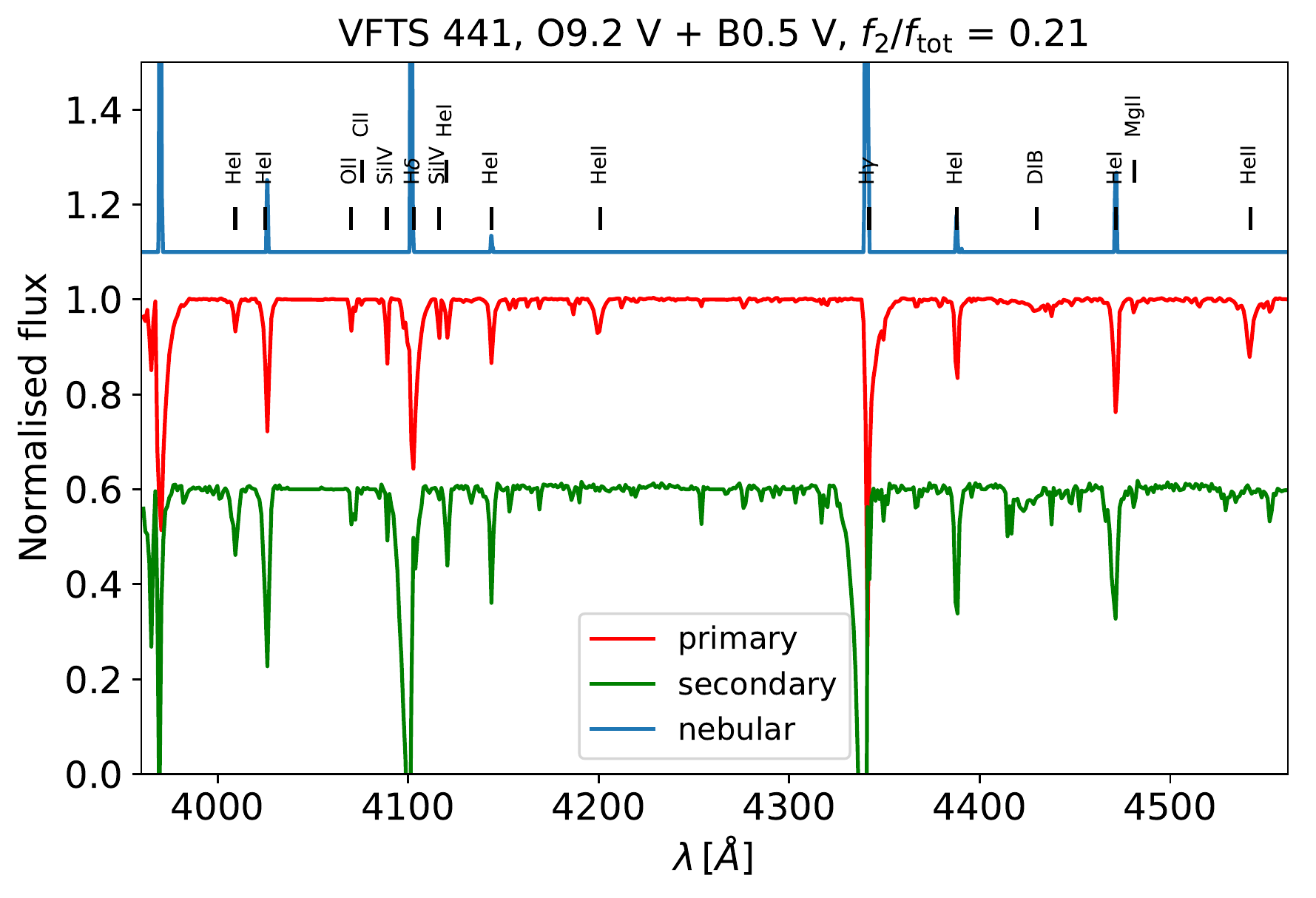}
\caption{As Fig.\,\ref{fig:VFFTS64_DISSPEC}, but for VFTS~441}
\label{fig:DISSPECTRA_VFTS441}
\end{figure}

\begin{figure}
\centering
\includegraphics[width=.5\textwidth]{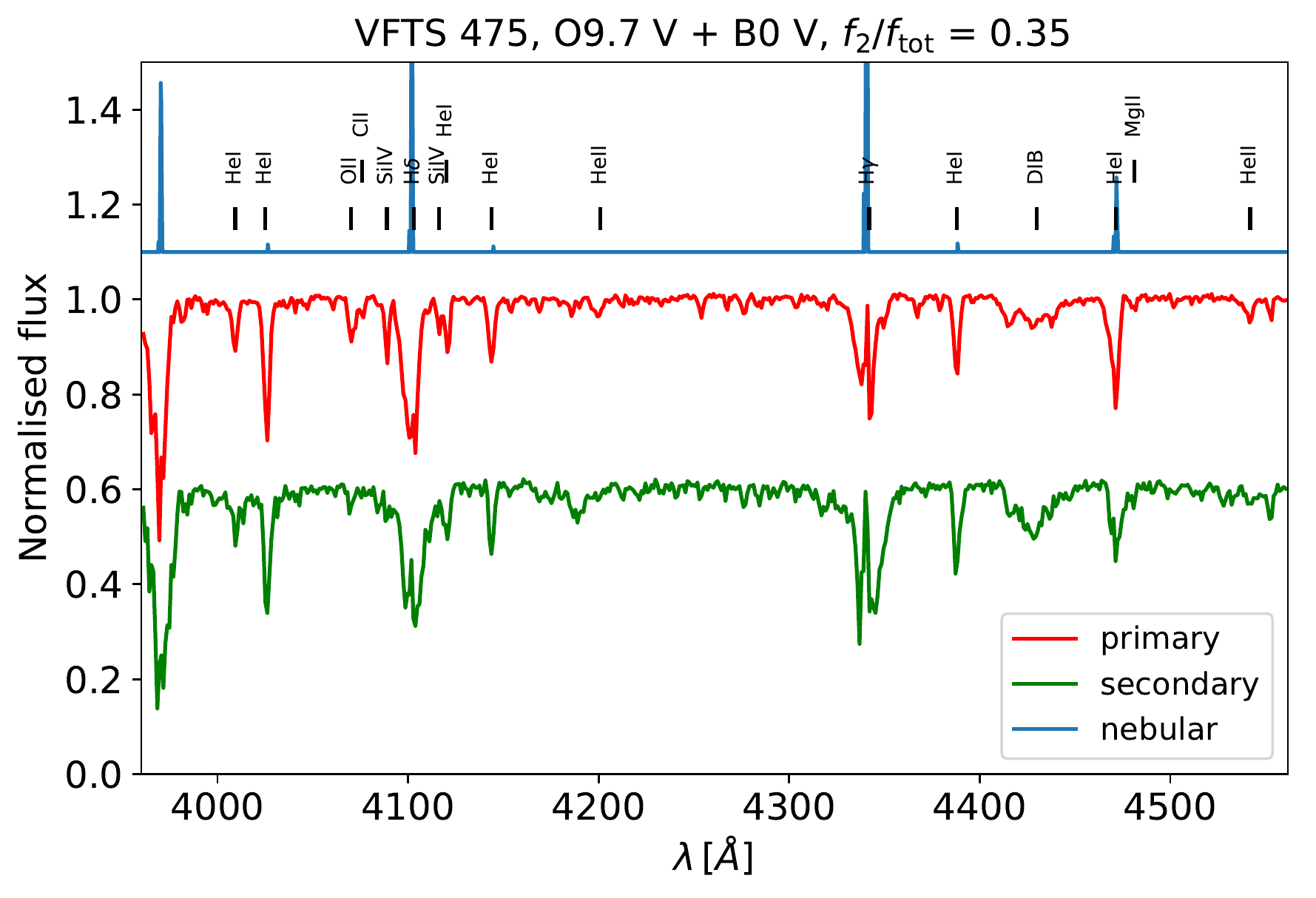}
\caption{As Fig.\,\ref{fig:VFFTS64_DISSPEC}, but for VFTS~475}
\label{fig:DISSPECTRA_VFTS475}
\end{figure}

\begin{figure}
\centering
\includegraphics[width=.5\textwidth]{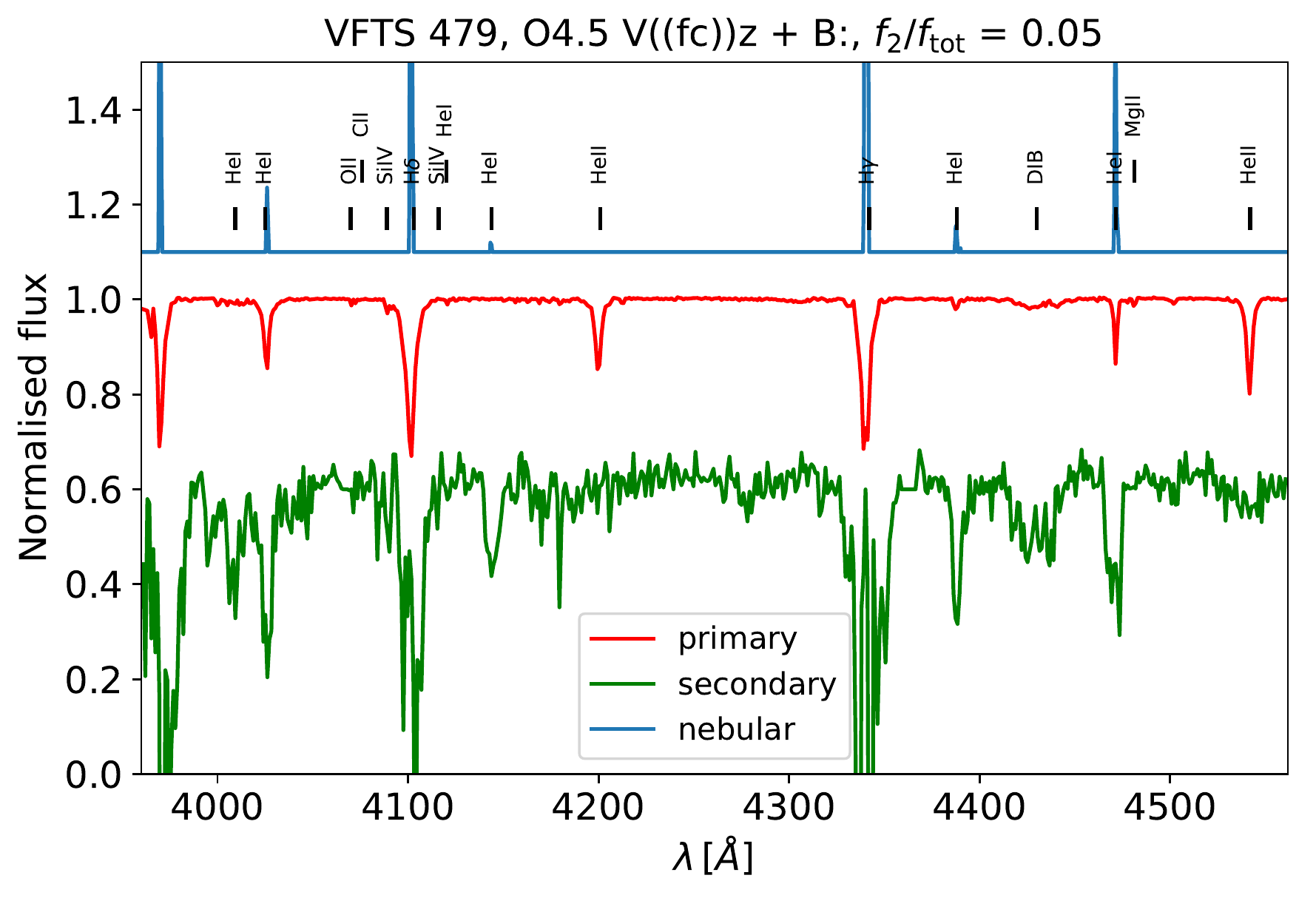}
\caption{As Fig.\,\ref{fig:VFFTS64_DISSPEC}, but for VFTS~479}
\label{fig:DISSPECTRA_VFTS479}
\end{figure}

\begin{figure}
\centering
\includegraphics[width=.5\textwidth]{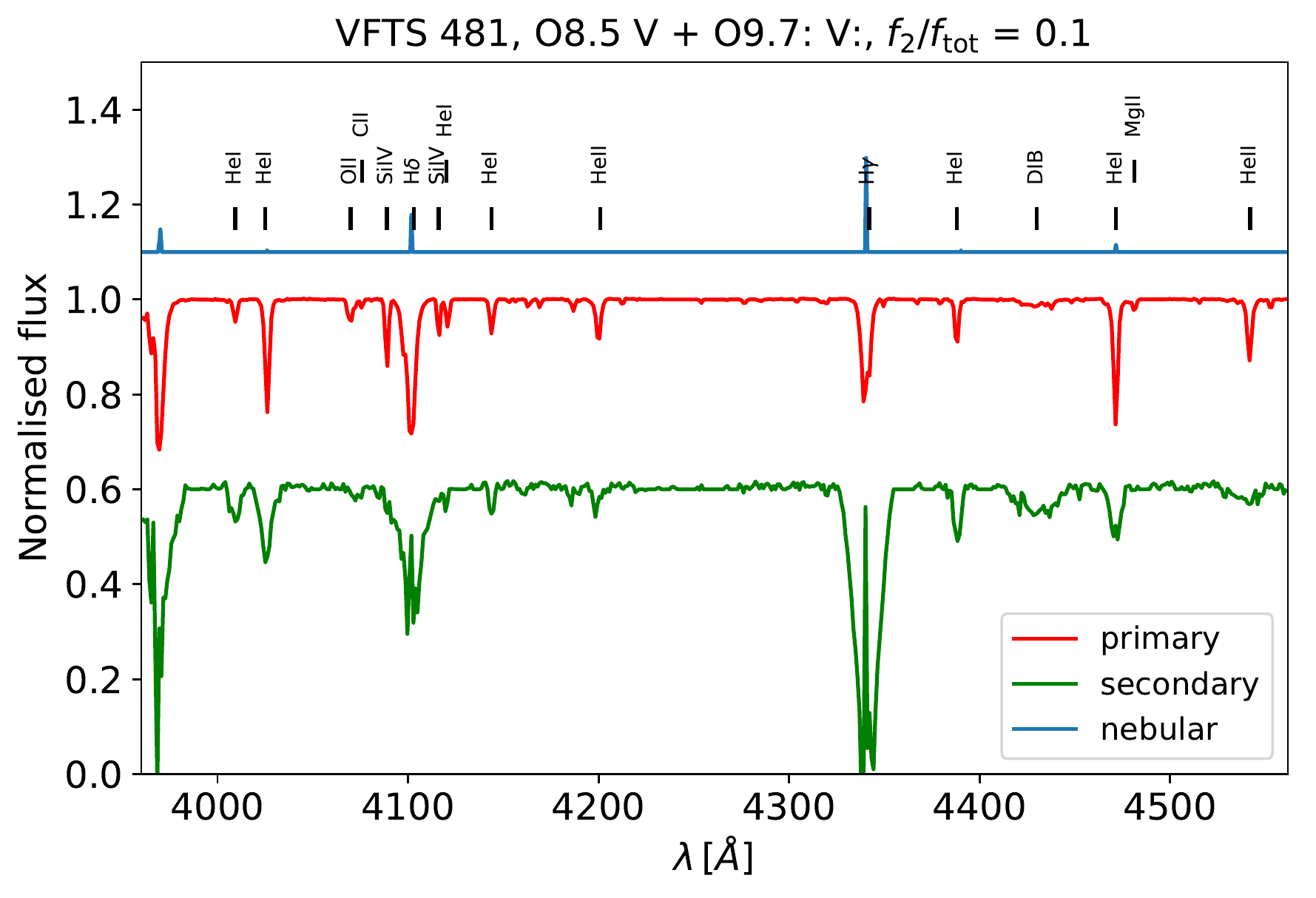}
\caption{As Fig.\,\ref{fig:VFFTS64_DISSPEC}, but for VFTS~481}
\label{fig:DISSPECTRA_VFTS481}
\end{figure}

\begin{figure}
\centering
\includegraphics[width=.5\textwidth]{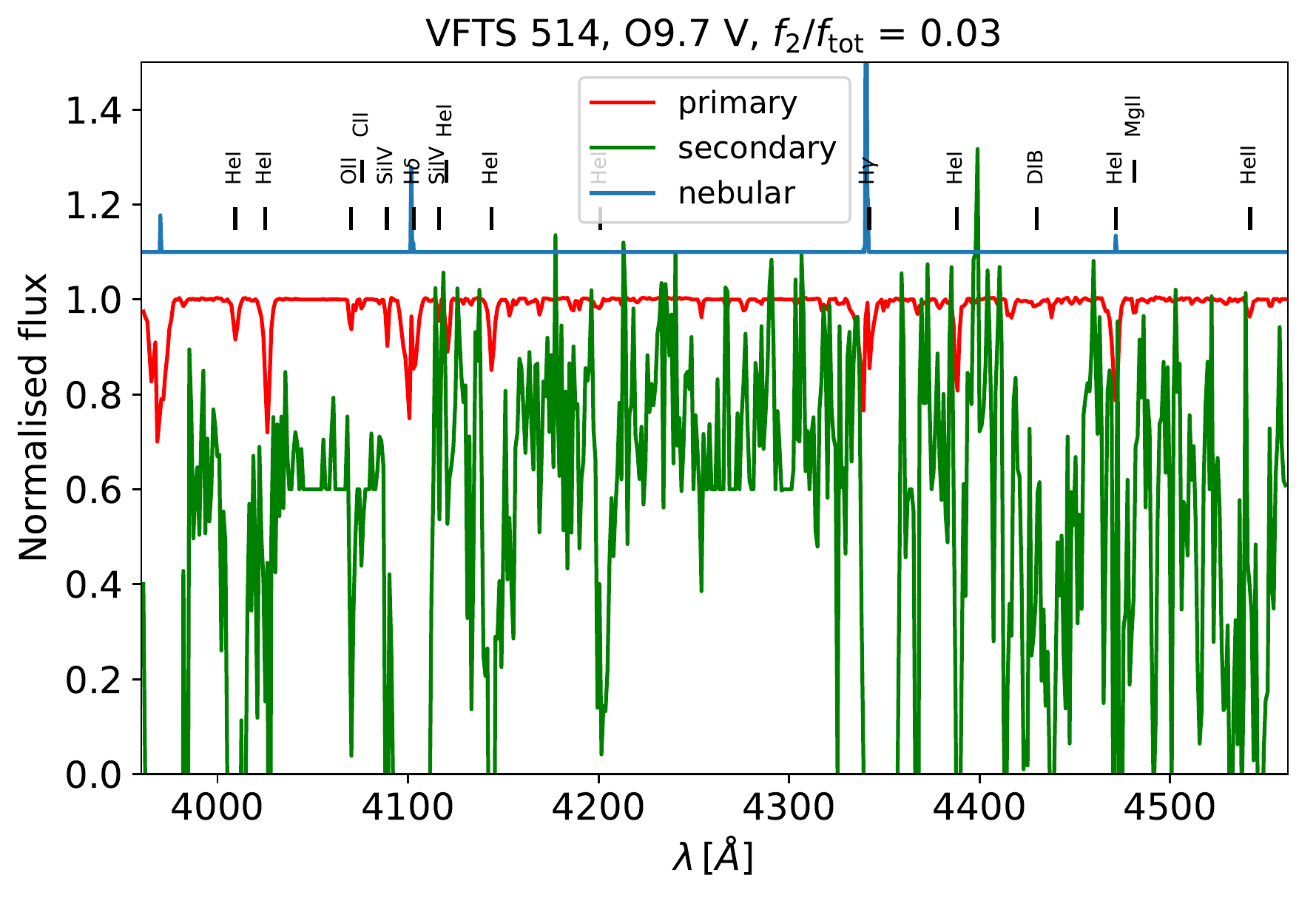}
\caption{As Fig.\,\ref{fig:VFFTS64_DISSPEC}, but for VFTS~514}
\label{fig:DISSPECTRA_VFTS514}
\end{figure}

\begin{figure}
\centering
\includegraphics[width=.5\textwidth]{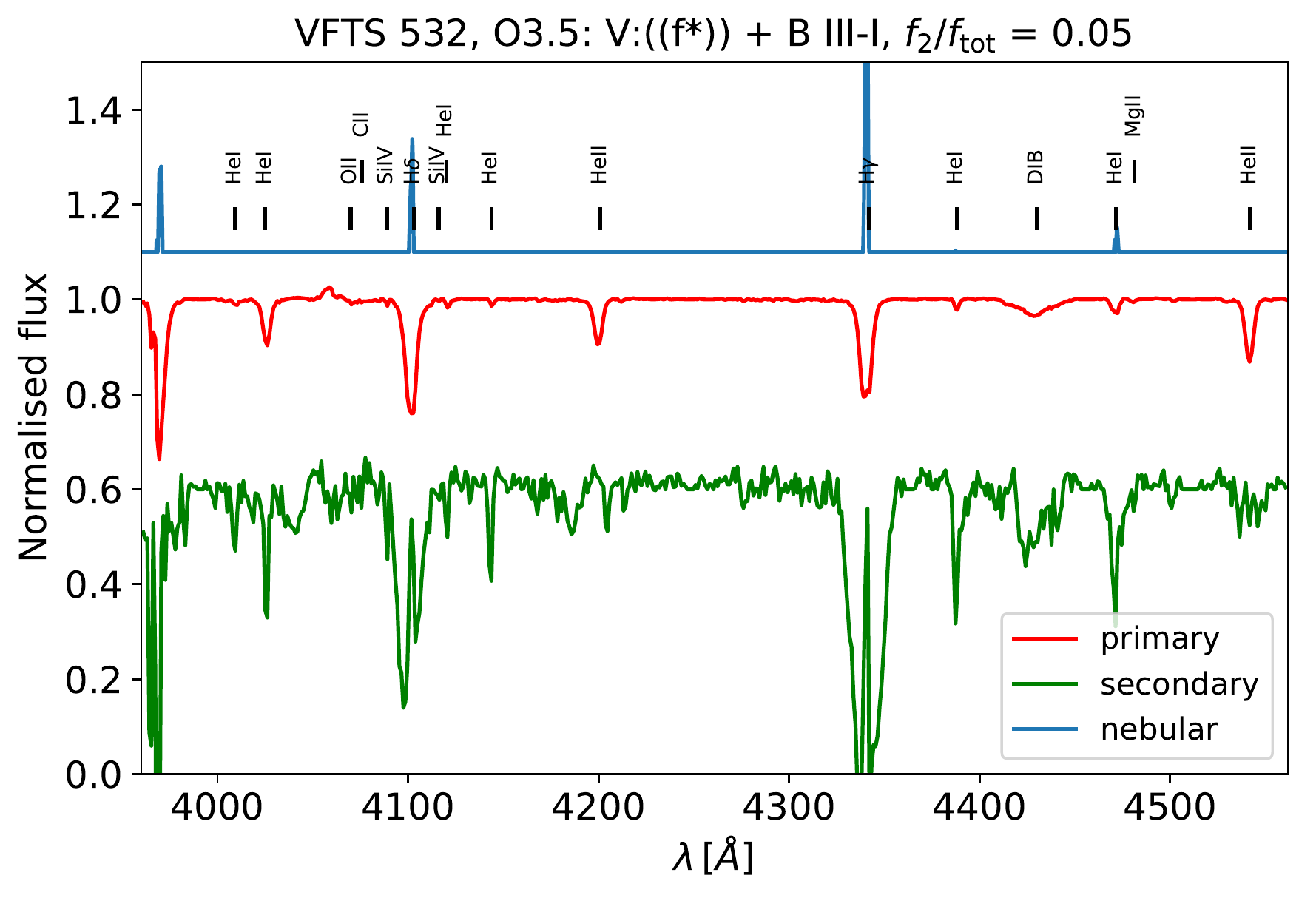}
\caption{As Fig.\,\ref{fig:VFFTS64_DISSPEC}, but for VFTS~532}
\label{fig:DISSPECTRA_VFTS532}
\end{figure}

\begin{figure}
\centering
\includegraphics[width=.5\textwidth]{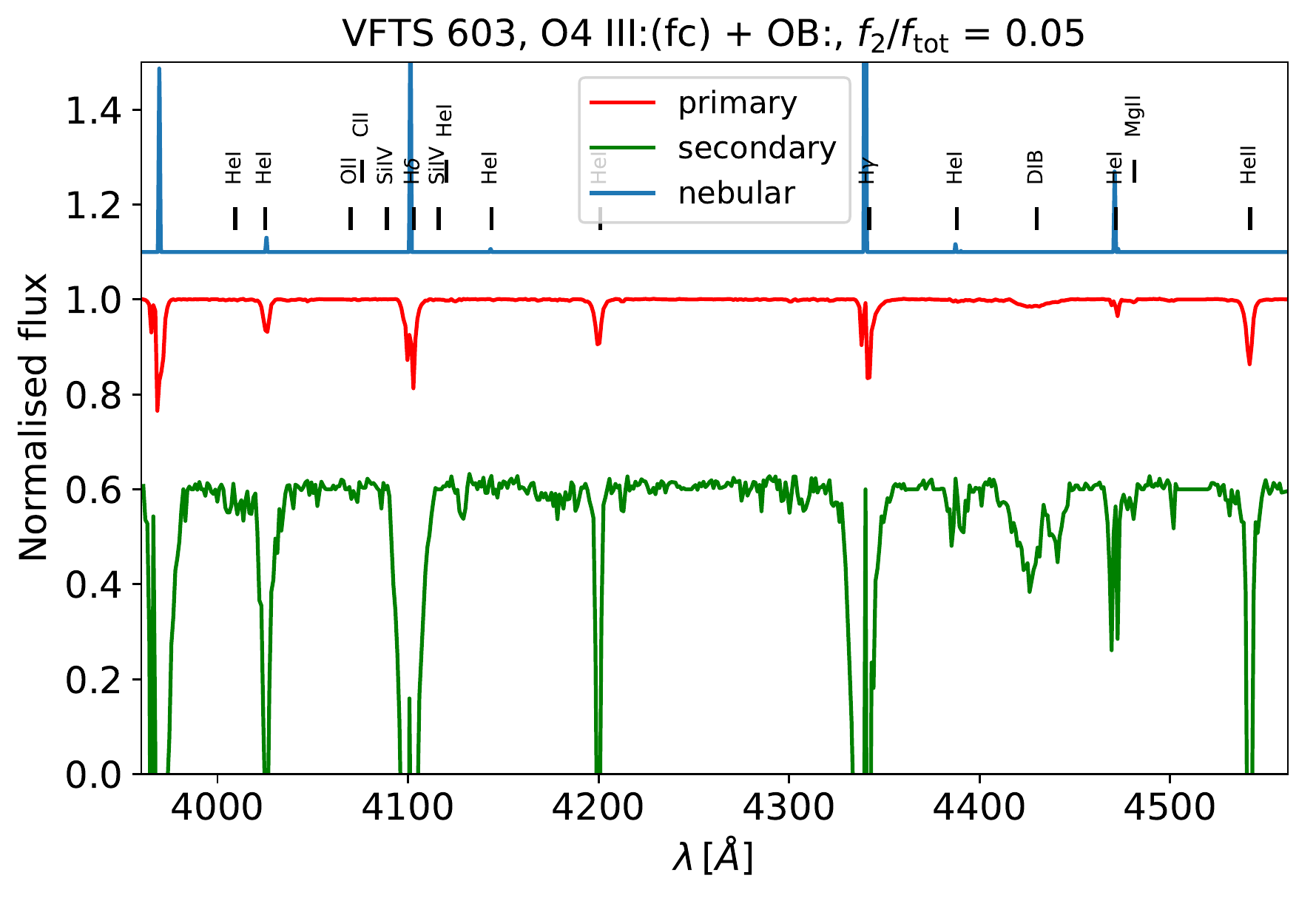}
\caption{As Fig.\,\ref{fig:VFFTS64_DISSPEC}, but for VFTS~603}
\label{fig:DISSPECTRA_VFTS603}
\end{figure}

\begin{figure}
\centering
\includegraphics[width=.5\textwidth]{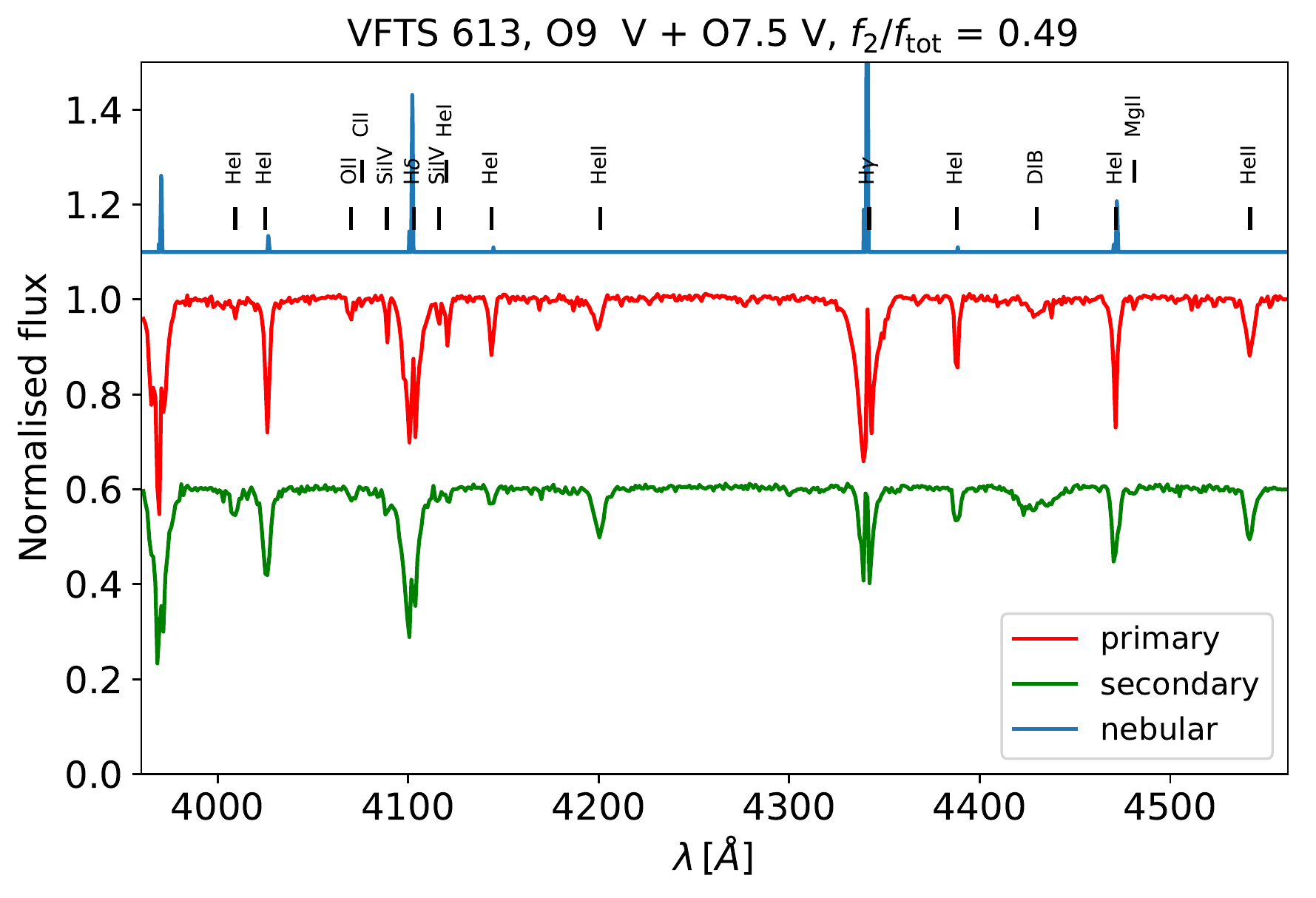}
\caption{As Fig.\,\ref{fig:VFFTS64_DISSPEC}, but for VFTS~613}
\label{fig:DISSPECTRA_VFTS613}
\end{figure}

\begin{figure}
\centering
\includegraphics[width=.5\textwidth]{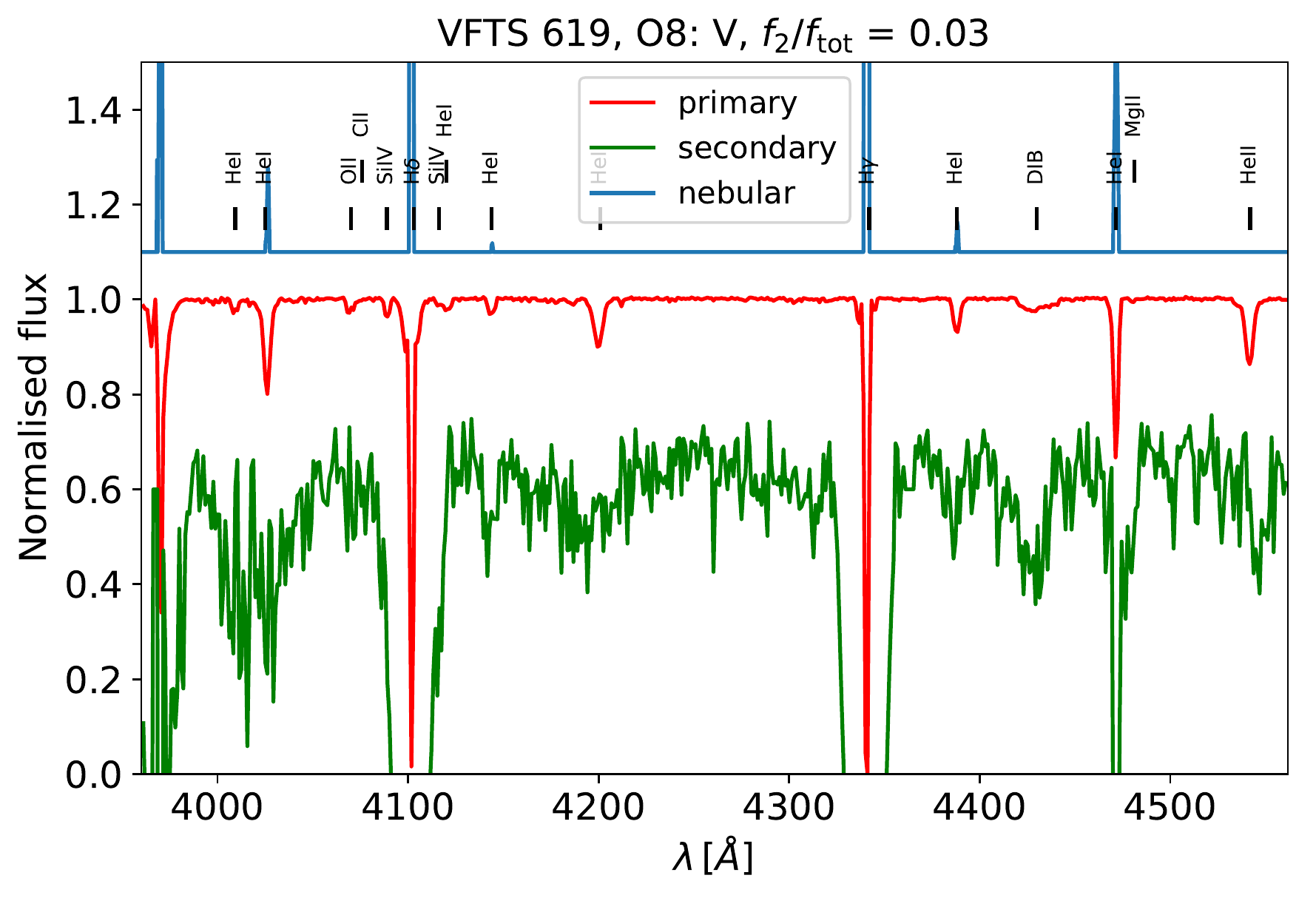}
\caption{As Fig.\,\ref{fig:VFFTS64_DISSPEC}, but for VFTS~619}
\label{fig:DISSPECTRA_VFTS619}
\end{figure}

\begin{figure}
\centering
\includegraphics[width=.5\textwidth]{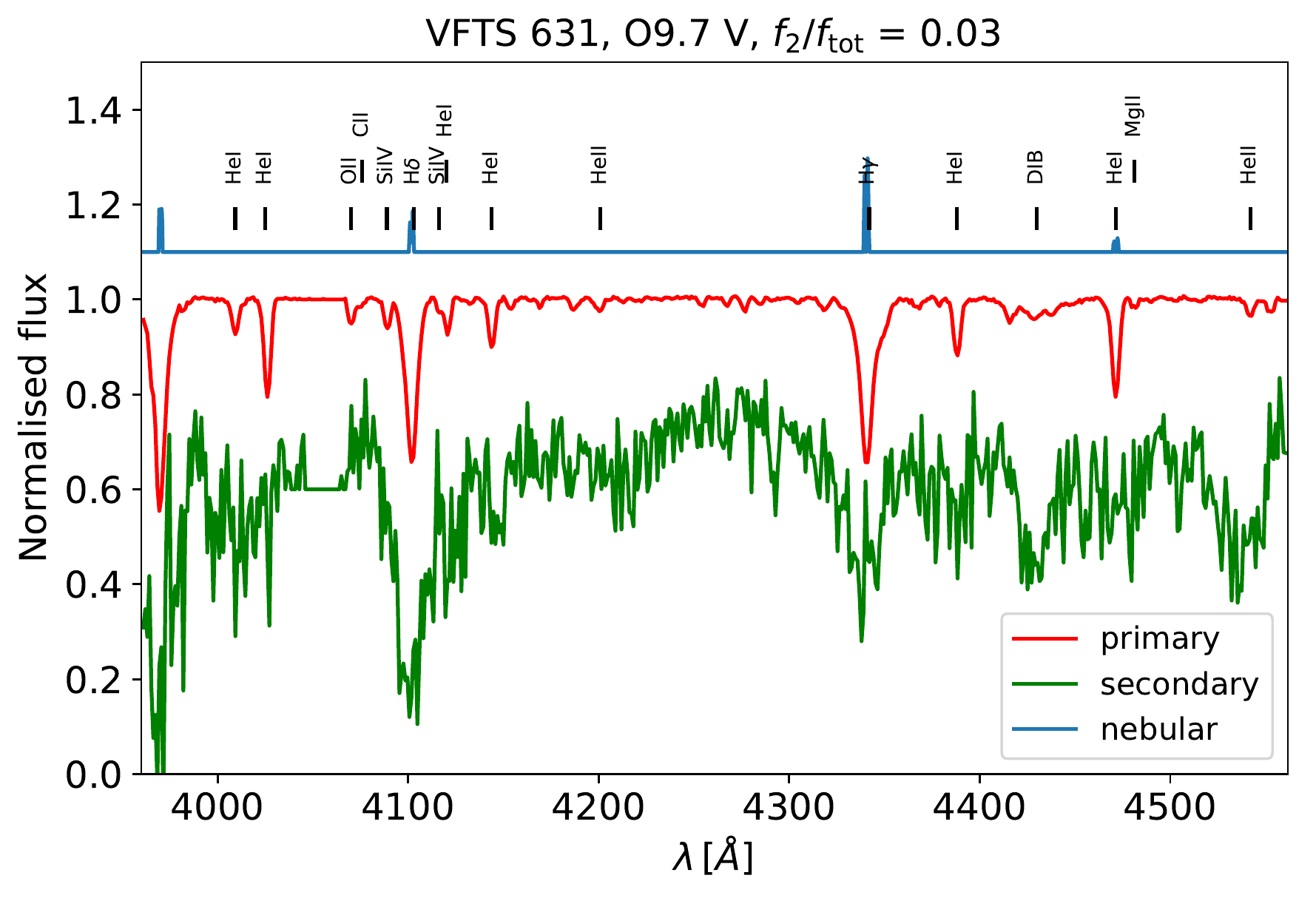}
\caption{As Fig.\,\ref{fig:VFFTS64_DISSPEC}, but for VFTS~631}
\label{fig:DISSPECTRA_VFTS631}
\end{figure}

\begin{figure}
\centering
\includegraphics[width=.5\textwidth]{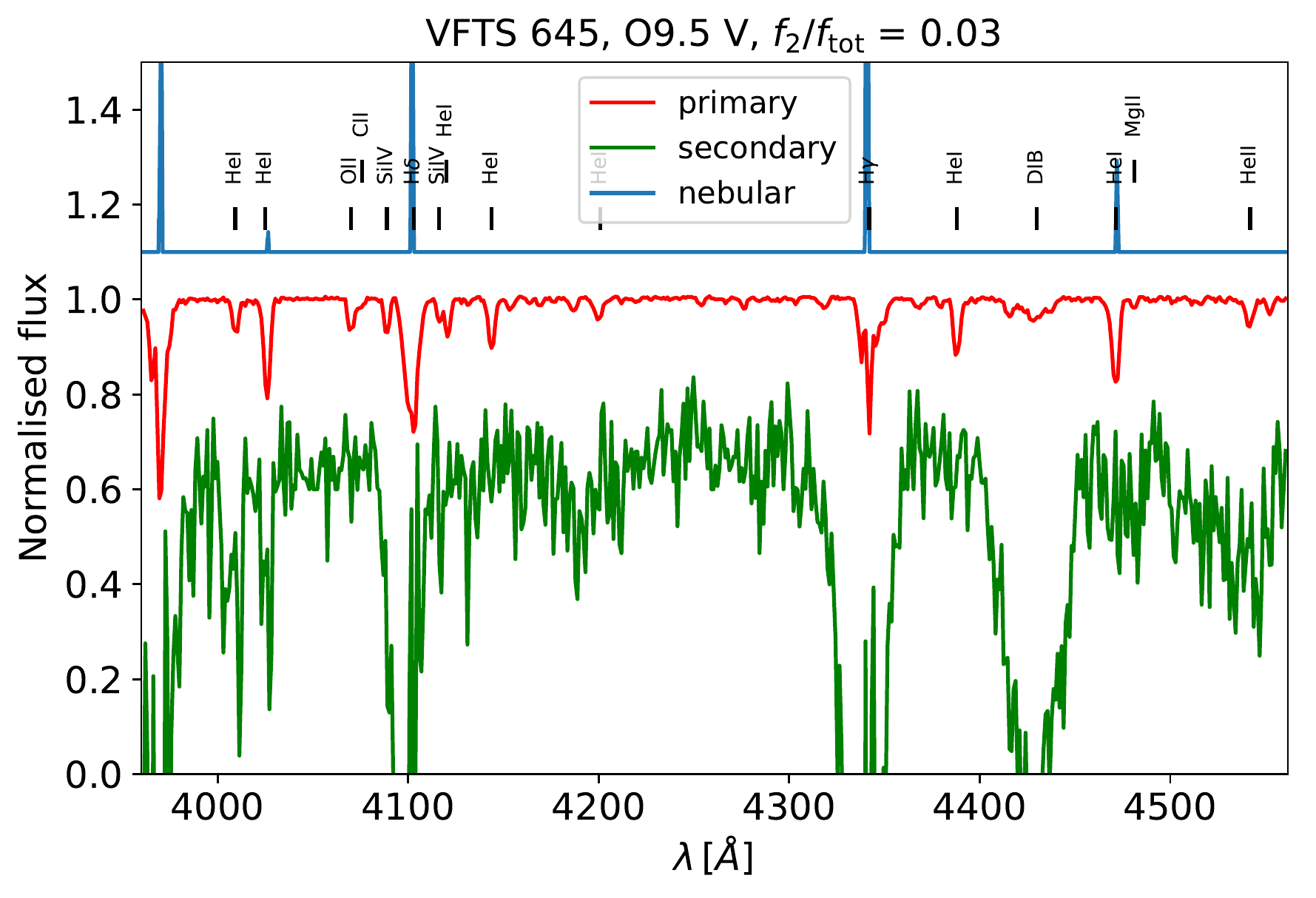}
\caption{As Fig.\,\ref{fig:VFFTS64_DISSPEC}, but for VFTS~645}
\label{fig:DISSPECTRA_VFTS645}
\end{figure}

\begin{figure}
\centering
\includegraphics[width=.5\textwidth]{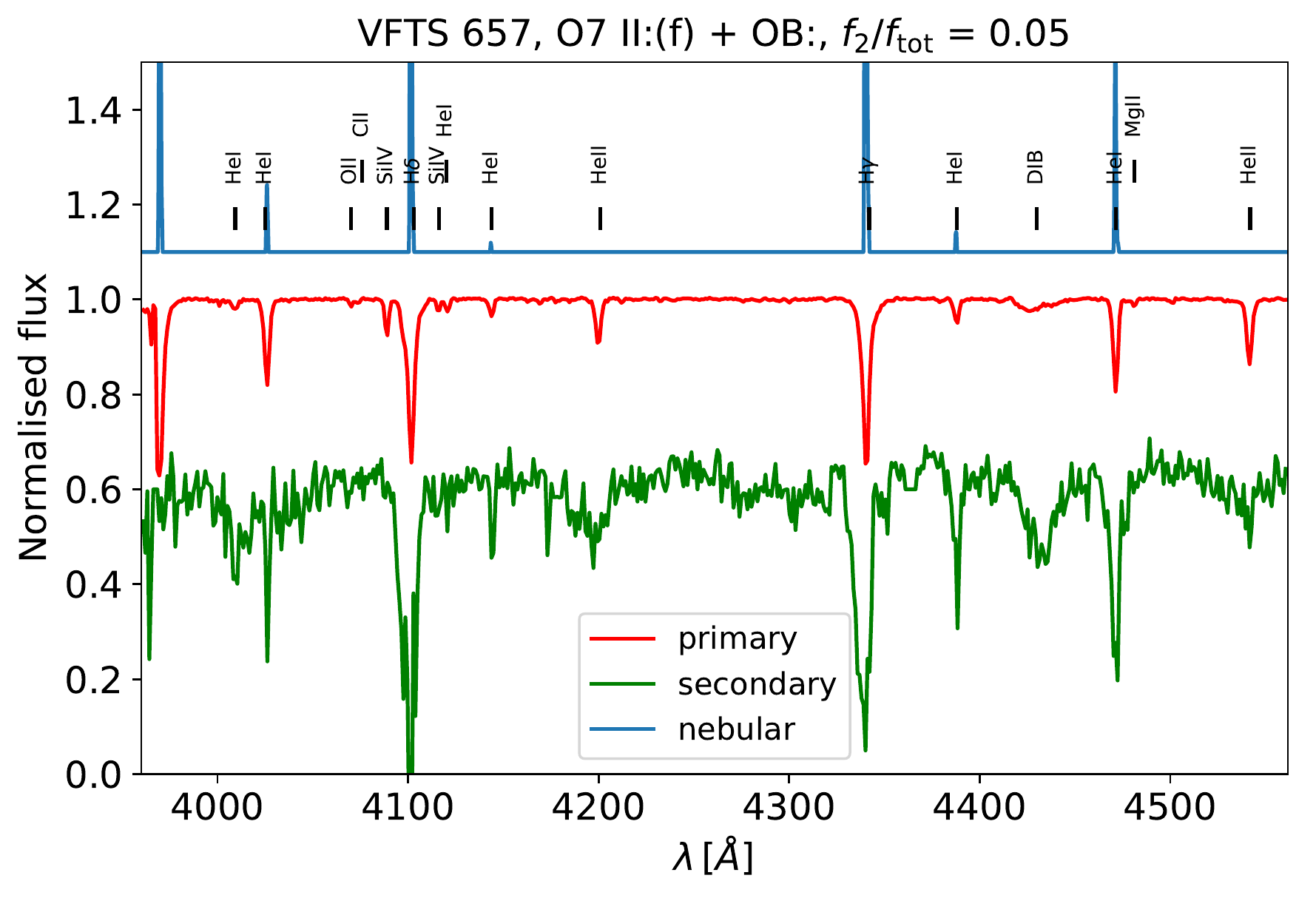}
\caption{As Fig.\,\ref{fig:VFFTS64_DISSPEC}, but for VFTS~657}
\label{fig:DISSPECTRA_VFTS657}
\end{figure}

\begin{figure}
\centering
\includegraphics[width=.5\textwidth]{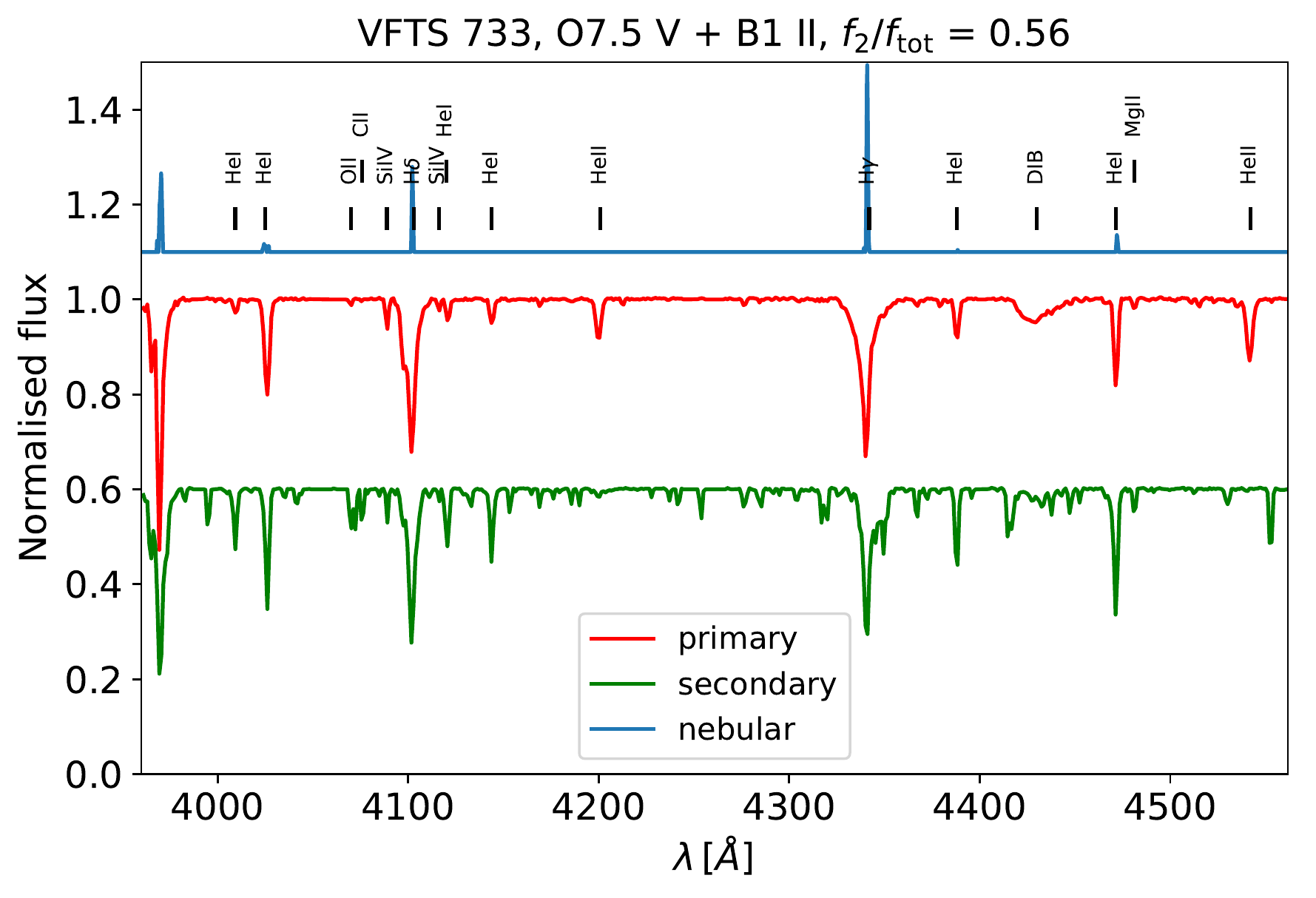}
\caption{As Fig.\,\ref{fig:VFFTS64_DISSPEC}, but for VFTS~733}
\label{fig:DISSPECTRA_VFTS733}
\end{figure}

\begin{figure}
\centering
\includegraphics[width=.5\textwidth]{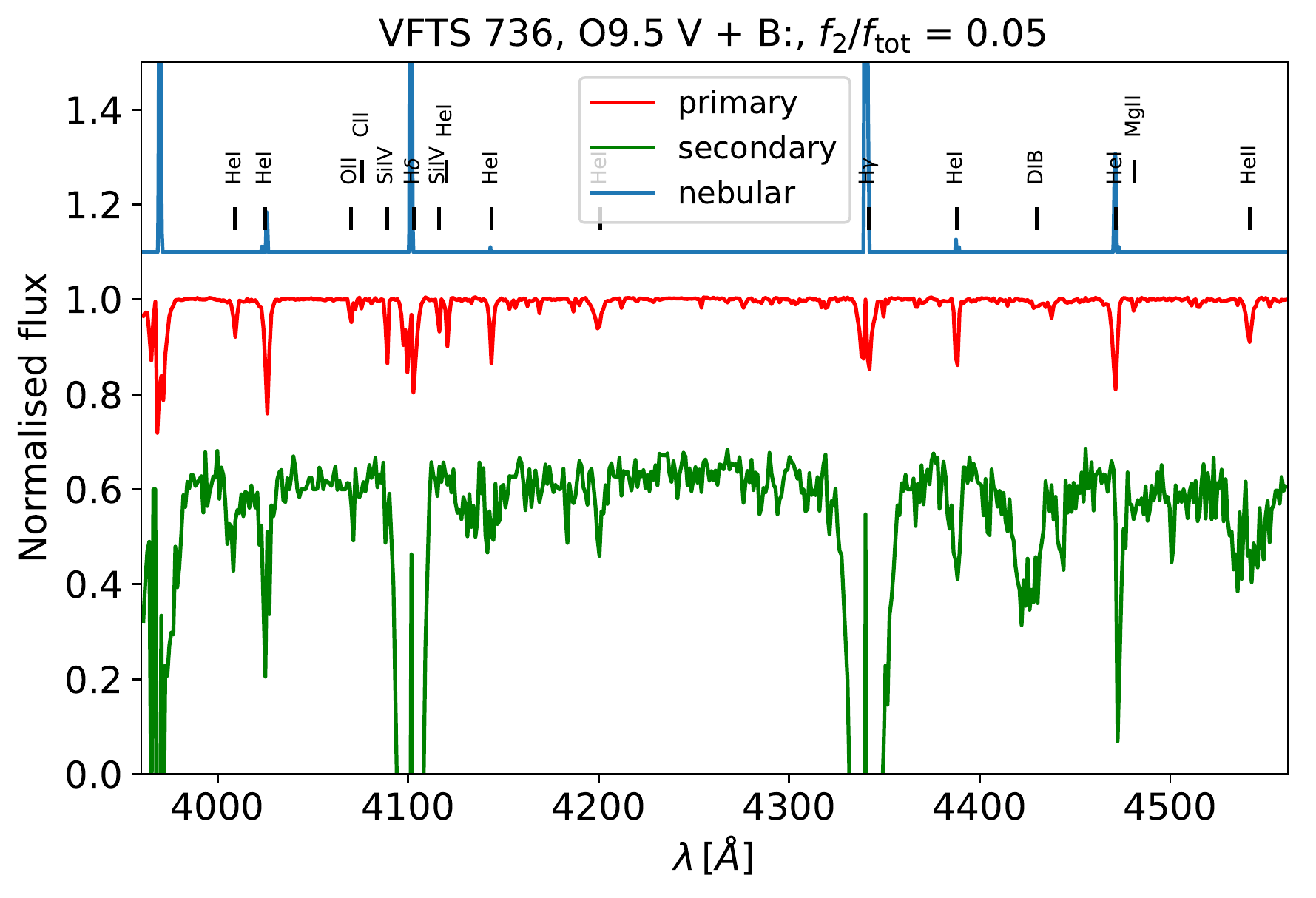}
\caption{As Fig.\,\ref{fig:VFFTS64_DISSPEC}, but for VFTS~736}
\label{fig:DISSPECTRA_VFTS736}
\end{figure}

\begin{figure}
\centering
\includegraphics[width=.5\textwidth]{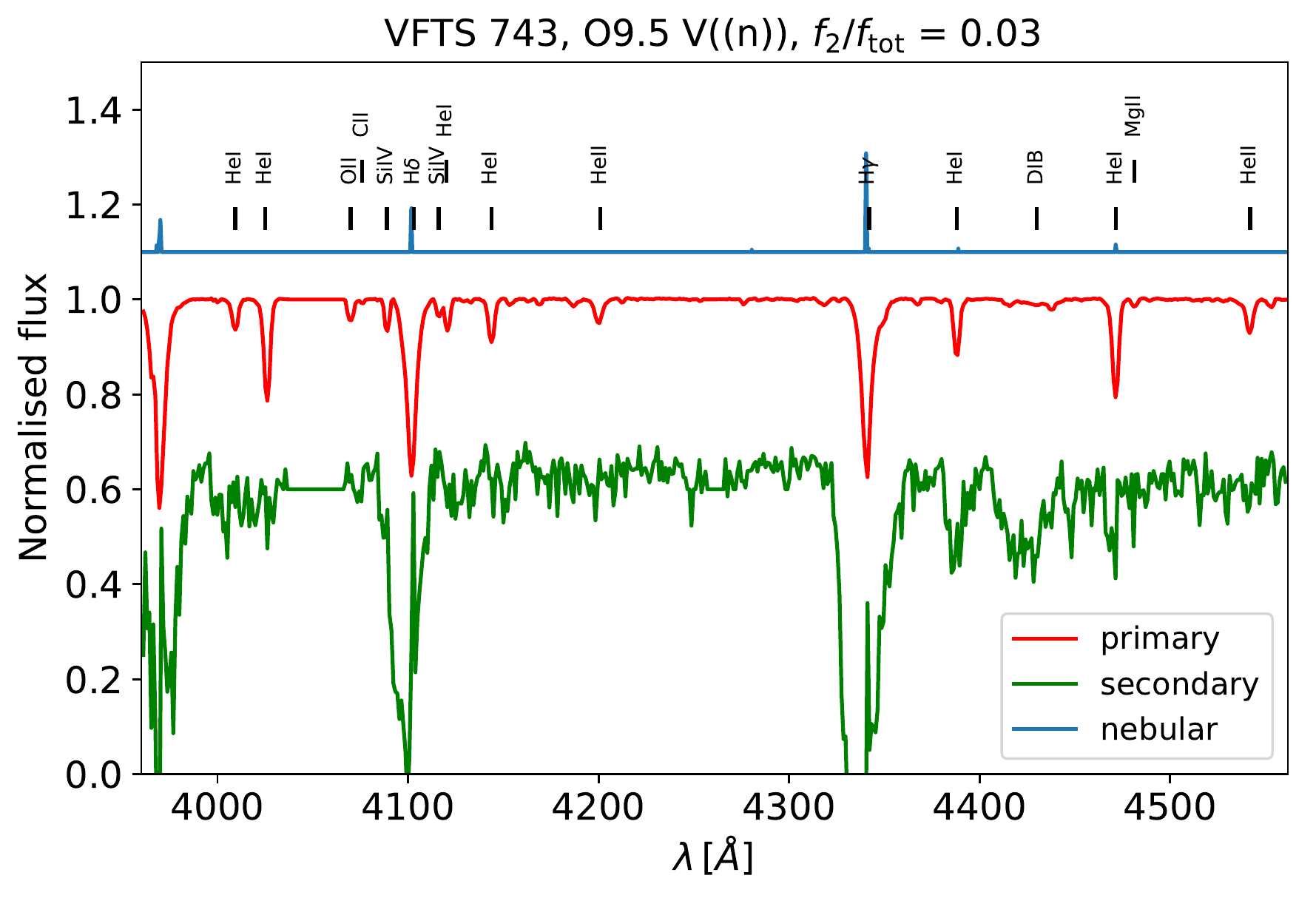}
\caption{As Fig.\,\ref{fig:VFFTS64_DISSPEC}, but for VFTS~743}
\label{fig:DISSPECTRA_VFTS743}
\end{figure}

\begin{figure}
\centering
\includegraphics[width=.5\textwidth]{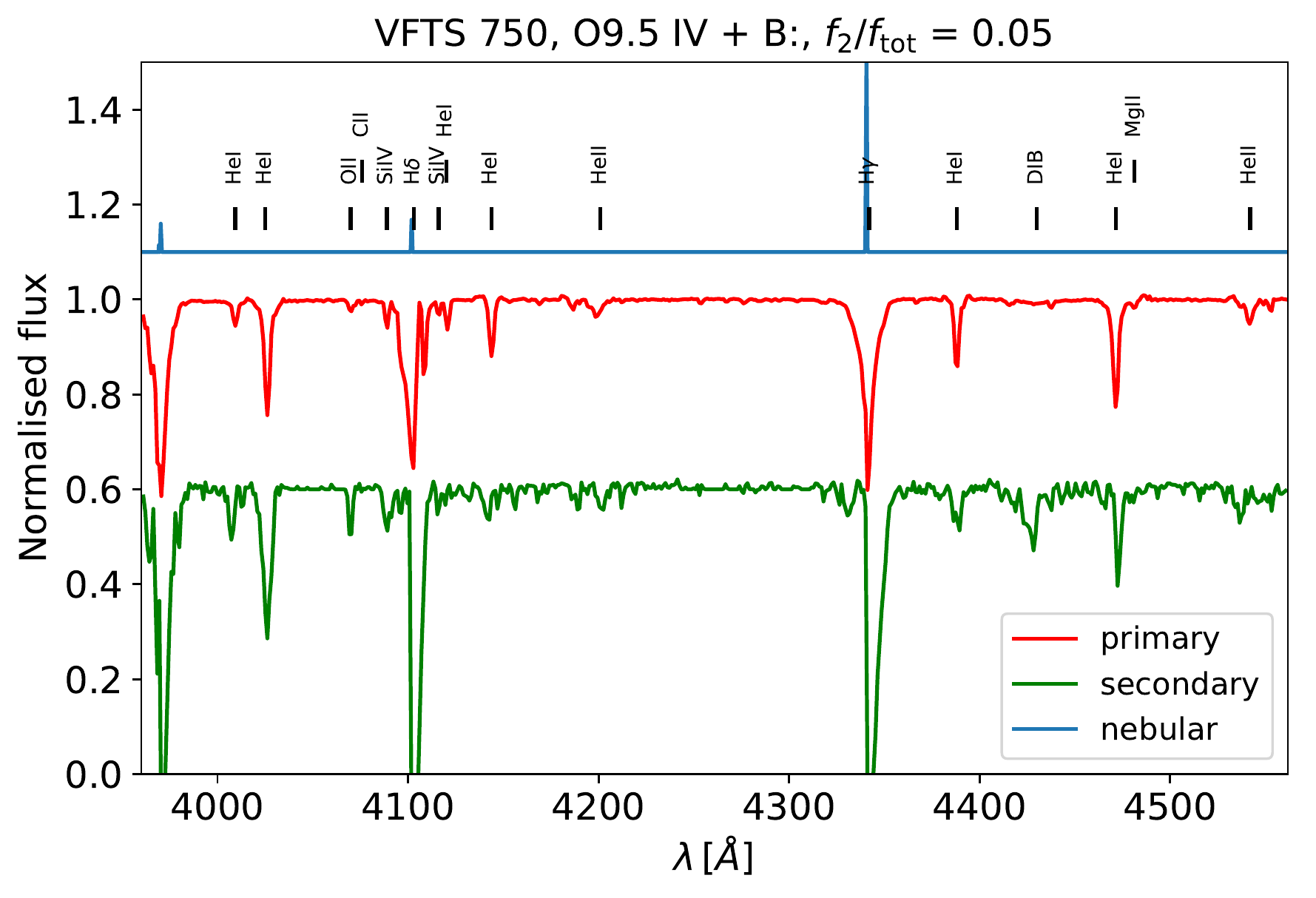}
\caption{As Fig.\,\ref{fig:VFFTS64_DISSPEC}, but for VFTS~750}
\label{fig:DISSPECTRA_VFTS750}
\end{figure}

\begin{figure}
\centering
\includegraphics[width=.5\textwidth]{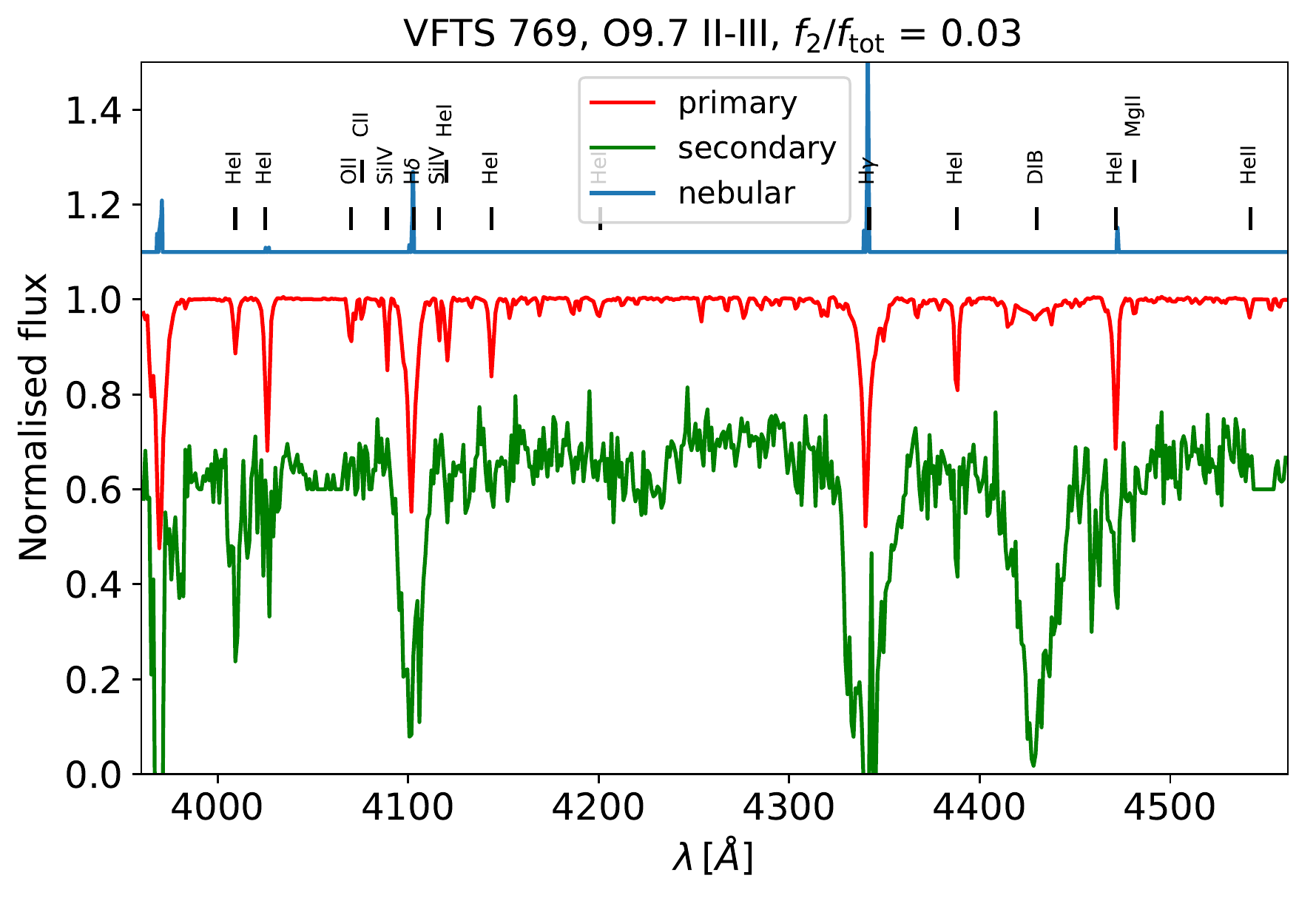}
\caption{As Fig.\,\ref{fig:VFFTS64_DISSPEC}, but for VFTS~769}
\label{fig:DISSPECTRA_VFTS769}
\end{figure}

\begin{figure}
\centering
\includegraphics[width=.5\textwidth]{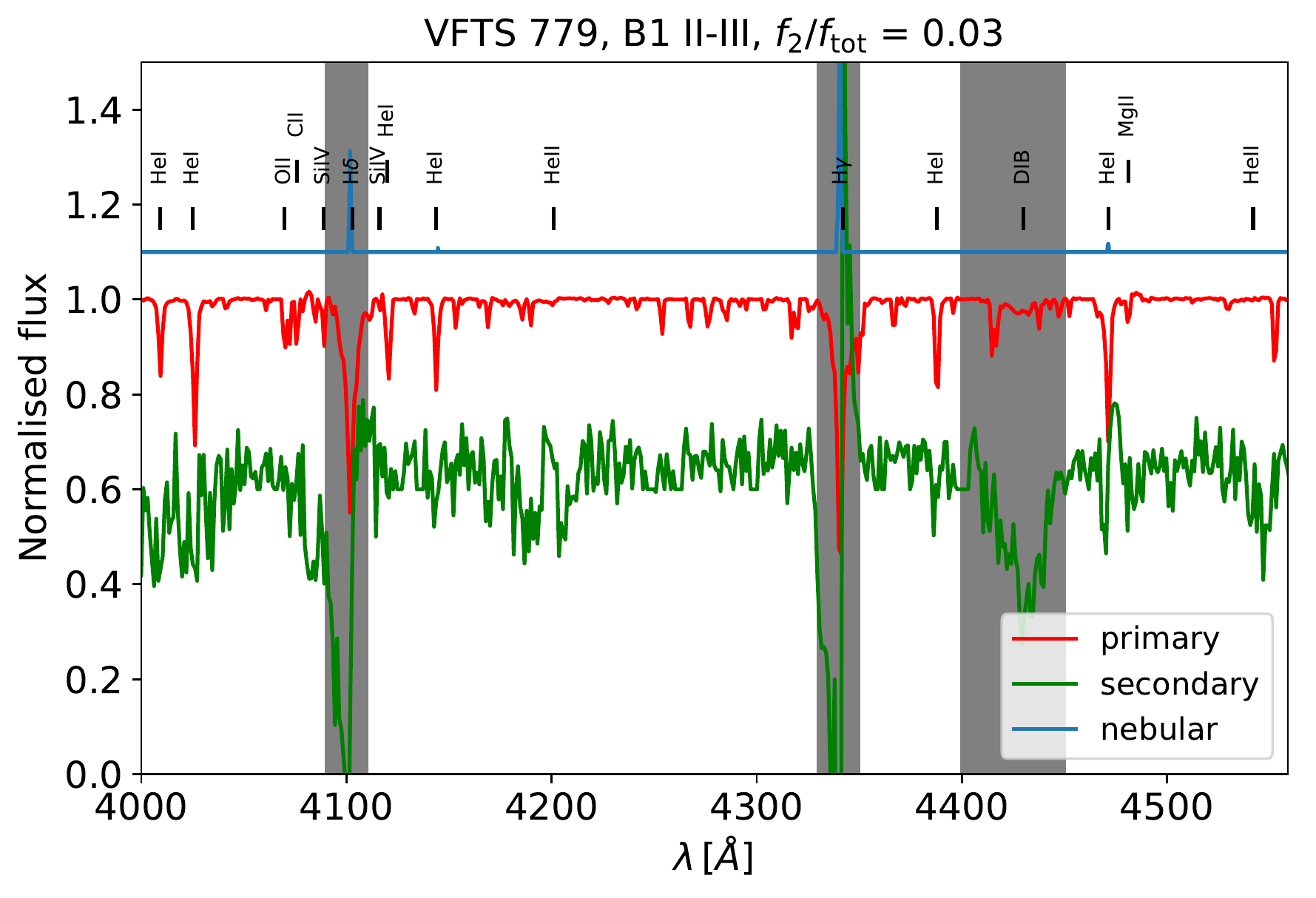}
\caption{As Fig.\,\ref{fig:VFFTS64_DISSPEC}, but for VFTS~779}
\label{fig:DISSPECTRA_VFTS779}
\end{figure}

\begin{figure}
\centering
\includegraphics[width=.5\textwidth]{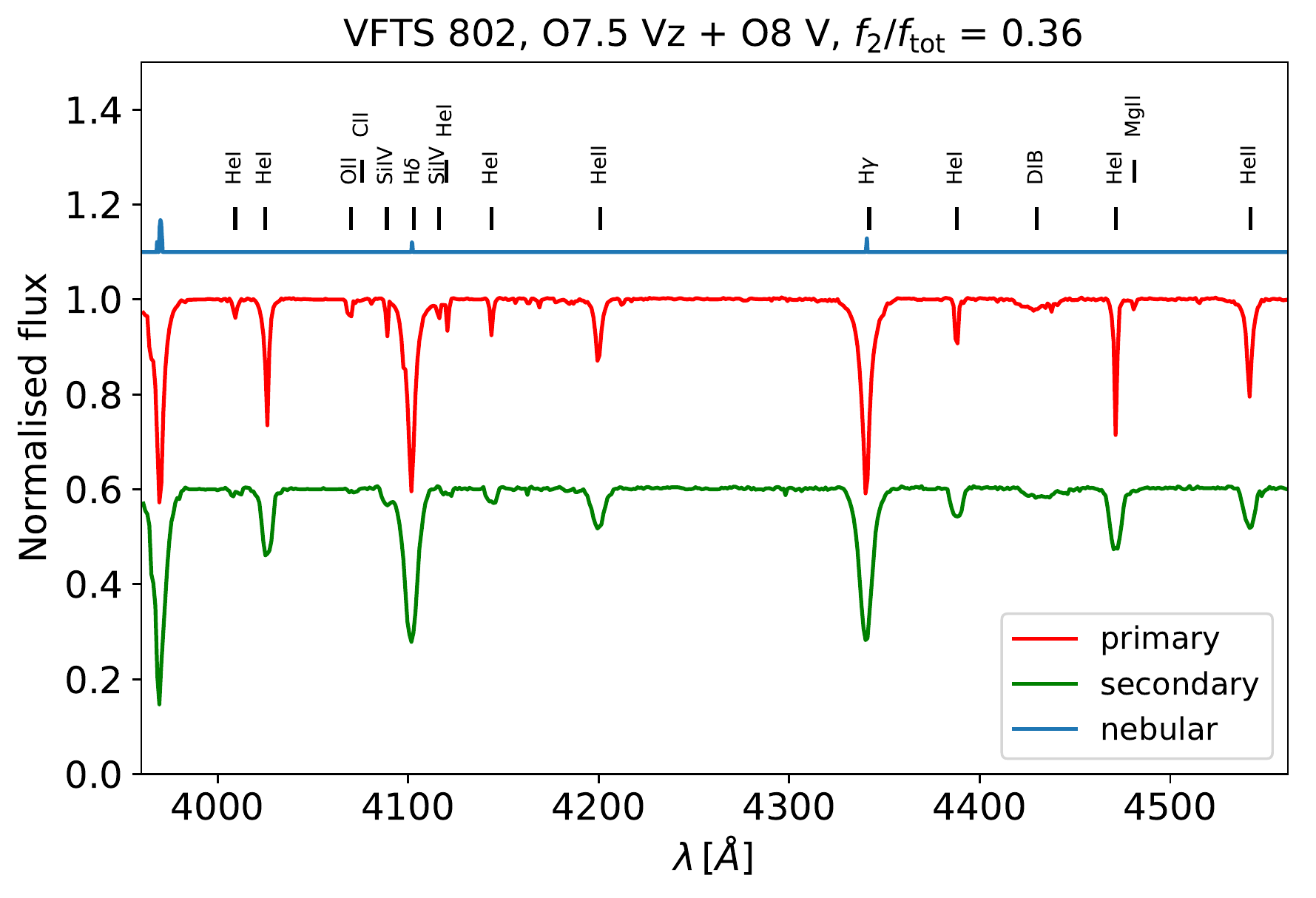}
\caption{As Fig.\,\ref{fig:VFFTS64_DISSPEC}, but for VFTS~802}
\label{fig:DISSPECTRA_VFTS802}
\end{figure}

\begin{figure}
\centering
\includegraphics[width=.5\textwidth]{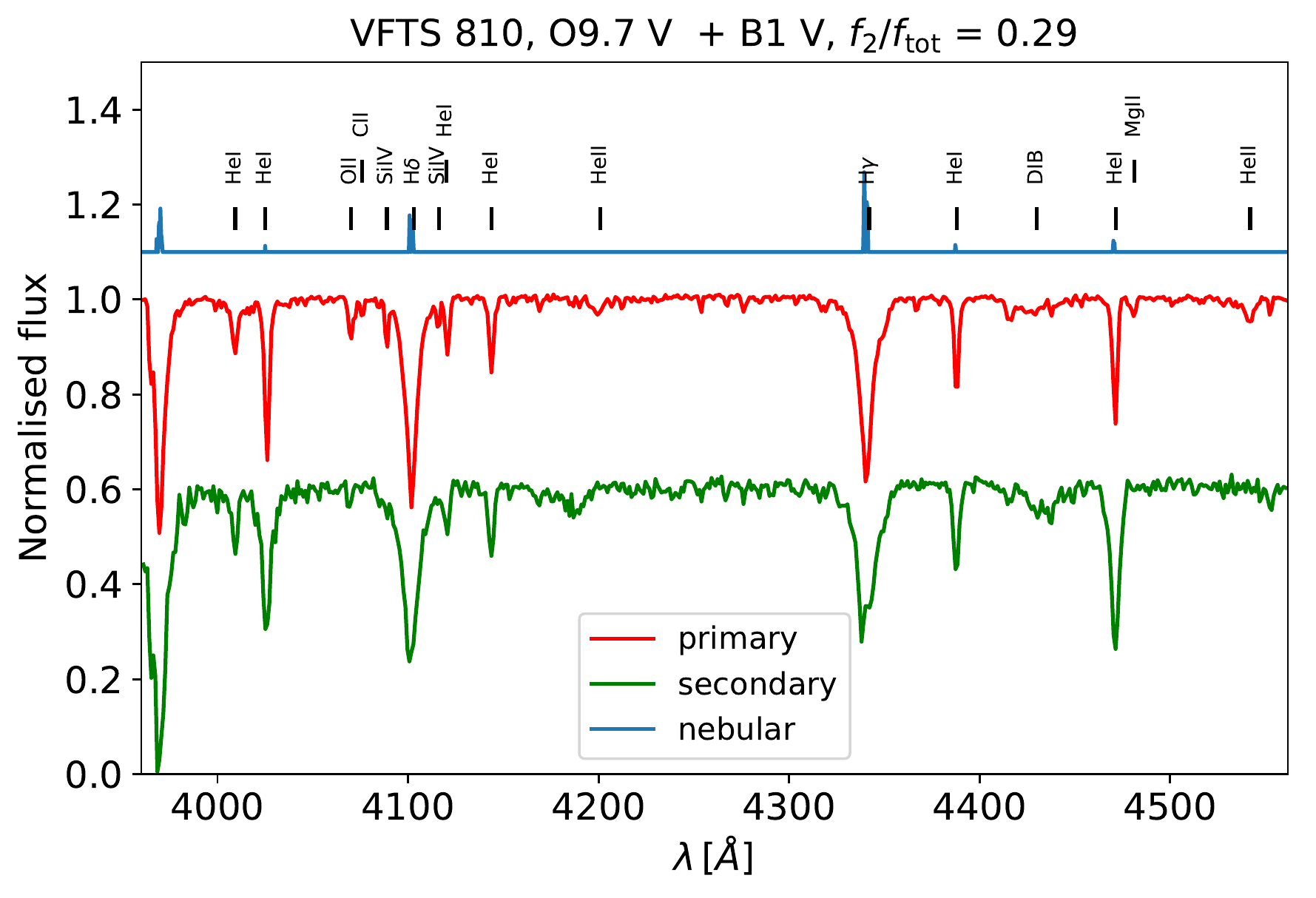}
\caption{As Fig.\,\ref{fig:VFFTS64_DISSPEC}, but for VFTS~810}
\label{fig:DISSPECTRA_VFTS810}
\end{figure}

\begin{figure}
\centering
\includegraphics[width=.5\textwidth]{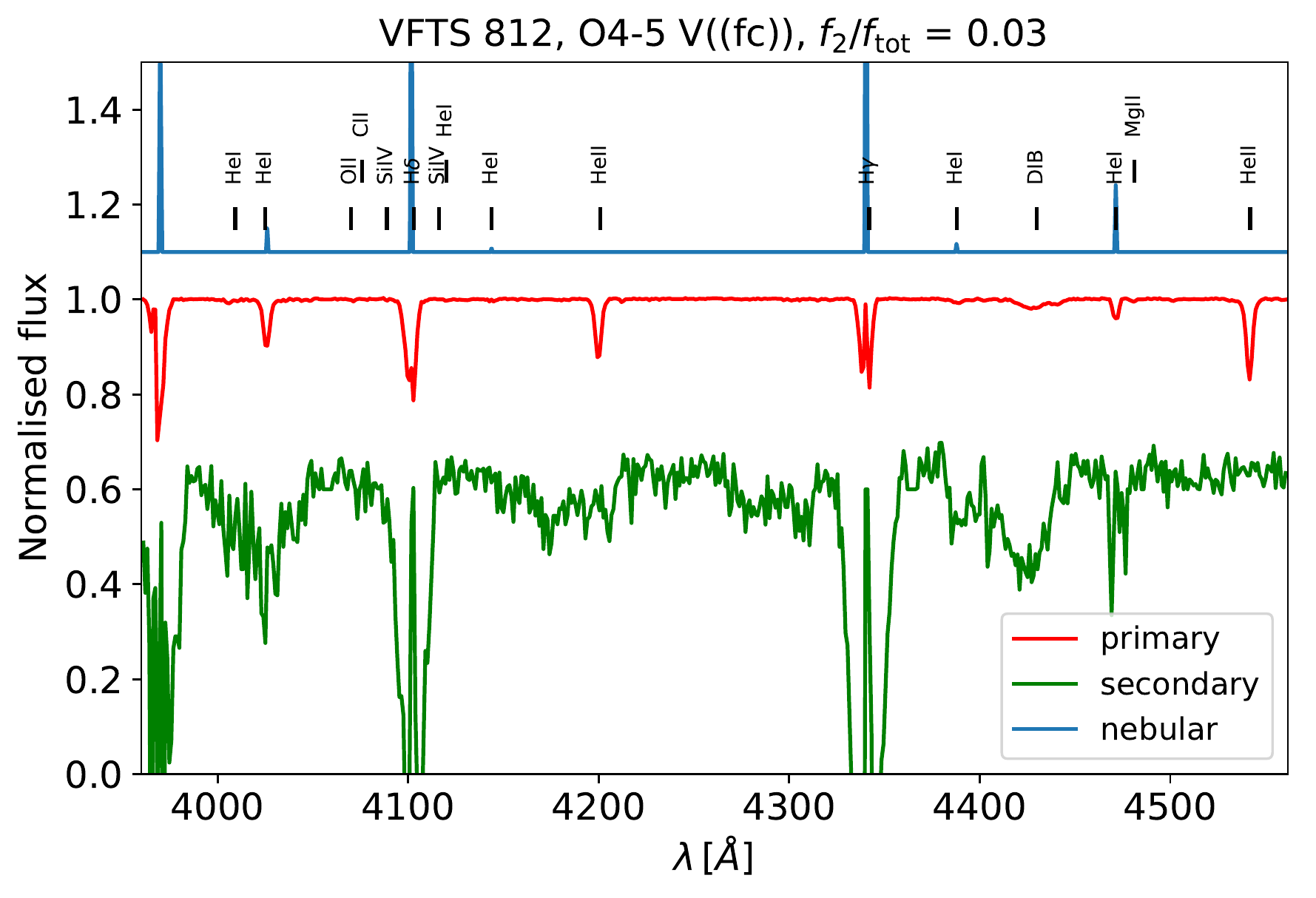}
\caption{As Fig.\,\ref{fig:VFFTS64_DISSPEC}, but for VFTS~812}
\label{fig:DISSPECTRA_VFTS812}
\end{figure}

\begin{figure}
\centering
\includegraphics[width=.5\textwidth]{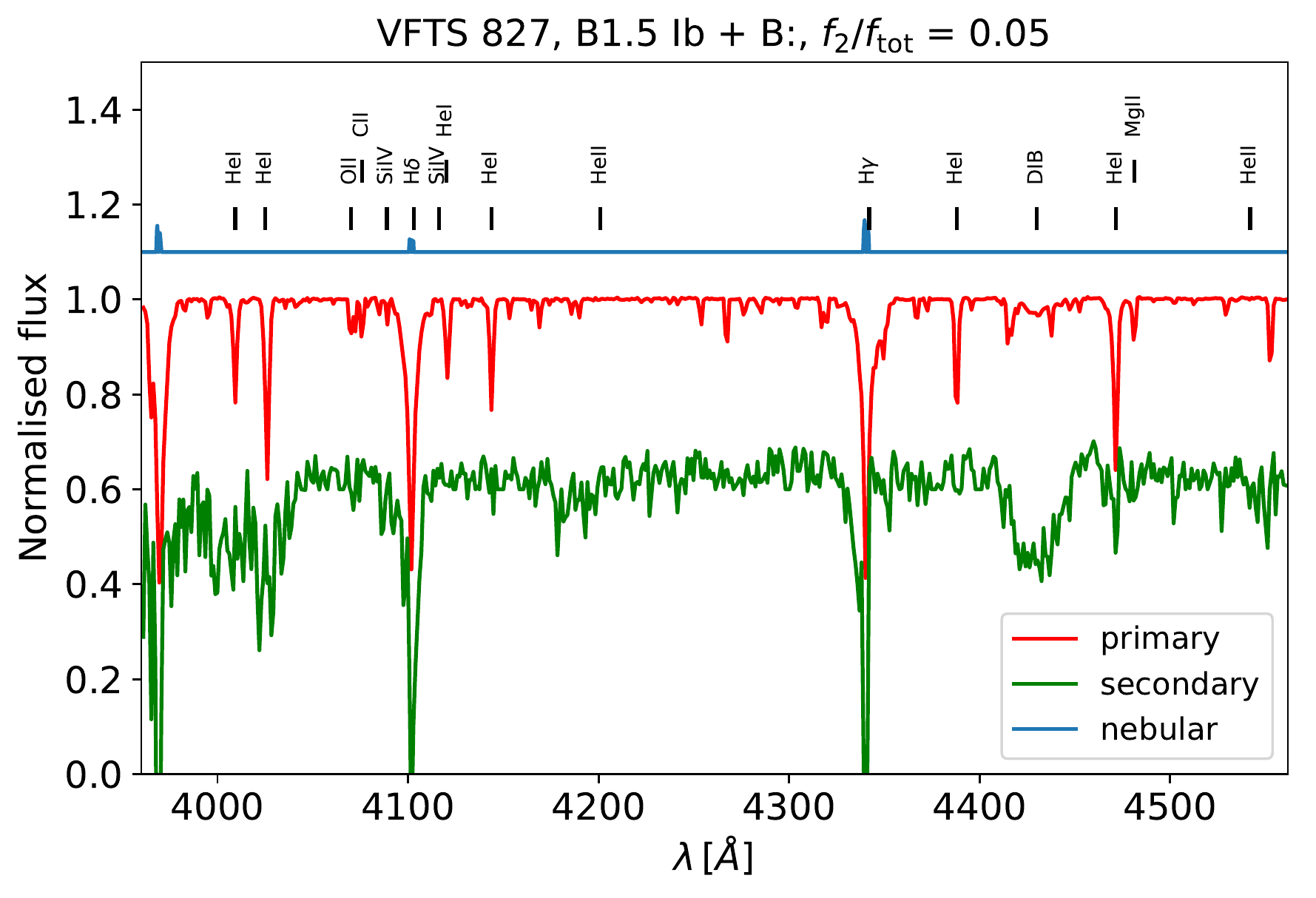}
\caption{As Fig.\,\ref{fig:VFFTS64_DISSPEC}, but for VFTS~827}
\label{fig:DISSPECTRA_VFTS827}
\end{figure}

\begin{figure}
\centering
\includegraphics[width=.5\textwidth]{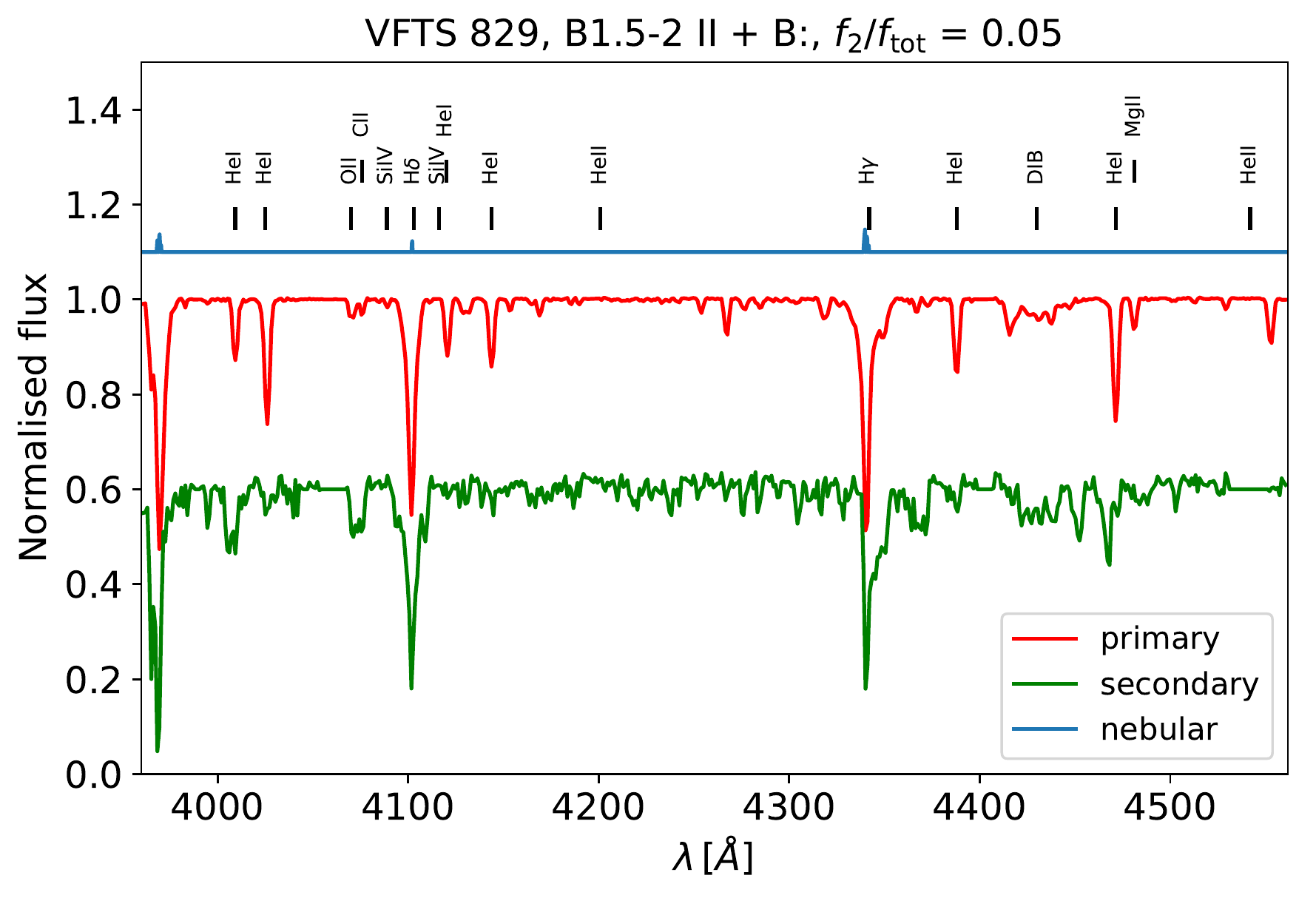}
\caption{As Fig.\,\ref{fig:VFFTS64_DISSPEC}, but for VFTS~829}
\label{fig:DISSPECTRA_VFTS829}
\end{figure}

\begin{figure}
\centering
\includegraphics[width=.5\textwidth]{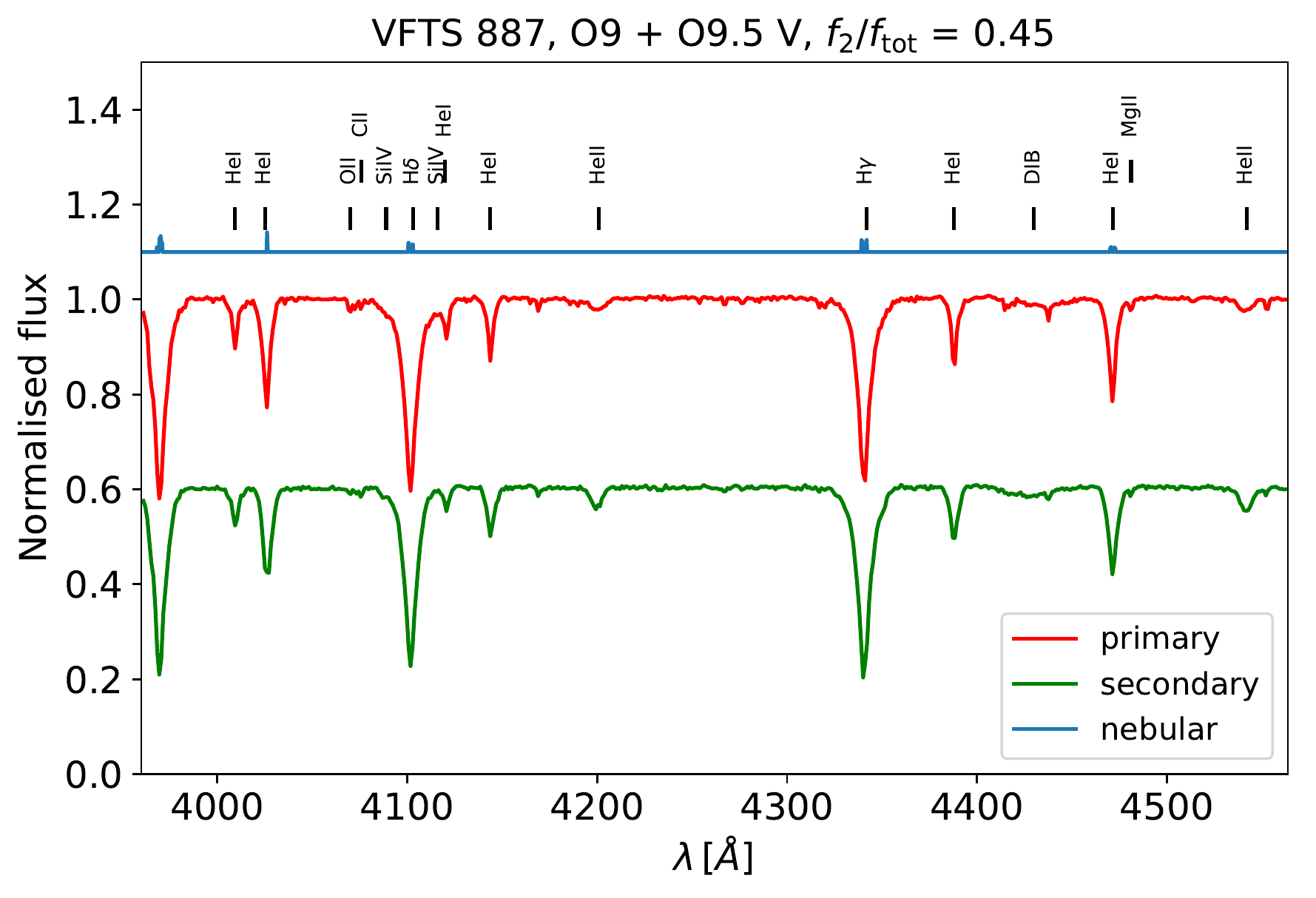}
\caption{As Fig.\,\ref{fig:VFFTS64_DISSPEC}, but for VFTS~887}
\label{fig:VFFTS887_DISSPEC}
\end{figure}

This section and its Figs.\,\ref{fig:VFFTS64_DISSPEC}-\ref{fig:VFFTS887_DISSPEC} provides the disentangled spectra for each of our targets.  

\clearpage

\section{Rotation velocities and minimum inclinations}
\label{sec:vsini}

Table\,\ref{tab:vsinis} provides $\varv \sin i$ values measured from the disentangled spectra using FWHM calibrations provided by \citet{Ramirez-Agudelo2015}, as well as minimum inclinations using Eq.\,(\ref{eq:icrit2}). We note that this relies on the assumption that the orbital axis is aligned with the rotational axis of the stars, which may not be valid for binaries with compact companions. Values should be considered as accurate to within $50-100\,$\kms.


\begin{table}
\centering
\caption{Projected rotational velocities and minimum inclinations}
\resizebox{.5\textwidth}{!}{\begin{tabular}{lcccc}\hline \hline
VFTS  & SpT & $\varv \sin i_1\,$ [\kms] &  $\varv \sin i_2\,$ [\kms] & $i_{\rm crit}$ \\ 
\hline \ 
64 & O7.5 II(f) + (O7~V+O7~V) & 55 & - & 6\\ 
64A & O6 + O6 & - & 274 & 33\\ 
73 & O9.5 III + B: & 90 & - & 10\\ 
86 & O9 III((n)) + O8~V & 80 & 138 & 16\\ 
93 & O9.2 III-IV + B2~V & 54 & 78 & 9\\ 
120 & (O9.5 IV: + B0~V) + (B + B) & - & - & - \\ 
120A & O9.5 IV + B0~V & - & - & - \\ 
120B & B + B & - & - & - \\ 
171 & O8 II-III(f) + B: & 55 & 112 & 13\\ 
184 & O6.5 Vnz + OB: & 249 & 302 & 37\\ 
191 & O9.5 V + B1~V: & 65 & 238 & 28\\ 
201 & O9.7 V  + B1~V & 96 & 133 & 15\\ 
225 & B0.7-1III-II & 58 & - & 7\\ 
231 & O9.7 IV:(n)  + B1~V & 198 & 265 & 32\\ 
243 & O7 V(n)((f)) & 129 & - & 15\\ 
256 & O7.5-8 V((n))z + B0~V: & 120 & - & 14\\ 
277 & O9 V + B4-5 & 72 & 315 & 39\\ 
314 & O9.7 IV:(n)  + B2-5~V & 171 & 296 & 36\\ 
318 & O8.5 + O8.5~V & 102 & 112 & 13\\ 
329 & O9.7 II-III(n) + B1~V & 196 & 265 & 32\\ 
332 & O9.2 II-V + O9~II-V & 64 & 42 & 7\\ 
333 & O8 II-III((f)) + O6~V & 56 & 94 & 11\\ 
350 & O8 V + O9.5~V & 57 & 143 & 17\\ 
386 & O8 + B1~V & 160 & 126 & 19\\ 
390 & O5-6 V(n)((fc))z + B0~V & 120 & 367 & 47\\ 
404 & O3.5 V(n)((fc)) + O4.5 & 100 & 117 & 14\\ 
409 & O4 V((f))z + B: & 71 & - & 8\\ 
429 & O7.5-8 V + B1~V & 82 & 189 & 22\\ 
440 & O6-6.5 II(f) + O9~V & 59 & 111 & 13\\ 
441 & O9.5 V + B1 V & 68 & 66 & 8\\ 
475 & O9.7 III + B0 V & 122 & 146 & 17\\ 
479 & O4-5 V((fc))z + B: & 72 & 133 & 15\\ 
481 & O8.5 III + B1 V & 60 & 177 & 21\\ 
514 & O9.7 III & 78 & - & 9\\ 
532 & O3 V(n)((f*))z + B2~III & 144 & 91 & 17\\ 
603 & O4 III(fc) + O4~V & 83 & 45 & 10\\ 
613 & O8.5 Vz + O7~V & 108 & 151 & 18\\ 
619 & O7-8 V(n) & 148 & - & 17\\ 
631 & O9.7 III(n) & 178 & - & 21\\ 
645 & O9.5 V((n)) & 124 & - & 14\\ 
657 & O7-8 II(f) + B2-3~V & 69 & 71 & 8\\ 
702 & O8 V(n) + OB + (OB+OB) & - & - & - \\ 
702A & O8 V(n) + OB & - & - & - \\ 
702B & OB + OB & - & - & - \\ 
733 & O7.5~V + B0~V & 86 & 57 & 10\\ 
736 & O9.5 V + B: & 56 & - & 6\\ 
743 & O9.5 V((n)) & 116 & - & 13\\ 
750 & O9.5 IV + B: & 92 & 210 & 25\\ 
769 & O9.7 II-III & 71 & - & 8\\ 
779 & B1 II-Ib & 55 & - & 6\\ 
802 & O7.5 Vz + O8~V & 60 & 269 & 33\\ 
810 & O9.7 V  + B1~V & 108 & 148 & 17\\ 
812 & O4-5 V((fc)) & 90 & - & 10\\ 
827 & B1.5 Ib & 62 & - & 7\\ 
829 & B1.5-2 II & 119 & - & 14\\ 
887 & O9 + O9.5~V & 128 & 225 & 27\\ 
\hline
\end{tabular}}
\label{tab:vsinis}\end{table}

\section{Binary detection probability}
\label{sec:DetProb}

\begin{figure}
\centering
\includegraphics[width=.5\textwidth]{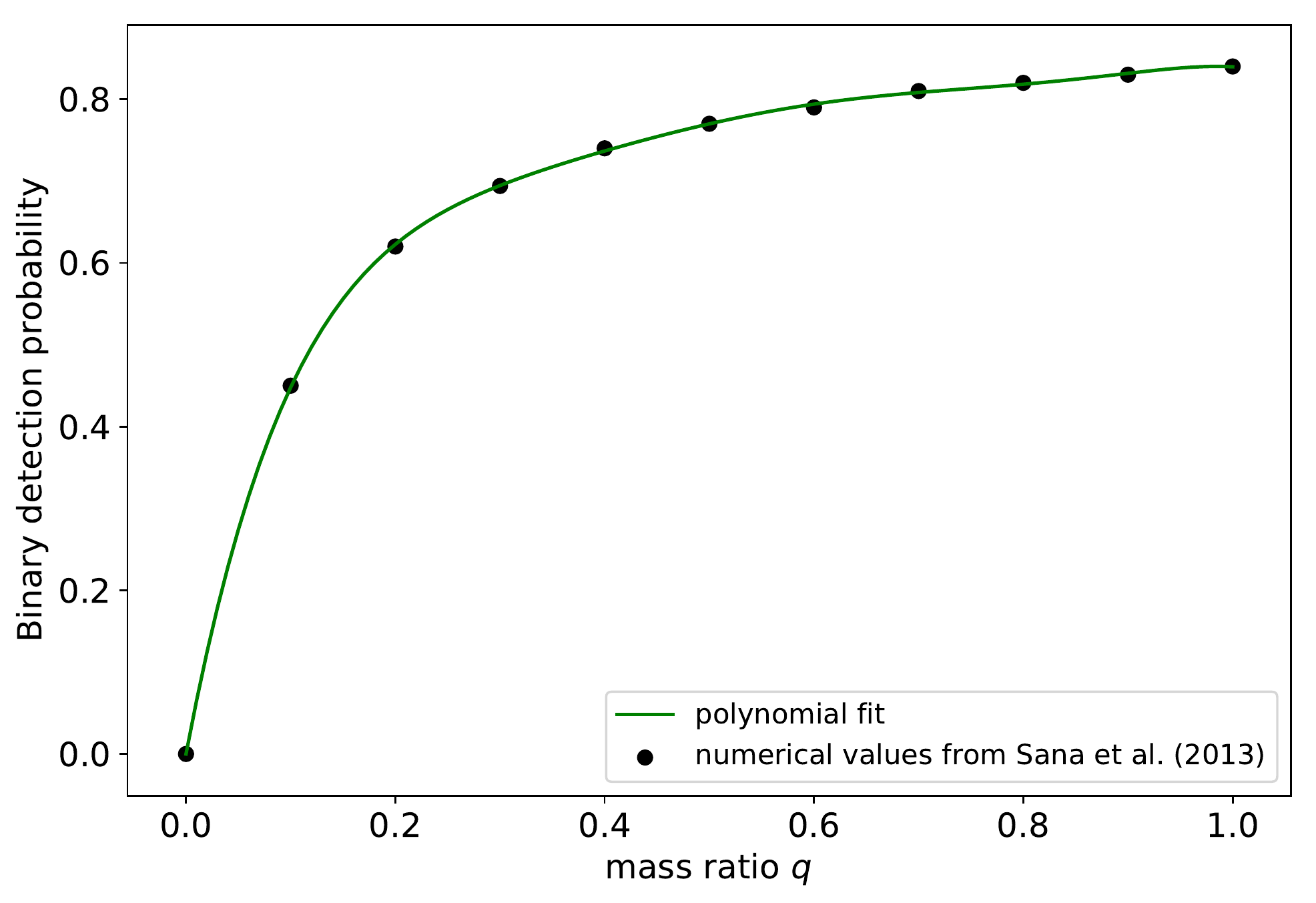}
\caption{Sixth order Polynomial fit to the numerical binary detection probability provided by \citet{Sana2013}.  }
\label{fig:BinDetProb}
\end{figure}

We evaluated the numerical binary detection probability, $f_{\rm det}(q)$, provided by \citet{Sana2013}, at ten individual points, and extended the function to $f_{\rm det}(0) = 0$. We find that a sixth order polynomial is needed to adequately fit the data and exhibit a smooth behaviour in the range $0 \le q \le 1$:

\begin{equation}
    f_{\rm det}(q) = a\,q^6 +  b\,q^5 +  c\,q^4 +  d\,q^3 +  e\,q^2 + f\,q,
\end{equation}
where a = -17.178762, b = 62.839749, c= -92.161434, d=69.536844, e=-28.963718, and f=6.7669361. The fit reaches an accuracy of 0.5\% or better in the range $0 \le q \le 1$ (Fig.\,\ref{fig:BinDetProb}).

\end{appendix}

\end{document}